\documentclass[
    aps,
    prd,
    10pt,
    amsmath,
    amssymb,
    tightenlines,
    preprint,
    nofootinbib,
    groupedaddress,
    superscriptaddress,
    floatfix
]{revtex4-2}

\usepackage{enumerate}
\usepackage{graphicx}
\usepackage{dcolumn}
\usepackage{bm}
\usepackage{microtype}

\usepackage{xcolor}
\usepackage{cancel}
\usepackage{setspace}
\usepackage{environ}
\usepackage{fontawesome5}
\usepackage{etaremune}
\usepackage{tikz}
\usepackage{slashed}
\usepackage[normalem]{ulem}
\usepackage{mathtools}
\usepackage{mathrsfs}
\usepackage{booktabs}

\usepackage[margin=1in]{geometry}

\definecolor{navyblue}{rgb}{0.0,0.0,0.5}

\definecolor{orcidlogocol}{rgb}{0.65,0.807,0.223}

\definecolor{mulberry}{rgb}{0.5,0,0.5}
\definecolor{elderberry}{rgb}{0.3,0,0.7}
\definecolor{raspberry}{RGB}{255,87,51}
\definecolor{gooseberry}{RGB}{0,171,79}
\definecolor{blueberry}{RGB}{75,134,247}
\definecolor{winterberry}{RGB}{255,135,0}

\definecolor{DarkSlateNavy}{rgb}{0.106,0.106,0.227}
\definecolor{DeepPurple}{rgb}{0.294,0.180,0.514}
\definecolor{HypotheticBlue}{rgb}{0.216,0.494,0.722}
\definecolor{HypotheticGreen}{rgb}{0.302,0.686,0.290}
\definecolor{HypotheticOrange}{rgb}{1.000,0.498,0.000}

\graphicspath{{./}{./}}

\newcommand{\calL}{\mathcal{L}}
\newcommand{\calO}{\mathcal{O}}

\newcommand{\asf}{\mathfrak{a}}
\newcommand{\asfew}{\mathfrak{a}_{\mathrm{ew}}}

\newcommand{\omegaEW}{\omega_{\mathrm{ew}}}

\newcommand{\fSN}{f_{\mathrm{SN}}}

\newcommand{\vpq}{V_{\cancel{\text{PQ}}}}
\newcommand{\vmono}{V_{\mathrm{mono}}}

\newcommand{\M}{\mathcal{M}}

\newcommand{\MUV}{M_{\mathrm{UV}}}

\newcommand{\Tew}{T_{\mathrm{ew}}}
\newcommand{\Tosc}{T_{\mathrm{osc}}}
\newcommand{\Trelax}{T_\mathrm{relax}}
\newcommand{\Troll}{T_\mathrm{roll}}
\newcommand{\Tswitch}{T_\mathrm{switch}}

\newcommand{\epsphi}{\epsilon_\Phi}
\newcommand{\epscirc}{\epsilon_\mathrm{circ}}
\newcommand{\epscirctd}{\tilde\epsilon_\mathrm{circ}}

\newcommand{\Hew}{H_{\mathrm{ew}}}

\newcommand{\GeV}{\;\mathrm{GeV}}
\newcommand{\MeV}{\;\mathrm{MeV}}

\newcommand{\keV}{\;\mathrm{keV}}

\newcommand{\eV}{\;\mathrm{eV}}

\newcommand{\Mpl}{M_{\mathrm{pl}}}
\newcommand{\Heq}{H_{\mathrm{eq}}}

\newcommand{\LQCD}{\Lambda_{\text{QCD}}}

\newcommand{\orcid}[1]{%
    $,$%
    \href{https://orcid.org/#1}{%
        \textcolor{orcidlogocol}{%
            \footnotesize\faOrcid
        }%
    }%
}

\NewEnviron{eq}{%
    \begin{align}
        \begin{split}
            \BODY
        \end{split}
    \end{align}
}

\makeatletter

\newcommand{\amarki}{\faWineGlass}
\newcommand{\amarkii}{\faCoffee}

\newcommand{\amarkiv}{\faBeer}

\def\@fnsymbol#1{{%
    \ifcase#1
    \or \amarkiv
    \or \amarkii
    \or \amarki
    \else \@ctrerr
    \fi
}}

\makeatother

\usepackage[
    colorlinks=true,
    linktoc=all,
    linkcolor=elderberry,
    citecolor=winterberry,
    urlcolor=gooseberry,
    pagebackref=true
]{hyperref}

\makeatletter
\let\original@NAT@sort@cites\NAT@sort@cites
\renewcommand*{\NAT@sort@cites}[1]{%
    \Hy@backout{#1}%
    \original@NAT@sort@cites{#1}%
}
\makeatother

\makeatletter
\renewcommand*{\backref}[1]{%
  #1%
  \@ifundefined{@bibstop}{}{%
    \csname bibitem@\@bibstop\endcsname
    \global\let\@bibstop\@undefined
  }%
}
\makeatother

\backrefparscanfalse

\makeatletter

\let\BibitemShutOriginal\BibitemShut

\renewcommand{\BibitemShut}[1]{%
    \BibitemShutOriginal{#1}%
    \backrefprint
}

\makeatother

\usepackage[nameinlink,noabbrev]{cleveref}

\crefname{section}{Sec.}{Secs.}
\Crefname{section}{Sec.}{Secs.}

\crefname{table}{Tab.}{Tabs.}
\Crefname{table}{Tab.}{Tabs.}

\crefname{appendix}{App.}{Apps.}
\Crefname{appendix}{App.}{Apps.}

\crefname{figure}{Fig.}{Figs.}
\Crefname{figure}{Fig.}{Figs.}

\crefname{equation}{Eq.}{Eqs.}
\Crefname{equation}{Eq.}{Eqs.}

\crefname{eqnarray}{Eq.}{Eqs.}
\Crefname{eqnarray}{Eq.}{Eqs.}

\crefname{align}{Eq.}{Eqs.}
\Crefname{align}{Eq.}{Eqs.}

\usepackage{xpatch}

\makeatletter
\xpatchcmd{\@ssect@ltx}
  {\addcontentsline{toc}{#1}{\protect\numberline{}#8}}
  {}
  {}
  {\PackageWarning{revtex-toc}{Could not patch starred section commands}}
\makeatother

\begin{document}
\count\footins=1000
\raggedbottom

\title{Technically Natural Baryogenesis with the QCD Axion}

\author{Pouya Asadi}
\email{pasadi@ucsc.edu}
\affiliation{Department of Physics, University of California Santa Cruz, Santa Cruz, CA 95064, U.S.A.}
\author{David Cyncynates}
\email{davidcyn@ictp.it}
\affiliation{Department of Physics, University of Washington, Seattle, WA 98195, U.S.A.}
\affiliation{International Centre for Theoretical Physics, Strada Costiera 11, 34151 Trieste, Italy}
\affiliation{Istituto Nazionale di Fisica Nucleare (INFN), Sezione di Trieste, I-34127 Trieste, Italy}
\author{Stefania Gori}
\email{sgori@ucsc.edu}
\affiliation{Department of Physics, University of California Santa Cruz, Santa Cruz, CA 95064, U.S.A.} 
\affiliation{Department of Physics, University of California, Berkeley, CA 94720, U.S.A.} 
\affiliation{Lawrence Berkeley National Laboratory, Berkeley, CA 94720, U.S.A.} 

\date{\today}

\begin{abstract}
The strong CP problem, dark matter, and the observed baryon abundance may all be connected dynamically through the QCD axion, though whether this connection is realized in Nature depends on additional degrees of freedom. We show that the required dynamics can arise from technically natural ingredients if multiple axions mix through the QCD anomaly. We review the connection between axion rotation and the baryon asymmetry, revisit the axiogenesis and kinetic-misalignment proposals, and derive new limits on their dynamics. Motivated by these considerations, naturalness and axion quality, we construct a realization in which mixing with additional axions both enhances the effective QCD axion field range through monodromy and dissipates excess rotational energy through mixing with a dark confining gauge sector, thereby avoiding overclosure. We study the resulting effective dynamics in several illustrative limits, uncovering phenomena that may arise more generally in multi-axion theories. We numerically evolve the full coupled system and show that it can simultaneously reproduce the observed baryon asymmetry and dark matter abundance within a predictive region of parameter space testable by near-future direct and indirect searches. Depending on the model parameters, either the QCD axion or the new heavy axion constitutes the dominant dark matter component. Successful baryogenesis favors a QCD axion decay constant near its current astrophysical lower bound, while the heavy axion lies within reach of forthcoming astrophysical and cosmological observations.
\end{abstract}

\maketitle

\newpage

\tableofcontents

\section{Introduction and Conclusion}
\label{sec:intro_conc}

The Standard Model (SM) of particle physics provides a very successful description of microscopic phenomena wherever it has been directly tested, yet leaves numerous fundamental properties of our Universe unexplained. Among these, we focus on three that may possess an interesting connection: the smallness of CP violation in the strong interactions, the origin of the baryon asymmetry, and the nature of dark matter (DM). The absence of observable CP violation in the strong interactions motivates the Peccei–Quinn mechanism~\cite{Peccei:1977hh,Peccei:1977ur}. The resulting pseudo-Nambu–Goldstone boson, the QCD axion~\cite{Weinberg:1977ma,Wilczek:1977pj}, can itself constitute dark matter~\cite{Preskill:1982cy,Abbott:1982af,Dine:1982ah}, thereby linking the solution to the strong CP problem to the origin of the cosmological DM abundance. Furthermore, the present-day closeness of the baryon and DM energy densities~\cite{Planck:2018vyg} motivates cogenesis scenarios in which the two abundances share a common cosmological origin~\cite{Nussinov:1985xr,Barr:1990ca,Barr:1991qn,Kaplan:1991ah}. The QCD axion-gluon coupling connects its dynamics to SM processes involved in $B+L$ violation, raising the possibility that the QCD axion itself may provide the common dynamical origin of the baryon and DM abundances.

Over the last several decades, the connection between QCD-axion dynamics and baryogenesis has been explored in a number of distinct settings. Early work explored axion-driven electroweak baryogenesis, though these scenarios generally require specialized dynamics such as a delayed or cold electroweak phase transition~\cite{Kuzmin:1992up, Servant:2014bla, Krauss:1999ng, Konstandin:2011ds}. A qualitatively different realization was subsequently proposed in the framework of axiogenesis~\cite{Co:2019wyp}. Here the Peccei–Quinn field itself undergoes Affleck–Dine rotation~\cite{Affleck:1984fy}, carrying a PQ charge that is transferred to the baryon asymmetry through the SM anomalies. Unlike the electroweak-baryogenesis scenarios above, axiogenesis does not require a first-order electroweak phase transition. If a sufficiently large axion rotation persists to the electroweak scale, the time-dependent axion background acts as an effective chemical potential~\cite{Cohen:1987vi,Cohen:1988kt,Servant:2014bla,Domcke:2020kcp,Dasgupta:2018eha}: strong sphalerons communicate the rotation to SM fermion asymmetries, which are converted into a net baryon asymmetry by electroweak sphalerons. The same rotation subsequently determines the axion relic abundance through kinetic misalignment~\cite{Co:2019jts}.

The implementation of Ref.~\cite{Co:2019wyp} introduces no additional low-energy degrees of freedom beyond those already present in standard KSVZ or DFSZ axion constructions~\cite{Kim:1979if,Shifman:1979if,Zhitnitsky:1980tq,Dine:1981rt}. The axion rotation is generated from a large initial displacement of the radial mode, the saxion, away from its vacuum expectation value. At large field values, higher-dimensional PQ-violating operators provide a torque as the saxion begins to relax, generating the PQ charge required for baryogenesis. Once these operators become negligible, the subsequent axion velocity is controlled by the approximately conserved PQ charge and the evolution of the saxion in its potential.

Retaining a sufficiently large axion velocity until the electroweak epoch places competing demands on this evolution. Once the saxion settles near its vacuum expectation value, conservation of PQ charge implies $\dot{\theta}\propto \asf^{-3}$, rapidly diluting the rotation required for baryogenesis. The onset of this regime must therefore be delayed, favoring a radial potential that remains parametrically flat over the relevant field range. At the same time, the PQ-violating operators responsible for generating the initial charge must be important at large saxion field values while becoming sufficiently suppressed near the vacuum to preserve the strong CP solution~\cite{Holman:1992us,Kamionkowski:1992mf,Barr:1992qq}.

Maintaining a coherent rotation introduces an additional constraint. Oscillations of the saxion about its evolving equilibrium can parametrically amplify inhomogeneous modes and fragment the homogeneous field~\cite{Fedderke:2025sic}. Efficient damping of the radial motion can suppress this instability, but the interactions responsible for such damping generically induce finite-temperature corrections to the saxion potential. The same interactions introduced to preserve the homogeneous rotation can therefore work against the flat radial potential required to retain a sufficiently large axion velocity.

If a sufficiently large and coherent rotation survives until electroweak sphaleron freeze-out, its subsequent evolution presents a separate challenge. The SM plasma exerts only a weak friction on the QCD axion, so the PQ charge required to generate the observed baryon asymmetry largely survives to lower temperatures. When the QCD potential becomes important, this charge is converted into an axion relic abundance that exceeds the observed dark matter density in the minimal setup. Successful axiogenesis must therefore preserve a large axion rotation until the electroweak phase transition while efficiently depleting its energy afterward. Several papers have proposed mechanisms that address these conditions, including the original implementation~\cite{Co:2019wyp,Co:2019jts} and subsequent work~\cite{Co:2020jtv,Co:2022kul,Co:2025lcs,Asadi:2025cvm}.

In this paper, we propose that the dynamics required to produce the observed baryon and dark matter abundances arise from QCD-axion mixing without invoking saxion dynamics, while preserving axion quality and introducing only technically natural dimensionful scales. The resulting framework is predictive, with viable parameter space (see \cref{fig:intro_roadmap} below) within reach of future direct and astrophysical searches, as anticipated in the companion paper Ref.~\cite{Asadi:2025cvm}.

\cref{sec:review} and \cref{sec:IC_preliminaries} pedagogically review the axion-coupled SM Boltzmann equations and derive consistency conditions for the generation and preservation of axion rotation in the minimal axiogenesis model. We find that initial conditions for which dimension twelve or higher operators dominate the saxion potential are required under the minimal assumptions of our setup. This lower bound persists independent of intervening matter- or kination-dominated epochs that conserve the bath entropy and PQ charge. We also find that efficient charge generation without tuning favors PQ-violating operators of comparable dimension to the dominant radial term. We further show that the subsequent radial evolution imposes bounds on the renormalizable part of the potential. Preserving the required rotation while allowing the saxion to relax before electroweak phase transition restricts its couplings to a finite interval. Finally, we consider interactions introduced to damp saxion oscillations, as are necessary to avoid fragmentation. We find that they can generate thermal corrections generically in tension with the required flatness.

We motivate a multi-axion realization of axiogenesis in \cref{sec:recap}, and then turn in \cref{sec:monodromy} to the origin of the axion velocity required at the electroweak phase transition. We introduce axion monodromy, in which the axion potential extends aperiodically over many fundamental domains, as the source of the axion velocity at the electroweak phase transition. Starting from ordinary axion-misalignment initial conditions, the required energy is stored as potential energy while the axion remains pinned by Hubble friction and therefore does not redshift with the expansion of the Universe. It is released only when the axion begins to roll, generating the angular motion required for baryogenesis. The monodromy mechanism does not rely on saxion dynamics and therefore also avoids the associated fragmentation and damping constraints~\cite{Fedderke:2025sic}. From a bottom-up perspective, a small monodromy scale is technically natural, since the axion shift symmetry is restored as the monodromy potential is taken to zero. Monodromies admit a technically natural four-dimensional effective description~\cite{Kaloper:2008fb,Kaloper:2011jz} and may arise from string compactifications with sufficiently rich structure~\cite{McAllister:2008hb,McAllister:2014mpa,McAllister:2016vzi}.

Using a phenomenological parametrization of the monodromy potential, we determine the conditions on its shape and on the initial axion misalignment required to reproduce the observed baryon asymmetry. We do so by solving the full set of coupled axion--Boltzmann equations governing the evolution of the axion together with the SM fermion asymmetries. Once the axion has entered its oscillatory regime by the electroweak phase transition, the baryon asymmetry exhibits a broad response as a function of the initial misalignment and monodromy scale, modulated by the axion phase at the electroweak transition. The observed abundance can therefore be obtained over a substantial range of these parameters, fixing each only to within roughly an order of magnitude.

We then examine to what extent the required monodromy potential can arise in a consistent ultraviolet (UV) construction. We consider its stability against membrane nucleation, quantum-gravity effects, and additional heavy degrees of freedom, as well as the consistency of the large winding numbers required to obtain the observed baryon asymmetry. The topology of the compactification may limit the available monodromy budget, and currently no known compactification has an Euler characteristic large enough to support the required winding. We also discuss whether the small monodromy scales required by the cosmological dynamics can be realized in the UV. As an illustrative example, we argue that a warped double Klebanov--Strassler throat geometry~\cite{Klebanov:2000hb,Hebecker:2015tzo,Hebecker:2018yxs} may generate a sufficiently flat monodromy potential, although we do not claim a complete UV embedding.

In \cref{sec:damping_from_dark_sphalerons}, we introduce a new confining SU$(N_d)$ gauge group in thermal contact with the SM as a mechanism for dissipating the axion energy and avoiding overclosure, following the strategy of the companion paper~\cite{Asadi:2025cvm}. Sphaleron transitions in the deconfined dark plasma provide a finite-temperature friction term that damps the axion motion. When the dark gauge group confines, however, it also generates an additional instanton potential. Consistency with the solution to the strong CP problem therefore requires the QCD and dark potentials to act along non-aligned directions in axion field space. We study the resulting two-axion system and show how the damping inherited by the QCD-coupled direction depends on axion mixing and on the relative importance of the QCD and dark instanton potentials.

Relative to the companion paper, which considered an SU$(N_d)$ theory with light dark quarks necessary for consistency with kinetic misalignment initial conditions, we show that a pure SU$(N_d)$ gauge theory without additional light matter can itself provide sufficient friction. We also study the associated monodromy--cosine crossing that occurs as the confining potential turns on. When the monodromy direction has already relaxed to within a single period, the crossing is well described by linear adiabatic evolution; if it remains displaced by many windings, the crossing becomes nonlinear and the final partition of energy between the two axion modes is effectively probabilistic. This partition determines how much of the original monodromy energy is deposited into the modes that subsequently redshift and dissipate.

\begin{figure}[t]
    \centering
    \resizebox{0.8\textwidth}{!}{
    \includegraphics[scale=1]{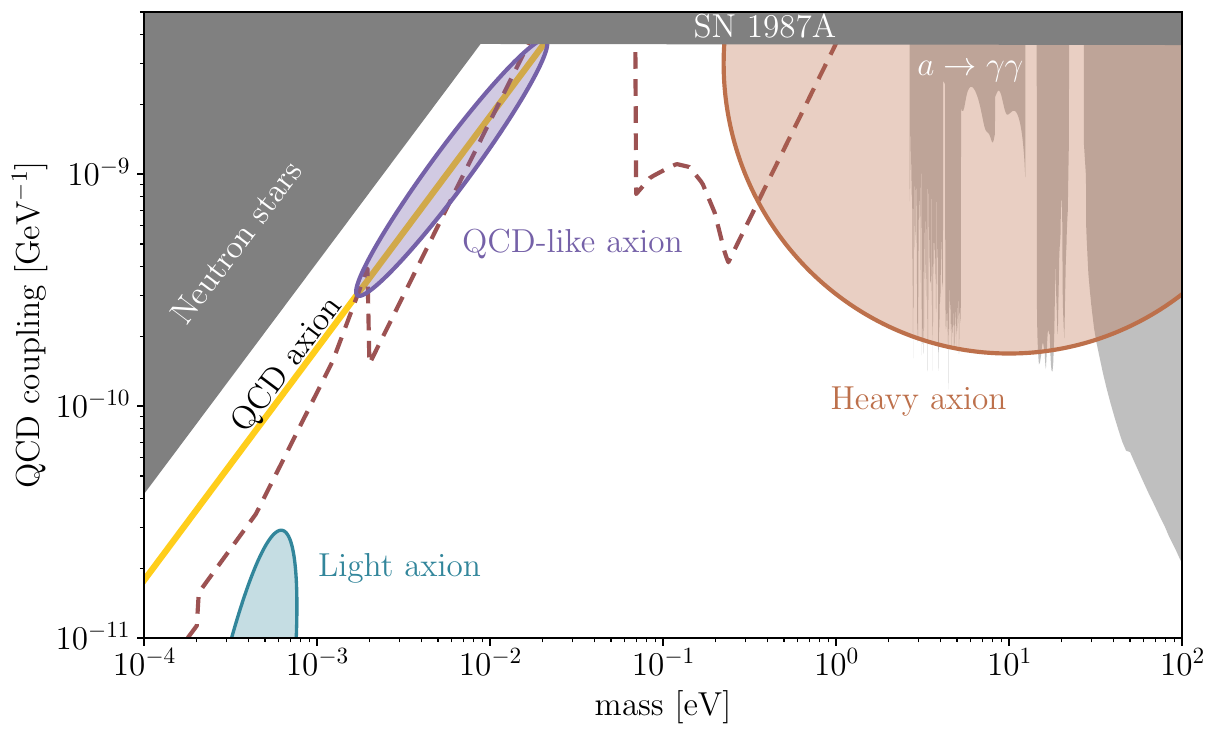}
    }
    \caption{Schematic parameter space for the QCD (\textcolor{elderberry}{purple}), heavier dark-aligned (\textcolor{HypotheticOrange}{orange}), and lighter monodromy-aligned (\textcolor{blueberry}{blue}) axions. Colored regions show representative viable targets, gray shading denotes existing constraints, and dashed red curves indicate projected experimental reach.}
    \label{fig:intro_roadmap}
\end{figure}

In \cref{sec:three_axions}, we combine the monodromy and dark-sphaleron ingredients into a complete model of axiverse baryogenesis. Since the construction contains three independent sources of axion potential: QCD, the dark confining group, and monodromy, a generic realization that preserves the strong CP solution requires three axion directions coupled to these sources in non-degenerate linear combinations. Despite this three-field structure, the cosmological evolution largely factorizes into a sequence of effectively two-axion systems: monodromy first generates the velocity required for baryogenesis, dark sphalerons dissipate the resulting axion energy, and subsequent dark and QCD crossings determine how the surviving energy is partitioned among the late-time mass eigenstates. We develop analytic approximations for this evolution and solve the full system numerically, using exact evolution through the electroweak epoch and WKB methods at later times. We validate these results with fully resolved crossing simulations for representative points, allowing us to identify parameter regions that simultaneously reproduce the observed baryon asymmetry and avoid overproducing dark matter. 
We illustrate these regions schematically in \cref{fig:intro_roadmap}.

An important source of uncertainty is the behavior of the dark-sphaleron friction at temperatures well above confinement but not yet in the semiclassical regime. To bracket this uncertainty, we introduce two friction prescriptions, based respectively on extrapolations of the lattice results of Refs.~\cite{Bonanno:2023thi,Bonanno:2023ljc,Kotov2018,BarrosoMancha:2022mbj} and on the high-temperature semiclassical asymptotic expressions of Refs.~\cite{Moore:2010jd,Bodeker:1998hm}, with the latter cut off before confinement. We find viable regions in both cases but qualitatively different late-time relic compositions.

In \cref{sec:phenomenology}, we study the observational consequences of both the confining dark sector and the resulting axion spectrum. Requiring the dark-sector relics to decay before BBN constrains the dark confinement scale and motivates heavy dark quarks that couple the dark sector to the SM, leading in turn to possible collider signatures in the quirk regime~\cite{Kang:2008ea}. The axion sector is imminently testable: much of the viable parameter space lies close to current SN1987A cooling bounds~\cite{Raffelt:1993ix,Carenza:2019pxu,Carena:1988kr,Brinkmann:1988vi,Raffelt:1996wa,Carenza:2023lci,Lella:2023bfb}, while thermally produced axions are constrained by cosmology~\cite{Graf:2010tv,Salvio:2013iaa,DEramo:2021lgb,DEramo:2021psx,Caloni:2022uya,Bouzoud:2024bom} and can provide an observable subdominant relic even when their misalignment abundance is strongly depleted. Axion decays to photons offer complementary sensitivity to the heavier state~\cite{Gemmell:2026xaa,Saha:2025any,Yin:2024lla,Regis:2020fhw,Todarello:2023hdk,Wang:2023imi,Grin:2006aw,Todarello:2024qci,Carenza:2023qxh,Nakayama:2022jza,Porras-Bedmar:2024uql,Bernal:2022xyi,Wadekar:2021qae,Liu:2023nct,Capozzi:2023xie,Bolliet:2020ofj,Cadamuro:2011fd}, while haloscopes can probe whichever eigenstate constitutes the dominant dark matter component~\cite{Baryakhtar:2018doz,BREAD:2021tpx}. In the lattice-extrapolated friction scenario, the QCD axion typically makes up most of the dark matter at masses above the canonical misalignment target, with the heavier state remaining as a thermally produced subcomponent. For the high-temperature semiclassical asymptotic friction prescription, the relic composition can instead be reversed, with the heavier dark-aligned axion constituting most of the dark matter. Finally, in \cref{sec:outlook} we outline future theoretical and phenomenological avenues for exploring this scenario. 

The appendices provide further details on the coupled axion--SM Boltzmann equations~(\cref{app:boltzmann,app:linear_sys}); the running couplings and masses~(\cref{app:running}), topological susceptibility, and sphaleron rate used in our calculations~(\cref{app:topological_susceptibility}); and the axion relic abundance from kinetic misalignment~(\cref{app:relic_abundance_from_adiabatic_invariant}). They also present limiting-case analyses of the axion--SM Boltzmann equations (\cref{app:rapid_oscillation_YB}) and coupled axion equations of motion (\cref{app:mode_mixing_decoupling}), studies of axion fragmentation from self-interactions (\cref{app:Monodromy_oscillation_fragmentation}) and monodromy crossing (\cref{app:Monodromy_Crossing_Fragmentation}), and details of our numerical scan of the full parameter space (\cref{app:fast_three}), including applications of large language models (LLMs) (\cref{app:semantic}).

Our construction realizes a cogenesis mechanism, linking the production of the baryon asymmetry and axion dark matter within the same cosmological evolution. Their relative abundance remains sensitive to the microscopic parameters and initial conditions, so the mechanism does not by itself explain the numerical coincidence of the two densities. Understanding whether additional structure can account for their observed ratio remains an important open question.

The technical naturalness of the low-energy construction raises the possibility that the mechanism can be realized in a UV completion without introducing new tunings. Multiple axions, additional confining gauge sectors, and axion monodromies all have known realizations in string compactifications, making the string axiverse~\cite{Witten:1984dg,Arvanitaki:2009fg} a natural setting in which to search for such a completion. Realizing the large monodromy charge required for baryogenesis remains the principal unresolved challenge, although no general obstruction to realizing the required winding is known. Constructing an explicit compactification that realizes the full mechanism therefore remains an important target for future work.

If the axiverse picture is realized in Nature, the multiplicity of axions emerging from the UV may provide the underlying structure responsible for several seemingly disparate features of the Universe we observe. The framework developed here illustrates how the resulting dynamics can solve multiple open problems of the SM (dark matter, baryon asymmetry, strong CP), and how these connections can make the cosmological evolution and experimental signatures predictive.

\section{Gauge anomalies, friction, and fermion asymmetries}\label{sec:review}

In this section we review aspects of sphalerons and the vacuum structure of non-abelian gauge theories relevant to our study. We show that both the Boltzmann equations for fermions in a homogeneous time-varying axion background and the friction terms in the axion equation of motion arise from the same underlying microphysics. Much of this section reviews existing results from the literature. It provides context and forms the basis for constructing our axiogenesis model in subsequent sections.

Let us consider the interaction
\begin{align}
    \mathcal{L} \supset \frac{\alpha}{8\pi}\,\theta\, G^{\alpha\beta} \tilde{G}_{\alpha\beta} ,
    \label{eq:L_initial}
\end{align}
where $G$ is the field strength of an $\mathrm{SU}(N_x)$ gauge group, $\tilde{G}^{\alpha\beta} = \frac{1}{2} \varepsilon^{\alpha\beta\gamma\rho} G_{\gamma\rho}$ is its dual, $\alpha$ denotes the corresponding gauge coupling, and $\theta = a/f_a$ is the dimensionless, $2\pi$-periodic axion-like field. The vacuum structure of such a gauge theory is labeled by discrete Chern-Simons (CS) numbers; non-perturbative configurations called sphalerons \cite{Klinkhamer:1984di} mediate transitions between vacua with different CS numbers.

The thermal expectation value $\langle G\tilde{G} \rangle_T$ is the unifying object that connects two different aspects of the dynamics: 
\begin{enumerate}
\item Through the anomalies, it affects the generation of fermion number asymmetries by biasing sphaleron transitions (see Sec. \ref{subsec:sph_boltzman}). 
\item It enters the axion’s equation of motion as an effective friction term (see Sec. \ref{subsec:EOM_friction}). 
\end{enumerate}
For comprehensive reviews, see Refs.~\cite{Rubakov:1996vz,Riotto:1998bt,Srednicki:2007qs}, and for numerical studies, see e.g. Refs.~\cite{Ambjorn:1995xm,Moore:1996qs}. The full thermodynamic derivation of relevant quantities for our setup can be found in \cref{app:boltzmann}; here we summarize the key results. 

A non-abelian gauge group SU$(N_x)$ has an infinite number of vacua, each characterized by a CS number, $N_{\mathrm{CS}}$. This number is associated with the divergence of the CS current, $j_{\mathrm{CS}}^\beta$
\begin{equation}
    \partial_\beta j_{\mathrm{CS}}^\beta = \frac{\alpha}{8\pi} G_{\mu\nu}^a \tilde{G}^{\mu\nu, a} \ ,~~~ N_{\mathrm{CS}}=\int{d^3 x ~j_{\mathrm{CS}}^0(t,\bar x)} \equiv n_\mathrm{CS} V.
    \label{eq:changeCS}
\end{equation}
A time-dependent $\theta$ acts as an effective chemical potential for the CS current of the gauge field \cite{Kuzmin:1992up,Servant:2014bla,Dasgupta:2018eha} 
\begin{equation}
    \mu_{\mathrm{CS}} = - \dot\theta \, .
    \label{eq:muNC_text}
\end{equation}
As a result, a rotating axion field in the early Universe ($\dot\theta \neq 0$) can bias transitions between different CS vacua. Through the chiral anomaly, changes in the CS number are directly linked to changes in the net asymmetries of Weyl fermions charged under the corresponding gauge group 

\begin{equation}\label{eq:NQNCS_main_body}
    \delta n_i = \mathcal{N}_i^x \times \delta n_{\mathrm{CS}} \,,
\end{equation}
where $n_i$ is the net asymmetry of fermion species $i$ charged under the gauge group, and $\mathcal{N}_i^x$ denotes the number of internal degrees of freedom $g_i$ of the fermion $i$, divided by its number of degrees of freedom arising from the gauge group SU($N_x$). For example, for a fermion in the fundamental representation of SU$(N_x)$, $\mathcal{N}_i^x=g_i/ N_x$. Consequently, the rotation of an axion(-like) particle in the early Universe can act as a source for generating fermion number asymmetries \cite{Co:2019wyp}.

The existing fermion asymmetry also backreacts on the sphaleron rate, resulting in a wash-out effect on the change of the CS number. A Weyl fermion asymmetry contributes to the free energy density and since sphaleron transitions change $\mu_i$, they modify the fermion contribution to the total free energy density. The rate of a sphaleron transition is exponentially sensitive to the change in free energy, see \cref{app:boltzmann} for more details.

Combining these two effects, we find that when the fermions are in the fundamental representation of the gauge group, the variation of the CS number can be written as
\begin{align}\label{eqn:N_CS}
   \partial_t n_{\mathrm{CS}} =  \frac{\alpha}{8\pi}\langle\tilde G G\rangle_T=\frac{\Gamma_{\mathrm{sph}}}{2T}\left(\mu_{\mathrm{CS}} -  \frac{6}{T^2} \sum_i \frac{{\cal N}_i^x}{g_i }n_i\right)\,
\end{align}
where 
$\langle \rangle_T$ denotes the thermal average, and 
$\Gamma_{\mathrm{sph}}$ is the average sphaleron rate per unit volume at the temperature $T$. The first equality in \cref{eqn:N_CS} is a basic fact about non-abelian gauge fields,\footnote{In writing this equality, we neglect the spatial variation in the corresponding CS current.} while the second equality is derived in \cref{app:boltzmann}.\footnote{At high temperature, the topological susceptibility is negligible and is not included in this study. As will be discussed in the upcoming sections, these terms become relevant at lower temperatures and play a pivotal role in the evolution of the axion field.}

Up to the proportionality constant $\mathcal{N}_i^x$ (see \cref{eq:NQNCS_main_body}), \cref{eqn:N_CS} also determines the time evolution of the various fermion asymmetries.
Crucially, since $\dot{\theta}$ is the only source of C and CP violation, all fermion asymmetries ($n_i$) in \cref{eqn:N_CS} will be proportional to $\dot{\theta}$, which implies
\begin{align}\label{eq:thermal_thetadot}
    \langle\tilde G G\rangle_T \propto \dot{\theta}\,.
\end{align}
As we will discuss in the upcoming sections, the same thermal average $\langle\tilde G G\rangle_T$ appears in the potential of the axion field $a$ and, thus, its equation of motion in the early Universe, acting as a friction term. 
In the next two subsections, we elaborate further on the interplay of the axion evolution and fermion asymmetries in the early Universe connected via sphaleron dynamics.

\subsection{Rotating Axions, Sphalerons, and Baryon Asymmetry in the Standard Model}
\label{subsec:sph_boltzman}

We begin our discussion focusing on the fermion asymmetries generated in the SM. Both non-perturbative and perturbative effects arising from the SM gauge and Yukawa interactions contribute to the evolution of fermion asymmetries. The perturbative terms can be thought of as \textit{chirality-flipping} processes. Due to the smallness of CP violation and the specific patterns of fermion number violation in the SM, these perturbative processes contribute only as wash-out terms in the Boltzmann equations for the various fermion asymmetries (see \cref{app:boltzmann} for details). Including finite-temperature effects, the resulting Boltzmann equations governing the evolution of various SM fermion asymmetries in the electroweak-symmetric phase, $T>T_{\rm{ew}}$,
can be written as 
\begin{equation}
\label{eq:final_boltz}
\begin{aligned}
    \dot{n}_{q_g} + 3H n_{q_g} &= 3\kappa_{{\rm ch},u}\min(\alpha_3 ,y_{u_g}^2)T\left( -\frac{n_{q_g}}{6} - \frac{n_{\bar{u}_g}}{3} - \frac{n_H}{4}  \right) \\ 
    &+ 3\kappa_{{\rm ch},d}\min(\alpha_3 ,y_{d_g}^2)T\left( -\frac{n_{q_g}}{6} - \frac{n_{\bar{d}_g}}{3} + \frac{n_H}{4}  \right) \\
    &+ \frac{9}{2} \frac{\Gamma_w}{T^3} \left( - \frac{c_W}{3} \dot{\theta} T^2 - \sum_{g'} (n_{q_{g'}} + n_{\ell_{g'}}) \right) \\
    &+ 2 \frac{\Gamma_c}{T^3} \left( - \frac{1}{2} \dot{\theta} T^2 - \sum_{g'} ( n_{q_{g'}} + n_{\bar{u}_{g'}} + n_{\bar{d}_{g'}}  ) \right)\, + W_{q_g} \,, \\
    \dot{n}_{\ell_g} + 3H n_{\ell_g} &= 2\kappa_{{\rm ch},e}\min(\alpha_2 ,y_{e_g}^2) T\left( -\frac{n_{\ell_g}}{2} - n_{\bar{e}_g} + \frac{n_H}{4}  \right) \\
    &+ \frac{3}{2} \frac{\Gamma_w}{T^3} \left( - \frac{c_W}{3} \dot{\theta} T^2 - \sum_{g'} (n_{q_{g'}} + n_{\ell_{g'}}) \right) \, + W_{\ell_g} \,,\\
    \dot{n}_{\bar{u}_g} + 3H n_{\bar{u}_g} &= 3\kappa_{{\rm ch},u}\min(\alpha_3 ,y_{u_g}^2)T\left( -\frac{n_{q_g}}{6} - \frac{n_{\bar{u}_g}}{3} - \frac{n_H}{4}  \right) \\
    &+  \frac{\Gamma_c}{T^3} \left( - \frac{1}{2} \dot{\theta} T^2 - \sum_{g'} ( n_{q_{g'}} + n_{\bar{u}_{g'}} + n_{\bar{d}_{g'}} )  \right)\,, \\
    \dot{n}_{\bar{d}_g} + 3H n_{\bar{d}_g} &= 3\kappa_{{\rm ch},d}\min(\alpha_3 ,y_{d_g}^2)T\left( -\frac{n_{q_g}}{6} - \frac{n_{\bar{d}_g}}{3} + \frac{n_H}{4}  \right) \\
    &+  \frac{\Gamma_c}{T^3} \left( - \frac{1}{2} \dot{\theta} T^2 - \sum_{g'} ( n_{q_{g'}} + n_{\bar{u}_{g'}} + n_{\bar{d}_{g'}})  \right)\,, \\
    \dot{n}_{\bar{e}_g} + 3H n_{\bar{e}_g} &= 2\kappa_{{\rm ch},e}\min(\alpha_2 ,y_{e_g}^2) T\left( -\frac{n_{\ell_g}}{2} - n_{\bar{e}_g} + \frac{n_H}{4}  \right) \,, 
\end{aligned}
\end{equation}
where, in our convention, all fermions are left-handed. 
$g^{(')}=1,2,3$ is the fermion generation index, $y_{f_g}$ is the SM Yukawa coupling of fermion $f$ in generation $g$, $n_i$ denotes the asymmetric abundance of particle species $i$, $n_H$ the Higgs asymmetry density, $H$ is the Hubble parameter, $\alpha_{3 \, (2)}$ refers to the strong (weak) fine structure constant, and $\kappa_{{\rm ch},u,d,e}$ refer to $\calO(1)$ numerical coefficients in the rate of chirality-flipping processes, which we take to be 1 for the remainder of the manuscript \cite{Boyarsky:2020cyk,Boyarsky:2020ani,Berghaus:2020ekh}.\footnote{In Refs.~\cite{Boyarsky:2020cyk,Boyarsky:2020ani}, this prefactor is calculated in special cases and was shown to be $\mathcal{O}(1)$.} $c_W$ is the model-dependent weak anomaly coefficient,\footnote{We assume that the $\theta$ term couples directly to the strong sector via $G\tilde{G}$, and only couples to the weak sector through the usual anomaly-induced interactions.} normalized as 
\begin{align}
    \mathcal{L} \supset \theta \frac{\alpha_2}{8\pi}c_WW_{\mu\nu}^a\tilde W^{\mu\nu~a}.
\end{align}
$W_{q_g/\ell_g}$ denotes contributions from off-diagonal SU(2)$_L$ interactions and cancels when summed over all generations. Since we are only interested in the total asymmetries after summing over the three generations, $W_{q_g,\ell_g}$ do not appear in any of our final expressions. The perturbative wash-out terms are proportional to $n_i/g_i$, where $g_i$ counts the degrees of freedom of species $i$. $\Gamma_{w \, (c)}$ denotes the sphaleron rates of the weak (color) gauge group, which has a simple parametric form at high temperatures \cite{Moore:2000ara,Moore:2010jd} 
\begin{align}
    \Gamma_c \approx 3^5 \alpha_3^5 T^4\, ,~~~\Gamma_w \approx 2^5 \alpha_2^5 T^4 \,.
    \label{eq:Gamma_exp}
\end{align}
The derivation of the numerical prefactors accompanying the sphaleron contributions $\Gamma_{w,c}$ in \cref{eq:final_boltz} is provided in \cref{app:boltzmann}. We discuss further details of this linear system of equations in \cref{app:linear_sys}. Note that the fine-structure constant $\alpha_i$ ought to be evaluated at temperature $T$ in these equations, see \cref{app:running} for details. We use the more accurate expression (including $\mathcal{O}(1)$ prefactors) for $\Gamma_\mathrm{sph}$ in \cref{app:topological_susceptibility} in our numerical calculations, as well as the lattice extrapolated expression described in \cref{subsec:sphaleron_rate_summary}.

\begin{figure}
    \centering
    \includegraphics[width=0.35\linewidth]{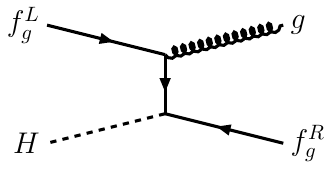}
    \hspace{0.6in}
    \includegraphics[width=0.35\linewidth]{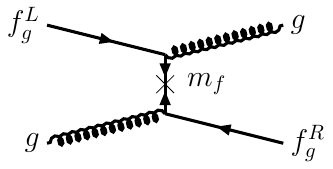}
    \caption{Representative diagrams giving rise to fermion chirality-flip for $T > T_{\mathrm{ew}}$, i.e. electroweak-symmetric phase (\textbf{left}), and for $T < T_{\mathrm{ew}}$, i.e. in the symmetry-broken phase (\textbf{right}). $g$ denotes a gluon or an EW gauge boson. These processes give rise to the perturbative wash-out terms in the Boltzmann equations in \cref{eq:final_boltz}. Properly including the finite temperature effects and resumming IR divergences gives rise to the rate in \cref{eq:final_boltz}, see the text and Refs.~\cite{Boyarsky:2020cyk,Boyarsky:2020ani} for details.  }
    \label{fig:ch_flip}
\end{figure}

A few words are in order regarding the contribution of perturbative chirality-flipping processes in \cref{eq:final_boltz}. In the left panel of \cref{fig:ch_flip}, we show representative diagrams that induce chirality flips for a SM fermion for $T>T_{\mathrm{ew}}$. The linear-response formalism has been used to compute chirality flip from such $2\to2$ scatterings \cite{Boyarsky:2020cyk,Boyarsky:2020ani}, where it was emphasized that finite-temperature effects and a systematic regularization of IR divergences are crucial for a proper treatment.\footnote{In Refs.~\cite{Boyarsky:2020cyk,Boyarsky:2020ani} it is argued that finite-temperature effects and conservation laws suppress $1\to1$ chirality-flipping processes, while $2\to1$ processes have a similar contribution as $2\to2$ scattering processes in the thermal bath. For simplicity, we only discuss the $2\to2$ processes here to provide intuition about the underlying physics.} In particular, one might have naively expected the rate associated with the diagrams in the left panel of \cref{fig:ch_flip} to scale as
\begin{equation}
    \Gamma_\mathrm{ch}^{T>T_{\mathrm{ew}}} \sim \alpha \, y_f^2 \, T .
\label{eq:2to2_chirality_naive}
\end{equation}
However, this treatment neglects the thermal corrections to the dispersion relations of different particles. The fermion thermal mass regulates the rate for $T > T_{\mathrm{ew}}$. In this case, the finite temperature mass gets a contribution from the gauge interactions and a contribution from chirality-breaking coupling to Higgs (e.g. see Refs.~\cite{Weldon:1982bn,Gatoff:1989uz,Ayala:2003pv}):
\begin{equation}
    m_f^2 (T \neq 0) \sim \alpha T^2 + y_f^2 T^2 \approx T^2 \times \max ~(\alpha,y_f^2) \ .
    \label{eq:mass_finite_T}
\end{equation}
Using this, the chirality-flip rate becomes:\footnote{More precisely, the symmetric-phase chirality flip rate for fermion $f$ can be written as
$$
\Gamma_{{\rm ch},f}^{T > T_\mathrm{ew}}=c_f(T)\,y_f^2\,T,
$$
where $c_f(T)$ contains additional gauge- and top-Yukawa-coupling dependence, logarithmic corrections, and Landau-Pomeranchuk-Migdal-resummed contributions \cite{Bodeker:2019ajh}. We neglect this detailed dependence because, for the slowly varying trajectories relevant to the final baryon asymmetry, $\Gamma_{\rm ch}$ is much larger than both the Hubble rate and the variation rate of the axion source. The interactions therefore impose chemical-equilibrium conditions, whose stationary solution is independent of their overall normalization. This approximation is not intended as a precision treatment of the rapidly oscillating regime.}
\begin{equation}
    \Gamma_\mathrm{ch}^{T>T_{\mathrm{ew}}} \sim \alpha \, y_f^2 \, \frac{T^3}{m_f^2(T\neq 0)} \sim T \times \min (\alpha,y_f^2)\,.
    \label{eq:2to2_chirality_highT}
\end{equation}
This discussion also shows that if the fermion $f$ is charged under multiple gauge fields in the bath, the contribution of the gauge field with the largest fine structure constant dominates. For a non-abelian gauge field, e.g. SU($N_c$), the total chirality-flip rate receives an extra factor of $N_c$, accounting for the multiplicity of fermion colors. Putting these ingredients together, and introducing an $\mathcal{O}(1)$ coefficient $\kappa_{\mathrm{ch}}$ to parameterize our uncertainty associated with finite-temperature effects and IR regularization, we obtain the chirality-flip rates summarized in \cref{eq:final_boltz}. 

Furthermore, it is evident that the strong sphalerons do not generate a net baryon number as the baryon number is non-anomalous under this symmetry. Any non-zero baryon asymmetry is generated by the electroweak sphalerons. Additionally, in the early Universe, the system quickly reaches a steady state where the comoving asymmetry of each species becomes constant; i.e., the right-hand side of the Boltzmann equations vanishes. This yields a system of linear equations in $n_X$; the resulting coefficient matrix multiplying the $n_X$ vector must be singular for a consistent solution to exist. This gives rise to a non-trivial consistency check on the numerical prefactors in the equations and we find that our \cref{eq:final_boltz} passes this check. 

In addition to being degenerate, this system is underdetermined, in part because there are two exact U(1) symmetries whose charges cannot be determined by Boltzmann equations: $B - L$ and hypercharge. Therefore, to fully determine the system, we must likewise specify these two numbers as well. Assuming the Universe is neutral under these U(1)'s, we obtain the two additional equations:
\begin{align}
    \sum_g \frac{1}{3}  ( n_{q_g} - n_{\bar{u}_g} - n_{\bar{d}_g}) - (n_{\ell_g} - n_{\bar{e}_g}) = 0 \,,\\ 
    \label{eq:B-L_zero}
    \frac{1}{2} n_H + \sum_g \frac{1}{6} n_{q_g} -\frac{2}{3} n_{\bar{u}_g} + \frac{1}{3} n_{\bar{d}_g} -\frac{1}{2} n_{\ell_g} + n_{\bar{e}_g} &= 0 \,.\nonumber
\end{align}
While these equations are not enough to determine each fermion asymmetry uniquely, they allow us to find a unique solution to the net comoving baryon asymmetry of the Universe 
\begin{equation}
    Y_B =  \frac{45}{2\pi^2 g_{\star S}} \frac{12 }{79} \, \left(\frac18~\frac{\sum_g (7 y_{u_g}^{-2} + 5 y_{d_g}^{-2})}{\frac{T^4}{\Gamma_{c}} + \sum_g (y_{u_g}^{-2} + y_{d_g}^{-2})} -  c_W\right) \, \frac{\dot{\theta}}{T} \simeq  \frac{45}{2\pi^2 g_{\star S}} \frac{12 }{79} \, (0.83 -  c_W) \, \frac{\dot{\theta}}{T} \,,
    \label{eq:fin_B_val}
\end{equation}
where we have taken $\min\{y_{u_i,d_i}^2,\alpha_3\} = y_{u_i,d_i}^2$, the Yukawa couplings computed at the EW scale. (This applies to all but the top quark.) $g_{\star S}$ is the entropy degrees of freedom of the bath. See Refs.~\cite{Domcke:2020kcp,Co:2020xlh} for similar formulas. The weak anomaly coefficient $c_W$ depends on the specific UV completion of the model, e.g. see Ref.~\cite{DiLuzio:2020wdo} for a review. To remain agnostic about the UV details, and since it only affects the final asymmetry by an $\mathcal{O}(1)$ factor, we neglect $c_W$ in our numerical study. A specific UV completion with $c_W = 0$ is the KSVZ model \cite{Kim:1979if,Shifman:1979if}. 
 
Crucially, we find that the asymmetry is controlled by the smallest Yukawa couplings in the SM, i.e. the first generation quarks, as also pointed out in Refs.~\cite{Co:2019wyp,Co:2020xlh}. As evident from \cref{eq:fin_B_val}, in the small Yukawa coupling limit, the final baryon asymmetry is independent of the sphaleron rate and its uncertainty. Furthermore, as expected, since $\dot{\theta}$ is the only source of C and CP violation, the asymmetry is proportional to $\dot{\theta}$. 

As the Universe cools, electroweak sphalerons become suppressed and the generated $B+L$ asymmetry is frozen in. The freeze-out temperature has been estimated to be $T_{\mathrm{ew}} \approx 130\,\mathrm{GeV}$~\cite{DOnofrio:2014rug}. The observed baryon asymmetry \cite{Planck:2018vyg} of the Universe corresponds to a yield of
\begin{align}
    Y_B \approx 8.7 \times 10^{-11}\,,
\end{align}
which requires the axion velocity at freeze-out to be
\begin{align}\label{eqn:velocity_boundary_condition}
    \omegaEW \equiv \dot\theta(T_{\mathrm{ew}}) \approx 3.9\,\mathrm{keV}\,.
\end{align}
Fixing the axion velocity at electroweak sphaleron freeze-out to reproduce the correct baryon asymmetry simultaneously determines the axion abundance. This will be the focus of the next section.

\subsection{Sphaleron Friction and Axion Evolution in the Standard Model}
\label{subsec:EOM_friction}

Having determined the criterion for the axion to produce the observed baryon asymmetry, we now discuss the resulting axion dynamics and its relic abundance, and in particular the friction due to the strong sphalerons. From \cref{eq:L_initial}, we can write down the QCD axion equation of motion in the background of the SM plasma,\footnote{We have omitted the electroweak topological term from the axion equation of motion for brevity and because it contributes neither to the friction nor the potential.} 
\begin{align}\label{eqn:axion_EOM}
    \ddot\theta + 3H \dot\theta  - \frac{\nabla^2}{\asf^2}\theta= \frac{\alpha_c}{8\pi f_a^2} \tilde{G}_c G_c    \,,
\end{align}
where $\asf$ is the scale factor and $G_c$ is the QCD field strength. As shown in \cref{eqn:N_CS}, the thermal average $\langle \tilde{G_c}G_c \rangle_T$ is proportional to a linear combination of $\mu_{\mathrm{CS}}$ and the fermion asymmetries $n_i$, all of which are proportional to $\dot\theta$ (see \cref{eq:muNC_text} and \eqref{eq:fin_B_val}). Thus, after taking a thermal average and neglecting the spatial variation of $\theta$ on length scales $\lesssim 1/\alpha T$, i.e. smaller than typical soft processes in the plasma such as sphaleron transitions~\cite{Arnold:1996dy}, \cref{eqn:axion_EOM} can be rewritten as~\cite{Laine:2016hma} 
\begin{align}\label{eqn:axion_EOM_2} 
    \ddot\theta + \left(3H + \frac{\Upsilon_c}{f_a^2} \right)\dot\theta - \frac{\nabla^2}{{\mathfrak a}^2}\theta + \frac{1}{f_a^2}\chi_c\sin\theta= 0\,,
\end{align}
where the friction coefficient 
\begin{align}\label{eqn:friction_definition}
    \Upsilon_c \equiv - \, \dot n_{\rm CS}/\dot\theta\,
\end{align}
encodes the dissipative effect of QCD sphalerons and arises from the change in the CS number (see \cref{eqn:N_CS}).\footnote{Our definition of the gauge friction coefficient $\Upsilon_c$ is different from the conventional parametrization which includes a factor of $1/f_a^2$~\cite{Berghaus:2019whh,Berghaus:2020ekh}. In the context of the present work, which in subsequent sections will invoke multiple confining sectors and axions, our choice to keep the axion and gauge field parameters separate will be more convenient.} The temperature-dependent QCD topological susceptibility $\chi_c$ is related to the local curvature of the axion potential and can be neglected at high temperatures, e.g. when the axion rotation sources the baryon asymmetry, but it should be included in the study of perturbative $\theta$ fluctuations and its equation of motion. Further detail of the definition of the susceptibility and explicit formulas are included in \cref{app:topological_susceptibility}. 

Using \cref{eqn:N_CS} and solving the system of Boltzmann equations in \cref{eq:final_boltz}, we can calculate $\Upsilon_c$ in the EW symmetric phase for $T>T_{\mathrm{ew}}$:
\begin{align}\label{eqn:Upsilon_c_Symmetric}
    \Upsilon_c^{(T \,> \, T_{\mathrm{ew}})} = \frac{\Gamma_{c}}{2 T}\left(1+\frac{\Gamma_c}{T^4}\sum_g(y_{u_g}^{-2} + y_{d_g}^{-2})\right)^{-1}\,.
\end{align}
The term in the parentheses is dominated by the light quark Yukawas, as observed in Refs.~\cite{Co:2019wyp,Domcke:2020kcp}. This can be attributed to the fact that the perturbative wash-out term in \cref{eq:final_boltz} for each fermion is proportional to its Yukawa, so the lighter fermions feel a smaller wash-out effect. Thus, thanks to their feeble coupling to the Higgs, the up and down quarks build up large chiral populations that contribute to the chemical potential felt by the Chern-Simons number density in \cref{eqn:N_CS}, blocking further net sphaleron transitions.\footnote{Sphaleron transitions still occur at the rate $\Gamma_c$ --- the effect of the chemical potentials is to bias the \textit{net} sphaleron transitions. When the chiral asymmetry of the light quarks balances the axion's chemical potential, no further average change of the winding number occurs.}

When the Yukawa coupling is sufficiently small as in the SM, the second term in \cref{eqn:Upsilon_c_Symmetric} dominates, and the expression simplifies to 
\begin{equation}
    \Upsilon_c^{(T \,> \, T_{\mathrm{ew}})} \,\sim\, \frac{1}{2}\, y_{u_1}^2\, T^3 \,,
    \label{eq:Upsilon_scale_Symmetric}
\end{equation}
making the $T^3$ scaling manifest. From this scaling, it is evident that, as the Universe expands, the sphaleron friction decreases faster than Hubble in a radiation-dominated Universe.

After the Higgs develops a non-zero vacuum expectation value for $T<T_{\rm{ew}}$, the SM fermions acquire their Dirac masses. As the temperature drops lower, the top quark and the Higgs scalar decay away and decouple from the dynamics, leaving the remaining light quarks.  Those quarks that do remain build up chiral populations that suppress sphaleron transitions. Once background Higgs bosons have decayed away, Yukawa-mediated scatterings in the left diagram of \cref{fig:ch_flip} are no longer active, and chirality-flipping is instead dominated by mass insertions, i.e. the right diagram in \cref{fig:ch_flip}. We can replicate the calculation for deriving \cref{eqn:Upsilon_c_Symmetric} in the broken phase from \cref{eq:final_boltz} by making the substitution $y_i\to \sqrt{\alpha}\ m_i/T$ to account for the scaling of the rate of chirality flipping for particles with a vector-like mass \cite{Boyarsky:2020cyk,Boyarsky:2020ani}. The final result is 
\begin{align}\label{eqn:Upsilon_c_Broken}
    \Upsilon_c^{(T \,< \, T_{\mathrm{ew}})} = \frac{\Gamma_{c}}{2 T}\left(1+\frac{\Gamma_c}{\alpha_3 T^2}\sum_g(m_{u_g}^{-2} + m_{d_g}^{-2})\right)^{-1}\,.
\end{align}
Once again, the small first generation quark masses dominate the sum, leading to the low temperature scaling 
\begin{equation}
    \Upsilon_c^{(T \,< \, T_{\mathrm{ew}})} \,\sim \, \frac{\alpha_3}{2} \, m_{u_1}^2 \,T \,.
    \label{eq:Upsilon_scale_Asymmetric}
\end{equation}
Notably, were this scaling to continue to arbitrarily low temperatures, eventually the sphaleron friction would exceed Hubble, exponentially damping away the axion velocity. This happens for\footnote{Here we neglect the running of the QCD fine structure constant $\alpha_3$ below $T \approx \Lambda_{\rm QCD}$ and set $\alpha_3 = 1$.}
\begin{equation}
    T\lesssim 0.28 ~{\rm{MeV}} \left(\frac{5\times 10^7~{\rm{GeV}}}{f_a}\right)^2\,.
\end{equation}
This equation shows that the SM friction $\Upsilon_c/f_a^2$ is always subdominant to Hubble throughout the regime of validity of our calculation, $T>\Lambda_c\approx 200\MeV$, and for values of $f_a$ allowed by the SN1987A bound~\cite{Lucente:2022vuo,Springmann:2024ret},
\begin{align}
    f_a\geq \fSN \sim 5\times 10^7~{\rm{GeV}}\,;
\end{align}
see Sec. \ref{sec:pheno_astro} for a discussion of this rather conservative bound. 

\begin{figure}
    \centering
    \includegraphics[width=\columnwidth]{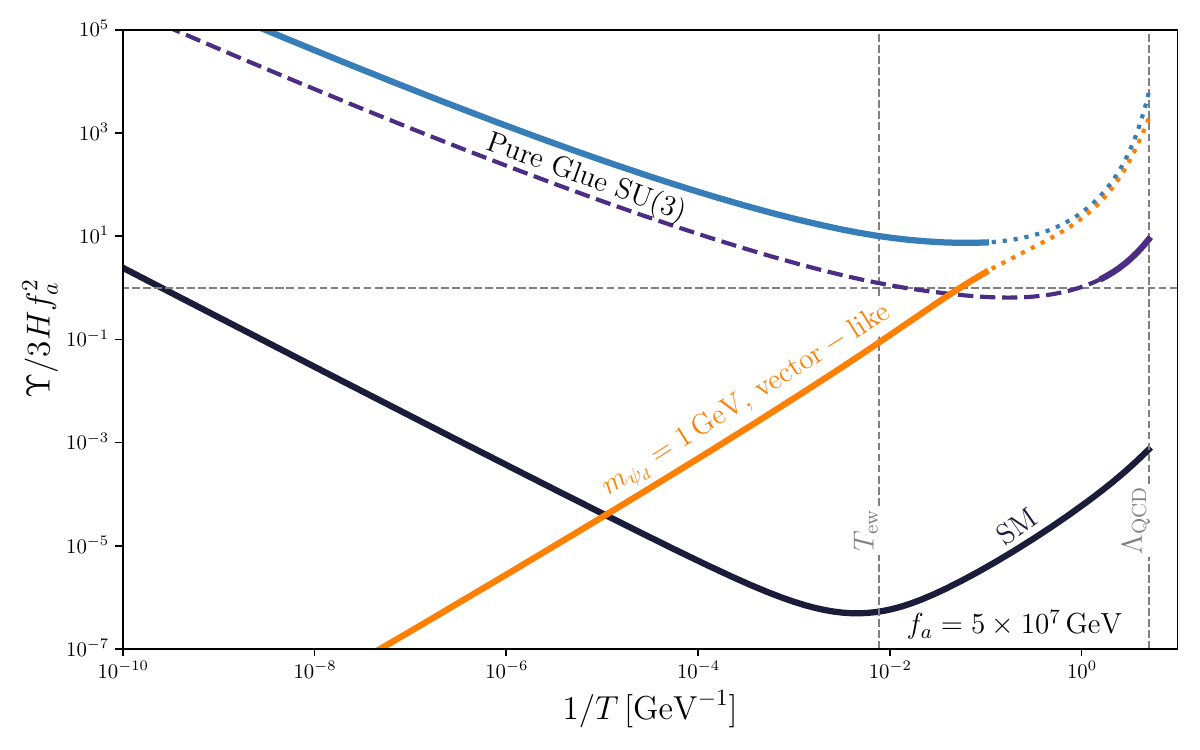}
    \caption{Sphaleron friction relative to Hubble damping, assuming $f_a=5\times10^7\GeV$ and radiation domination, for several choices of matter content charged under a confining $\mathrm{SU}(3)$ gauge group with $\Lambda=\Lambda_{\rm QCD}$. The \textcolor{DarkSlateNavy}{black} curve shows the SM friction, dominated by the small Yukawa couplings and masses of the light quarks. The \textcolor{HypotheticBlue}{blue} and \textcolor{DeepPurple}{purple} curves show pure-glue estimates based respectively on the high-temperature asymptotic expression and the lattice-motivated extrapolation discussed in \cref{subsec:sphaleron_rate_summary}. Dotted and dashed portions mark regions where these estimates are expected to break down most severely or are not directly measured, respectively. The \textcolor{HypotheticOrange}{orange} curve shows a confining $\mathrm{SU}(3)$ dark sector with a lightest vectorlike dark quark of mass $1\GeV$, illustrating that quarks suppress the friction only when they are light enough to build up a substantial chiral population. Above the horizontal dashed line, sphaleron friction exceeds Hubble damping and axion motion damps rapidly.}
    \label{fig:Friction_Sampler}
\end{figure}

The SM result is illustrated by the black curve in \cref{fig:Friction_Sampler}. In the electroweak-symmetric phase, $\Upsilon_c\propto T^3$, so $\Upsilon_c/H$ decreases as the Universe cools during radiation domination. In the electroweak-broken phase, $\Upsilon_c\propto T$, so $\Upsilon_c/H$ grows as the Universe cools. Nevertheless, because the SM contribution is controlled by the small Yukawa couplings and masses of the light quarks, it remains below Hubble damping throughout the regime in which QCD sphalerons are active. Thus, SM strong sphalerons do not efficiently damp the axion rotation relevant for axiogenesis.

While we have focused on QCD, the friction results in \cref{eqn:Upsilon_c_Symmetric} and \eqref{eqn:Upsilon_c_Broken} carry over straightforwardly to new confining dark sectors. In particular, for a theory containing only vectorlike fermions, whose masses can arise from bare mass terms, the friction continues to decrease relative to Hubble with increasing temperature even when the fermions are relativistic. This behavior is shown by the orange curve in \cref{fig:Friction_Sampler}. When the lightest dark quark is light enough relative to the temperature to build up a substantial chiral population, the effective chemical potential of the axion rotation is partially canceled by this population, and the effective friction is suppressed. Once this quark becomes Boltzmann suppressed, the theory approaches the pure-glue limit and the sphaleron friction becomes more efficient. Thus, the presence of vectorlike quarks changes the temperature dependence by suppressing the friction only over the range in which the quark is thermally populated.

The first term in parentheses in both \cref{eqn:Upsilon_c_Symmetric} and \eqref{eqn:Upsilon_c_Broken} corresponds to the contribution from the thermal population of gauge-bosons. Equivalently, for a pure dark Yang--Mills sector coupled to the axion, the friction coefficient is
\begin{equation}\label{eq:YM}
    \Upsilon_d=\frac{\Gamma_d}{2T}\, .
\end{equation}
This provides the maximal friction attainable from a confining dark sector of this type: the inclusion of fermions, whether chiral or vectorlike, only suppresses the friction relative to the pure Yang--Mills result, as is evident from \cref{eqn:Upsilon_c_Symmetric} and \eqref{eqn:Upsilon_c_Broken}. The blue curve in \cref{fig:Friction_Sampler} shows the corresponding pure-glue estimate using the high-temperature sphaleron rate. As the gauge coupling grows toward the confinement scale, this estimate rises rapidly and can exceed Hubble damping, but this near-confinement regime is precisely where the high-temperature expression is least reliable.

The purple curve in \cref{fig:Friction_Sampler} previews the lattice-motivated extrapolation discussed in the next subsection. We include it here to indicate the size of the theoretical uncertainty associated with the sphaleron rate near confinement. The main conclusion of \cref{fig:Friction_Sampler} is therefore qualitative: SM QCD friction is far too small to affect the axion rotation, while a new confining sector, either pure Yang--Mills or with sufficiently heavy vectorlike matter, may plausibly provide friction large enough to damp the axion motion, though the size of this damping depends sensitively on the sphaleron rate near confinement, which we now discuss.

\subsection{Summary of the sphaleron rate and friction}
\label{subsec:sphaleron_rate_summary}

The sphaleron rate is best understood in the hot, weakly coupled regime where the 't~\!Hooft parameter $\lambda \equiv 4\pi \alpha N$ is small, and the various length scales become parametrically separated. In particular, hard thermal excitations $\sim T^{-1}$ are much smaller than the Debye (electric) screening length $\sim \lambda^{-1/2}T^{-1}$, which is itself shorter than the magnetic screening length scale $\sim \lambda^{-1}T^{-1}$~\cite{Linde:1980ts,Gross:1980br}. Non-perturbative physics, including sphaleron transitions, occur at the magnetic scale, and can therefore be described after integrating out the hard and electric modes~\cite{Bodeker:1998hm,Bodeker:2002gy}. The resulting sphaleron rate has been studied on the classical Minkowski lattice for $N=2,3,5$ and $8$~\cite{Moore:2010jd} and scales as~\cite{Bodeker:1998hm}
\begin{equation}\label{eqn:rate_asymptotic}
    \Gamma_{\rm sph}\propto0.21\left(\frac{\lambda}{4\pi}\right)^5 T^4\left[\frac32\log\left(\frac{1}{\lambda}\right) + \mathcal{O}(1)\right]\equiv\Gamma_{\rm high},
\end{equation}
with additional dependence on the Debye mass and gauge group given in Ref.~\cite{Moore:2010jd} and \cref{app:topological_susceptibility}. In the particular case of  $\mathrm{SU}(2)_L$, the weak coupling limit is applicable down to the electroweak phase transition, allowing the electroweak sphaleron rate to be well understood~\cite{Kuzmin:1985mm,Arnold:1987mh,Ambjorn:1990pu,Bodeker:1995pp,Ambjorn:1995xm,Ambjorn:1997jz,Moore:1997cr,Moore:1997sn,Arnold:1996dy,Bodeker:1998hm,Arnold:1998cy,Bodeker:1999gx,Moore:1998zk,Arnold:1999ux,Arnold:1999uy}. 

At temperatures approaching confinement, the hierarchy of thermal scales collapses, and classical Minkowski lattice simulations no longer suffice. Nevertheless, in recent years, interest has developed in the QCD sphaleron rate because of its relevance to chirality flipping during electroweak baryogenesis~\cite{McLerran:1990de,Giudice:1993bb}, the chiral magnetic effect in heavy ion collisions~\cite{Fukushima:2008xe}, and because of its connection to axion physics~\cite{Notari:2022ffe,Bonanno:2023thi}, spurring on the development of a number of techniques to extract reliable non-perturbative information near confinement.

In Ref.~\cite{Altenkort:2020axj}, the Euclidean topological charge density two-point function, itself being calculable on the lattice, was used to constrain the sphaleron rate through the Kubo relation~\cite{Kubo:1957mj}. Since the sphaleron rate is determined by the small-frequency behavior of the corresponding spectral density, while the analytic continuation method in Ref.~\cite{Altenkort:2020axj} relies on perturbative input for the spectral density at large frequency, this method yields a lower bound rather than a direct determination. In the case of pure-glue ${\rm SU}(3)$, they find $\Gamma_{\rm sph}/T^4 >  0.024$ at $T = 1.5 T_c$, where $T_c\simeq 290\MeV$ is the critical deconfinement temperature of pure-glue ${\rm SU}(3)$ Yang--Mills~\cite{Borsanyi:2022xml,Giusti:2025fxu}.

Ref.~\cite{BarrosoMancha:2022mbj} proposes another technique, where the thermal occupancy of Euclidean configurations straddling the field-space hypersurface between adjacent winding number domains is used to infer the sphaleron rate. This picture assumes sphalerons remain robust, high-action objects near confinement, sufficiently distinct from ordinary thermal fluctuations. The method is applied to pure-glue ${\rm SU}(3)$ Yang--Mills at temperatures between $1.3T_c$ and $20T_c$, with $T_c\simeq 290\MeV$ the deconfinement temperature. As emphasized in Ref.~\cite{BarrosoMancha:2022mbj}, the approach is subject to several theoretical limitations, including the treatment of a dynamical prefactor characterizing correlation between sphaleron transitions, the large-volume limit, and the inability to access temperatures below $1.3T_c$.

Separately, Refs.~\cite{Kotov2018,Bonanno:2023ljc,Bonanno:2023thi} calculate the sphaleron rate using the Kubo relation to express it in terms of the spectral density of the Euclidean topological charge density correlator. Extracting the sphaleron rate then amounts to solving an inverse problem for the spectral density in terms of the Euclidean correlator. To approximate this inversion, Ref.~\cite{Kotov2018} applies the Backus--Gilbert method~\cite{Backus:1968svk}, while Refs.~\cite{Bonanno:2023ljc,Bonanno:2023thi} apply the HLT method~\cite{Hansen:2019idp}, a modification of Backus--Gilbert which allows the systematic uncertainties associated with the inverse problem to be quantified and included in the total error estimate. Ref.~\cite{Bonanno:2023ljc} introduces and tests this method in pure-glue ${\rm SU}(3)$, finding agreement with Ref.~\cite{BarrosoMancha:2022mbj}. Ref.~\cite{Bonanno:2023thi} then applies this method to obtain the first non-perturbative determination of $\Gamma_{\rm sph}$ in $2+1$ flavor QCD with physical quark masses, for temperatures between $1.48T_{\rm pc}$ and $3.68T_{\rm pc}$, where $T_{\rm pc}\simeq155\MeV$ is the chiral crossover pseudocritical temperature~\cite{HotQCD:2014kol}.

\begin{figure}
    \centering
    \includegraphics[width=\linewidth]{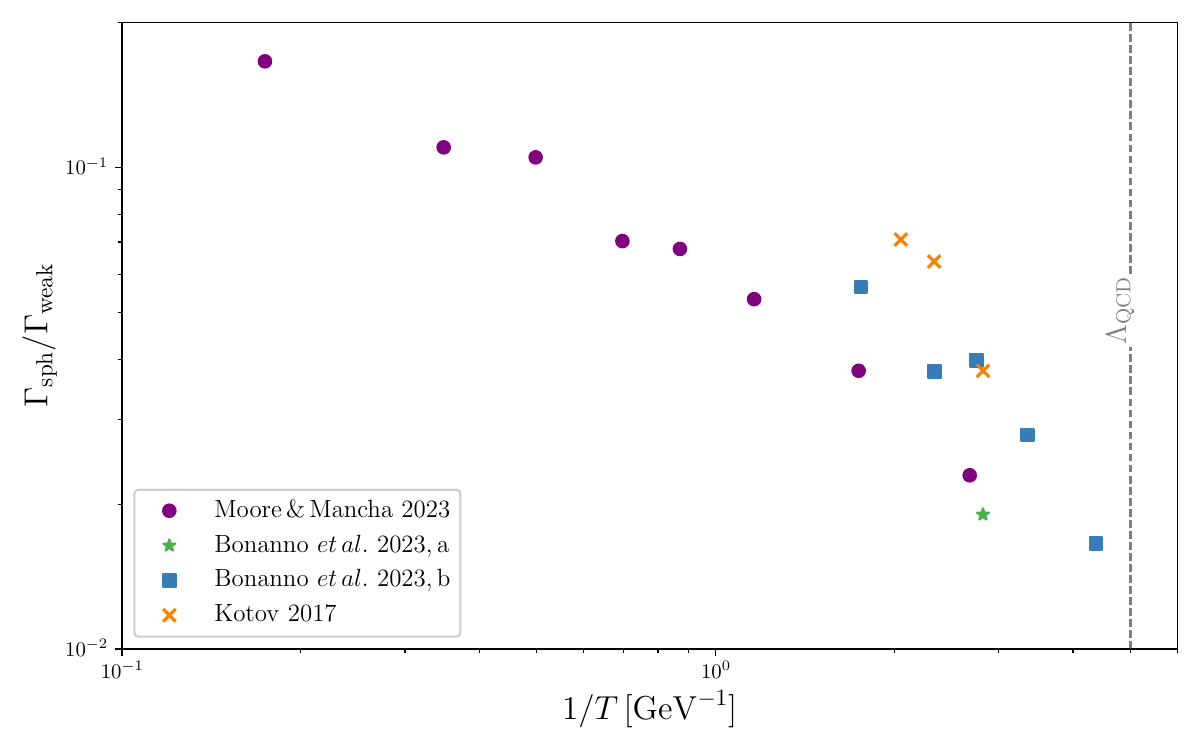}
    \caption{The ratio of near-confinement calculations of the sphaleron rate in $\mathrm{SU}(3)$ gauge theories to the weak-coupling asymptotic expression provided in Ref.~\cite{Moore:2010jd} and \cref{app:topological_susceptibility}. The \textcolor{blueberry}{blue} squares correspond to the full $2+1$ flavor QCD calculation of Ref.~\cite{Bonanno:2023thi}, while the remaining points are pure-glue results. The \textcolor{DeepPurple}{purple} circles correspond to the method of Ref.~\cite{BarrosoMancha:2022mbj}, while the \textcolor{winterberry}{orange} crosses are calculated in Ref.~\cite{Kotov2018} using the Backus--Gilbert method. The \textcolor{gooseberry}{green} star of Ref.~\cite{Bonanno:2023ljc} is derived using the HLT method in pure glue.}
    \label{fig:Sphaleron_Rate}
\end{figure}

In \cref{fig:Sphaleron_Rate}, we illustrate the ratio of the sphaleron rate at low temperatures as calculated using the methods discussed above and its weak coupling limit in \cref{eqn:rate_asymptotic}, evaluated for pure ${\rm SU}(3)$ using the conventions of \cref{app:topological_susceptibility}. In the near-confinement regime shown, the available results and bounds lie well below a direct extrapolation of the weak-coupling expression in \cref{eqn:rate_asymptotic}. The correct interpolation between the near-confinement temperatures, where lattice results exist, and high temperatures, where the asymptotic weak-coupling calculation of \cref{eqn:rate_asymptotic} is accurate, is currently unknown.

For a dark confining force with a confinement scale in the $\mathcal{O}({\rm GeV})$ range, the temperature regime with the largest uncertainty spans much of the region between the electroweak and QCD phase transitions. Further, while the weak-coupling $N$-dependence has been studied, to our knowledge no near-confinement lattice determinations exist for $\mathrm{SU}(N)$ with $N>3$. As the sphaleron rate plays a central role in determining the axion dynamics discussed in this paper, we parametrize the uncertainty in two ways.

First, we use the fit to the lattice data from Ref.~\cite{Bonanno:2023thi}:
\begin{align}
    \frac{\Gamma_{\rm sph}(T)}{T^4}=\left[\frac{A_0}{\log(T^2/T_{{\rm pc},d}^2)+\log(B_0^2)}\right]^5\,,
\end{align}
with $A_0 = 2.96(51)$ and $B_0 = 4.34(1.67)$. When applying this fit to the dark sector, we identify $T_{{\rm pc},d}$ with the corresponding dark confinement scale $\Lambda_d$.  Although this fit was obtained for $2+1$ QCD, the difference relative to pure Yang--Mills appears to be only ${\cal O}(1)$. Moreover, the asymptotic form in \cref{eqn:rate_asymptotic}, together with the more complete expression in \cref{app:topological_susceptibility}, suggests that the sphaleron rate should not vary parametrically across different ${\rm SU}(N)$ gauge groups. We therefore use this fit as a representative benchmark for the dark sector, while treating the normalization of the rate near confinement as an ${\cal O}(1)$ theoretical uncertainty.

Second, we define the dimensionless threshold
\begin{equation}\label{eq:Rth}
    R_{\rm th} \equiv \frac{T_{\rm th}}{T_c}\,,
\end{equation}
and we take the parametrization
\begin{equation}\label{eq:Gamma_Rth}
    \Gamma_{\rm sph}(T)=\Gamma_{\rm high}(T)\,\Theta\left(\frac{T}{T_c}-R_{\rm th}\right)\,,
\end{equation}
where $\Gamma_{\rm high}$ is the asymptotic expression whose limiting form is given in \cref{eqn:rate_asymptotic} and reproduced including ${\cal O}(1)$ factors in \cref{app:topological_susceptibility}. Thus, for temperatures below $T_{\rm th}$, this parametrization removes the sphaleron friction. This sharp cutoff is not intended as a model of the physical near-confinement rate, which lattice calculations find to be nonzero, but as a simple way of exposing the sensitivity of our results to this theoretical uncertainty. In particular, it exposes the extent to which our mechanism depends on early friction near the electroweak phase transition or late friction near confinement. We therefore scan over $R_{\rm th}$ when presenting results in \cref{sec:axion_signatures}.

\section{Minimal axiogenesis}\label{sec:IC_preliminaries}

We now review the early-Universe evolution of the axion and its radial mode in a minimal setup that can give rise to axiogenesis. We divide the discussion into epochs above and below the electroweak scale $T_{\rm{ew}}$, while assuming
\begin{align}
    \dot\theta(T_{\rm{ew}}) = \omegaEW \, ,
    \label{eq:velocity_IC_EW}
\end{align}
so as to generate the observed baryon asymmetry today. 
The goal of this discussion is to highlight several challenges facing the minimal realization of axiogenesis that motivate and inform the construction of our model in the following sections.

In \cref{subsec:axion_relic}, we will focus on temperatures below the electroweak phase transition, assuming that the axion’s radial mode $\sigma$ has relaxed to its vacuum expectation value, $\sigma = f_a$, and compute the axion relic abundance. We use the adiabatic approximation, assuming the axion rotation is rapid compared to the evolution of the Hubble rate and axion mass. Our results agree with those of Refs.~\cite{Co:2019wyp,Co:2019jts}. Our different approach for calculating the rotating axion relic abundance generalizes straightforwardly to multiple axions (with non-negligible sources of friction). The resulting relic abundance highlights the minimal axiogenesis overproduction problem: QCD axion velocities large enough to account for the observed baryon asymmetry overproduce dark matter \cite{Co:2019wyp}.

In \cref{sec:kinetic_misalignment}, we study the coupled evolution of the axion and its radial mode in a minimal axiogenesis setup with a field-theoretic axion at temperatures well above the electroweak phase transition. By analyzing the equations of motion, we determine the scaling of the energy density in this system. We find that, under the assumptions of our minimal setup, generating an axion velocity sufficient to explain the observed baryon asymmetry requires a high-dimensional radial potential, with efficient PQ-charge generation favoring PQ-breaking operators of comparable dimension. The required suppression of lower-dimensional PQ-breaking operators is closely related to the axion quality problem. The PQ-preserving quartic coupling must also be parametrically suppressed, even when the quadratic term dominates the renormalizable potential. We also comment on dynamical challenges associated with kinetic misalignment in the early Universe in \cref{subsec:redshifting_consistency,subsec:saxion_damping}. Much of this discussion builds on the existing literature, e.g.\ see in particular Ref.~\cite{Eroncel:2024rpe}, and serves to put into context the model of axiogenesis in the upcoming sections.

\subsection{Below $T_{\rm{ew}}$: axion equation of motion and relic abundance}
\label{subsec:axion_relic}

We now turn our attention to the QCD axion relic abundance, assuming that some dynamics have generated the velocity of the axion at the electroweak scale, and will show that generating the correct baryon asymmetry and axion dark matter are at odds with each other in this minimal axiogenesis setup, as already noted in the original proposal~\cite{Co:2019wyp}. We attribute this tension to the negligible friction that the QCD axion feels in the SM bath.

Below $T_{\rm{ew}}$, the axion evolves through three distinct regimes. (i) For $\Tosc<T<T_{\rm{ew}}$, where $\Tosc$ denotes the temperature at which the axion kinetic energy and the maximum of its potential energy become equal, the QCD susceptibility $\chi_c$ is negligible. (ii) For $\LQCD<T<\Tosc$, the QCD susceptibility becomes important and depends strongly on the temperature. (iii) For $T<\LQCD$, the susceptibility remains important but can be approximated by its zero-temperature value. 

We discuss these regimes in turn, beginning with the determination of $\Tosc$. This temperature can be found by solving
\begin{align}\label{eqn:Tosc}
    \frac12f_a^2\dot\theta^2(\Tosc) = 2\chi_c(\Tosc) \, .
\end{align}
Above this temperature, $\Tosc<T<\Tew$, the QCD susceptibility can be neglected and the axion field obeys the following equation of motion:\footnote{In \cref{fig:Friction_Sampler}, we showed that the QCD friction $\Upsilon_c$ is negligible at this temperature as well; we simply include it here to develop the full solution to this equation of motion, with an eye toward extensions in which the friction term is not negligible.}
\begin{align}\label{eq:EOM1}
    0&=\ddot\theta + \left(3H + \frac{\Upsilon_c}{f_a^2}\right)\dot\theta\,.
\end{align}
This equation can be directly integrated to obtain 
\begin{align}\label{eqn:rolling_adiabatic}
    \dot\theta^2 = \omegaEW^2\left(\frac{\asfew}{\asf}\right)^6\exp\left\{-\int_{T}^{T_{\rm{ew}}}{\rm d}\log T'\, \gamma_{\rm roll}(T')\right\}\,,\hspace{1cm}T_{\rm{ew}}>T>\Tosc\,,
\end{align}
where the instantaneous temperature power-law index is
\begin{align}
    \gamma_{\rm roll} = 2j\frac{\Upsilon_c}{H f_a^2}\,, \quad j \equiv 1 + \frac13\frac{d\log g_{\star S}}{d\log T}\,.
\end{align}
In this regime, the energy density of the axion is made of its kinetic energy and redshifts as $\asf^{-6}$. This implies that $\Tosc$ can also be found, after neglecting the small QCD friction, via solving 
\begin{align}\label{eqn:Tosc_Estimate}
    \frac12 f_a^2\omegaEW^2 \frac{g_{\star S}^2(\Tosc)}{g_{\star S}^2(T_{\rm{ew}})}\left(\frac{\Tosc}{T_{\rm{ew}}}\right)^6= 2\chi_c(\Tosc)\,.
\end{align}
This scaling is valid as long as the associated radial mode of the axion is settled at its minimum---an assumption we maintain in this section for $T < \Tew$. 

In both regimes, $\LQCD<T < \Tosc$ and $T<\LQCD$, the axion kinetic energy is less than the potential's maximum, and thus the axion no longer rolls freely over the hills and valleys of its potential---instead it oscillates around its potential minimum (hence the notation $\Tosc$). From this point on, the equation of motion is more accurately given by
\begin{align}
    0&=\ddot\theta + \left(3H + \frac{\Upsilon_c}{f_a^2}\right)\dot\theta + \frac{\chi_c}{f_a^2}\theta\,,
    \label{eq:EOM2}
\end{align}
where we have neglected the anharmonicity in the potential. Anharmonic effects are relevant only during the brief interval in which the axion kinetic and potential energies are comparable. This interval is short because the potential energy grows rapidly as $T^{-8}$, while the kinetic energy simultaneously redshifts as $T^{6}$. Indeed, in Ref.~\cite{Co:2019jts}, it is argued that solving the full homogeneous equation of motion, accounting for the anharmonic terms in the potential, introduces a factor of 2 boost to the final axion relic abundance compared to analytically solving \cref{eq:EOM2} and matching onto the rolling solution of \cref{eq:EOM1}. 

Solving \cref{eq:EOM2} using the WKB approximation for the axion oscillations, we find
\begin{align}\label{eqn:oscillating_adiabatic}
    \langle\theta^2\rangle \approx 2\times2\left(\frac{\asf_{\rm osc}}{\asf}\right)^3\exp\left\{-\int_T^{\Tosc}{\rm d}\log T'\,\gamma_{\rm osc}(T')\right\}\hspace{1cm}T < \Tosc\,,
\end{align}
where the overall normalization of \cref{eqn:oscillating_adiabatic} includes one factor of 2 determined by matching the energy density at $\Tosc$, and a second factor of 2 from Ref.~\cite{Co:2019jts}. The instantaneous temperature power-law index in the oscillatory regime is
\begin{align}
    \gamma_{\rm osc} = j\frac{\Upsilon_c}{H f_a^2} - \frac12 \frac{d\log\chi_c}{d\log T}\,,
\end{align}
where the first term accounts for the dissipation due to sphaleron transitions, while the second term accounts for the time-variation of the potential. In particular, the form of the $\chi_c$ term encodes the usual $\theta \propto m_a^{-1/2}$ scaling in the WKB approximation.

\begin{figure}
    \centering
    \includegraphics[width=\columnwidth]{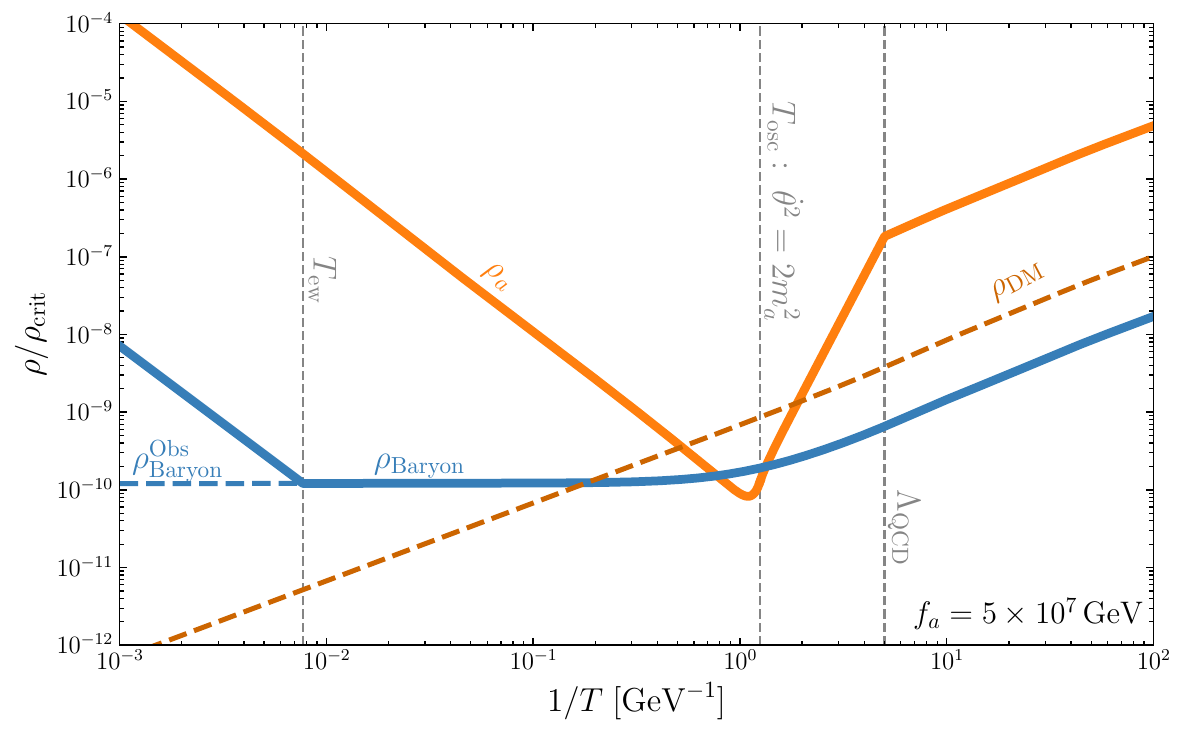}
    \caption{
    The energy density of the axion (solid \textcolor{orange}{orange}) and SM baryon asymmetry with $c_W = 0$ (solid \textcolor{blueberry}{blue}) relative to the critical density $\rho_{\rm crit} = 3\Mpl^2 H^2$ in the original axiogenesis model. At high temperatures, the QCD axion dilutes as $\rho_a\propto T^6$ (assuming the axion's radial mode is frozen). At the temperature of electroweak sphaleron freeze-out $T_{\mathrm{ew}}$, the baryon asymmetry decouples from the rolling axion, matching onto the observed baryon asymmetry (dashed \textcolor{blueberry}{blue}). At $T \sim T_\mathrm{osc}$, QCD susceptibility is no longer negligible and the axion potential increases proportional to $T^{-8}$, giving rise to an increase in its net energy density. Once the QCD topological susceptibility stops growing at $T = \Lambda_{c}\approx 200\MeV$, the axion energy density dilutes like cold matter. For comparison, we plot the energy density of a relic with the observed dark matter abundance (dashed, \textcolor{orange}{orange}). By requiring the QCD axion to produce the observed baryon asymmetry, it overproduces the dark matter, even for the most optimistically small axion decay constant $f_a = \fSN$. 
    }
    \label{fig:axiogenesis}
\end{figure}

The axion mode amplitude continues to dilute due to the expansion of the Universe, while also gaining energy from the growing QCD topological susceptibility. During this period, its energy density is accurately expressed in terms of the cycle-averaged amplitude, 
\begin{align}
    \rho_a = \frac12 f_a^2 \dot\theta^2 + \chi_c (1 - \cos\theta) \;\approx\; \frac12 f_a^2 \langle \dot\theta^2 \rangle + \frac12 \chi_c \langle \theta^2 \rangle\approx \chi_c\langle\theta^2\rangle \,, 
\end{align}
where in the last equality we have used equipartition.

Given the negligible QCD friction $\Upsilon_c$, \cref{eqn:oscillating_adiabatic} implies that for $T < T_\mathrm{osc}$, the axion energy density evolves as 
\begin{align}
    \label{eq:density_scale_below_QCD}
    \rho_a \approx 4\chi_c(T) \left(\frac{\chi_c(\Tosc)}{\chi_c(T)}\right)^{1/2} \frac{g_{\star S}(T)}{g_{\star S}(\Tosc)}\left(\frac{T}{\Tosc}\right)^3 = 2\chi_c^{1/2}(T) f_a \omega_{\rm ew} \frac{g_{\star S}(T)}{g_{\star S}(T_{\rm{ew}})}\left(\frac{T}{T_{\rm{ew}}}\right)^3 \,.
\end{align}
The first equality follows from the WKB approximation and the matter-like redshifting of the axion energy density, as described by \cref{eqn:oscillating_adiabatic}. In the second equality, we restrict to $T\leq\Tosc$ and use \cref{eqn:Tosc_Estimate} to eliminate $\chi_c(\Tosc)$.

Putting the numbers to the equation above and assuming the axion velocity at $T_{\rm{ew}}$ is such that we generate the observed baryon asymmetry in the Universe, we find
\begin{align}
    \frac{\Omega_a}{\Omega_{\rm cdm}} \approx 49\times \left(\frac{f_a}{\fSN}\right)\,,
\end{align}
that, up to effects of anharmonicity and nonadiabaticity, is entirely independent of the rate at which the QCD instanton potential turns on. While this appears as an accidental cancellation of temperature dependence when viewed from the perspective of the equations of motion, the result must be independent of the detailed temperature dependence of $\chi_c(T)$ because the conserved PQ charge in the rolling regime matches onto the approximately conserved axion number in the oscillatory regime. We review this calculation, as originally employed in Ref.~\cite{Co:2019jts}, in detail in \cref{app:relic_abundance_from_adiabatic_invariant}.

The evolution of the axion relic abundance normalized to the SM bath is plotted as a function of temperature in \cref{fig:axiogenesis}. In making this plot, we use the initial condition $\omega_{\rm{ew}}\equiv\dot{\theta} \, (T_{\rm{ew}}) = 3.9~{\rm{keV}}$ needed to explain the observed baryon asymmetry today. We find that this velocity gives rise to too much axion energy density today and overcloses the Universe, illustrating the need for additional dissipation channels. Although the QCD sphaleron process in principle supplies dissipation, for decay constants that satisfy the SN1987A bound, $f_a \gtrsim 5\times 10^7\GeV$~\cite{Lucente:2022vuo,Springmann:2024ret}, the sphaleron-induced friction in QCD, $\Upsilon_c/f_a^2$, is always negligible (see \cref{fig:Friction_Sampler}), and is not capable of dissipating away the axion velocity---resulting in an overabundant axion relic. In the upcoming sections, we argue that this overclosure problem can be avoided with \textit{new sources of friction}.

So far, we have assumed that the axion radial mode is frozen at its minimum while the axion carries a nonzero angular momentum. We next study the coupled equations of motion of the axion and its \textit{radial mode} in a minimal field-theoretic axion setup.

\subsection{Above $T_{\rm{ew}}$: kinetic misalignment and PQ breaking}\label{sec:kinetic_misalignment}

We now turn to the axion dynamics above $\Tew$ and the generation of a sufficiently large post-inflationary axion velocity. 
We review how such initial conditions can arise through the Kinetic Misalignment Mechanism of Ref.~\cite{Co:2019jts}, and then compute the resulting axion velocity in an explicit model, highlighting both the underlying dynamics and the associated tensions.

Let us consider a model in which the axion descends from the phase of a complex scalar, and in conjunction with a partner CP-even scalar field, $\sigma$:
\begin{align}
    \Phi\equiv \frac{1}{\sqrt{2}}\sigma e^{i\theta}\,.
\end{align}
The radial mode sits in a PQ-conserving potential $V_\sigma (\sigma)$, while the system also receives an explicit PQ-breaking potential $\vpq \, (\sigma,~\theta)$. 
The coupled equations of motion are given by
\begin{align}
    \label{eq:EOMsigma}
    0&=\ddot\sigma +3 H\dot\sigma+\partial_\sigma V_\sigma(\sigma) + \partial_\sigma \vpq(\sigma,~\theta) - \sigma\dot\theta^2\,,\\
    \label{eq:EOMtheta}
    0&=\ddot\theta + \left(3 H + 2\frac{\dot\sigma}{\sigma}\right)\dot\theta + \frac{1}{\sigma^2} \partial_\theta \vpq (\sigma,~\theta)\,,
\end{align}
with $V_\sigma^\prime\equiv\partial_\sigma V_\sigma$. 
In writing these equations we have neglected the small QCD friction at high temperatures.
We now examine these equations more closely.

Let us first study the equations of motion without a PQ-breaking potential $\vpq$.
From \cref{eq:EOMtheta}, one can immediately read off the ``angular momentum'' conservation law:
\begin{align}
    0&=\frac{\partial}{\partial t}\left(\asf^3\sigma^2\dot\theta\right)\,,
\end{align}
which we use to define the conserved global charge density which dilutes with comoving volume
\begin{align}
    \label{eq:conserved_ntheta}
    n_\theta\equiv \sigma^2\dot\theta\propto \asf^{-3}\,,
\end{align}
in terms of which the radial equation of motion becomes
\begin{align}
    0&=\ddot\sigma + 3 H\dot\sigma + V_\sigma'(\sigma) - \frac{n_\theta^2}{\sigma^3}\,.
    \label{eq:EOMsig2}
\end{align}
The last term encodes the angular momentum barrier set up by a non-zero $\dot\theta$.
The balance between the bare radial potential $V_\sigma$ and the angular momentum barrier $\frac12n_\theta^2/\sigma^2$ gives rise to a dynamical equilibrium for the radial mode:
\begin{align}
    V_\sigma'(\sigma_{\rm eq}) = \frac{n_{\theta}^2}{\sigma_{\rm eq}^3}\,.
    \label{eq:eqb_V_KE}
\end{align}

\cref{eq:EOMsig2} can be solved to show that the minimum of the radial mode effective potential is described by an effective mass
\begin{align}
    m_{\rm eff}^2 = V_\sigma''(\sigma_{\rm eq}) +3\frac{n_\theta^2}{\sigma_{\rm eq}^4}= V_\sigma''(\sigma_{\rm eq}) + \frac{3}{\sigma_{\rm eq}}V_\sigma'(\sigma_{\rm eq})\,.
    \label{eq:meff}
\end{align}
When $m_{\rm eff}$ is large compared with $H$, the radial field can perform damped oscillations about the slowly evolving equilibrium.
Ref.~\cite{Fedderke:2025sic} shows that these oscillations, if undamped, may lead to a parametric resonance instability involving both the axion and radial mode. At saturation, the instability causes the formation of coherent counter-rotating regions: on the domain boundaries between these regions pairs of axion strings nucleate, undergoing periodic production and annihilation, and the system radiates a significant portion of its energy in the form of scalar radiation. If this resonance persists long enough to saturate, the picture of the coherently rotating axion field breaks down, possibly spoiling kinetic misalignment. Therefore, dissipation channels that efficiently damp the $\sigma$ oscillations are necessary to maintain analytical control over kinetic misalignment. We consider the implication of one possible dissipation channel in \cref{subsec:saxion_damping}.

Assuming that the radial oscillations are efficiently damped, the radial field adiabatically tracks the equilibrium set by the balance between the radial potential and the angular momentum barrier. The redshifting of the resulting rotating configuration is then controlled by the functional form of $V_\sigma$. Assuming that the potential is a monomial in $\sigma$,
\begin{align}\label{eqn:perturbative_potential_saxion}
    V_\sigma(\sigma) \propto \sigma^m,
    \qquad m \geq 2\,,
\end{align}
the equilibrium condition, together with conservation of $n_\theta\propto \asf^{-3}$, implies that the total energy density of the rotating field redshifts as \cite{Turner:1983he}
\begin{align}
    \rho_{\rm tot}\propto\asf^{-3(1+w_\sigma)}\,,\qquad w_\sigma=\frac{m-2}{m+2}\,.
    \label{eq:EOS_radial}
\end{align}
Here $w_\sigma$ is the effective equation of state of the rotating field.\footnote{The adiabatic rotating solution is the ``spintessence'' regime from Refs.~\cite{Boyle:2001du,Johnson:2008se}. More generally, for any bounded orbit of a homogeneous complex field in a U(1)-symmetric monomial potential, the orbit-averaged virial relation implies the cycle-averaged equation of state is given by \(\overline{w}_\sigma=(m-2)/(m+2)\), independently of the orbit's eccentricity and including the purely radial limit~\cite{Turner:1983he}. This statement assumes that Hubble expansion and dissipation are slow on an orbital timescale.}
Along the equilibrium trajectory, the angular kinetic energy and radial potential energy satisfy
\begin{align}
    \frac{1}{2}\sigma_{\rm{eq}}^2\dot\theta_{\rm{eq}}^2=\frac{m}{2}V_\sigma(\sigma_{\rm{eq}})\,,
    \label{eq:eqb_cond}
\end{align}
so that the condensate has the same effective equation of state as a rapidly oscillating real scalar in the potential of \cref{eqn:perturbative_potential_saxion}, although in the present case the kinetic energy is carried predominantly by the axion rotation rather than by radial oscillations. 
While the radial field adiabatically tracks this equilibrium, its subsequent evolution is described by the following scaling relations.

The radial field value and axion rotational velocity consequently redshift as
\begin{align}\label{eqn:kmm_redshifting}
    \sigma\propto\asf^{-\frac{3}{2}(1-w_\sigma)}\,,\qquad\dot\theta\propto\asf^{-3w_\sigma}\,,\qquad{(\rm centrifugal\,\,equilibrium)}\,.
\end{align}
Two cases are particularly illustrative: for the quadratic potential $m = 2$, $\rho_{\rm tot}$ dilutes like matter ($\rho_{\rm tot}\propto \asf^{-3}$),
while for the quartic potential $m = 4$, $\rho_{\rm tot}$ dilutes like radiation ($\rho_{\rm tot}\propto \asf^{-4}$). 
In the former (latter) case, $\dot{\theta} \propto \asf^0$ ($\dot{\theta} \propto \asf^{-1}$).
This result should be contrasted with the $\dot{\theta} \propto \asf^{-3}$ scaling assumed in \cref{sec:review}, that is valid in the case where the radial mode is fixed at its minimum. 
This is illustrative of the more general point that the full dynamics of the UV model can qualitatively change the redshifting of the energy density in the axion system and its rotational velocity. 

The circular approximation can break down for sufficiently steep saxion potentials. During radiation domination, \cref{eqn:kmm_redshifting} implies
\begin{align}
    \frac{\dot\theta_{\rm eq}}{H} \propto \asf^{-\frac{m-10}{m+2}}\,.
\end{align}
As we show below, the initial angular velocity generated by a subdominant PQ-breaking potential is of order $H_{\rm roll}$ or smaller. Therefore, for $m>10$ the rotation never becomes parametrically rapid compared with Hubble expansion. Consequently, as the conserved PQ charge density redshifts, the centrifugal equilibrium \cref{eq:eqb_V_KE} moves inward, but the radial field eventually fails to track it.

In the absence of additional radial dissipation, the centrifugal force becomes subdominant and the field approaches the scaling solution
\begin{align}\label{eq:steep_saxion_scaling}
    \sigma\propto\asf^{-\frac{4}{m-2}},\qquad\frac{\dot\theta}{H}\propto\asf^{-\frac{m-10}{m-2}}\,,\qquad m > 10\,,
\end{align}
where the second proportionality follows from conservation of angular momentum. This scaling behavior is analogous to that of a real scalar in a steep monomial potential~\cite{Ema:2015dza}, where the radial force is balanced by Hubble friction, and $\sigma$ decreases more slowly than in the circular approximation.

Axion rotation can be generated by a PQ-violating potential when the radial field, initially displaced from its minimum, begins to roll in the early Universe~\cite{Co:2019wyp,Co:2019jts}. 
In the absence of PQ violation, the potential exerts no torque in the angular direction. Consequently, a field starting with vanishing angular velocity evolves purely radially, and no coherent axion rotation, $\dot{\theta}\neq 0$, is generated.

To clarify the dynamics of PQ charge generation,  
we now focus on a toy potential and cosmological history that can be solved analytically \cite{Liddle:1998xm,Ema:2015dza}. 
Specifically, we assume a matter-dominated early Universe
and a potential 
\begin{align}
    V(\sigma,\theta) = V_{\sigma}(\sigma,6) + V_{\cancel{\text{PQ}}}(\sigma,\theta,n)\,,
    \label{eq:tot_V_sigma_PQ}
\end{align}
with $V_{\cancel{\text{PQ}}}(\sigma,\theta,n)$ the PQ-breaking part,
\begin{align}
    V_{\cancel{\text{PQ}}}(\sigma,\theta,n) = \frac12C_n \frac{\Phi^n}{M_{\rm UV}^{n-4}} + {\text{h.c.}} = |C_n|\frac{\sigma^n}{2^{n/2}M_{\rm UV}^{n - 4}}\cos(n\theta + \arg C_n)\, ,
\end{align}
The PQ conserving radial potential is given by 
\begin{align}\label{eqn:Stabilizing}
    V_{\sigma}(\sigma,m) = B_m\frac{\sigma^{m}}{M_{\rm UV}^{m - 4}}\,.
\end{align}
We assume $n\geq 6$ and that either $B_6\gg |C_6|$ (for $n=6$) or $\sigma\ll M_{\rm UV}$ (for $n>6$) so that we can safely treat the PQ-violating potential as a perturbation compared to the radial potential. We will use this setup as an example to illustrate a broader point: the initial $\dot{\theta}$ when it starts rolling is smaller than the Hubble rate at that instant.

After a long period of inflation, $\sigma$ is generically displaced from its present-day vacuum expectation value. From large field initial conditions, the radial mode rolls towards its vacuum expectation value at smaller field values, while the axion receives a small push from the PQ-violating operator. As the radial mode decreases, the irrelevant PQ-violating potential becomes less important.

We denote the initial radial field displacement at the onset of rolling by $\sigma_0\equiv\sigma(T_{\rm roll})$. The radial mode starts rolling at the time determined by the condition 
\begin{equation}
    3H \approx \sqrt{V_\sigma'(\sigma_0)/\sigma_0}\, ,
    \label{eq:rolling_condition}
\end{equation}
yielding
\begin{align}
    t_{\rm roll}\approx \sqrt{\frac{2}{3}}\frac{M_{\rm UV}}{B_6^{1/2}}\frac{1}{\sigma_0^2}\,.
\end{align}
We can then solve the radial mode equation of motion in \cref{eq:EOMsigma}. In doing so we neglect the PQ-violating potential and the axion rotation as they are both negligible compared to the radial mode potential. We find that $\sigma$ evolves approximately according to the power law~\cite{Liddle:1998xm,Ema:2015dza}:\footnote{The pseudo-scaling solution relevant for the degree 6 monomial potential in matter domination exhibits this power law for a long but finite period of time before pausing and then resuming a long time later \emph{ad infinitum}~\cite{Ema:2015dza}. Since the power law holds for several orders of magnitude, we neglect this feature since other terms in the potential will spoil the pseudoscaling long before it is relevant.}
\begin{align}\label{eqn:sigma_background}
    \sigma\approx \sqrt{\frac{2}{3}}\frac{M_{\rm UV}}{B_6^{1/2}}\frac{1}{t\sigma_0}\,.
\end{align}

Using this background $\sigma$ evolution, $\theta$'s equation of motion \cref{eq:EOMtheta} can be rewritten as
\begin{align}\label{eqn:theta_kmm}
    0&=\theta'' + \frac{2}{3}\frac{n^2|C_n|}{2^{n/2}B_6} \left(\frac{\sigma_0}{M_{\rm UV}}\right)^{n - 6}\tau^{2-n}\theta\,,
\end{align}
where now $'$ represents derivatives with respect to $\tau = t/t_{\rm roll}$, we have replaced the full cosine potential with its approximation about the minimum, and we have absorbed the $\arg C_n$ phase into the definition of $\theta$.

Solving this equation with the initial condition $\theta(t_{\rm roll}) = \theta_0$ and $\dot\theta(t_{\rm roll}) = 0$, and to the leading order in $|C_n|$, we find an expression for the asymptotic axion charge 
\begin{align}\label{eqn:conserved_charge}
    |n_\theta| = \sqrt{\frac{2}{3}}\frac{|C_n|}{2^{n/2}B_6^{1/2}}\frac{n^2}{n-3}\theta_0M_{\rm UV}^3\left(\frac{\sigma_0}{M_{\rm UV}}\right)^{n-2}\frac{\asf_{\rm roll}^3}{\asf^3}\,,
\end{align}
where $\asf_{\rm roll}$ is the scale factor when $t = t_{\rm roll}$. 
Extrapolating this asymptotic charge back to $t=t_\mathrm{roll}$, we can write\footnote{Note that $n_{\theta,\rm roll}$ need not exactly match $n_\theta(T_{\rm roll})$, since the transient PQ violating force may still be active at that time.}
\begin{align}\label{eq:nthetaroll}
    n_{\theta,\rm roll} = \frac{\epsilon_{\rm circ}}{\sqrt{20}}\sigma_0^2 H_{\rm roll}\,,
\end{align}
with the charge-generation efficiency
\begin{align}\label{eq:epsilon_circ}
    \epsilon_{\rm circ} = \frac{\sqrt{20} n^2}{n-3}\frac{|C_n|}{2^{n/2}B_6}\theta_0\left(\frac{\sigma_0}{M_{\rm UV}}\right)^{n-6}\,.
\end{align}
Clearly, in the small PQ-violating limit ($|C_6| \ll B_6$ or $\sigma_0\ll M_{\rm{UV}}$), we have $\epsilon_{\rm circ} \ll 1$, thus
\begin{align}\label{eq:nth_smaller_H}
    n_{\theta,\rm roll} \ll \sigma_0^2H_{\rm roll} \Longrightarrow \dot{\theta}_{\rm roll} \ll H_{\rm roll} \,,
\end{align}
which is the main takeaway from this example. We note that, here, $\dot\theta_{\rm roll}$ is the velocity inferred by extrapolating the asymptotic conserved charge back to $T_{\rm roll}$, as opposed to the dynamical initial condition $\dot\theta(T_{\rm roll}) = 0$.

This result can be intuitively generalized to any monomial $V_\sigma$ provided the PQ-violating potential can be treated as a perturbation of $V_\sigma$. With the perturbative PQ-violating potential, we can show \cite{Co:2019jts}
\begin{align}
    n_{\theta,\rm roll} = \left.\epsilon_{\rm dyn} \frac{\partial_\theta V_{\cancel{\rm PQ}}}{\sqrt{\partial_\sigma ^2V_{\sigma}}}\right|_{\sigma = \sigma_0,\,\theta = \theta_0}\, ,
    \label{eq:nth_roll}
\end{align}
with $\epsilon_\mathrm{dyn} \sim \mathcal{O}(1)$.\footnote{This quantity relates the PQ-breaking to the resulting PQ-charge \cite{Co:2019jts}. For the example studied above, one can show
$$\epsilon_{\rm dyn} = \frac{2\sqrt{5}}{n - 3}\, .$$
}
Let us define a counterpart quantity for the radial direction
\begin{align}
    n_{\sigma,\rm roll} \equiv \frac{\sigma}{m} \left.\frac{\partial_\sigma V_{\sigma}}{\sqrt{\partial_\sigma^2 V_{\sigma}}}\right|_{\sigma = \sigma_0}= \left.\frac{V_{\sigma}}{\sqrt{\partial_\sigma^2 V_{\sigma}}}\right|_{\sigma = \sigma_0}\,.
    \label{eq:nsigma_roll}
\end{align}

Then the ratio of the gradient of the potential in axion and in the radial mode is quantified by
\begin{align}
    \epsilon_{\rm circ} \equiv \frac{|n_{\theta,\rm roll}|}{n_{\sigma,\rm roll}}\,,
    \label{eq:ecirc_general_def}
\end{align}
which reduces to \cref{eq:epsilon_circ} for the specific example studied above. 

The charge-generation efficiency $\epscirc$ determines the angular kinetic energy associated with the initial displacement $\sigma_0$. Using \cref{eq:ecirc_general_def,eq:nsigma_roll}, we find
\begin{align}\label{eq:general_thetadot_H}
    \frac{n_{\theta,\rm roll}^2}{2\sigma_0^2}=\frac{\epscirc^2}{2m(m-1)}V_\sigma(\sigma_0,m)\approx\frac{9\epscirc^2}{2m^2(m-1)}\sigma_0^2H_{\rm roll}^2\,.
\end{align}
where $\dot\theta_{\rm roll}\equiv n_{\theta,\rm roll}/\sigma_0^2$ is the angular velocity extrapolated from the asymptotic PQ charge, and the example of this section gives $|\dot\theta_{\rm roll}|\ll H_{\rm roll}$. Consequently,
\begin{align}\label{eq:fin_thetadot_bound}
    \frac{|n_{\theta,\rm roll}|}{\sigma_0^2}=|\dot\theta_{\rm roll}|\approx\frac{3\epscirc}{m\sqrt{m-1}}H_{\rm roll}\,.
\end{align}
where the second equality follows from the monomial potential and the rolling condition.

In the parametric analysis that follows, we focus on a deliberately optimistic benchmark, $n_{\theta,\rm roll}\sim\sigma_0^2H_{\rm roll}$. We adopt the prompt-damping approximation, assuming that PQ charge is generated around the onset of radial motion at $\sigma(T_{\rm roll})=\sigma_0$, and that PQ-preserving interactions subsequently damp the radial motion and bring the field to its circular equilibrium at $\sigma_{\rm circ}$ on a timescale much shorter than a Hubble time. We therefore neglect the cosmological expansion between the onset of rolling and circularization, identifying the temperature and Hubble rate during these stages with $T_{\rm roll}$ and $H_{\rm roll}$.

The condensate therefore reaches its circular equilibrium before significant cosmological redshifting occurs, so that, up to $\mathcal{O}(1)$ factors,
\begin{equation}
    \sigma_{\rm circ}\sim\sigma_0\,,\qquad\dot{\theta}_{\rm circ}\sim H_{\rm roll}\,,\qquad\rho_\Phi(T_{\rm roll})\sim V_\sigma(\sigma_0)\sim\sigma_0^2H_{\rm roll}^2\,.
\end{equation}
In the next section, we study the subsequent redshifting of this initial velocity and derive the consistency conditions that a minimal axiogenesis setup must satisfy.

\subsection{Consistency conditions for a successful axiogenesis}
\label{subsec:redshifting_consistency}

In this section, we develop an analytic estimate of the cosmological evolution of a rapidly circularized axion--saxion system carrying a conserved PQ charge, and derive the conditions under which it can support successful axiogenesis. We work primarily in a radiation-dominated background and consider a two-stage radial potential consisting of a renormalizable and a non-renormalizable piece. We find that the simultaneous requirements of (i) \textit{successful baryogenesis}, (ii) \textit{inflationary initial conditions}, and (iii) \textit{EFT consistency} are highly restrictive. Within radiation domination, these conditions require a UV radial mode potential with dimension $m>10$ and a parametrically flat quartic term in the same potential, and they strongly constrain the axion decay constant to be near its present astrophysical lower bound, $f_a \gtrsim 5\times 10^7\GeV$ \cite{Lucente:2022vuo,Springmann:2024ret}, particularly for smaller $m$. Several structural features of this result persist for more general cosmological backgrounds between $\Troll$ and $\Tew$. 

We assume the potential of the system is made up of a non-renormalizable part like \cref{eqn:Stabilizing} and a renormalizable piece whose importance only grows as $\sigma$ approaches its minimum from above 
\begin{equation}
    V_\sigma \sim B_m \frac{\sigma^m}{\MUV^{m-4}} + \lambda_\ell f_a^{4-\ell} |\sigma^2 - f_a^2|^{\ell/2} \, ,
    \label{eq:potential_Vsogma_Vrenorm}
\end{equation}
We will set the dimensionless coupling $B_m = 1$ for the rest of the analysis. As we are mainly concerned with $\sigma \gg f_a$, we will replace $\sigma^2-f_a^2\to\sigma^2$ in \cref{eq:potential_Vsogma_Vrenorm}.

\begin{figure}
    \centering
    \includegraphics[width=\linewidth]{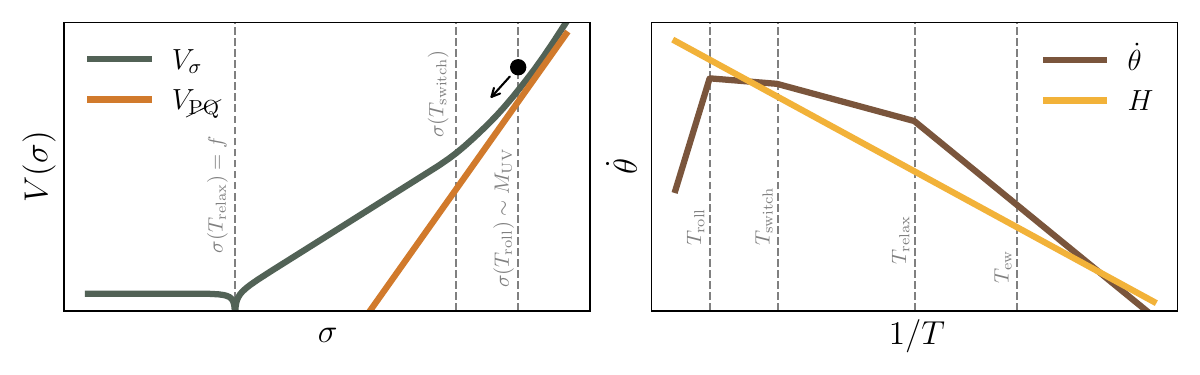}
    \caption{Evolution of the radial mode $\sigma$ in its potential (\textbf{left}) and the corresponding axion velocity $\dot\theta$ (\textbf{right}). The radial mode begins rolling appreciably at $T=\Troll$, generating PQ charge during a transient interval around this epoch, after which the PQ-violating contribution is assumed to become negligible. At large field values, the non-renormalizable term $V_\sigma\propto\sigma^m$ dominates, while below $T=\Tswitch$ the renormalizable contribution $V_\sigma\propto\sigma^\ell$ takes over; see \cref{eq:potential_Vsogma_Vrenorm}. The radial mode finally settles near $\sigma\simeq f_a$ at $T=T_{\rm relax}$. Correspondingly, $\dot\theta$ redshifts according to the dominant radial potential, changes its scaling at $T_{\rm switch}$ once the renormalizable part of the potential takes over, and redshifts as $\asf^{-3}$ once $\sigma$ has settled at its minimum. Changes in the steepness of the radial potential affect both the velocity reached at $T_{\rm relax}$ and the duration of the subsequent rapid redshifting. As shown in the text, these effects compensate in the final $\dot\theta(\Tew)$, rendering it insensitive to several details of the intermediate radial evolution.
}
    \label{fig:above_Tew_evolution}
\end{figure}

At $T=T_{\rm roll}$, the saxion begins rolling from its initial displacement $\sigma_0$, generating PQ charge through the explicit PQ-breaking interaction. We assume that this interaction becomes negligible after charge generation, so that the subsequent evolution conserves the comoving PQ charge. Under the prompt-damping approximation, radial motion is rapidly damped and the field approaches its circular equilibrium at $\sigma_{\rm circ}$, with negligible cosmological expansion during this process (see \cref{eq:eqb_cond}). The subsequent evolution of $\sigma$ and $\dot\theta$ is summarized in \cref{fig:above_Tew_evolution}. Initially, $\Phi$ evolves in the presence of both PQ-conserving and PQ-violating contributions to the potential, as in \cref{eq:tot_V_sigma_PQ}. We assume that after the transient charge-generation interval around the onset of rolling $T = \Troll$, the PQ-violating contribution becomes negligible. We assume for $\Tswitch \leq T \leq \Troll$ the non-renormalizable term from the potential in \cref{eq:potential_Vsogma_Vrenorm} dominates. Below $\Tswitch$, the renormalizable piece dominates. Eventually, the radial mode settles at its minimum $f_a$ at $T \sim \Trelax$. Below this temperature, $\sigma$ is approximately fixed and conservation of the axion angular momentum implies $\dot\theta\propto\asf^{-3}$.\footnote{The effect of a temperature-dependent radial vacuum expectation value is studied in Ref.~\cite{Chun:2024gvp}.}

At the onset of rolling, the scalar energy density is dominated by the radial potential evaluated at the initial displacement $\sigma_0$. Using the rolling condition in \cref{eq:rolling_condition}, we have
\begin{equation}\label{eq:sigma_roll_1}
    \rho_\Phi(T_{\rm roll})\sim V_\sigma(\sigma_0)\sim\sigma_0^2H_{\rm roll}^2\,.
\end{equation}
Here $\rho_\Phi(T_{\rm roll})$ denotes the scalar energy density before the subsequent dissipation of radial motion. Defining the initial energy fraction by $\rho_\Phi(T_{\rm roll})\equiv\epsilon_\Phi(T_{\rm roll})\rho_{\rm rad}$, we obtain
\begin{equation}\label{eq:sigma_roll_2}
    \sigma_0\sim\sqrt{\epsilon_\Phi(T_{\rm roll})}\,M_{\rm pl}\,.
\end{equation}
Thus, either the axion constitutes a very small fraction of the total energy density at $\Troll$, or radial mode must begin with a Planckian field excursion. This presents a potential challenge for minimal axiogenesis: perturbative control may be lost at such large field values, and quantum-gravity effects may become important. At the same time, a smaller energy density fraction may suppress the resulting axion velocity.

Our conclusions can mostly be derived directly from the conservation of the PQ charge after $\Troll$ and in terms of the initial conditions at that time. Below, we cover different possible scenarios and show that in each case, minimal axiogenesis requires non-generic potentials.

In the remainder of this subsection, we will express bounds in terms of the following quantity, which expresses the efficiency of PQ charge generation
\begin{align}\label{eq:ecirc_tilde_def}
    \epscirctd\equiv\frac{|n_{\theta,\rm roll}|}{\sigma_0^2H_{\rm roll}}\,.
\end{align}
Observe $\tilde{\epsilon}_{\rm circ} = 3\epsilon_{\rm circ}/m\sqrt{m-1}$ for the monomial potential $V_\sigma(\sigma,m)$ considered above. In our numerical estimates, we take the deliberately optimistic benchmark value $\tilde{\epsilon}_{\rm circ}=1$.

\subsection*{Non-renormalizable Potential Dominates}

Let us first assume a radiation-dominated background cosmology with the non-renormalizable term in \cref{eq:potential_Vsogma_Vrenorm} dominating the initial radial mode displacement. 
Conservation of the comoving PQ charge, combined with a negligible variation in the number of relativistic degrees of freedom and entropy between $\Troll$ and $\Tew$, and assuming the radial mode $\sigma$ has reached its minimum $f_a$ by $\Tew$, implies
\begin{align}\label{eq:PQ_cons_data_1}
    \dot{\theta}(\Tew)=\epscirctd H_{\rm roll}\left(\frac{\sigma_0}{f_a}\right)^2\left(\frac{\Tew}{\Troll}\right)^3\,.
\end{align}
Combining this result with \cref{eq:sigma_roll_2,eq:rolling_condition} yields
\begin{align}\label{eq:PQ_cons_data_2}
    \dot{\theta}(\Tew)\sim g_\star^{3/4} \frac{\Tew^3}{f_a^2} \epscirctd \epsphi^{\frac{10-m}{8}}(\Troll)\left(\frac{\MUV}{\Mpl}\right)^{\frac{m-4}{4}} \,,
\end{align}
where $g_\star$ denotes the number of relativistic degrees of freedom in the thermal bath. 
This relation may also be solved for $\epsphi$ and combined with \cref{eq:sigma_roll_2} to yield an expression for $\sigma_0/\MUV$:
\begin{align}\label{eq:sigroll_over_MUV}
    \frac{\sigma_0}{M_{\rm UV}}\sim\left(\frac{g_\star^3\epscirctd^4M_{\rm UV}^6\Tew^{12}}{\dot{\theta}^4(\Tew)f_a^8\Mpl^6}\right)^{\frac{1}{m-10}}\,.
\end{align} 
Consistency of the EFT treatment requires $\sigma_0/\MUV < 1$, thereby imposing bounds on $f_a$.

It is worth emphasizing that, in deriving \cref{eq:PQ_cons_data_2}, we assumed a radiation-dominated Universe at $T=\Troll$. Similarly, in deriving \cref{eq:PQ_cons_data_1}, we assumed that the scale factor satisfies $\asf\propto T^{-1}$ at $T=\Troll$ and $T=\Tew$, where $T$ denotes the temperature of the radiation bath. Therefore, all the results presented below apply to any background cosmology satisfying these conditions and, in particular, are independent of the detailed cosmological evolution between $\Troll$ and $\Tew$. For example, Ref.~\cite{Co:2021lkc} argues that minimal axiogenesis models naturally give rise to an epoch of matter domination followed by a kination-dominated epoch between $\Troll$ and $\Tew$. Our analysis applies directly to such a cosmological history because it does not depend on the intermediate evolution.

Successful axiogenesis requires the radial field to remain displaced throughout inflation, ensuring that its post-inflationary evolution begins at the large field value required near $\Troll$. The field value at which the local curvature of the radial-mode potential becomes comparable to Hubble friction during inflation is
\begin{align}\label{eq:sig_osc_def}
    \sigma_{\rm osc}\sim M_{\rm UV}\left(\frac{H_I^2}{m B_m M_{\rm UV}^2}\right)^{\frac{1}{m-2}}\,,
\end{align}
where $H_I$ denotes the inflationary Hubble scale. If the field is initially displaced beyond $\sigma_{\rm osc}$, the potential curvature overcomes Hubble friction, causing it to roll during inflation and reducing its amplitude to approximately $\sigma_{\rm osc}$ or below. Consequently, the field value relevant for successful axiogenesis is bounded by
\begin{align}\label{eqn:roll_bound}
    \sigma_0\lesssim\sigma_{\rm osc}\, .
\end{align}

Combining \cref{eq:sigma_roll_2,eqn:roll_bound} with \cref{eq:PQ_cons_data_2}, and taking $B_m\sim1$, yields
\begin{equation}\label{eq:fin_bound_fa}
    f_a \,\mathrel{\begin{Bmatrix}<\\\sim\\>\end{Bmatrix}} \,
    g_\star^{3/8} \, \epscirctd^{1/2} \, \frac{T_{\mathrm{ew}}^{3/2}}{\omegaEW^{1/2}} \, \frac{M_{\mathrm{UV}}^{\frac{m-4}{m-2}}H_I^{\frac{10-m}{4(m-2)}}}{\Mpl^{3/4}}\qquad
    \begin{cases}
        m<10\\
        m=10\\
        m>10
    \end{cases}\,.
\end{equation}
Note that there is no dependence on the IR potential (the $\sigma^\ell$ term in \cref{eq:potential_Vsogma_Vrenorm}). The three cases arise because $m=10$ is the critical exponent in radiation domination: at this value, the dependence of $\dot{\theta}(\Tew)$ on the initial condensate fraction changes sign. 

Together with the current supernova bounds on $f_a$ \cite{Lucente:2022vuo,Springmann:2024ret}, \cref{eq:fin_bound_fa} excludes the region $m<10$. Even for the largest allowed inflationary Hubble scale, $H_I=5\times10^{13}\GeV$~\cite{Planck:2018vyg,BICEP:2021xfz}, and $M_{\rm UV}\sim\Mpl$, the upper bound from \cref{eq:fin_bound_fa} (for $m < 10$) is
\begin{align}\label{eq:fa_supernova_exclusion}
    f_a\lesssim10^6~{\rm GeV}\,,
\end{align}
well below the allowed QCD-axion window.

The same conclusion follows from EFT consistency, see \cref{eq:sigroll_over_MUV}. For the EFT expansion in powers of $\sigma/M_{\rm UV}$ to be under control, the ratio $\sigma_0/\MUV$ must be smaller than unity. For $\dot\theta (\Tew) \sim \omegaEW$ (for successful axiogenesis), setting $\sigma_0/\MUV < 1$ for $m<10$ requires $M_{\rm UV}\gtrsim \Mpl$, where quantum-gravitational effects can no longer be neglected. Thus, within the radiation-dominated cosmology assumed here, and if a non-renormalizable term $\sigma^m$ from \cref{eq:potential_Vsogma_Vrenorm} dominates the radial field potential at $T\sim\Troll$, successful minimal axiogenesis requires $m>10$. Similarly, for $m=10$, \cref{eq:fin_bound_fa} implies that for $\MUV = \Mpl$, we need 
\begin{equation}
    f_a \sim 4 \, \times 10^6 \, \mathrm{GeV} \, ,
    \label{eq:bound_fa_m=10}
\end{equation}
which is again outside the phenomenologically viable range of $f_a$.

Even for $m>10$, the setup is quite constrained. While the lower bound in \cref{eq:fin_bound_fa} in this case is not constraining, other consistency conditions introduce stringent bounds. We have assumed the non-renormalizable term in \cref{eq:potential_Vsogma_Vrenorm} dominates the initial displacement $\sigma_0$. Furthermore, successful axiogenesis requires that when the radial mode relaxes at its minimum $f_a$, presumably at some $\Trelax > \Tew$, we should have $\dot\theta (\Trelax) > \omegaEW$. These two (relatively) loose conditions give rise to the following bounds on $\lambda_\ell$:
\begin{align}\label{eq:range_lambda}
    \left(\frac{\omega_{\rm ew}}{f_a}\right)^2<\lambda_\ell<\left(\frac{\sigma_0}{\MUV}\right)^{m-4}\left(\frac{\sigma_0}{f_a}\right)^{4-\ell}\,.
\end{align}
The lower bound ensures that the radial restoring force near $f_a$ is sufficiently large to accommodate the required angular velocity. If the field follows its circular equilibrium until relaxation, this also ensures $T_{\rm relax}\gtrsim T_{\rm ew}$. For $m>10$, however, the radial field can enter the slower scaling regime, and relaxation by the electroweak epoch imposes an additional dynamical requirement that we do not discuss further here. To derive the lower bound in \cref{eq:range_lambda}, we used the equilibrium condition \cref{eq:eqb_cond} to find an expression for the axion velocity at $\Trelax$: 
\begin{align}\label{eq:dotthetafa_eq}
     f_a^2\dot\theta^2(\Trelax)\sim \lambda_\ell f_a^4\,,
\end{align}
which yields:
\begin{align}\label{eq:thetafa_result}
    \dot{\theta}(\Trelax) \sim \sqrt{\lambda_\ell} f_a \,.
\end{align}
The upper bound enforces $\sigma(\Troll)>\sigma(\Tswitch)$, where $\Tswitch$ refers to the temperature when the renormalizable and the non-renormalizable contributions in \cref{eq:potential_Vsogma_Vrenorm} are comparable---this condition on $\sigma$ ensures the non-renormalizable potential dominates at $\Troll$ and our previous equations are valid.

For example, when $\ell=4$, \cref{eq:range_lambda} implies
\begin{align}\label{eq:lambda4_upper}
    \lambda_4\lesssim\left[\frac{\epscirctd^{2/3}}{26}\frac{M_{\rm UV}}{\Mpl}\left(\frac{5\times10^7\,{\rm GeV}}{f_a}\right)^{4/3}\right]^{\frac{6(m-4)}{m-10}}\,.
\end{align}
For $M_{\rm UV}\sim \Mpl$, $m=12$, $\epscirctd\sim1$, and $f_a\simeq5\times10^7 {\rm GeV}$, this gives
\begin{align}
    \lambda_4\lesssim10^{-34}\,.
\end{align}
Thus, even where the EFT treatment and initial condition constraints for a proper inflation and axiogenesis are formally consistent, minimal axiogenesis requires a very flat IR potential. 

Indeed, the interval in \cref{eq:range_lambda} is empty over a substantial part of the parameter space, unless the following inequality is satisfied: 
\begin{equation}
    \frac{\sigma_0}{\MUV}>\left[\left(\frac{\omega_{\rm ew}}{f_a}\right)^2\left(\frac{f_a}{\MUV}\right)^{4-\ell}\right]^{\frac{1}{m-\ell}} .
    \label{eq:consistency_bounds_lambda}
\end{equation}
Substituting \cref{eq:sigroll_over_MUV} and imposing the baryogenesis condition yields an upper bound on $f_a$. For $\MUV=\Mpl$, the resulting upper limits are displayed in \cref{tab:fa_max_lambda_consistency}. Combined with the lower bound on $f_a$ from supernova cooling, these results show that the region satisfying all ordering conditions is narrow.

\begin{table}[t]
\centering
\begin{tabular}{c|c|c}
    \hline\hline
    $m$
    & $f_a^{\rm max}\,[{\rm GeV}],\ \ell=2$
    & $f_a^{\rm max}\,[{\rm GeV}],\ \ell=4$
    \\
    \hline
    $12$ & $6.7\times10^{7}$ & $2.8\times10^{7}$ \\
    $14$ & $4.2\times10^{8}$ & $9.4\times10^{7}$ \\
    $16$ & $1.5\times10^{9}$  & $2.3\times10^{8}$ \\
    $20$ & $8.7\times10^{9}$ & $7.4\times10^{8}$ \\
    \hline\hline
\end{tabular}
\caption{ 
Maximum axion decay constant for which the interval in \cref{eq:range_lambda} is nonempty in a radiation-dominant background. We take $g_\star=106.75$, $\epscirctd=1$, $\MUV=\Mpl=2.435\times10^{18}\GeV$, $\Tew=130\GeV$, and $\omega_{\rm ew}=3.9\keV$. For $f_a>f_a^{\rm max}$, no value of $\lambda_\ell$ satisfies all the assumed temperature-ordering conditions. Comparing these results with the current lower bound, $f_a \gtrsim 5\times 10^7\,\GeV$ \cite{Lucente:2022vuo,Springmann:2024ret}, we conclude that the radial potential must be very steep at the initial condition.
}
\label{tab:fa_max_lambda_consistency}
\end{table}

Although the quadratic potential can accommodate an order-one coupling $\lambda_2$, its dominance at large field values requires strong suppression of the quartic interaction. Including a contribution $\lambda_4(\sigma^2-f_a^2)^2$, quadratic dominance below the crossover with the non-renormalizable potential requires
\begin{align}
    \lambda_4\lesssim\lambda_2\left(\frac{f_a}{\sigma_{\rm switch}}\right)^2\sim\left(\lambda_2\frac{f_a^2}{M_{\rm UV}^2}\right)^{\frac{m-4}{m-2}}.
\end{align}
Thus, even when the quadratic coupling is not small, the quartic interaction must be parametrically suppressed to realize the assumed hierarchy of radial operators.

To summarize, under the assumptions of a radiation-dominated background and the non-renormalizable term in \cref{eq:potential_Vsogma_Vrenorm} dominating at $T \sim \Troll$, minimal axiogenesis requires a high-power UV radial potential with $m>10$, together with a flat IR potential. Even then, the simultaneous inflationary, EFT, and temperature-ordering constraints restrict $f_a$ to a narrow range near its present astrophysical lower bound.

As discussed in the previous section, generating an $\mathcal{O}(1)$ value of $\epscirctd$ requires the dimension $n$ of the leading PQ-violating operator to be comparable to that of the dominant PQ-preserving operator, $m$; see \cref{eq:epsilon_circ,eq:ecirc_tilde_def}. If $n\gg m$, the generated PQ charge is parametrically suppressed unless the initial field value is tuned close to $\MUV$, where control of the EFT expansion becomes questionable. Our conclusion that the setup considered here requires $m>10$ therefore implies that efficient charge generation also requires a high-dimensional PQ-violating operator. This, in turn, requires a mechanism that forbids or strongly suppresses all lower-dimensional PQ-violating operators—precisely the axion quality problem.

\subsection*{Renormalizable Potential Dominates}

If instead the renormalizable contribution dominates the radial potential at the initial displacement $\sigma_0$, we can derive analogous consistency conditions for PQ-charge generation. In this case, the dominant PQ-preserving operator has dimension $\ell\leq4$. The argument of the previous section then suggests that efficient charge generation requires the leading PQ-violating operator to have dimension $n$ not much larger than $\ell$, and hence not much larger than four, or else $\epscirctd$ would be $(\sigma_0/\MUV)^{n-\ell}$-suppressed, though this can be compensated by a sufficiently flat potential. A low-dimensional source of explicit PQ breaking would shift the axion vacuum enough to spoil the solution to the strong CP problem. Nevertheless, we analyze this setup for completeness.

Repeating the steps that led to \cref{eq:PQ_cons_data_2} in this scenario, we find
\begin{equation}\label{eq:dottheta_IR}
    \omega_{\rm ew}\sim g_\star^{3/4}\frac{T_{\rm ew}^3}{f_a^2} \epscirctd\epsphi^{(10-\ell)/8} \lambda_\ell^{-1/4} \left( \frac{f_a}{M_{\rm pl}} \right)^{(\ell-4)/4}\,.
\end{equation}
We can solve this equation for $\epsphi$ and, as before, use that expression to find constraints on the model in this scenario. In this analysis we assume the SM comoving entropy is conserved, i.e.\ no additional entropy injection.

Even though the non-renormalizable term in the radial potential is negligible, demanding that the radial field excursion is below $\MUV$ (EFT consistency) imposes $\sigma_0 < \MUV$; considering the very loose limit of $\MUV \sim \Mpl$, this results in:\footnote{Same inequality arises from demanding the field excursion is small enough that we can neglect poorly-understood quantum gravity effects.}
\begin{equation}\label{eq:bound_VIR_Mpl}
    \sigma_0< \Mpl \Longrightarrow \lambda_\ell < \frac{g_*^3\epscirctd^4T_{\rm ew}^{12}}{\omegaEW^4f_a^2M_{\rm pl}^6}\left( \frac{\Mpl}{f_a}\right)^{10-\ell} \sim2.4 \times 10^{-73}\left(\frac{\Mpl}{f_a} \right)^{10-\ell}\left(\frac{5\times10^7\GeV}{f_a}\right)^2\,.
\end{equation}
Even this rather loose condition introduces a fairly stringent bound on $\lambda_\ell$, especially for $\ell=4$ (assuming $\epscirctd \sim 1$):
\begin{align}
    \ell=2: \quad \lambda_2&\lesssim7.4\times10^{12}\,\left(\frac{5\times10^7\,{\rm GeV}}{f_a}\right)^{10} \,,\\
    \ell=4: \quad\lambda_4&\lesssim3.1\times10^{-9}\,\left(\frac{5\times10^7\,{\rm GeV}}{f_a}\right)^8 \,.
\end{align}
These equations suggest, unless $f_a$ is very close to its lower bound of $\sim 5 \times 10^7$~GeV, the potential has to be very flat.

The renormalizable potential dominating at $\Troll$ also introduces another constraint:
\begin{equation}
    \lambda_\ell \,> \,\left(\frac{\sigma_0}{\MUV}\right)^{m-4}\left(\frac{\sigma_0}{f_a}\right)^{4-\ell} \, .
    \label{eq:IR_dominating_constraint}
\end{equation}
Using the expression for $\sigma_0$, \cref{eq:sigma_roll_2}, and \cref{eq:dottheta_IR} to solve for $\epsphi$, we can find constraints on $\lambda_\ell$. For $m< 10$, this introduces a lower bound on $\lambda_\ell$. Even for the extreme case of $\MUV \sim \Mpl$ this lower bound will be orders of magnitude larger than 1---in tension with perturbativity bounds. For $m > 10$, we instead find a tight upper bound on $\lambda_\ell$, see \cref{tab:lambda-upper-bounds-IR}. These results show that the case of the renormalizable term in the potential dominating at $T \sim \Troll$ is not generic and requires flat potentials and high powers of $m$. 

\begin{table}[t]
    \centering
    \begin{tabular}{c | c | c}
        \hline\hline
        $m$ & $\ell=2$ & $\ell=4$ \\
        \hline
        $12$ & $7.2\times10^{-22}$ & $9.7\times10^{-35}$ \\
        $14$ & $7.3\times10^{-5}$  & $5.5\times10^{-22}$ \\
        $16$ & $3.4\times10^{1}$   & $9.8\times10^{-18}$ \\
        \hline\hline
    \end{tabular}
    \caption{Upper bounds on $\lambda_\ell$ derived from demanding the renormalizable potential dominates at $\Troll$ and for $m>10$. We assumed $f_a=5\times10^7\,{\rm GeV}$, $M_{\rm UV}=M_{\rm pl}$, and $\epscirctd=1$---the bounds become more stringent with smaller $\MUV$ or larger $f_a$. }
    \label{tab:lambda-upper-bounds-IR}
\end{table}

Even if the quadratic contribution dominates from the onset of rolling, the quartic interaction must be strongly suppressed. Requiring $\lambda_4\sigma_0^4\lesssim\lambda_2f_a^2\sigma_0^2$ gives $\lambda_4\lesssim\lambda_2(f_a/\sigma_0)^2$, which is parametrically small for the near-Planck-scale initial saxion excursions considered here.

\subsection{Circularization via thermal dissipation}
\label{subsec:saxion_damping}

The preceding analysis assumed that radial oscillations are promptly damped after charge generation, allowing the system to reach its circular equilibrium before appreciable expansion takes place. Undamped elliptical oscillations may otherwise excite a parametric resonance that fragments the condensate, as has been demonstrated for specific scalar potentials in Ref.~\cite{Fedderke:2025sic}. For $m > 10$, the radial field instead approaches a non-oscillatory scaling solution, which may likewise damp the oscillations, though whether this avoids fragmentation after the transition to the subsequent IR potential is a model-dependent question that we leave to future work. Here, we now investigate a simple model of interactions with the thermal bath and study to what extent they can provide the necessary damping while preserving the assumed evolution.

In a KSVZ-like scenario, the saxion interacts with vectorlike quarks through
\begin{align}\label{eq:saxion_yukawa}
    \mathcal{L}_{\rm damp}\supset-y_\psi\Phi\,\bar\psi_L\psi_R+\mathrm{h.c.}\,,\qquad m_\psi^2(\sigma)=\frac{1}{2}y_\psi^2\sigma^2\,,
\end{align}
where we take $y_\psi>0$. For one colored Dirac fermion in thermal equilibrium and $m_\psi\ll T$, the leading thermal correction and the damping rate considered in Ref.~\cite{Co:2019jts} are
\begin{align}\label{eq:saxion_thermal_mass}
    \Delta V_T=\frac12m_{\rm th}^2\sigma^2\,,\qquad m_{\rm th}^2=\frac14y_\psi^2T^2\,,
\end{align}
and
\begin{align}\label{eq:saxion_damping_rate}
    \Gamma_\sigma=c_\Gamma y_\psi^2T\,,
\end{align}
where $c_\Gamma$ is an order one constant.

Assuming negligible transfer of PQ charge to the bath, these interactions add $\Gamma_\sigma\dot\sigma+m_{\rm th}^2\sigma$ to \cref{eq:EOMsig2}. In order that the vacuum potential dominate over the in-medium potential, one requires $m_{\rm th}^2\ll V_\sigma'/\sigma$. Using \cref{eq:saxion_thermal_mass} and the rolling condition in \cref{eq:rolling_condition} gives
\begin{align}\label{eq:saxion_prompt_damping_bound}
    \frac{\Gamma_\sigma}{H_{\rm roll}}\ll36c_\Gamma\frac{H_{\rm roll}}{\Troll}\,.
\end{align}
For $c_\Gamma\sim {\cal O}(1)$ and $\Troll\gg H_{\rm roll}$, this is incompatible with prompt dissipation, which requires at least $\Gamma_\sigma\gtrsim H_{\rm roll}$. Efficient damping via \cref{eq:saxion_yukawa} therefore modifies the restoring force responsible for the assumed vacuum-driven onset.

Let us instead suppose the thermal correction dominates the radial potential. The circular-equilibrium condition in \cref{eq:eqb_V_KE} then gives
\begin{align}\label{eq:thermal_circular_equilibrium}
    |\dot\theta|=m_{\rm th}\,,\qquad \sigma_{\rm eq}^2=\frac{|n_\theta|}{m_{\rm th}}\,.
\end{align}
Within the prompt-damping approximation, the radial field evolves from its initial displacement $\sigma_0$ to the circular equilibrium $\sigma_{\rm eq}(\Troll)$ before appreciable cosmological expansion. Using the definition of $\epscirc$ in \cref{eq:ecirc_general_def} (as opposed to $\epscirctd$ of the previous subsection), conservation of PQ charge during this process implies
\begin{align}
    \sigma_{\rm eq}(\Troll)=\sqrt{\frac{\epscirc}{2}}\sigma_0\,,
\end{align}
where we have assumed that the thermal quadratic potential dominates throughout circularization. With conserved PQ charge and bath entropy, and neglecting changes in the relativistic degrees of freedom, 
\begin{align}
    m_{\rm th}\propto T\propto\asf^{-1} \quad \implies \quad  \sigma_{\rm eq},|\dot\theta|\propto\asf^{-1}\,,
\end{align}
reproducing the quartic scaling of \cref{eqn:kmm_redshifting}. Matching the thermal restoring force to $4\lambda_{4,T}\sigma^3$ defines a constant effective quartic coupling,
\begin{align}\label{eq:thermal_effective_quartic}
    \lambda_{4,T}\equiv\frac{m_{\rm th}^2}{4\sigma^2} = \frac{m_{\rm th}^3}{4 |n_\theta|}=\frac{y_\psi^3\Tew^3}{32f_a^2\omegaEW}\,,
\end{align}
where in the second equality we used $\dot\theta = m_{\rm th}$ and in the last equality we used PQ charge conservation, conserved radiation bath entropy, and $\sigma(\Tew)=f_a$. Under these assumptions, \cref{eq:thermal_effective_quartic} is independent of the evolution between $T_{\rm roll}$ and $\Tew$ and does not require thermal domination to persist until $\Tew$.

If the thermal mass controls the onset of rolling, then $m_{\rm th}(\Troll)\sim 3H_{\rm roll}$. Consequently, the field-excursion requirement $\sigma_0<\MUV$ gives, assuming prompt circularization $\Gamma_\sigma(T_{\rm roll})\gg H_{\rm roll}$ and radiation domination at $T_{\rm roll}$,
\begin{align}\label{eq:thermal_quartic_excursion_bound}
    \lambda_{4,T}\lesssim\frac{1}{32}\epscirc^3\left(\frac{\pi^2g_\star(\Troll)}{10}\right)^3\left(\frac{\Tew^3}{f_a^2\omegaEW}\right)^4\left(\frac{\MUV}{\Mpl}\right)^6\,,
\end{align}
reproducing the $\lambda_4\propto f_a^{-8}$ scaling in \cref{eq:bound_VIR_Mpl}.

Further, the requirement that the fermions remain thermally populated while damping circularizes the orbit, $m_\psi\lesssim T$, implies
\begin{align}\label{eq:thermal_quartic_light_fermions}
    \lambda_{4,T}\gtrsim\frac{y_\psi^4}{16\epscirc}\quad\implies\quad y_\psi\lesssim\frac{\epscirc}{2}\frac{\Tew^3}{f_a^2\omegaEW}\,.
\end{align}
Assuming that the thermal effective potential determines the onset of oscillations, the damping rate satisfies
\begin{align}\label{eq:thermal_dominated_prompt_damping}
    \frac{\Gamma_\sigma}{H_{\rm roll}}\simeq6c_\Gamma y_\psi\lesssim3c_\Gamma\epscirc\frac{T_{\rm ew}^3}{f_a^2\omega_{\rm ew}}\,.
\end{align}
Since $\Tew^3/f_a^2\omegaEW\ll 1$ in the parameter range of interest, this channel cannot provide the assumed prompt circularization even when the thermal potential dominates.

This analysis shows, within the benchmark KSVZ axion considered here, that if the vacuum potential dominates, efficient damping necessarily modifies the onset of radial motion, whereas if the thermal potential dominates, the damping is too slow to produce prompt circularization. We emphasize that this is not a general no-go theorem, though a similar tension may arise in other axion constructions whenever the interactions responsible for dissipation also generate significant thermal corrections to the radial potential.

\section{Motivation for multi-axion axiogenesis}
\label{sec:recap}

The preceding sections show that minimal single-axion axiogenesis asks the same field to satisfy several competing requirements. The axion must rotate fast enough near the electroweak phase transition to generate the observed baryon asymmetry, while the radial dynamics must preserve a sufficiently large and spatially homogeneous rotation until that time. The same rotation must not produce an excessive late-time axion abundance, and the UV theory must generate appreciable PQ violation in the early Universe without spoiling the solution to the strong CP problem today. None of these requirements is by itself a no-go theorem, but satisfying them simultaneously leaves a narrow range of viable cosmological histories.

The most direct quantitative tension concerns the axion relic abundance. For the minimal electroweak anomaly coefficient, producing the observed baryon asymmetry requires an angular velocity of order $3.9\keV$ at the electroweak sphaleron freeze-out temperature, see \cref{eq:fin_B_val}. The SM plasma exerts negligible friction on the QCD axion, so the corresponding PQ charge survives after baryogenesis. When the QCD potential later becomes important, this charge produces an axion abundance that generically exceeds the observed dark-matter density. One possible way around this conclusion is therefore to introduce new dynamics that damps the QCD axion rotation after baryogenesis but before the QCD potential becomes important. One proposed example uses dark magnetic monopoles and the Witten effect~\cite{Witten:1979ey} to drain the rotational energy through transitions among dyonic states~\cite{Co:2025lcs}.

Preserving the required angular velocity until the electroweak epoch places strong restrictions on the radial dynamics as well. Conservation of PQ charge implies
\begin{align}
    n_\theta\propto \asf^{-3}\,,\qquad n_\theta=\sigma^2\dot\theta\,.
\end{align}
Once the radial field approaches its vacuum value, $\sigma\simeq f_a$, the decrease of $\sigma$ can no longer compensate for the dilution of $n_\theta$, and the angular velocity begins to redshift as $\dot\theta\propto\asf^{-3}$. An initial displacement of the radial field can delay this rapid dilution. We considered the possibility that the radial potential is dominated by high dimension operators, 
where centrifugal equilibrium gives
\begin{align}
    \sigma\propto\asf^{-p_\sigma}\,,\qquad\dot\theta\propto\asf^{2p_\sigma-3}\, ,
    \qquad p_\sigma > 0 \,,
\end{align}
thereby delaying the rapid-$\dot\theta$-redshifting epoch.
Maintaining the required velocity until $\Tew$ demands both a Planckian initial radial displacement and a non-generic radial field potential with high powers of $\sigma$, decay constants very close to the current $5 \times 10^7$~GeV lower bound, and very small couplings. The constraints on the potential are tightened not only thanks to the constraints on $\dot\theta(\Tew)$, but also the EFT consistency conditions and constraints arising from inflationary initial conditions.

On the other hand, from the EFT perspective of the axion-saxion model considered here, there is no symmetry reason for the quartic to be so small, and tuning away the zero-temperature quartic does not necessarily remove the problem at finite temperature. Interactions with the primordial plasma capable of damping radial oscillations can generate additional thermal contributions to the $\sigma$ potential. For a thermal mass of the form $\sim T^2\sigma^2$, the temperature dependence during radiation domination again gives $\sigma\propto\asf^{-1}$ and $\dot\theta\propto\asf^{-1}$, reproducing the same redshifting as the quartic theory ($\ell=4$ in the notation of the previous section). Nevertheless, interactions with the plasma seem necessary: undamped radial oscillations can parametrically amplify inhomogeneous modes, fragmenting the rotating field and generating regions with opposite axion velocity~\cite{Fedderke:2025sic}. Thus, the interactions introduced to preserve a homogeneous rotation may also regenerate the radial scaling that makes the required velocity difficult to maintain.

The UV origin of the initial torque presents a separate but related tension. Kinetic misalignment requires explicit PQ breaking that is strong enough at large field values to accelerate the axion, while the strong CP problem requires the same breaking to be negligible near the present-day vacuum. In EFT constructions, a sufficiently large discrete symmetry can forbid low-dimensional PQ-violating operators while permitting a higher-dimensional operator that becomes important when $\sigma\sim M_{\rm UV}$. A large ${\mathbb Z}_N$ symmetry can enforce this pattern and may itself descend from an extra-dimensional gauge symmetry~\cite{Babu:2002ic,Dias:2002gg}. For the surviving higher-dimensional operator to provide an appreciable torque, however, $\sigma$ must typically probe field values near the cutoff, where the expansion of the scalar potential into higher-dimensional operators is least parametrically controlled.
 
\textbf{Our goal in the remainder of this paper is therefore to determine to what extent axiogenesis can be realized within a technically natural theory.} To this end, we introduce two new ingredients: \emph{axion monodromy}, which can generate the large axion velocity required to source the baryon asymmetry, and a \emph{new confining sector}, whose sphaleron dynamics can dissipate it at later times. For a single axion, however, both mechanisms introduce explicit PQ breaking that is incompatible with a high-quality solution to the strong CP problem. \emph{Multiple axions} resolve this tension by allowing the cosmologically important sources of PQ breaking to act along directions distinct from the QCD direction, while preserving a light axion combination that relaxes the strong CP phase.

The role of multiple axions can be understood geometrically. In general, multiple axions can couple to the QCD anomaly, defining a linear combination that we call the \textit{QCD direction}. Other sources of axion potential, such as the monodromy potential or instantons from a new confining force, couple to linear combinations of axions as well. These directions may overlap with, but need not be parallel to, the QCD direction. Additional axion potentials spoil quality only if together they prevent the axions from adjusting to cancel the QCD angle; a single contribution is therefore dangerous only if it is parallel to the QCD direction. A sizable monodromy potential can therefore drive motion with a nonzero QCD velocity while preserving a separate light direction that solves the strong CP problem.

Axion monodromy offers two important benefits over kinetic misalignment as the origin of this velocity. First, the axion energy is stored in the monodromy potential until near the electroweak phase transition, when the axion begins rolling and converts this potential energy into kinetic energy. This avoids the long period of redshifting that must be overcome in kinetic misalignment. Second, the axion radial mode does not participate in the dynamics, avoiding the associated fragmentation instability~\cite{Fedderke:2025sic}, although instabilities of the axion itself must still be considered.

Multiple axions arise naturally in a variety of extensions of the SM and are especially well motivated in string compactifications~\cite{Witten:1984dg,Arvanitaki:2009fg}. Likewise, axion monodromy and extended field ranges have been realized in explicit string constructions~\cite{Silverstein:2008sg,McAllister:2008hb}. These observations make the axiverse an attractive UV setting for the mechanism developed here. At the same time, the cosmological evolution we require demands a large monodromy charge and we are not aware of an explicit compactification that realizes the required field range. Establishing such a realization while controlling backreaction over the trajectory remains an important question for future work. We therefore treat the multi-axion theory below as an effective description rather than assuming a specific axiverse UV completion.

The remainder of the paper is organized around the aforementioned three main ingredients:
\begin{enumerate}
    \item \textbf{Axion monodromy}: In \cref{sec:monodromy}, we show how the monodromy potential generates the axion velocity required at electroweak sphaleron freeze-out, determine the field range over which it must remain controlled, and discuss specific challenges for (string theory) UV completions.
    \item \textbf{Sphaleron damping}: In \cref{sec:damping_from_dark_sphalerons}, we study the subsequent evolution below the electroweak scale and show that friction from an additional confining sector can dissipate the axion kinetic energy, thereby preventing overclosure. We also explain how uncertainties in the SU($N$) sphaleron rate propagate into significant uncertainties in the viable parameter space.
    \item \textbf{Multi-axion dynamics}: In \cref{sec:three_axions}, we study the combined dynamics of the multi-axion system and describe the numerical techniques used to solve the equations of motion.
\end{enumerate}
Finally, in \cref{sec:phenomenology}, we examine the allowed parameter space and the resulting phenomenology.

\section{Angular momentum from axion monodromy}\label{sec:monodromy}

We now propose an alternative mechanism for generating axion rotation in the early Universe. Rather than relying on a large initial displacement of the radial mode, we consider an axion displaced over \textit{many windings along a branch of a monodromy potential}, while the radial mode remains stabilized near its vacuum value. The resulting nonperiodic potential provides a force that accelerates the axion as it rolls toward its minimum. Keeping the radial mode fixed also avoids the resonance and fragmentation associated with radial oscillations in kinetic misalignment.

A monodromy potential acting directly along the QCD axion direction would spoil the shift symmetry required to solve the strong CP problem. In a multi-axion field space, however, the monodromy can act along a different but overlapping direction, accelerating the QCD-coupled combination while preserving a light direction capable of relaxing the QCD angle. We defer the discussion of the multi-axion construction to the following sections and first study the monodromy-driven dynamics in isolation.

We show that a broad class of monodromy potentials can generate the angular velocity required for axiogenesis. Their essential advantage is that the axion can remain nearly at rest until close to the electroweak phase transition, with the energy required for its later motion stored in its potential. Monodromy therefore avoids the prolonged redshifting of rotation generated shortly after inflation, as occurs in kinetic misalignment.

The required flatness of the monodromy potential is also technically natural from an EFT perspective. Taking the monodromy scale to zero restores the axion shift symmetry, so quantum corrections to the axion potential are controlled by the same small symmetry-breaking parameter. This contrasts with the quartic coupling of the radial field in kinetic misalignment, whose vanishing does not restore an additional symmetry of the effective theory.

In \cref{subsec:phenomenological_monodromy} we review the development of axion monodromy and present the phenomenological parametrization we use in the subsequent sections. In \cref{sec:monodromy_for_natural_axiogenesis} we use this phenomenological monodromy parametrization to study the dynamics of the axion and the SM fermion asymmetries and derive the necessary conditions for successful baryogenesis. In \cref{subsec:monodromy-discussion} we discuss these conditions from the perspective of UV model building. While string theoretic constructions offer a promising avenue, no explicit construction yet achieves the requisite number of windings.

\subsection{Phenomenological monodromy potential}
\label{subsec:phenomenological_monodromy}

The central idea of monodromy is that a periodic parameter can act nontrivially on an integer-labeled family of states or branches of a multi-valued parameter. Shifting the parameter by one period returns the theory to itself, but only after relabeling $k \to k+1$, so that the branch energies obey $E_k(\theta + 2\pi) = E_{k+1}(\theta)$. The collection of branches respects the periodicity, but the physics of any fixed branch need not. If transitions between branches are slow, a state that remains on a single branch therefore explores a potential that is unbounded by the fundamental period, while the underlying theory retains only the discrete shift symmetry.

This structure can already be seen in the Schwinger model of electrodynamics in $1+1$ dimensions of spacetime~\cite{Schwinger:1962tp}.\footnote{We thank Bobby Acharya for bringing this historical note to our attention.} There, the monodromy structure emerges because the gauge field has no propagating degrees of freedom: the gauge field is instead characterized by its integer-labeled electric flux plus a continuous parameter $\theta$, with energy~\cite{Lawrence:2012ua}
\begin{align}
    E_n(\theta)\sim \frac{e^2}{2}\left(n+\frac{\theta}{2\pi}\right)^2\,.
\end{align}
The angle $\theta$ shifts the electric field continuously, so that the flux depends on the combination $n+\theta/2\pi$. Thus the collection of fluxes is exactly periodic as required, even though the flux along any fixed branch is not. Transitions between neighboring branches occur through charged-pair production, which screens the electric field by one unit $n \to n \pm 1$.

A closely related four-dimensional phenomenon is the Witten effect \cite{Witten:1979ey}. In a non-Abelian gauge theory containing magnetic monopoles, varying the $\theta$ angle causes the electric charge of a monopole to vary continuously, $\propto \theta/(2\pi)$. Varying $\theta$ through one complete rotation necessarily brings the full theory back to where it started, while a monopole that originally had zero electric charge now carries one unit of electric charge. This seeming contradiction is resolved by a tower of dyonic states, each separated by one unit of electric charge: a $2\pi$ shift of $\theta$ returns the full spectrum to itself by permuting the states within this tower.

The possible multibranched structure of the vacuum energy, more directly relevant to axion potentials, appeared shortly thereafter in studies of large-$N$ QCD~\cite{DiVecchia:1980yfw,Witten:1980sp}. There, it was realized that, in the chiral effective theory, for any given value of the $\theta$ parameter there exist ${\cal O}(N)$ local minima described by the potential
\begin{align}
    E(\phi,\theta)=f_\pi^2\left[-\mu^2\cos\phi+\frac{a}{2N}(\phi-\theta)^2\right]\,,
\end{align}
for a theory with a single light quark flavor and corresponding would-be Goldstone mode $\phi$. Here, $\mu$ is the quark-mass-induced mass parameter, while $a$ characterizes the size of the anomaly-induced contribution to the potential. One of these minima is the true vacuum, while the others are metastable. Tunneling between the various vacua ultimately allows the system to relax to the lowest-energy configuration through bubble nucleation, although in the $N\to\infty$ limit each configuration becomes exactly stable.

As $\theta$ is fundamentally $2\pi$ periodic, the apparent breaking of this periodicity is resolved by thinking of the energy landscape as multibranched, with each of the branches labeled by an integer $k$~\cite{Witten:1998uka}:
\begin{align}
    E_k(\theta)&=N^2h\left(\frac{\theta+2\pi k}{N}\right)\,,&E_{\rm vac}(\theta)&=
    \min_k E_k(\theta)\,,
\end{align}
where $h$ is a smooth function. A fundamental shift of the $\theta$ angle permutes the branches,
\begin{align}
    E_k(\theta+2\pi)=E_{k+1}(\theta)\,,
\end{align}
so that the collection of branches viewed as a whole is exactly $2\pi$ periodic, as demanded by the fundamental shift symmetry, even though the energy along a fixed branch is not. In finite-$N$ realizations, a given branch of the potential returns to its original configuration only after $\theta\to\theta+2\pi N$, while in the $N\to\infty$ limit this period diverges and each branch becomes effectively nonperiodic. Upon promoting $\theta$ to a dynamical axion, this multibranched vacuum energy becomes an axion monodromy potential. 

The modern string-theory development of monodromy began in the context of large-field inflation~\cite{Silverstein:2008sg}. A D4-brane wrapping a compact one-cycle lengthens and grows heavier as it is carried repeatedly around another compact direction of the internal space. Although the brane returns to the same position after each circuit, it does not return to the same configuration. Holding the remaining moduli fixed, this process may be repeated over many windings, yielding a nonperiodic potential for the D4-brane position modulus. Accounting for the field-dependent canonical normalization of this modulus, the large-field potential in the construction of Ref.~\cite{Silverstein:2008sg} scales as 
\begin{align}\label{eq:monod_V_23}
    V(\varphi)\propto \varphi^{2/3}\,.
\end{align}

Shortly thereafter, a string-theory realization directly involving closed-string axions was developed~\cite{McAllister:2008hb}. Closed-string axions arise by integrating higher-form gauge potentials over compact cycles in the extra dimensions. When a space-filling fivebrane wraps the corresponding internal cycle, winding the axion changes the flux carried by the brane, so that the brane's energy grows monotonically rather than periodically. The resulting potential takes the schematic form
\begin{align}\label{eq:PotMono1}
    V(\theta)\propto \sqrt{\ell^4+\theta^2}\,,
\end{align}
where $\ell$ is the volume of the wrapped cycle in string units. This potential is linear at large axion displacement but it becomes a constant plus a quadratic term near its minimum. Thus, the potential interpolates between an ordinary mass term near the origin and a flattened, sub-quadratic potential at large displacement. The closed-string axion may therefore evolve through many fundamental periods along an asymptotically linear branch.

At approximately the same time, a four-dimensional realization of axion monodromy was developed, again for the purpose of realizing large-field inflation~\cite{Kaloper:2008fb,Kaloper:2011jz}. In this construction, an axion couples to the four-form field strength of a three-form gauge field. Since a three-form has no local propagating degrees of freedom in four dimensions, the theory is divided into sectors labeled by a quantized flux $q_n$ that is locally fixed in the absence of charged membranes. Integrating out the four-form yields a collection of quadratic branches,
\begin{align}
    V_n(\theta)=\frac{1}{2}\left(q_n+\mu f_a\theta\right)^2\,,
\end{align}
whose minima are separated by one fundamental axion period. The discrete axion shift symmetry acts simultaneously on $\theta$ and the flux label $n$, thereby permuting the branches. The full theory is therefore periodic even though the potential on any fixed branch is not. The axion can consequently evolve through many periods along a single quadratic branch, while transitions between branches remain possible through membrane nucleation, which changes the four-form flux by one unit.

Other power-law potentials were later explored~\cite{Dong:2010in}, illustrating that the axion potential tends to flatten as heavy fields adjust to more energetically favorable configurations in the background of the axion winding. Since then, additional mechanisms giving rise to axion monodromy have been discovered, including F-term potentials generated by background fluxes or torsion~\cite{Marchesano:2014mla}. Type-IIB constructions were also developed in which an axion descending from the NS--NS two-form couples to the R--R field strength~\cite{McAllister:2014mpa}. In these theories the stabilized moduli shift in an axion-dependent way, and the resulting backreaction flattens the potential: at large field displacement one finds $V \sim \theta^{\,p}$ with effective powers $p = 2/3,\, 4/3,\, 2$, and $3$. 

\begin{figure}
    \centering
    \resizebox{\columnwidth}{!}{\includegraphics[width=0.5\linewidth]{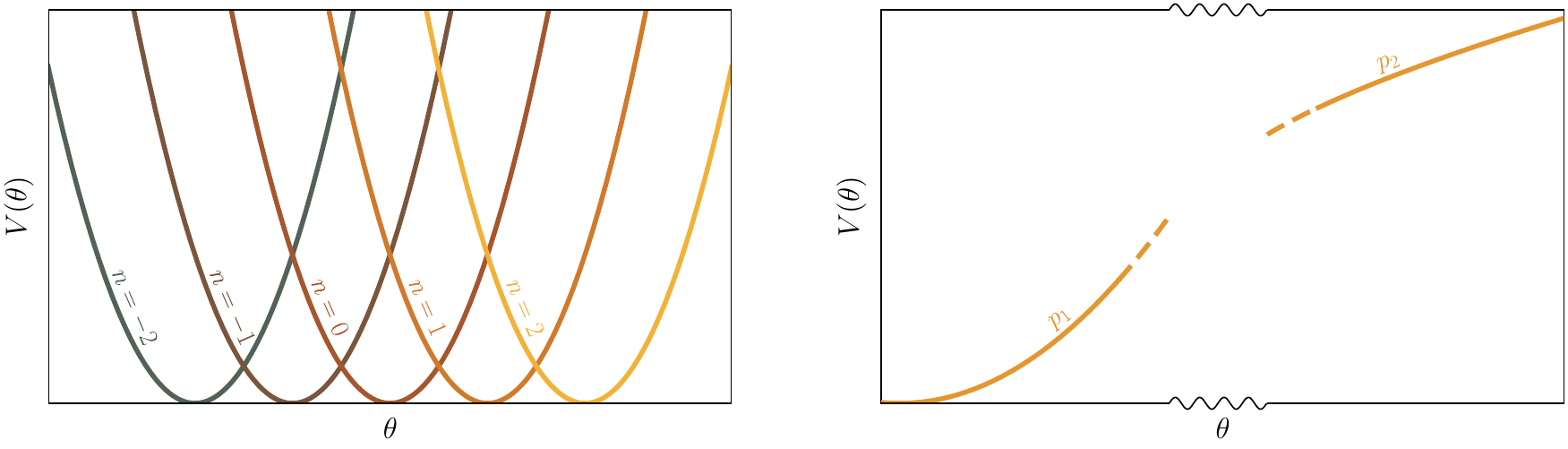}}
    \caption{\textbf{Left:} The multi-branch potential of the axion in a generic monodromy construction. Different curves indicate different branches of the potential labeled by the integer $n$. While the multi-branched potential shows a discrete shift symmetry, if the axion stays on a single branch, it feels a non-trivial non-periodic potential that appears as a PQ-violating contribution in the effective 4D theory. \textbf{Right:} In the most general monodromy constructions, as the result of backreaction from new fields in extra dimensions, a single branch of the potential can have different functional dependence at small (the left side of the plot) and large (the right side of the plot) field excursions, i.e. different values of $p$ in the schematic form of \cref{eq:mono-potential}. In particular, the potential can become flatter ($p_2 < p_1$) as we go to larger axion field excursions, enabling a wide variety of effective PQ-violating potentials from monodromy. }
    \label{fig:monodromy_potential}
\end{figure}

Given the varied microscopic origins of axion monodromy and the resulting range of potential shapes, we parameterize the monodromy potential phenomenologically as a function that is quadratic near the minimum of each branch but approaches a model-dependent power law at large field displacement. One convenient smooth parameterization is
\begin{equation}
    {\vmono}_{,n}(\theta) \equiv \frac{\mu^2 f_a^2}{  p} \left(\alpha^2 +(\theta - 2\pi n)^2\right)^{p/2}\, ,
    \label{eq:mono-potential}
\end{equation}
see the left panel of Fig. \ref{fig:monodromy_potential}. In this potential, $n$ labels the branch, $\alpha$ is the width of the quadratic region, $p$ is the asymptotic power, and $\mu$ sets the curvature at the minimum, $m^2 = \mu^2\alpha^{\,p-2}$. This potential satisfies $V_{\rm mono,n}(\theta + 2\pi) = V_{\rm mono,n-1}(\theta)$, so the branches are permuted by a shift of one period. The case $p=1$, $\alpha^2 = \ell^4$ reproduces the fivebrane potential in \cref{eq:PotMono1}. 
When transitions between branches can be neglected, we work on a fixed branch and suppress the label $n$. Periodic contributions to the axion potential, including those generated by instantons, may then be added separately.

\subsection{Monodromy for axiogenesis}
\label{sec:monodromy_for_natural_axiogenesis}

In this section, we show how the monodromy potential in \cref{eq:mono-potential} can facilitate axiogenesis. Like the radial potential in kinetic misalignment, the monodromy potential must be parametrically flat. Here, however, this flatness is technically natural and serves a qualitatively different purpose: it delays the onset of axion motion until near the electroweak phase transition.

In the setup reviewed in \cref{sec:IC_preliminaries}, the axion acquires an initial angular kick from a PQ-violating potential at a large radial field excursion. As the radial mode moves toward lower values, this PQ-violating potential becomes negligible, after which the angular momentum is conserved. Once the radial mode is stabilized at its minimum, conservation of comoving angular momentum implies $\dot{\theta} \propto \asf^{-3}$, see \cref{eq:EOMtheta,eq:conserved_ntheta}. Such rapid redshifting typically suppresses the rotational velocity well before the electroweak phase transition, making it challenging to obtain the required $\dot\theta$ at $\Tew$. In the original axiogenesis proposal~\cite{Co:2019wyp}, this issue is addressed by assuming that the radial mode evolves in a very flat potential with an initial condition far from its minimum. The slow relaxation of the radial field delays the regime in which angular momentum conservation enforces $\dot{\theta} \propto \asf^{-3}$, thereby allowing a sizable rotational velocity to persist until $\Tew$.

Monodromy provides a different way to delay this redshifting. As long as the axion rolls along a single branch, it experiences a PQ-violating potential over an extended field range. While Hubble friction dominates, the axion remains nearly stationary and its potential energy is not diluted. If the monodromy potential is sufficiently shallow, the axion begins rolling only at late times, near the electroweak phase transition. The extended field range provided by monodromy then allows the axion to acquire a sufficiently large rotational velocity to drive baryogenesis.

In contrast to the flatness of the radial potential, the required flatness of the monodromy potential is \textit{technically natural}. When the monodromy is taken to vanish ($\mu\to 0$), the axion shift symmetry is restored, so the small scale $\mu$ controlling the potential \cref{eq:mono-potential} is stable against radiative corrections. In \cref{subsec:monodromy-discussion}, we discuss possible UV origins of such shallow monodromies and the challenges associated with realizing the required field range, leaving a complete UV construction for future work.

Successful baryogenesis requires the monodromy dynamics to occur on the appropriate timescale. If $\mu$ is too small, the axion does not acquire the required rotational velocity by $\Tew$, while if $\mu$ is too large, the axion reaches the minimum of the monodromy potential and begins oscillating too early, leading to an exponential suppression of the baryon abundance~\cite{Jeong:2018jqe}. In what follows, we show that there is a range of $\mu$ for which the axion is rolling sufficiently rapidly near $\Tew$. We also show that the mechanism can remain viable if the axion begins oscillating shortly before the electroweak phase transition.

To demonstrate the viability of monodromy-driven axiogenesis, we first study the cosmological evolution of a single QCD axion in the potential of \cref{eq:mono-potential}. This simplified setup isolates the dynamics responsible for generating the baryon asymmetry, setting aside both the subsequent axion overabundance and the axion quality problem. We address the former through additional dissipation in \cref{sec:damping_from_dark_sphalerons}, and in \cref{sec:three_axions} we embed these dynamics in a multi-axion theory compatible with a high-quality solution to the strong CP problem.

We now quantify for what parameters the monodromy potential can generate the required axion velocity. Neglecting the QCD potential and friction, which are irrelevant around the electroweak epoch, the axion evolves according to:\footnote{We also assume that the axion remains on the $n=0$ branch of the monodromy potential. See Sec. \ref{subsec:monodromy-discussion} for a discussion of the validity of this assumption.}
\begin{align}\label{eq:single_axion_monodromy_EOM}
    0&=\ddot\theta + 3 H \dot\theta + \mu^2\theta(\alpha^2 + \theta^2)^{p/2 - 1}\,.
\end{align}
We study this evolution during radiation domination.

For a simple analytical estimate, consider the early evolution at large initial displacement, $\theta_0^2\gg\alpha^2$, where the potential gradient can be approximated by its value at $\theta_0$. During radiation domination, \cref{eq:single_axion_monodromy_EOM} then gives
\begin{equation}
    \theta(t) \simeq \theta_0 - \frac{1}{5}\mu^2 t^2 \theta_0^{p-1}\,,
    \label{eq:analytic_sol_theta}
\end{equation}
at small values of $t$. Requiring the axion not to reach the minimum before the electroweak phase transition gives parametrically
\begin{equation}
    \mu \lesssim \theta_0^{1-p/2} \Hew\,,
    \label{eq:analytic_sol_tHESFO}
\end{equation}
where $\Hew$ is the Hubble rate at $\Tew$. Saturating this bound gives the largest velocity attainable at $\Tew$ while the axion is still rolling on its first descent,
\begin{equation}
    \dot{\theta}(\Tew)\sim \theta_0 \Hew\,.
    \label{eq:analytic_sol_thetadot}
\end{equation}

Remarkably, the estimate in \cref{eq:analytic_sol_thetadot} is independent of the power $p$. Achieving the required velocity, $\dot{\theta}(T_{\rm EW})\sim 3.9~{\rm keV}$, therefore requires an initial winding
\begin{equation}
    N_i = \frac{\theta_0}{2\pi} \sim 2.6\times 10^7 \,,
    \label{eq:analytic_sol_winding}
\end{equation}
independently of the asymptotic form of the monodromy potential. We discuss the feasibility of such a large field excursion in the following subsection.

Combining \cref{eq:analytic_sol_tHESFO} and \eqref{eq:analytic_sol_winding} gives the approximate upper bound
\begin{align}\label{eqn:monodromy_size}
    \mu \lesssim (1.6\times 10^8)^{1-\frac{p}{2}}\times 2.4\times 10^{-5}\eV\,.
\end{align}
For the quadratic case, $p=2$, considered in Refs.~\cite{Kaloper:2008fb,Kaloper:2011jz}, this gives $\mu\lesssim 2.4\times 10^{-5}\,\eV$; such small values of $\mu$ are technically natural because the axion shift symmetry is restored in the limit $\mu\to0$. This estimate assumes that the axion remains on its first descent through $\Tew$. As we now show numerically, successful axiogenesis can persist for substantially larger values of $\mu$, even when the axion begins oscillating before the electroweak phase transition.

We numerically solve the coupled axion and Boltzmann equations for several monodromy potentials, fixing the large initial winding required by \cref{eq:analytic_sol_winding}. The numerical evolution allows us to follow the transition from the rolling regime into the oscillatory regime and determine when cancellations between successive oscillations suppress the baryon asymmetry.

\begin{figure}
    \centering
    \includegraphics[width=\linewidth]{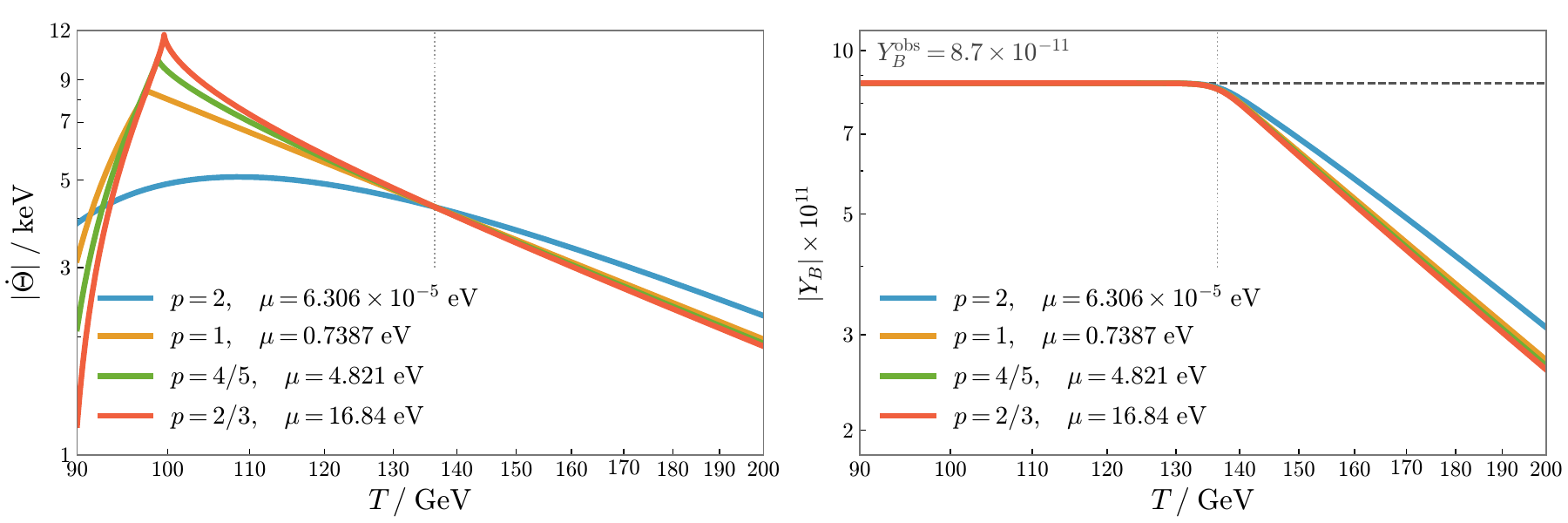}
    \caption{Numerical solution for the axion velocity $\dot{\theta}$ (\textbf{left}) and the corresponding baryon yield $|Y_B|$ (\textbf{right}) for the monodromy potentials \cref{eq:mono-potential} with $p = 2,1,4/5,2/3$ and $N_i$ given by \cref{eq:analytic_sol_winding}. For each $p$, the mass parameter $\mu$ is chosen such that we reproduce the observed baryon asymmetry; these values of $\mu$ match the analytic estimate from \cref{eqn:monodromy_size} to within a factor of $\sim 2$.  The nearly equal axion velocities at $T\simeq136\,{\rm GeV}$ illustrate that, for these slowly varying trajectories, the final baryon yield predominantly samples $\dot{\theta}$ at temperatures slightly above the conventional electroweak sphaleron freeze-out temperature $\Tew$. This offset reflects the rapid decrease of the sphaleron rate, which makes the time available for further baryon-number relaxation much shorter than a Hubble time. The axions reach their potential minima only at $T\sim 100\,{\rm GeV}$, after sphaleron freeze-out, making the final yield essentially insensitive to the shape of the potential near its minimum in these cases.}
    \label{fig:monodromy_evolution}
\end{figure}

In \cref{fig:monodromy_evolution}, we numerically solve the axion equation of motion \cref{eq:single_axion_monodromy_EOM} together with the full set of SM fermion asymmetry Boltzmann equations, see \cref{eq:final_boltz}, to find the baryon asymmetry in this toy setup.\footnote{We also include self-consistent backreaction of the plasma, though in practice it is negligible.} We present the evolution of $\dot{\theta}$ as a function of temperature for several monodromy potentials with different monomial functional forms. We have chosen the mass parameters so that the resulting baryon yield matches the observed value $|Y_B| \approx 9\times 10^{-11}$, and we find that it closely matches the analytically estimated values in \cref{eq:analytic_sol_tHESFO}. For all these benchmark points, the axion is still in the rolling phase and has not reached the bottom of its potential and oscillations have not started. \cref{fig:monodromy_evolution} demonstrates that a range of monodromy potentials (with different values of $p$) can sustain a sufficiently large axion rotational velocity at the electroweak scale. This is important, given that the detailed structure of the potential at large field excursions is model-dependent and may exhibit flattening of the monodromy potential \cite{Dong:2010in}.

\begin{figure}
    \centering
    \includegraphics[width=1\linewidth]{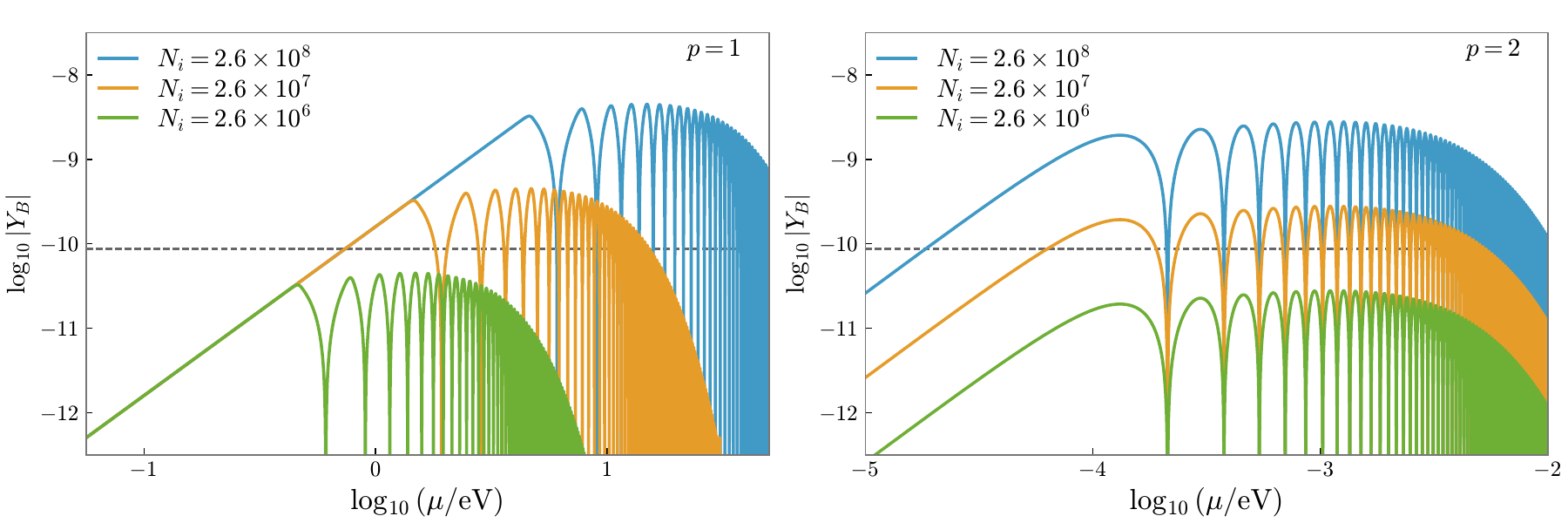}
    \caption{Baryon yield as a function of the monodromy potential mass scale $\mu$ for several choices of $N_i$, with $\alpha = 1$ held fixed. The \textbf{left} and \textbf{right} figures correspond to the linear and quadratic monodromy potentials, i.e.\ $p=1$ or $p=2$, respectively. In the quadratic case (\textbf{right}), the time at which oscillations commence is determined only by the mass scale, while the initial misalignment determines only the maximum velocity achieved when the axion reaches the bottom of its potential. On the other hand, in the linear potential (\textbf{left}) the maximum axion velocity achieved is determined by both the mass scale and the initial misalignment. For small enough values of $\mu$, the axion is still rolling down its potential at $\Tew$; as $\mu$ increases, eventually the axion reaches and starts oscillating at the bottom of its monodromy potential. When $\mu$ is large enough that the axion is deep in the oscillation regime at $\Tew$, the baryon asymmetry follows the exponentially decreasing $\Gamma_w$ and becomes suppressed.  
}
    \label{fig:monodromy_evolution_YB}
\end{figure}

Figure \ref{fig:monodromy_evolution_YB} illustrates the solution to the same system of equations, but now scanning over the mass parameter $\mu$ and for various choices of initial winding number and monodromy power law $p$. The orange curves correspond to the initial field excursion $\theta_0$ from \cref{eq:analytic_sol_winding}. At masses smaller than \cref{eq:analytic_sol_tHESFO}, the axion continues to roll monotonically as the electroweak sphaleron process freezes out. Increasing the mass parameter correspondingly increases the velocity of the axion field at the electroweak scale up to the point that the axion reaches the bottom of its monodromy potential before $\Tew$. 

For values of $\mu$ larger than in \cref{eq:analytic_sol_tHESFO} the axion reaches the minimum of its potential and begins oscillating before electroweak sphaleron freeze-out; it may still generate the observed baryon asymmetry, though for masses beyond a certain value the baryon yield is suppressed. On one hand, larger masses increase the axion velocity and thus cause a linear increase in the baryon yield $\dot\theta\propto \mu$. On the other hand, each half oscillation of the axion yields a baryon abundance of the opposite sign. As $\Gamma_w$ decreases, the baryon yield can not exactly track the instantaneous axion velocity, causing each half oscillation to partly cancel with the other half. The faster the axion oscillates, the more precise this cancellation~\cite{Jeong:2018jqe}, resulting in the exponential fall-off of $|Y_B|$ observed at large $\mu$. There is approximately an order of magnitude over $\mu$ where the linear enhancement and the exponential suppression approximately cancel out, leading to a plateau in the maximum achievable baryon abundance as a function of $\mu$. We illustrate the dynamics, and in particular the proportionality $Y_B\propto \Gamma_w$ in the oscillatory regime, in \cref{app:rapid_oscillation_YB}. 

\begin{figure}
    \centering
    \includegraphics[width=1\linewidth]{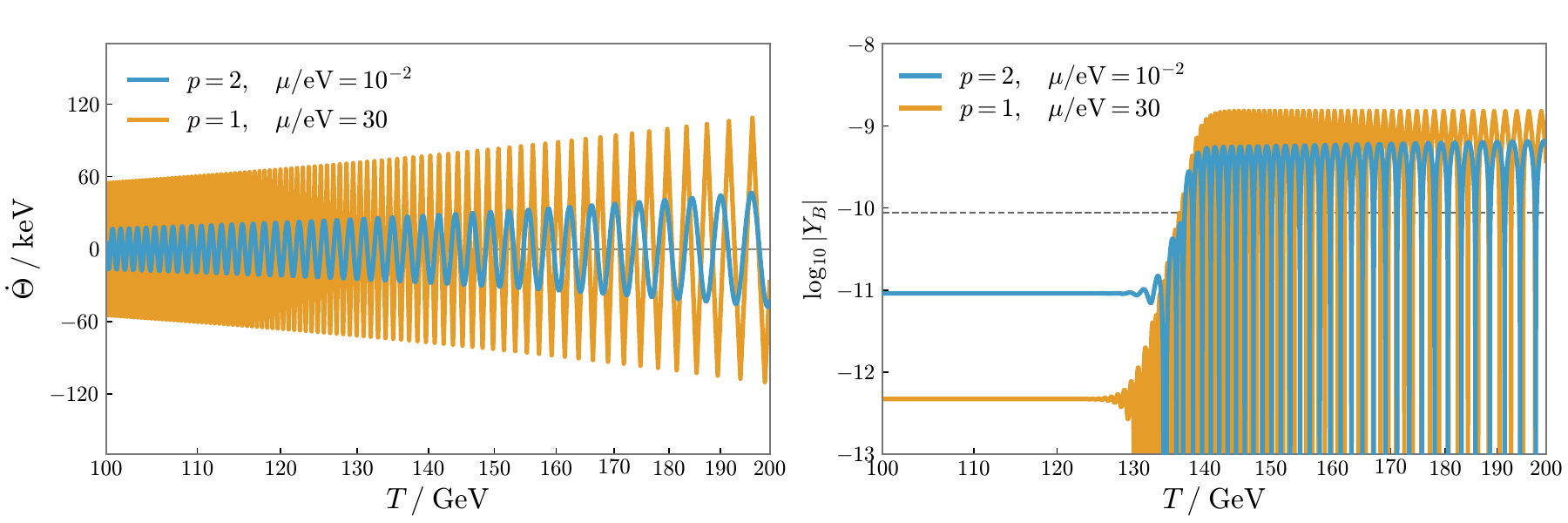}
    \caption{The axion velocity (\textbf{left}) and corresponding baryon yield (\textbf{right}) as a function of temperature in the quadratic $p = 2$ (\textcolor{blueberry}{blue}) and linear $p = 1$ (\textcolor{winterberry}{orange}) potentials, with $N_i$ given by the analytic estimate in \cref{eq:analytic_sol_winding}, where the monodromy mass scale $\mu$ is taken large enough that the axion oscillates rapidly at the time of electroweak phase transition. The baryon yield is suppressed when $\Gamma_w$ is small enough that it can no longer track the axion velocity, causing subsequent oscillations to partially cancel one another.
    }
    \label{fig:monodromy_evolution_osc}
\end{figure}

In \cref{fig:monodromy_evolution_osc} we present the time evolution of the axion for two benchmark sets of parameters deep in the oscillatory phase during electroweak sphaleron freeze-out, illustrating the partial cancellation that takes place in the rapidly oscillating regime.

\subsection*{Fragmentation}
The monodromy potential studied in this section generically introduces axion self-interactions that can cause certain momentum modes to grow, with the exception of $p = 2$. If these modes become nonperturbatively large, then they can spoil the homogeneity of the axion, possibly invalidating the axion-Boltzmann system of equations we used to infer the baryon abundance. We provide a detailed treatment of the growth of these modes in \cref{app:Monodromy_oscillation_fragmentation}, and summarize the main results here. 

We focus on the specific range of exponents $1 \leq p \leq 2$. There, we find that the dominant instability band corresponds to wavenumbers on the order of $k \sim \mu \theta_0^{(p-2)/2}\sqrt{2 - p}$. This instability can be approximately understood from the equation of state of the homogeneous oscillating scalar field $w_p = (p-2)/(p+2)$, corresponding to an imaginary speed of sound when $p < 2$~\cite{Turner:1983he,Cembranos:2015oya}. The resulting modes cause the axion to fragment after several oscillations
\begin{align}
    N_{\rm osc}\sim \frac{7.3}{2-p}\,,\qquad 1 <p < 2\,,
\end{align}
assuming the initial long wavelength perturbation amplitude is ${\cal O}(10^{-5})$.

Other power laws exhibit instability, though the resonances are qualitatively different. In particular, for $p < 1$, the instability is more efficient due to periods during the oscillation when the perturbations become tachyonic. $p > 2$ on the other hand exhibits weaker instability and the instability band does not extend to zero momentum. The full construction of \cref{sec:three_axions} specializes to quadratic monodromy $p=2$ (see Refs.~\cite{Kaloper:2008fb,Kaloper:2011jz}), for which the self-fragmentation studied here is absent, while fragmentation associated with the periodic confinement potentials occurs only after electroweak sphaleron freeze-out (see \cref{subsec:monodromy_crossing,app:Monodromy_Crossing_Fragmentation}); such fragmentation therefore does not affect the generated baryon asymmetry, although it may modify the subsequent axion energy partition and relic abundance. 

\subsection{Feasibility of large monodromy charges}
\label{subsec:monodromy-discussion}

The large winding number in \cref{eq:analytic_sol_winding} required to produce the observed baryon asymmetry raises several concerns regarding the consistency of monodromy for axiogenesis: (i) possible uncontrolled corrections to the potential at large field excursions, (ii) topological consistency of a large initial winding number $N_i$ itself, and (iii) rapid tunneling between different branches of the potential. The purpose of this section is to assess whether the ingredients required by the infrared theory can coexist in a controlled UV setting, rather than to present a complete UV construction. We examine a particular 10D string geometry that illustrates how some of these requirements may be addressed, and use it to assess whether the parametric hierarchies required by the infrared theory are compatible with known UV structures.  Though this geometry is suggestive, establishing an explicit model satisfying all of the required conditions remains an open problem.

To illustrate the essential features, whenever we want to refer to an explicit monodromy model, we will use the model of Refs.~\cite{Kaloper:2008fb,Kaloper:2011jz}, while noting that related realizations have been developed in, e.g., Refs.~\cite{Silverstein:2008sg,McAllister:2008hb,Dong:2010in,McAllister:2014mpa}. In the setup of Refs.~\cite{Kaloper:2008fb,Kaloper:2011jz}, a scalar field, $a \equiv f_a \theta$, is coupled to a 3-form gauge potential $A_3$ with 4-form field strength $F_4 = dA_3$,
\begin{equation}
    \mathcal{L} \supset - \frac{1}{48} F_{\alpha \beta \gamma \sigma}F^{\alpha \beta \gamma \sigma} + \frac{\mu}{24}\, a  \,\epsilon^{\alpha \beta \gamma \sigma}F_{\alpha \beta \gamma \sigma} \,.
    \label{eq:KS_Lag1}
\end{equation}
Since the 3-form contributes no propagating degrees of freedom in four dimensions, it can be integrated out, yielding the axion potential
\begin{equation}
    \vmono = \frac{1}{2} (q + \mu a)^2\,,
    \label{eq:KS_pot}
\end{equation}
where $q$ labels the quantized background electric field for the 3-form gauge field. In UV completions, the 3-form can descend from higher-dimensional forms after compactification, so $q$ is often related to a quantized internal flux. Within the 4D theory, membranes charged under the three-form can nucleate and change $q$ by integer multiples of the minimal membrane charge $e$, so that $q_n=n \, e$. Demanding the full potential to be invariant under this shift implies
\begin{equation}
    2\pi f_a\mu=k e,\qquad k\in\mathbb Z,    
\end{equation}
where $k$ is the integer axion–three-form coupling level. We assume the minimal choice $k=1$, for which $e=2\pi f_a\mu$ and hence $q_n=2\pi n f_a\mu$. Under $a\to a+2\pi f_a$, the flux label is shifted by one unit, thereby permuting the branches.

Within the 4D theory, membranes charged under the 3-form can nucleate and shift $q$ by discrete units of the membrane charge, leading to the branch structure characteristic of axion monodromy. This implies that $q$ is quantized in units of $2\pi f_a\mu$~\cite{Kaloper:2011jz}. This quantized flux is the origin of the multi-branch monodromy potential in this construction, see \cref{fig:monodromy_potential}. The theory remains invariant under a discrete axion shift symmetry only if the 4-form flux $q$ also shifts by a corresponding number of units under the same action. The potential in \cref{eq:KS_pot} has the functional form of the monodromy potential in \cref{eq:mono-potential} with $p=2$. 

In the benchmark scenarios considered in the previous section, achieving the required rotational velocity relies on a large initial axion field excursion $a_i \equiv f_a \,\theta_0$. The largeness of this field excursion can only be assessed within particular UV completions, where scales associated with new degrees of freedom are known. If $a_i$ exceeds any of these scales, it can change the character of the monodromy potential or invalidate the effective theory entirely. Below we review some of the dangerous effects that may arise in large field excursions.

\subsection*{Quantum gravity effects}
A theoretical concern limiting the axion field excursion is whether $a$ traverses a distance that exceeds $\Mpl$. In this case, quantum-gravity effects are conjectured to become relevant, for example through backreaction or the appearance of additional light degrees of freedom~\cite{Ooguri:2006in,Grimm:2018ohb}. On the other hand, the axion itself is a fundamentally periodic field, and its shift symmetry can protect the theory from generic polynomial corrections even over large field excursions~\cite{McAllister:2008hb,Valenzuela:2016yny}. A super-Planckian axion excursion is therefore not by itself sufficient for a breakdown of the effective theory.

Nevertheless, the axion decay constants required for successful axiogenesis dynamics tend to range between $5\times 10^7\GeV$ and $10^9\GeV$, as will be shown in the upcoming sections. As a result, even for the large monodromy charges needed in our setup, see \cref{eq:analytic_sol_winding}, the required field excursions are never super-Planckian. The more significant UV concern in our setup is instead the large number of windings and the corresponding accumulated monodromy charge, which we discuss below.

\subsection*{Heavy degrees of freedom} 

If the initial field displacement $a_i$ is large compared to the characteristic scale controlling the monodromy sector, typically set by background fluxes, heavy fields, or moduli that participate in generating the monodromy potential, backreaction from these degrees of freedom can become important. Such backreaction does not necessarily invalidate the axion effective theory: the heavy modes can play a non-trivial role even while remaining integrated out. If the heavy modes remain fixed at their vacuum expectation values, the axion potential remains stiff; for example, the quadratic potential of \cref{eq:KS_pot} stays quadratic to large field excursions. If instead the heavy modes adjust their expectation values in response to fluxes or charges generated by the axion at large winding number, they can relax the potential, which from the perspective of the single axion theory leads to an effective flattening at large field values \cite{Dong:2010in}; see the right panel in \cref{fig:monodromy_potential}.

As illustrated in Ref.~\cite{Dong:2010in}, the precise large-field behavior is model-dependent. We have therefore analyzed several monomial parameterizations of the monodromy potential and shown that successful baryogenesis can be achieved across a range of functional forms, see \cref{fig:monodromy_evolution,fig:monodromy_evolution_YB}, suggesting that the mechanism is not highly sensitive to the detailed large-field structure of a particular construction. A precise delineation of the viable parameter space within a given UV-complete model, however, requires an explicit treatment of the backreaction and its interplay with moduli stabilization. 
We leave such an analysis to future work.

\subsection*{Topological constraints} 

Aside from the absolute initial displacement, another physically relevant scale is the number of times the field winds relative to its fundamental period $2\pi f_a$, i.e.\ the monodromy charge $N_i \equiv a_i/(2\pi f_a)$. A large field excursion in absolute units need not correspond to a large winding number if $f_a$ is large, while many windings may occur even for sub-Planckian excursions when $f_a \ll M_{\rm pl}$. Although the full theory is periodic when all branches are viewed together, distinct winding numbers along a fixed branch correspond to physically distinct configurations.

In explicit string realizations of axion monodromy, the effective winding number is often associated with accumulated D-brane or flux charge, which is subject to global consistency conditions. 
In such constructions, one may parametrically write
\begin{equation}
    N_i \lesssim N_{\rm D}\,,
    \label{eq:tadpole}
\end{equation}
where $N_{\rm D}$ denotes the available D-brane charge budget. In F-theory compactifications, this charge budget is bounded by the Euler characteristic $\chi$ of the associated Calabi--Yau fourfold,
\begin{equation}
    N_{\rm D} \lesssim \chi/24\,,
    \label{eq:euler}
\end{equation}
see, e.g., Refs.~\cite{Vafa:1995fj,Sethi:1996es}. Together, \cref{eq:tadpole}-\eqref{eq:euler} show that the maximal achievable winding number can be constrained by the topology of the compactification.

Values as large as $\chi \sim 10^6$ have been realized in explicit string constructions~\cite{Klemm:1996ts}, corresponding parametrically to winding numbers $N_i \lesssim 10^5$. No known theorem forbids larger Euler characteristics, but explicit constructions with significantly larger values are not currently known; see Ref.~\cite{Klemm:1996ts} for further discussion. The winding numbers required for successful axiogenesis therefore lie beyond those demonstrated in known constructions. Moreover, increasing the available charge budget is not by itself sufficient: smaller values of $N_i/N_{\rm D}$ generally improve control over the backreaction of the monodromy charge on the background geometry~\cite{McAllister:2016vzi}, while sufficiently large winding can shift the stabilized moduli and potentially destabilize the compactification. Whether the required $N_i$ can be realized consistently therefore depends both on the available charge budget and on the details of moduli stabilization.\footnote{Ref.~\cite{McAllister:2016vzi} argued that achieving large $N_i$ presents a significant obstacle for UV completions of the relaxion model~\cite{Graham:2015cka}. Although some of the challenges in UV completing the relaxion setup and our framework are similar, the central difficulty for the relaxion mechanism is that large initial windings suppress the periodic barrier responsible for stopping the relaxion, giving rise to a runaway relaxion. This erosion of the barrier potential is not relevant for our axiogenesis mechanism.}

In warped Type-IIB throats supported by D3-brane charge, and at fixed throat geometry, the charge is related parametrically to the throat radius by~\cite{McAllister:2016vzi}
\begin{equation}
    N_{\rm D} \propto \left(\frac{R}{l_{\rm str}}\right)^4\,,
    \label{eq:winding_ratio}
\end{equation}
where $l_{\rm str}$ denotes the fundamental string length.\footnote{The examples in Ref.~\cite{McAllister:2016vzi} consider D$n$-brane charges with odd $n$. By assuming that the 4-form descends from a higher-dimensional flux whose induced charge arises from such branes, we are effectively working within a Type IIB/F-theory framework. In other string compactifications, the relevant tadpole constraints involve different brane charges~\cite{Sethi:1996es,Giddings:2001yu,DeWolfe:2005uu,Collinucci:2008pf}.} The fourth-power scaling in \cref{eq:winding_ratio} illustrates the possibility of parametrically large charge budget from a modest hierarchy of scales, though whether the values required in \cref{eq:analytic_sol_winding} can be accommodated must be demonstrated in an explicit construction.

\subsection*{Tunneling} 

A final consistency requirement is that the axion remain on a single branch of the monodromy potential throughout its evolution. At large field displacements the potential difference between neighboring branches can grow, thereby enhancing the tunneling rate between them. In the higher-dimensional theory, branch changes are mediated by membrane nucleation. In four dimensions these appear as bubble nucleation events that shift the quantized four-form flux by one unit \cite{Kaloper:2008fb,Kaloper:2011jz}. The tunneling rate per unit volume is parametrically
\begin{equation}
    \frac{\Gamma}{V}\sim A\,e^{-S_E},
    \label{eq:tunneling1_new}
\end{equation}
with the Euclidean bounce action
\begin{equation}
    S_E \sim \frac{27\pi^2}{2}\,\frac{T^4}{(\Delta V)^3},
    \label{eq:monodromy_tunneling_new}
\end{equation}
in the flat-space, thin wall approximation, where $T$ is the 4D membrane tension and $\Delta V$ is the vacuum-energy difference between the two branches.\footnote{Transitions between neighboring branches are generically favored over jumps across many branches. Although $\Delta V$ increases roughly linearly with the number of branches crossed, the corresponding wrapped-brane tension typically grows as well, so the bounce action tends to increase rather than decrease for large jumps.} Hence, tunneling is exponentially suppressed whenever $T^{4/3}$ is parametrically larger than the scale that sets the monodromy potential, i.e. $\Delta V\sim \mu^2 f_a^2 N_i$, where the scaling with $N_i$ assumes a quadratic monodromy potential. The principal concern is therefore that the large monodromy charge required for axiogenesis enhances the energy gained by tunneling to a neighboring branch and may render the single-branch evolution unstable.

Large membrane tensions suppress the tunneling rate and may arise in a number of ways. For instance, $T$ grows with the internal volume of the cycle wrapped by the membrane, though taking the entire compactification volume to be very large also lowers the KK scale and can invalidate the four-dimensional effective description. Another possibility is warping: the monodromy potential and membrane tension can depend differently on the warp factor, allowing the potential to become parametrically small compared to the scale controlling the domain-wall tension.

Warping to this effect can be achieved in the string theory construction of Refs.~\cite{Hebecker:2015tzo,Hebecker:2018yxs}. In the double Klebanov--Strassler \cite{Klebanov:2000hb} throat geometry of Ref.~\cite{Hebecker:2015tzo}, the axion arises from the integral of $C_2$ over an $S^2$ that is shared between two throats and shrinks only in the IR tip of the throats. It is shown in Ref.~\cite{Hebecker:2015tzo} that the axion mass is generated in the IR and is thus warped down compared to local throat scales. In particular, Ref.~\cite{Hebecker:2018yxs} argues that the axion potential is warped down by 
\begin{equation}
    \frac{\Delta V}{N_i} \propto \left(r_{\mathrm{IR}}/r_{\mathrm{UV}}\right)^6 \equiv \omega_\mathrm{IR}^6\,,
    \label{eq:pot_warped}
\end{equation}
with the warping factor $\omega_\mathrm{IR} \ll 1$. At the same time, in this geometry, the domain wall is a D5 brane wrapped on a three-sphere localized at the IR end of the geometry, whose tension is expected parametrically to scale as:\footnote{We thank Jakob Moritz for clarifications on the results of Ref.~\cite{Hebecker:2018yxs}. }
\begin{align}
    T \propto \omega_\mathrm{IR}^3\,.
    \label{eq:tension_warped}
\end{align}
While an explicit calculation of the tension in this setup does not exist in the literature, if this scaling holds, it is enough for our purposes. Combined together, \cref{eq:tunneling1_new}--\eqref{eq:tension_warped} suggest
\begin{align}
    \frac{\Gamma}{V} \propto e^{-\frac{\omega_\mathrm{IR}^{-6}}{N_i^3}}\,,
\label{eq:action_warped}
\end{align}
illustrating that strong warping parametrically suppresses the tunneling rate in this double Klebanov-Strassler throat geometry. 

The double-throat construction discussed above should be viewed only as a plausible template for obtaining the hierarchy $\sqrt{N_i}\mu f_a \ll T^{2/3}$ in an explicit compactification to suppress tunneling between different monodromy branches. While enabling this hierarchy, this construction could also suffer from a strong backreaction in the large field excursion limit, hence limiting the attainable monodromy charge before the potential completely flattens \cite{Hebecker:2018yxs}. Establishing whether this geometry can provide the required $T^{2/3}/f_a\mu\sqrt{N_i}$ hierarchy, while evading detrimental backreaction effects, would require exploring the flattening of the potential at large field excursions, identifying the precise branch-changing wrapped brane, computing its tension in the full geometry, and analyzing the corresponding bubble nucleation rate. We leave such a study to future string model building works.

It is worth emphasizing that, if one assumes the electric weak gravity conjecture for the four-dimensional three-form appearing in axion monodromy, then there should exist a membrane whose tension is bounded by its charge in Planck units,
\begin{equation}
    T \;\lesssim\; e_3\, \Mpl\, ,
\end{equation}
up to order-one factors \cite{Arkani-Hamed:2006emk,Hebecker:2015zss,Harlow:2022ich}. In the Kaloper--Sorbo (KS) description of monodromy \cite{Kaloper:2008fb,Kaloper:2011jz}, the three-form charge quantum is set by the product of the monodromy parameter and the axion decay constant, $e_3 \sim \mu f_a$, so this gives the parametric estimate
\begin{equation}
    T \;\lesssim\; \mu\, f_a\, \Mpl\, .
    \label{eq_wgc}
\end{equation}
For codimension-one objects such as membranes, this version of the weak gravity conjecture is more subtle than the more familiar particle case, and \cref{eq_wgc} should therefore be interpreted as an order-of-magnitude guide rather than as a sharp theorem. Nevertheless, for the range of $\mu$ and $f_a$ relevant to our setup, the bound is sufficiently weak that it does not obviously obstruct exponentially-slow branch-changing transitions.

Taken together, these considerations suggest that the monodromy origin of $\dot{\theta} (\Tew)$ is not obviously excluded by standard UV concerns. However, a fully controlled realization requires an explicit compactification with a sufficiently large charge budget, stabilized moduli at large monodromy charge, and a full computation of the relevant branch-changing membrane tension. A fully explicit realization in a concrete higher-dimensional compactification is left for future work.

\medskip

To summarize the main results of this section, we have found no obvious UV obstruction to the axion evolving along a monodromy potential  and acquiring the rotational velocity required for baryogenesis. Realizing the required large monodromy charge in a controlled UV completion nevertheless imposes nontrivial constraints from backreaction, global charge budget of the compactification, and tunneling between potential branches. We have identified a candidate string geometry that illustrates how these constraints may be addressed, though whether all of the required conditions may be realized simultaneously in a fully explicit example remains an open question.

\section{Damping from dark sphalerons}
\label{sec:damping_from_dark_sphalerons}

In minimal axiogenesis, the axion velocity that generates the observed baryon asymmetry also overproduces axion dark matter. In this section, we illustrate how this tension can be relieved by dissipating the axion kinetic energy into a plasma through sphaleron transitions in a new confining sector. In contrast to the previous section, where the effect of the monodromy potential at the electroweak phase transition could be understood without involving the multi-axion field space, the rate of dissipation and the final relic abundance are intimately connected to the physics of axion mixing.

To quantify dissipation, we consider pairs of interacting axions. Generically, the interaction eigenstates, which determine the couplings of the axions to QCD and the dark gauge group, do not align with the mass eigenstates. Consequently, the relationship between the sphaleron friction coefficient and the actual dissipation felt by the oscillating states can be quite different from the naive single-axion expectation. We calculate the scaling of the dissipation in all relevant dynamical regimes in \cref{sec:WKB}. In our companion paper~\cite{Asadi:2025cvm}, we considered a simplified scenario in which we remained agnostic to the origin of the axion velocity in the early Universe, imposed $\dot\theta_{\rm QCD}(\Tew) = \omegaEW$, and showed how dissipation from a dark gauge group could avoid the overproduction inherent to single-axion axiogenesis. These model-agnostic initial conditions provide a useful benchmark, which we will apply here. 

The same two-axion framework also lets us address how the energy initially stored in the monodromy direction is partitioned among the coupled axion states as a confining potential develops. Since these states subsequently redshift and dissipate differently, this partition can affect the final relic abundance even though the redistribution itself is not dissipative. In \cref{subsec:monodromy_crossing}, we identify a regime in which the outcome depends sensitively on the phase of the monodromy oscillations, and develop a statistical description of the resulting energy distribution.

\subsection{Two-axion dissipation}\label{sec:WKB}

We now isolate the damping dynamics from the mechanism that generates the initial axion velocity. In \cref{sec:monodromy}, we showed how a monodromy potential can naturally produce a large axion velocity before the electroweak phase transition. Here we ask the complementary question: given such a large velocity, can dissipation reduce the axion kinetic energy enough to avoid overclosing the Universe?

To answer this question, we study a simplified two-axion effective theory with a new confining dark force. 
\begin{align}
    \label{eq:Ls_two_axion}
    {\cal L}_\theta=\frac12(\partial \phi_1)^2+\frac12(\partial \phi_2)^2+\frac{\alpha_d}{4\pi}\frac{\phi_2}{f_2} {\rm Tr}\tilde G_d G_d +  \frac{\alpha_c}{4\pi}\left(\frac{\phi_1}{f_1}+C\frac{\phi_2}{f_2}\right){\rm Tr}\tilde G_c G_c\,.
\end{align}
The role of the dark sector is twofold. First, sphaleron transitions in the dark confining gauge group provide a finite-temperature friction term that damps the axion motion. Second, the same dark gauge group confines at late times and generates an additional instanton potential. Since this potential is an additional source of PQ breaking, consistency with the strong CP solution requires a second axion direction: the QCD and dark gauge groups must couple to non-aligned linear combinations, so that the dark-sector phase can be relaxed without displacing the QCD axion from the strong CP minimum.

After canonically normalizing the axions and performing a Gram--Schmidt rotation on the two instanton charge vectors~\cite{Demirtas:2021gsq,Gendler:2023kjt,Baryakhtar:2026oun}, the general two-axion system can be written in a basis where the dark gauge group couples to a single axion direction, while QCD couples to both axions. The relevant low-energy parameters are the two decay constants $f_1$ and $f_2$, together with a mixing coefficient $C$ that measures the projection of the dark axion direction onto the QCD-coupled direction.

Let us now turn to the dynamics of the homogeneous part of the axion fields. We will assume that both the SM and the dark gauge group are at a common temperature $T$. To distinguish the homogeneous part of the axion field from the full spatially varying field, we define $\Theta_i(t)\equiv\langle\theta_i(t,\bm x)\rangle$, which then satisfy the following equations of motion
\begin{subequations}\label{eqn:two_axion_EOM}
\begin{align}
  0&=\ddot\Theta_1 + 3 H\dot\Theta_1 + \frac{\Upsilon_c}{f_1^2}(\dot\Theta_1 + C\dot\Theta_2) + \frac{\chi_c}{f_1^2}\sin(\Theta_1 + C\Theta_2)\,,
  \label{eqn:two_axion_EOM_1}\\
  0&=\ddot\Theta_2 + 3 H\dot\Theta_2 + C\frac{\Upsilon_c}{f_2^2}(\dot\Theta_1 + C\dot\Theta_2) + \frac{\Upsilon_d}{f_2^2}\dot\Theta_2 + C\frac{\chi_c}{f_2^2}\sin(\Theta_1 + C\Theta_2) + \frac{\chi_d}{f_2^2}\sin\Theta_2\,.
  \label{eqn:two_axion_EOM_2}
\end{align}
\end{subequations}
We use subscripts $c$ and $d$ to refer to a term in the axion effective theory arising from QCD or the new confining force respectively. $\chi$ denotes topological susceptibility while $\Upsilon$ denotes friction. In this basis, $\Theta_2$ is the axion direction directly damped by dark sphalerons, while $\Theta_1+C\Theta_2$ is the QCD-coupled direction whose velocity is fixed by baryogenesis. The boundary condition
\begin{align}
    \dot\Theta_1 + C\dot\Theta_2 = \omegaEW \qquad \text{at} \qquad T = T_{\mathrm{ew}}\,,
\end{align}
therefore fixes only the velocity of the QCD-coupled interaction eigenstate, leaving the relative energy stored in $\Theta_1$ and $\Theta_2$ unconstrained at electroweak sphaleron freeze-out. This is the basic reason that axion mixing is central to the damping problem.

As the Universe expands and cools, the two-axion system of \cref{eqn:two_axion_EOM} passes through three dynamical regimes: (i) both axions are effectively massless and rolling, (ii) one axion oscillates while the other remains massless and rolling, and (iii) both axions oscillate about the minimum of their potential. As in the single-axion case of \cref{subsec:axion_relic}, we use the WKB approximation to describe the evolution of the axion modes in time. The first regime does not arise for the monodromy-driven initial conditions studied in \cref{sec:monodromy} and later in \cref{sec:three_axions}, since in that case the monodromy direction always has a non-negligible mass. Nevertheless, we include it to connect with the dynamics of the companion paper Ref.~\cite{Asadi:2025cvm}.

The central object of interest is the characteristic polynomial associated with \cref{eqn:two_axion_EOM}. For the purposes of this WKB analysis, we neglect the QCD sphaleron friction, which is parametrically small compared to the dark sphaleron friction and Hubble in the regimes of interest. The form of the characteristic polynomial depends on which instanton potentials are dynamically important. 

\subsection*{(iii) Both axions oscillating}
At low temperatures and small axion velocities, both axions perform nearly harmonic oscillations about the minimum of their potential, and it is self-consistent to neglect anharmonic corrections. In this regime, after rescaling out the trivial Hubble redshifting, setting $\Upsilon_c=0$, and assuming solutions of the type $\Theta_i\propto e^{\lambda t}$, the system of \cref{eqn:two_axion_EOM} is governed by the roots of the fourth-order characteristic polynomial
\begin{align}\label{eqn:four}
    0 &= x^4 + \alpha x^3 + (1 + \delta^2)x^2 + \alpha\beta^2 x + \beta^2\delta^2 \,,
\end{align}
with
\begin{equation}\label{eqn:parameters}
\begin{aligned}
    x &= \frac{\lambda}{m_A}\,,& m_A^2 &\equiv \chi_c\left(\frac{1}{f_1^2} + \frac{C^2}{f_2^2}\right), \\
    \alpha &= \frac{\Upsilon_d}{f_2^2 m_A}\,,& \beta^2 &= \frac{f_2^2}{C^2f_1^2 + f_2^2}\,,& \delta^2 &= \frac{1-\beta^2}{C^2}\frac{\chi_d}{\chi_c}\,.
\end{aligned}
\end{equation}
Here $\alpha,\delta>0$, $0<\beta<1$, and we define $\bar\beta^2\equiv1-\beta^2$.

The roots of \cref{eqn:four} characterize the instantaneous frequencies and damping rates of the eigenstates. When the corresponding modes are underdamped, the roots come in complex conjugate pairs, which we denote by $\lambda_i^\pm$ with $\lambda_i^+=(\lambda_i^-)^*$. We write
\begin{align}
    \lambda_i^\pm = \gamma_i \pm {\rm i}\omega_i\,,
\end{align}
where $\gamma_i$ and $\omega_i$ are the instantaneous damping rate and oscillation frequency of the $i$th mode, respectively. Up to $\mathcal{O}(1)$ corrections from damping and from the slow rotation of the eigenmodes, the corresponding instantaneous normal-mode amplitudes scale as
\begin{align}\label{eqn:massive_adiabatic}
    Q_i(t)\propto \asf^{-3/2}\,\omega_i^{-1/2} \exp\left\{\int^t{\rm d}t'\,\gamma_i(t')\right\}\,,
\end{align}
as long as the eigenfrequencies remain nondegenerate and the adiabatic condition is satisfied. Schematically, this requires
\begin{align}
    \left|\frac{\dot\omega_i}{\omega_i^2}\right|\ll 1\,,\qquad|\dot\varphi|\ll |\omega_1-\omega_2|\,,
\end{align}
where $\varphi$ denotes the instantaneous rotation angle between the interaction basis and the eigenbasis. 

The roots of \cref{eqn:four} have simple expressions in certain limiting cases. First consider the large-$\delta$ limit, in which the dark instanton potential is parametrically larger than the QCD potential. The oscillation eigenstates are then closely aligned with $\Theta_1$ and $\Theta_2$. Expanding the roots of \cref{eqn:four}, we find
\begin{align}\label{eqn:roots_four_large_delta}
\notag
    \Theta_1:\qquad\frac{\lambda_1^\pm}{m_A}&= -\frac{\alpha \beta^2\bar\beta^2}{2\delta^4}\pm {\rm i}\,\beta\,,\\
    \Theta_2:\qquad\frac{\lambda_2^\pm}{m_A}&= -\frac{\alpha}{2}\pm{\rm i}\,\delta\,,
\end{align}
up to corrections suppressed by powers of $1/\delta$. Thus $\Theta_2$ undergoes rapid oscillations with large damping, while the damping of $\Theta_1$ is parametrically suppressed. This suppression has a simple physical interpretation: the light mode, $\Theta_1$, is mostly orthogonal to the direction directly damped by dark sphalerons, so it loses energy only through its small mixing with the heavy dark-sector mode.

The opposite limit, $\delta\ll1$, is more naturally described in terms of the QCD-coupled direction
\begin{align}\label{eqn:A}
    A = \Theta_1 + C\Theta_2
\end{align}
and the approximately orthogonal light direction
\begin{align}\label{eqn:B}
    B = \frac{f_2^2\Theta_2 - C f_1^2\Theta_1}{C^2f_1^2+f_2^2}\,.
\end{align}
The corresponding decay constants are
\begin{align}
    f_A^2 = \bigl(f_1^{-2} + C^2f_2^{-2}\bigr)^{-1}\,,\qquad f_B^2 = C^2f_1^2 + f_2^2\,.
\end{align}
In this limit, $A$ is the massive QCD axion direction, while $B$ is approximately massless when $\delta\ll1$. When both modes are underdamped, the roots are
\begin{align}
    A:\qquad\frac{\lambda_A^\pm}{m_A}&=-\frac{1}{2}\alpha\bar\beta^2\pm{\rm i}\,,\\
    B:\qquad\frac{\lambda_B^\pm}{m_A}&=-\frac{1}{2}\alpha\beta^2\pm{\rm i}\,\beta\delta\sqrt{1-\frac{\alpha^2\beta^2}{4\delta^2}}\,.
\end{align}
The first line shows that the QCD axion direction, $A$, is damped only by the projection of the dark sphaleron friction onto the $A$ direction. The second line shows that the light mode becomes overdamped when $\alpha\beta\gtrsim 2\delta$, in which case the complex conjugate pair should be replaced by two real roots.

The preceding limits assume that both eigenvalues are large enough for the corresponding fields to oscillate.

\subsection*{(ii) One axion oscillating} At earlier times, however, one eigenmode may be rolling while the other oscillates about its minimum. In this regime, the restoring force of the rolling mode may be neglected. There are two distinct limits, depending on whether the dark or QCD instanton potential provides the dominant restoring force. The oscillating mode again obeys the scaling \cref{eqn:massive_adiabatic}, while the velocity of the rolling mode scales as 
\begin{align} 
    \dot Q_{\rm roll}\propto \asf^{-3}\exp\left\{\int^t{\rm d}t'\,\gamma_{\rm roll}(t')\right\}\,. 
\end{align} 
\paragraph{Dark-dominated regime.} In the dark-dominated regime, $\delta\gg1$, the eigenstates are well approximated by $\Theta_1$ and $\Theta_2$. The dark instanton potential provides the restoring force for the oscillating $\Theta_2$ mode, while the much smaller QCD restoring force may be neglected for the rolling $\Theta_1$ mode, leading to
\begin{align}\label{eqn:three_large_delta} 
    \Theta_1:\qquad\frac{\lambda_{\rm roll}}{m_A}&=-\frac{\alpha\beta^2\bar\beta^2}{\delta^4}\,,\\ 
    \Theta_2:\qquad\frac{\lambda_{\rm osc}^\pm}{m_A}&=-\frac{\alpha}{2}\pm{\rm i}\,\delta\,. 
\end{align} The slight misalignment between the rolling mode and the dark-sphaleron direction allows the rolling mode to feel a small amount of dark sphaleron damping. In particular, this damping vanishes in the decoupling limit $\bar\beta\to0$, as required. 

\paragraph{QCD-dominated regime.} In the QCD-dominated regime where the dark instanton potential is negligible, $\delta\ll 1$, the characteristic polynomial reduces to \begin{align}\label{eqn:three_d} 
    0=x^4 + \alpha x^3 + x^2 + \alpha\beta^2 x\,. 
\end{align} 
This polynomial has four roots: one trivial zero root reflecting the exact shift symmetry of the massless mode, one purely real root encoding the decay rate of the rolling mode's velocity, and a complex conjugate pair corresponding to the damped oscillating mode. We denote the nonzero real root by \begin{align} \lambda_{\rm roll} = \gamma_{\rm roll}\,. \end{align} For weak damping in the dark sector, $\alpha\ll1$, the eigenstates are well approximated by $A$ and $B$, and the roots are \begin{align} A:\qquad\frac{\lambda_A^\pm}{m_A}&=-\frac{1}{2}\alpha\bar\beta^2\pm{\rm i}\,,\\ B:\qquad\frac{\lambda_{\rm roll}}{m_A}&=-\alpha\beta^2\,. \end{align} As expected, the damping of each mode is set by the corresponding projection of the dark sphaleron friction. On the other hand, if the damping is strong, $\alpha\gg1$, the physical eigenstates are more naturally expressed in the original $\Theta_1,\Theta_2$ basis. In this limit the roots of \cref{eqn:three_d} behave as \begin{align}\label{eqn:large_alpha_small_delta} 
    \Theta_1:\qquad\frac{\lambda_1^\pm}{m_A}&=-\frac{\bar\beta^2}{2\alpha}\pm {\rm i}\,\beta\,,\\
    \Theta_2:\qquad\frac{\lambda_{\rm roll}}{m_A}&=-\alpha\,. 
\end{align}

\subsection*{(i) Neither axion oscillating} Finally, at the earliest times, when both $\chi_c$ and $\chi_d$ are negligible, the characteristic polynomial simplifies to
\begin{align}
\label{eqn:two}
    0&=x^4+\alpha x^3\,.
\end{align}
This polynomial has one nontrivial root, which describes the decay of the velocity of $\Theta_2$, while the velocity of $\Theta_1$ redshifts only due to the expansion of the Universe:
\begin{align}
    \Theta_1:\qquad\lambda_1 &=0\,,\\
    \Theta_2:\qquad\lambda_2 &=-\frac{\Upsilon_d}{f_2^2}=-\alpha m_A\,.
\end{align}

Before moving on to the next discussion, we note one subtlety of the two-axion decoupling limit. Although taking one decay constant much larger than the other one can make the system appear reducible to a one-axion effective theory over intermediate timescales, the decoupling is dynamical rather than exact. For any finite ratio of decay constants, the initially excited mode generically has a small projection onto the more weakly damped mode. This component may be negligible at early times, but it can control the asymptotic evolution after the more strongly damped oscillatory mode has decayed away.
We illustrate this effect in a simple limiting example in \cref{app:mode_mixing_decoupling}.

\subsection{Monodromy--cosine crossing}
\label{subsec:monodromy_crossing}

The analysis above assumed that the two-axion system could be treated in the linearized WKB regime, so that the evolution is controlled by instantaneous eigenmodes and their damping rates. For monodromy-driven initial conditions, however, there is an additional possibility. The cosine potential may become important while the monodromy direction is still displaced by many windings, so that the crossing between the monodromy mode and the confining cosine cannot be described by expanding about a single minimum. This regime is particularly relevant to the lattice-extrapolated friction prescription of \cref{subsec:sphaleron_rate_summary}, for which the friction can be too weak to fully damp the monodromy-driven motion before the crossing. The crossing is then nonlinear, and the final partition of energy between the light and heavy axion modes depends sensitively on where the monodromy trajectory lies when the cosine potential turns on.\footnote{Related sensitivity to initial conditions at nonlinear axion level crossings appears
in the axion-roulette mechanism~\cite{Daido:2015bva,Daido:2015cba}, as well as axion friendship~\cite{Cyncynates:2021xzw,Cyncynates:2022wlq}.}

Though this effect is not itself dissipative, it determines how much of the initial monodromy energy is deposited into each of the modes that subsequently redshift and dissipate. This is particularly important in the weak-damping regime near the boundary of the allowed parameter space studied in \cref{sec:three_axions}, where the final relic abundance remains sensitive to the energy partition at crossing. We therefore study a simplified two-axion crossing problem in order to characterize the resulting energy partition and to determine when it can be treated probabilistically.

Here we study the system of equations relevant to the discussion of monodromy--cosine crossing in \cref{sec:three_axions}.  We start from the Lagrangian
\begin{align}
    {\cal L} = \frac12f_1^2(\partial\Theta_1)^2 + \frac12f_2^2(\partial\Theta_2)^2 - \frac12\Lambda_{\rm mon}^4(\Theta_1 + \Theta_2)^2 - \chi_{\rm conf}(1-\cos\Theta_2)\,.
\end{align}
Here $\chi_{\rm conf}$ denotes the topological susceptibility of the confining gauge group responsible for the cosine potential. In the applications below it may be either $\chi_c$ or $\chi_d$.

The equations of motion in radiation domination are given by
\begin{align}
    0&=\ddot \Theta_1 + \frac{3}{2t}\dot\Theta_1 + \frac{\Lambda_{\rm mon}^4}{f_1^2}(\Theta_1+\Theta_2)\,,\\
    0&=\ddot\Theta_2 + \frac{3}{2t}\dot\Theta_2 + \frac{\Lambda_{\rm mon}^4}{f_2^2}(\Theta_1+\Theta_2) + \frac{\chi_{\rm conf}}{f_2^2}\sin\Theta_2\,,
\end{align}
where we have dropped the friction terms, beyond the Hubble friction. The regime of interest is
\begin{align}
    \Theta_1(0)+\Theta_2(0)\gg 1\,,\qquad\Lambda_{\rm mon}^4(f_1^{-2}+f_2^{-2})\ll \frac{\chi_{\rm conf}(t\to\infty)}{f_2^2}\,.
\end{align}
At early times, when the confining potential is negligible, it is useful to organize the dynamics into a monodromy mode
\begin{align}
    X \equiv \Theta_1+\Theta_2\,,
\end{align}
and an orthogonal approximately massless mode
\begin{align}
    Y \equiv \frac{f_2^2\Theta_2-f_1^2\Theta_1}{f_1^2+f_2^2}\,.
\end{align}
Equivalently,
\begin{align}
    \Theta_1=\beta^2 X-Y\,,\qquad\Theta_2=Y+\bar\beta^2 X\,,
\end{align}
where
\begin{align}
    \beta^2\equiv \frac{f_2^2}{f_1^2+f_2^2}\,,\qquad\bar\beta^2\equiv \frac{f_1^2}{f_1^2+f_2^2}\,.
\end{align}
In these variables, the equations of motion are
\begin{align}\label{eq:two-axion-monodromy-crossing-eom}
    0&=\ddot X+\frac{3}{2t}\dot X+m_X^2X +\frac{\chi_{\rm conf}}{f_2^2}\sin\!\left(Y+\bar\beta^2 X\right)\,,\\
    0&=\ddot Y+\frac{3}{2t}\dot Y +m_Y^2\sin\!\left(Y+\bar\beta^2 X\right)\,,
\end{align}
with
\begin{align}
    m_X^2\equiv \Lambda_{\rm mon}^4(f_1^{-2}+f_2^{-2})\,, \qquad m_Y^2\equiv \frac{\chi_{\rm conf}}{f_1^2+f_2^2}\,.
\end{align}
For $\chi_{\rm conf}\simeq0$, the monodromy mode $X$ is a harmonic oscillator while $Y$ is an exactly massless direction.

If the oscillation amplitude of $X$ is already $\mathcal O(1)$ or smaller when the confining potential becomes important, the cosine may be expanded about the nearest minimum and the system remains in the linear regime. In this case, the usual adiabatic level crossing applies (up to typically small Landau-Zener dynamics)~\cite{Cyncynates:2023esj,Ho:2018qur,Dunsky:2025sgz}: energy initially stored in the heavy mode $X$ is transferred to the final heavy mode, which is predominantly $\Theta_2$, while energy stored in $Y$ is transferred to the final light mode, which is predominantly $\Theta_1$.

The regime of interest here is qualitatively different: if $|X|\gg1$ at the time of crossing, the full sinusoidal structure of the confining potential becomes important, and the linear approximation breaks down. To build intuition, consider the limiting case in which
\begin{align}
    \chi_{\rm conf}=0\quad{\rm for}\quad t<t_0\,,
    \qquad
    \chi_{\rm conf}\to\infty\quad{\rm for}\quad  t\geq t_0\,.
\end{align}
At the instant the cosine turns on, $t_0$, $\Theta_2$ is effectively trapped at the nearest multiple of $2\pi$, while $\Theta_1$ continues rolling in the monodromy potential. The subsequent partition of energy is therefore determined by where the monodromy trajectory is at the time the confining potential turns on. In the realistic problem, $\chi_{\rm conf}$ grows continuously to a finite value, and the final partition depends both on the value of the potential and on the kinetic energy carried by the monodromy mode at the time the two sectors mix.

It is therefore useful to define the crossing time $t_\times$ by comparing the monodromy energy stored in the $X$ mode to the barrier height of the cosine potential,
\begin{align}
    \rho_{\rm monodromy}(t_\times)=2\chi_{\rm conf}(t_\times)\,.
\end{align}

\begin{figure}
    \centering
    \includegraphics[width=0.75\columnwidth]{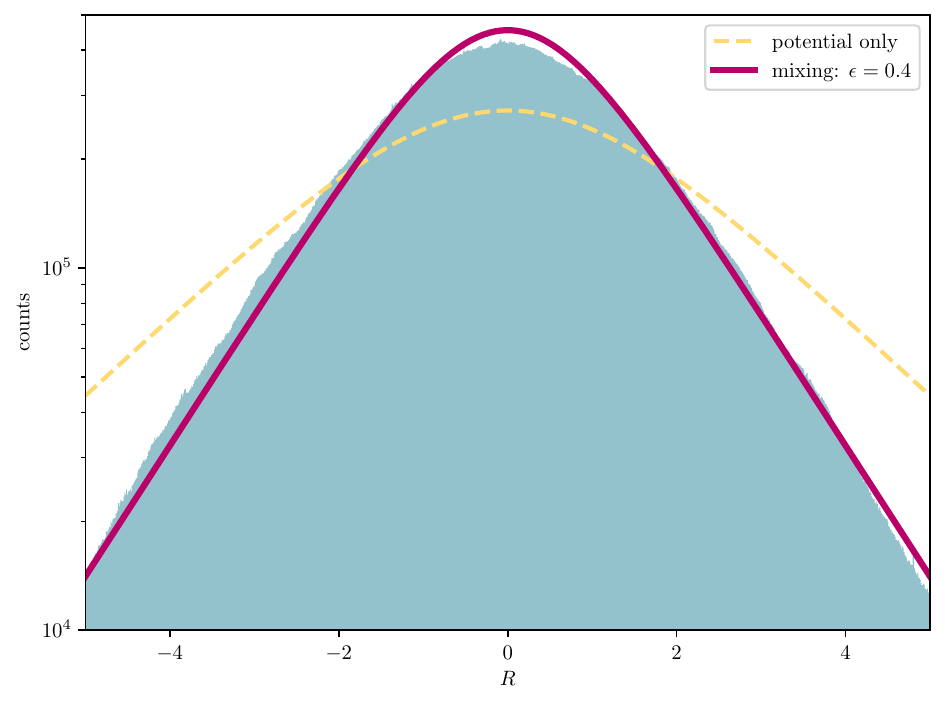}
    \caption{Counts of $R$ (defined in \cref{eq:R}) for the simulation of monodromy--cosine crossing as described in \cref{subsec:monodromy_crossing}. This histogram includes initial monodromy displacements $\theta_1(0) + \theta_2(0)$ uniformly sampled over the interval $[300,1000]$ (the histograms as a function of this initial displacement are visible in \cref{fig:final_ratio_sim}. Here we focus on the values of the displacement that lead to a fairly uniform distribution). On top of the histogram we plot two theoretical curves. The dashed yellow line corresponds to the analytical sudden-crossing estimate where the final partition of energies is determined by the value of the monodromy potential energy at the crossing. This predicted distribution is too broad. 
    We model this mismatch with the phenomenological exponent $\epsilon$ (see text for details). 
    Choosing $\epsilon = 0.4$ yields a good fit to the data, as illustrated by the purple curve.}
    \label{fig:ratio_hist}
\end{figure}

If the monodromy mode oscillates with amplitude $\bar X\equiv\sqrt{2\langle X^2\rangle}$, then the probability of finding $X=x$ at a random time is
\begin{align}
    p_X(x|\bar X)=\frac{1}{\pi\sqrt{\bar X^2-x^2}}\,,
    \qquad |x|<\bar X\,.
\end{align}
At fixed amplitude, knowing the instantaneous value of $X$ also fixes the kinetic energy. Neglecting nonlinear redistribution, this sudden-crossing estimate suggests that for the heavy, $\Theta_2$, and the light, $\Theta_1$, modes
\begin{align}
    \rho_{\rm heavy}(t_\times)\propto \bar X^2-X^2\,,\qquad \rho_{\rm light}(t_\times)\propto X^2\,.
\end{align}
Defining
\begin{align}\label{eq:R}
    R\equiv \log\frac{\rho_{\rm heavy}(t_\times)}{\rho_{\rm light}(t_\times)} =\log\frac{\bar X^2-X^2}{X^2}\,,
\end{align}
the corresponding probability distribution is
\begin{align}
    p_R(r)=\frac{1}{2\pi}{\rm sech}\!\left(\frac{r}{2}\right)\,.
\end{align}
Numerically, however, the probability distribution obtained with this sudden-crossing estimate is too broad, as illustrated by \cref{fig:ratio_hist} by the dashed yellow curve. 
The reason is that, in the realistic setting, the energy partition between the two states is not entirely determined by the value of the monodromy potential energy at crossing. Instead, nonlinearities in the cosine potential cause energy to mix between the heavy and light states, and thus move the system toward equipartition.

A simple phenomenological way to model the effect of nonlinear mixing is therefore to soften the sudden-crossing ratio by writing
\begin{align}\label{eqn:expR_mix}
    \exp R = \frac{\bar X^2-X^2}{X^2} \longrightarrow \left[\frac{\bar X^2-X^2}{X^2}\right]^{1-\epsilon}\,,
\end{align}
with the phenomenological parameter $0\leq \epsilon<1$. Here $\epsilon=0$ corresponds to the sudden-crossing estimate, while increasing $\epsilon$ pushes the system toward equipartition. Consequently, the distribution narrows:
\begin{align}\label{eqn:pRepsilon}
    p_R(r;\epsilon) = \frac{1}{2(1-\epsilon)\pi} {\rm sech}\!\left[\frac{r}{2(1-\epsilon)}\right]\,.
\end{align}
We find that $\epsilon\simeq0.4$ provides a good phenomenological description of the numerically observed distributions, as seen in \cref{fig:ratio_hist}.

We now turn to a numerical study of this system under a simplified set of assumptions. First, we fix $f_1=f_2\equiv f$, which is one of the main limitations of the present analysis. The common monodromy scale is then $\mu^2\equiv {\Lambda_{\rm mon}^4}/{f^2}$, so that the early-time monodromy frequency is $m_X=\sqrt{2}\mu$. Second, we take initial conditions
\begin{align}
    \dot\Theta_1(0)=\dot\Theta_2(0)=0\,, \qquad \Theta_1(0)=\theta_0 x\,, \qquad \Theta_2(0)=\theta_0(1-x)\,,
\end{align}
with $x\in[0,1]$ and $\theta_0\in[50,1000]$. 
Thus $X(0)=\theta_0$, while $x$ scans the initially flat direction. Third, we adopt the toy topological susceptibility
\begin{align}
    \chi_{\rm conf}(t)= \chi_0\min\left\{1,\left(\frac{t}{t_0}\right)^n\right\}\,,
\end{align}
with $n=4$ in all simulations. 
The choice of $n=4$ mimics the time dependence of the susceptibility during radiation domination, where the high-temperature scaling $\chi(T) \propto T^{-8}$ derived from the dilute instanton gas approximation (see Appendix \ref{app:topological_susceptibility}) translates to $\chi(t) \propto t^4$. The plateau at large values of $t$ captures the freezing of the susceptibility at its constant zero-temperature value once the Universe cools below the confinement scale.

\begin{figure}
    \centering
    \includegraphics[width=\columnwidth]{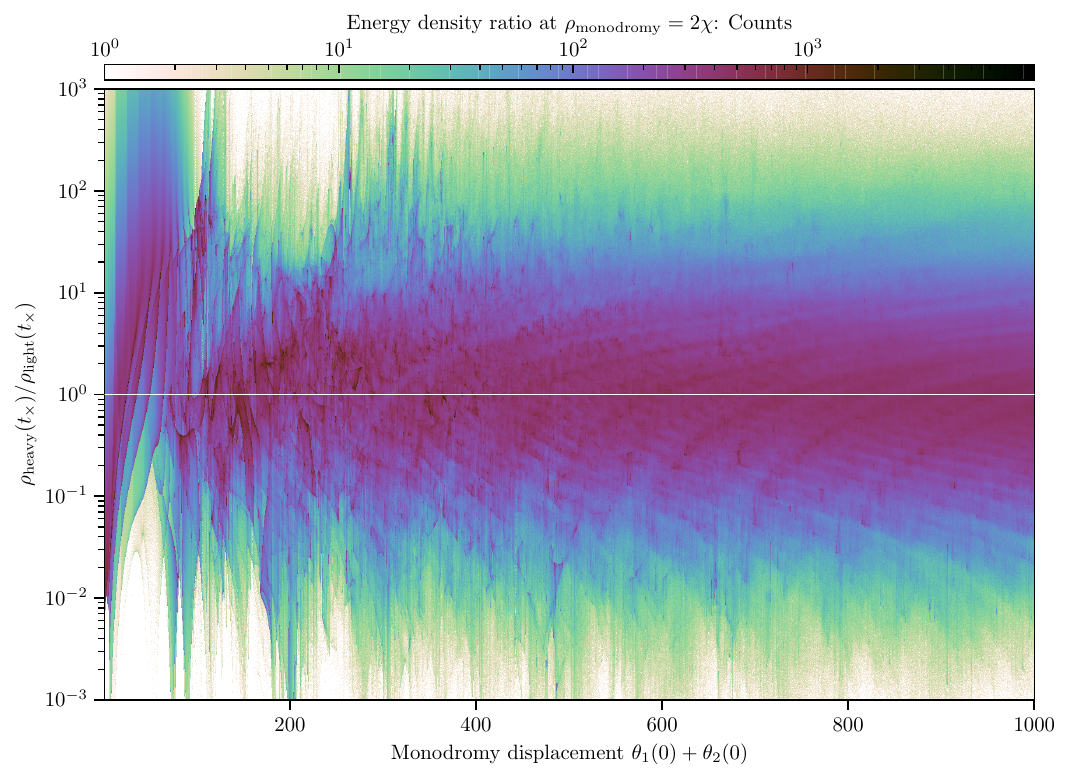}
    \caption{The ratio of energy densities between the final heavy and light states at the crossing time $t_\times$, plotted against the initial monodromy displacement $\theta_1(0) + \theta_2(0)$. Because the raw energies are measured at the end of the simulation ($t_f$), the values plotted here have been analytically 'rewound' to $t_\times$ by factoring out the WKB mass growth caused by the evolving potential $\chi(1 - \cos\theta_2)$. 
    At larger values of the initial monodromy displacement, we see that the complicated dependence of the energy density ratio on the initial conditions smears out into a continuous band. As demonstrated in Fig. \ref{fig:ratio_hist}, the integrated distribution of this large-displacement regime is well approximated by Eq. (\ref{eqn:pRepsilon}) for $\epsilon = 0.4$. 
    Solving for the monodromy displacement such that the time of $\rho_{\rm monodromy} = 2\chi$ is the same as the time when the unmixed masses are comparable, we find $\theta_1(0) + \theta_2(0)\sim \mathcal{O}(100)$, roughly matching the transition from an irregular distribution at small misalignments to a fairly uniform distribution at larger misalignments.}
    \label{fig:final_ratio_sim}
\end{figure}

The simulations start at $t=0.01/\mu$ and run until $t_f=10^3/\mu$. The integration step size is chosen as
\begin{align}
    \Delta t = 0.1\times \max\!\left\{\mu\theta_0(2\mu t)^{-3/4},\frac{\chi_{\rm conf}^{1/2}(t)}{f}\right\}^{-1},
\end{align}
so that the shortest instantaneous timescale is always resolved. The numerical values used are $\mu = {\rm meV}$, $f = 10^{12}\GeV$, $\chi_0 = 10^4\GeV^4$, and $t_0 = 5\times 10^{14}\GeV^{-1}$. The results of these simulations are presented in \cref{fig:final_ratio_sim,fig:ratio_hist}.

The practical importance of this energy partition is that the post-crossing evolution is not symmetric between the two modes. If most of the energy at $t_\times$ is deposited into the mode whose mass subsequently grows, then its energy density is enhanced by the usual WKB scaling as the confining potential turns on. If, instead, a smaller fraction of the energy is deposited into this mode, the final energy density of the system is correspondingly reduced. In this sense, monodromy--cosine crossing gives the system a way to lose energy without dissipation: the total energy is not erased by friction, but the nonlinear crossing can place less of the initial monodromy energy into the mode that later receives the adiabatic mass enhancement. Conversely, if the mode whose mass grows is also the mode that is most efficiently damped, depositing more energy into this mode can reduce the eventual relic abundance because the subsequent friction may be stronger than the WKB enhancement.

\subsection*{Fragmentation}

The homogeneous analysis above neglects the growth of spatial perturbations, which leads to fragmentation for many of the initial conditions studied. To assess its effect on the energy partition, we perform a limited set of lattice simulations in $2+1$ dimensions, described in \cref{app:Monodromy_Crossing_Fragmentation}. For the benchmark considered, the distribution of the post-crossing energy partition, including the energy stored in oscillons~\cite{Copeland:1995fq,Fodor:2008es,Amin:2014eta,Cyncynates:2021rtf}, is qualitatively similar to that obtained from the corresponding homogeneous evolution. These results provide preliminary support for the statistical description developed above, although a more extensive study, including a larger number of simulations and in $3+1$ dimensions, is needed to establish its accuracy.

The nonlinear structures and enhanced small-scale density fluctuations generated during crossing may also seed dark matter substructure, as occurs in other multi-axion scenarios~\cite{Daido:2015bva,Daido:2015cba,Eroncel:2022vjg,Cyncynates:2021xzw,Cyncynates:2022wlq}. We leave their subsequent evolution and observational implications to future work.

\subsection{Summary}
The two-axion theory provides the minimal effective description of dark-sphaleron damping consistent with the strong CP solution, as originally explored in the companion paper Ref.~\cite{Asadi:2025cvm}. Once the dark sector is introduced to dissipate the axion kinetic energy, its late-time confining potential represents an additional source of PQ violation, and a second axion is required so that the dark and QCD phases can both be relaxed. In the Gram--Schmidt basis this structure is especially transparent: the dark instanton couples to a single field direction, while the QCD axion is a misaligned linear combination of the two axions. The subsequent dynamics are then governed by a small set of instantaneous normal modes, whose damping rates and oscillation frequencies are determined by the roots of the characteristic polynomials in the regimes where neither, one, or both instanton potentials are important.

The main physical lesson is that dark sphalerons can efficiently damp the component of the axion motion aligned with the dark gauge group, while the amount of damping inherited by the QCD axion direction depends on the misalignment of the two instanton charge vectors and on the hierarchy between $\chi_c$ and $\chi_d$. Because the evolution is adiabatic over the parameter space of interest, the light and heavy mode energies can be evolved separately, with negligible Landau--Zener--type transfer between them. This gives a controlled way to track how an initially large axion velocity is reduced and how much energy remains in the final axion relics.

We also considered the related problem of monodromy--cosine crossing that may occur if the axion velocity at the electroweak phase transition originates from a monodromy, as in \cref{sec:monodromy}.
Although this monodromy--crossing effect is not itself dissipative, it determines how much of the initial monodromy energy is deposited into each of the modes that subsequently redshift and dissipate. When the confining cosine becomes important only after the monodromy direction has relaxed to within one period, the crossing is well described by the usual linear adiabatic evolution. However, if the monodromy direction is still displaced by many windings at crossing, the full nonlinear structure of the cosine potential becomes important. In this regime, the final energy partition is effectively probabilistic and is well described by a simple distribution. Since only the mode whose mass subsequently grows receives the associated WKB enhancement, this probabilistic partition can effectively reduce the final relic energy by placing less of the initial monodromy energy into that mode. It may also be that the mode whose mass is growing is also the one which is damped, in which case the opposite partition will work to suppress the net relic abundance.

So far, however, we have either imposed the initial velocity by hand or neglected damping. Combining the monodromy mechanism of \cref{sec:monodromy}, which supplies the required velocity, with the dark-sphaleron damping studied in this section introduces three independent sources of axion potential: QCD, the dark confining group, and the monodromy. Thus, a completely generic model consistent with the strong CP solution requires three distinct axion directions, coupled to these sources in non-degenerate linear combinations. In the next section we show that, for the parameter ranges of interest, this full three-axion system factorizes into the monodromy dynamics that generate the required axion velocity, the two-axion damping dynamics described above, and the nonlinear crossing dynamics that determine the initial partition of energy among the post-crossing modes.

\section{Complete effective theory and dynamics}
\label{sec:three_axions}

In the previous two sections, we introduced new ingredients to solve different challenges of the minimal axiogenesis setup. In \cref{sec:monodromy}, we showed how an axion rolling in a monodromy potential can generate a large rotational velocity at $T_{ew}$. In  \cref{sec:damping_from_dark_sphalerons}, we established the damping mechanism from a new confining dark force, which is necessary to deplete the excess kinetic energy in this system and avoid overclosure.

The reason for embedding the mechanism in a multi-axion framework is simple: the construction involves three independent sources of axion potential, or effective PQ breaking, namely QCD, the dark confining force, and monodromy. 
A generic model capable of resolving strong CP therefore contains at least three axion directions, coupled to these sources in non-degenerate linear combinations. This allows the monodromy potential to generate the required axion velocity, the dark sphalerons to dissipate the resulting kinetic energy, and the remaining light axion direction to relax the physical strong CP phase. We note, however, that the mechanism could also work with only two axions if, for example, the dark confining sector and the monodromy potential coupled to the same axion-field combination, though such an alignment appears less generic. The two axion limit nevertheless serves as a useful benchmark in parameter space, as illustrated by the companion paper Ref.~\cite{Asadi:2025cvm}.

Further, though the full dynamics involve three distinct axion field combinations, the evolution is often characterized by the dynamics of interacting pairs of axions. At early times, the monodromy direction generates the large axion velocity needed for baryogenesis. At later times, the damping and relic abundance are controlled by the two-axion normal-mode dynamics described in \cref{sec:damping_from_dark_sphalerons}. If the cosine potential from the new confining force becomes important while the monodromy direction is still displaced by many windings, the intermediate evolution also involves the nonlinear monodromy--cosine crossing discussed in \cref{subsec:monodromy_crossing}. The three-axion system therefore factorizes into a sequence of effectively two-axion systems studied above.

In the remainder of this section, we first describe the qualitative structure of the three-axion evolution in \cref{sec:three_axions_qualitative}, then develop analytic approximations in \cref{sec:three_axions_parameter_space_estimate}, and finally solve the full system numerically in \cref{sec:three_axions_numerical}. We then identify the parameter space in which the observed baryon asymmetry and dark matter abundance are simultaneously reproduced. 

\subsection{Dynamics of monodromy-driven baryogenesis}
\label{sec:three_axions_qualitative}

We begin by reviewing the ingredients of the full model. First, there is monodromy, which is responsible for generating the large axion velocity at the electroweak phase transition. For concreteness, we specialize to Kaloper--Sorbo (KS) monodromy~\cite{Kaloper:2008fb,Kaloper:2011jz}. This choice simplifies both the analytical calculation in \cref{sec:three_axions_parameter_space_estimate} and the numerics in \cref{sec:three_axions_numerical}, since the potential introduces only a single time scale and avoids introducing self interactions that would cause the axion to fragment before the electroweak phase transition. 

Second, we must include significant dissipation to avoid overclosure from the large axion energy stored in the monodromy. We take the dissipative sector to be a confining dark gauge sector at the same temperature as the SM, and treat it as pure SU($N_d$), since light charged fermions only suppress the induced friction (see \cref{fig:Friction_Sampler}). Ultimately, heavy quarks are necessary to dump the entropy of the dark sector into the SM before BBN; we discuss this process and the associated constraints in \cref{sec:depletion_of_DS}. We find that the dark confinement scale must be sufficiently large for the dark glueballs to decay efficiently, giving the lower bound $\Lambda_d \gtrsim 0.3\GeV$.

The confinement scale is also bounded from above. At high temperatures, dark sphaleron transitions dissipate axion kinetic energy, while at low temperatures dark instantons generate an additional axion potential. Once dark confinement becomes important relative to the monodromy potential, it shuts off axion mixing, so that only the heavy axion can continue to dissipate. The confinement scale is therefore constrained from both sides: from below by the requirement that the dark sector decay before BBN, and from above by the requirement that dissipation remain active for long enough to remove the excess axion energy.

As in the two-axion system of \cref{sec:damping_from_dark_sphalerons}, it is useful to choose a Gram--Schmidt basis in which the independent PQ-violating sources take a lower-triangular form~\cite{Demirtas:2021gsq,Gendler:2023kjt,Baryakhtar:2026oun}. We choose the ordering of the Gram-Schmidt basis states so that the fields are closely aligned with the late-time mass eigenstates, though we remind the reader that the results are basis invariant.  For the hierarchy of axion potentials
\begin{align}
\Lambda_{\rm KS}^4 \ll \chi_c(0) \ll \chi_d(0)\,,
\end{align}
we work in the basis where the dark instanton potential depends only on $\theta_3$, the QCD potential on the combination $\theta_2+C_{23}\theta_3$, and the KS monodromy potential on $\theta_1+C_{12}\theta_2+C_{13}\theta_3$, where all $C_{ij}$ are $\mathcal{O}(1)$ numbers. Thus, $\theta_1$ enters only the monodromy potential, $\theta_2$ enters the monodromy and QCD potentials, and $\theta_3$ enters all three. The corresponding dissipative terms act on the same field combinations.

In this parametrization, the Lagrangian is
\begin{align}
    {\cal L} = {\cal L}_{\rm SM} + {\cal L}_d + {\cal L}_\theta \,,
\end{align}
where ${\cal L}_d$ is the dark-sector Lagrangian consisting of dark gauge and fermion fields, and ${\cal L}_\theta$ contains the axions
\begin{align}
    \label{eq:Ls_three_axion}
    {\cal L}_\theta
    =
    \frac12 \sum_{i = 1}^3 (\partial \phi_i)^2
    +\frac{\alpha_d}{4\pi}\frac{\phi_3}{f_3} {\rm Tr}\tilde G_d G_d
    +  \frac{\alpha_c}{4\pi}\left(\frac{\phi_2}{f_2}+C_{23}\frac{\phi_3}{f_3}\right){\rm Tr}\tilde G_c G_c\, + \mathcal{L}_\mathrm{KS} \, ,
\end{align}
while the Kaloper--Sorbo monodromy term is given by
\begin{align}
    {\cal L}_{\rm KS}
    =
    \frac{\Lambda_{\rm KS}^2}{24}
    \left(\frac{\phi_1}{f_1} + C_{12}\frac{\phi_2}{f_2} + C_{13}\frac{\phi_3}{f_3} \right)
    \frac{\varepsilon^{\mu\nu\rho\sigma}}{\sqrt{-g}}
    F_{\mu\nu\rho\sigma}
     + \cdots\,,
\end{align}
where the $\cdots$ refer to terms involving only $F^{\mu\nu\rho\sigma}$ and the corresponding three-form gauge field $A^{\mu\nu\rho}$~\cite{Kaloper:2008fb,Kaloper:2011jz}.

The axion equations of motion may then generically be written in the Gram-Schmidt basis, where axion interactions assume a lower-triangular form:
\begin{equation}\label{eqn:full_three_axion_EOM}
\begin{aligned}
    0&=\ddot\theta_1 + 3H\dot\theta_1
    + \frac{\Lambda_{\rm KS}^4}{f_1^2}
    (\theta_1 + C_{12}\theta_2 + C_{13}\theta_3)\,,
    \\
    0&=\ddot\theta_2 + 3H\dot\theta_2
    +C_{12}\frac{\Lambda_{\rm KS}^4}{f_2^2}
    (\theta_1 + C_{12}\theta_2 + C_{13}\theta_3) + \frac{\Upsilon_c}{f_2^2}
    ( \dot\theta_2 + C_{23}\dot\theta_3)
    +\frac{\chi_c}{f_2^2}\sin(\theta_2 + C_{23}\theta_3)\,,
    \\
    0&=\ddot\theta_3 + 3H\dot\theta_3
    +C_{13}\frac{\Lambda_{\rm KS}^4}{f_3^2}
    (\theta_1+C_{12}\theta_2 + C_{13}\theta_3)+ C_{23}\frac{\Upsilon_c}{f_3^2}
    (\dot\theta_2 + C_{23}\dot\theta_3)
    +C_{23}\frac{\chi_c}{f_3^2}\sin(\theta_2 + C_{23}\theta_3)
    \\
    &\quad
    + \frac{\Upsilon_d}{f_3^2}\dot\theta_3
    + \frac{\chi_d}{f_3^2}\sin\theta_3\,.
\end{aligned}
\end{equation}

In what follows, we solve these equations for different choices of the free parameters in order to identify the regions of the parameter space that reproduce the observed DM abundance and SM baryon asymmetry today.
The parameter space of this model is spanned by the following quantities: (1--3) the axion decay constants $f_i$, (4--6) the mixing coefficients $C_{ij}$, (7) the dark confinement scale $\Lambda_d$, (8) the scale of the Kaloper--Sorbo monodromy potential $\Lambda_{\rm KS}$, and (9) the initial monodromy charge $N_{\rm mon}$. The initial displacements orthogonal to the monodromy (10--11) also contribute two free parameters, though the physics is relatively insensitive to their particular values.
In \cref{sec:phenomenology}, we will see that a viable model also requires a dark quark, whose mass $m_\Psi$ must be specified, though it does not play a dynamical role provided it is sufficiently heavy. 

Given the large number of free parameters in our model, a full scan of the complete parameter space is a daunting task. We will instead identify favorable directions in parameter space to 
produce viable examples that are consistent with existing constraints.
Even after reducing the parameter space in this way, the numerical integration is challenging. In particular, resolving the axion evolution from above the electroweak phase transition to below the QCD phase transition requires switching between WKB and exact integration for different combinations of normal modes. While this is technically feasible, it is not practical for even a reduced parameter-space scan.
Therefore, when reporting parameter regions, we use a combination of exact and WKB integration, supplemented by physically motivated heuristics near avoided mode crossings. We validate these approximations against resolved numerical integrations for a representative set of points (see the insets in \cref{fig:hard_cutoff_money,fig:monodromy_crossing_money} below).
The precise methods for both the resolved and heuristic integrations are discussed in \cref{sec:three_axions_numerical} and \cref{app:fast_three}.

In the following subsection, we first perform an analytical estimate to illustrate the approximate parameter space one should anticipate from the numerical scan. The important conceptual point is that although we have introduced three axions to satisfy three independent requirements, the dynamics of the axions themselves do not factorize. As a result, the system gives rise to a highly predictive parameter space, as illustrated in \cref{fig:hard_cutoff_money,fig:monodromy_crossing_money}.

\subsection{Example analytic three-mode evolution}
\label{sec:three_axions_parameter_space_estimate}

Here we perform an analytic estimate to illustrate typical parameter space of the full model, restricting ourselves to the simple case
\begin{equation}
    f_1 = f_2 = f_3\equiv f \, , \quad C_{12} = C_{13} = C_{23} = 1\,,
    \label{eq:simplify_f}
\end{equation}
which provides a useful benchmark for comparison with the numerics in \cref{sec:three_axions_numerical}.
We evaluate the friction using the high-temperature approximation with cutoff prescription as discussed in \cref{subsec:sphaleron_rate_summary} \footnote{The one-loop expression for $\alpha_d$ is given in \cref{app:running}; for analytic convenience, we drop the additive $1$ in this estimate.}
\begin{align}
    \Upsilon_d\approx \begin{cases}\frac12(N_d\alpha_d)^5T^3&T>T_{\rm{cut}}\,,\\
    0& T<T_{\rm{cut}}\,,
    \end{cases}
\end{align}
with $T_{\rm{cut}}=R_{\rm th}\Lambda_d$. We fix the monodromy potential scale $\Lambda_{\rm KS}$ and the initial monodromy charge so that the axion system begins oscillating at $\Tew$ and reproduces the observed baryon asymmetry, with the minimal possible displacement along the $\theta_{3\perp}$ direction (see \cref{sec:monodromy} and \cref{eq:analytic_sol_winding,eqn:monodromy_size}). The initial QCD angle $\theta_2 + C_{23}\theta_3$ is taken to be $0.3$ while the initial dark angle is taken to be $0$. These choices do not have a significant impact on our results. Finally, we take the cosmology to be radiation dominated and ignore the changing relativistic degrees of freedom. With these assumptions, we study the available parameter space in the $(\Lambda_d,f)$ plane, given a choice of $R_{\rm th}$.

\begin{figure}
    \centering
    \includegraphics[width=0.9\columnwidth]{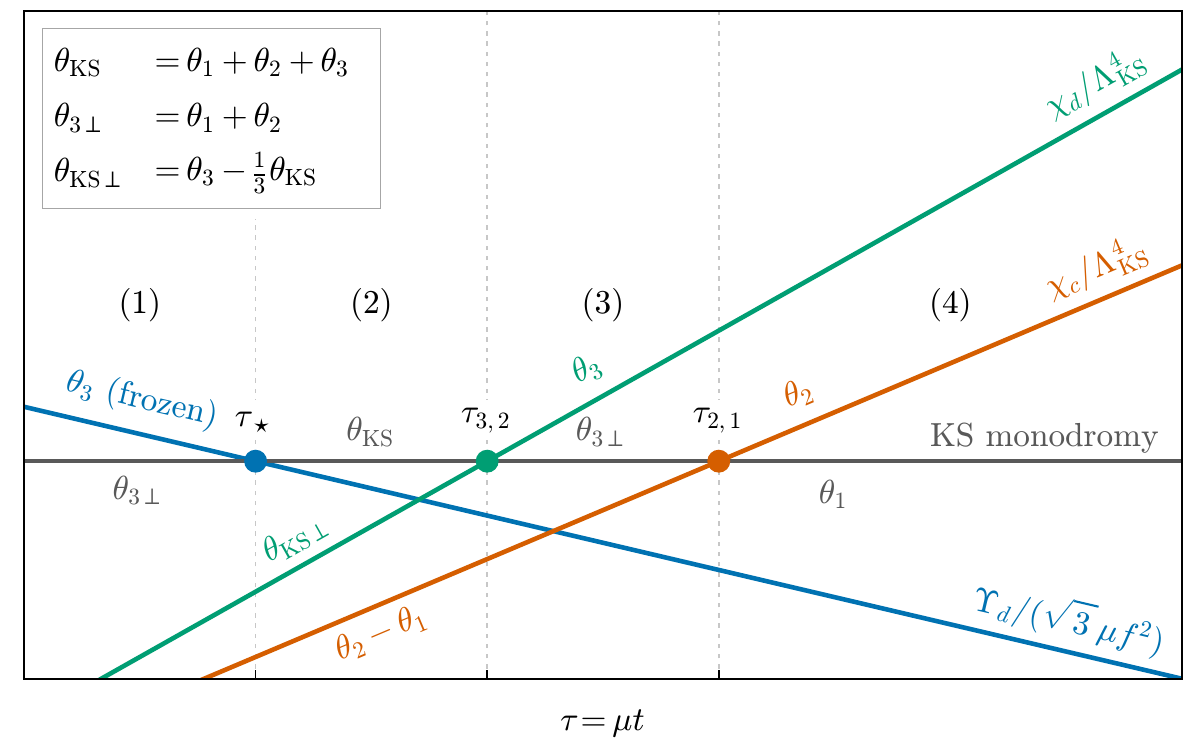}
    \caption{Schematic evolution of the three initial populations for the equal-decay constant unity $C_{ij}$ prescription of \cref{sec:three_axions_parameter_space_estimate}.
    The four time intervals (1)--(4) correspond to distinct epochs with different terms dominating the axion potentials, see the text.
    The colored lines correspond to the various contributions to the axion potential: the KS monodromy is gray, the dark instanton potential is \textcolor{gooseberry}{green}, the QCD instanton potential is \textcolor{red}{red}, and the dark sphaleron friction term is in \textcolor{blueberry}{blue}. The blue, red, and green lines are normalized to the KS potential for better comparison. 
    The three crossings are labeled by their respective times $\tau_{\star},\tau_{3,2},\tau_{2,1}$, and indicated by dots. 
    To the left and right of each dot, the axion eigenstate whose evolution is dominated by the corresponding potential/friction contribution is indicated.
    The linear combinations of axions relevant to different epochs are defined in the inset. 
    }
    \label{fig:three_axion_mode_evolution}
\end{figure}

Before presenting our analytic estimate, we summarize the physical picture on which it is based. The evolution approximately factorizes into a sequence of epochs involving at most two dynamically relevant axion combinations, with each epoch controlled by a subset of the terms in \cref{eqn:full_three_axion_EOM}. Their changing hierarchy is illustrated schematically in \cref{fig:three_axion_mode_evolution}. We measure time in units of the monodromy scale,
\begin{align}
    \tau \equiv \mu t\,,
    \qquad
    \mu \equiv \frac{\Lambda_{\rm KS}^2}{f}\,,
\end{align}
following the notation of \cref{sec:monodromy}.

To understand the sequential decoupling of these modes, it is instructive to track the system of equations in Eq. (\ref{eqn:full_three_axion_EOM}) through four distinct chronological epochs: (1) $\tau<\tau_\star$: initially, at temperatures well above confinement ($\chi_c = \chi_d = 0$), the large dark-sector friction $\Upsilon_d$ dominates the third line of Eq. (\ref{eqn:full_three_axion_EOM}), effectively freezing $\theta_3$ in place. With $\theta_3$ locked, the first two lines of Eq. (\ref{eqn:full_three_axion_EOM}) show that the remaining orthogonal combination $\theta_{3\perp} \equiv \theta_1 + \theta_2$ evolves exclusively under the influence of the KS monodromy potential. (2) $\tau_\star<\tau<\tau_{3,2}$: As the Universe cools and $\Upsilon_d$ decreases, $\theta_3$ is released. Because the instanton potentials remain negligible, all three fields in Eq. (\ref{eqn:full_three_axion_EOM}) are driven symmetrically by the KS restoring force. The oscillations thus reorganize into the fully KS-aligned direction $\theta_{\mathrm{KS}} \equiv \theta_1 + \theta_2 + \theta_3$, alongside the orthogonal, effectively massless combination $\theta_{KS\perp} \equiv \theta_3 - \frac{1}{3}\theta_{\mathrm{KS}}$. (3) $\tau_{3,2}<\tau<\tau_{2,1}$: At later times, the dark gauge group confines and the growing dark susceptibility $\chi_d$ turns on the $\sin\theta_3$ term in the third line of Eq. (\ref{eqn:full_three_axion_EOM}). This massive restoring force dominates the $\theta_3$ equation of motion, trapping the field and effectively decoupling it from the other axions as the new heavy mass eigenstate. The remaining monodromy energy is forced into the $\theta_1 + \theta_2$ subspace. (4) $\tau>\tau_{2,1}$: Finally, once the QCD susceptibility $\chi_c$ grows and dominates over the monodromy scale, the $\sin(\theta_2 + \theta_3)$ term in the second and third lines of Eq. (\ref{eqn:full_three_axion_EOM}) turns on. Because $\theta_3$ is already frozen by the deep dark potential, this term acts as a massive restoring force predominantly on $\theta_2$, decoupling it as the QCD axion mass eigenstate. By process of elimination, the first line of Eq. (\ref{eqn:full_three_axion_EOM}) dictates that $\theta_1$ is left as the sole remaining light mode governed by the monodromy potential. At late times, the interaction basis $(\theta_1, \theta_2, \theta_3)$ is therefore closely aligned with the mass eigenstates, whose masses are controlled predominantly by $\Lambda_{\mathrm{KS}}$, $\chi_c$, and $\chi_d$, respectively. 

We now elaborate on the evolution of each dynamical mode in different epochs.
(1) At very early times, 
both the QCD and dark topological susceptibilities are negligible. However, following the discussion in \cref{sec:WKB}, dark sphalerons act only along the $\theta_3$ direction, locking it in place as long as
\begin{align}
    \frac{\Upsilon_d}{f^2}\gtrsim\mu\,.
\end{align}
The field combination $\theta_{3\perp}$ can nevertheless move, and this motion causes small forced oscillations of $\theta_3$, leading to a small amount of dissipation. In total, the monodromy mode approximately satisfies
\begin{align}
    0&\approx\theta_{3\perp}''+\left(\frac{3}{2\tau} + \frac{1}{D(\tau)}\right)\theta_{3\perp}'+2\theta_{3\perp}\,,\qquad D(\tau)\gtrsim \sqrt{3}\,,
    \label{eq:EOM_epoch_1}
\end{align}
where $'$ denotes differentiation with respect to $\tau$ and the condition on $D(\tau)$ is chosen to continuously match onto the later weak damping regime (as discussed below). 
The sphaleron friction coefficient is 
\begin{align}
    D(\tau)\equiv\frac{\Upsilon_d}{\mu f^2}
=\frac{\gamma^2}{\tau^{3/2}\log^5(\zeta/\tau)}\Theta(\tau_{\rm th}-\tau)\,,
\end{align}
with
\begin{align}
    \label{eqn:D-Units}
    \gamma^2=\frac12\left(\frac{12\pi}{11}\right)^5\frac{\zeta^{3/2}\Lambda_d^3}{\mu f^2}\,,\qquad\zeta=\frac{3}{\pi\sqrt{g_\star}}\sqrt{\frac52}\frac{\Mpl\mu}{\Lambda_d^2}\,,
\end{align}
and $\tau_{\rm th}$ denotes the time when the sphaleron friction is cut off
\begin{align}
    \tau_{\rm th} \equiv \frac{\zeta}{R_{\rm th}^2}\,.
\end{align}
The onset of monodromy oscillations is then given by
\begin{align}
    \tau_{\rm osc}\approx\frac{3}{2\sqrt{2}}\,,
\end{align}
provided that $D(\tau_{\rm osc})\gtrsim \sqrt{3}$ such that the sphaleron friction is negligible at the oscillation time, if compared to Hubble [see \cref{eq:EOM_epoch_1}].
Fixing the timescale of the first oscillation gives $\Lambda_{\rm KS}\propto f^{1/2}$.

(2) Well before dark confinement, the dark sphaleron friction decreases as $T^3$. As the friction weakens, $\theta_3$ participates increasingly in the monodromy oscillation. In the weak-friction limit, the monodromy mode approximately satisfies 
\begin{align}
    \label{eqn:KS-EOM-Simple}
    0&\approx\theta_{\rm KS}''+\left(\frac{3}{2\tau}+\frac13D(\tau)\right)\theta_{\rm KS}'+3\theta_{\rm KS}\,,\qquad D(\tau)\lesssim \sqrt{3}\,,
\end{align}
The dimensionless energy-damping rate of the oscillating monodromy mode approaches $1/D$ for strong friction and $D/3$ for weak friction. We match these limits at $D(\tau_\star)=\sqrt3$, as described in \cref{sec:WKB}.
The matching time between the strong and weak damping limits is
\begin{align}
    \label{eqn:tau_osc_friction}
    \tau_\star&=\zeta\,\exp\!\left[\frac{10}{3}W_{-1}\!\left(-\frac{3^{9/10}\gamma^{2/5}}{10\zeta^{3/10}}
    \right)\right]\,.
\end{align}

In the WKB approximation, the amplitude of the monodromy mode after $\tau_\star$ therefore evolves as
\begin{align}
    \label{eqn:theta_3_WKB}
    \langle\theta_{\rm KS}^2\rangle &\approx\frac12\theta_{3\perp}^2(0)\left(\frac{\tau}{\tau_{\rm osc}}\right)^{-3/2}\exp\!\left[-\frac13\int_{\tau_\star}^{\min\{\tau,\tau_{\rm th}\}}{\rm d}\tau'\,D(\tau')-\int_{\tau_{\rm osc}}^{\tau_\star}{\rm d}\tau'\,D^{-1}(\tau')
    \right]\,,
\end{align}
where dilution with volume begins at the onset of oscillations. Here, we assume that the dark sphaleron friction shuts off at a time $\tau_{\rm th}$ before the first level crossing $\tau_{3,2}$, which we account for with the minimum in the integration limits. 

If the dark instanton potential barriers become comparable to the energy stored in the monodromy oscillations while the monodromy amplitude still spans many windings, the chaotic crossing dynamics discussed in \cref{subsec:monodromy_crossing} become relevant. When the monodromy has already unwound, the subsequent crossing can instead be treated as an avoided level crossing of approximately harmonic modes~\cite{Daido:2015cba,Cyncynates:2023esj,Dunsky:2025sgz}. In favor of determinism and to avoid treating these cases separately, we restrict to the regime in which the monodromy unwinds before the light mode $\theta_{\rm KS\perp}$ begins to relax appreciably.

(3) The avoided level crossing between the monodromy and dark confining mass eigenstates occurs at a time $\tau_{3,2}$ determined by
\begin{align}
    \chi_d\approx\Lambda_{\rm KS}^4\,.
\end{align}
For pure ${\rm SU}(N_d)$, the zero-temperature topological susceptibility is approximately
\begin{align}
    \chi_d(0)\approx0.16\,\Lambda_d^4\,,
\end{align}
where the numerical coefficient is nearly independent of $N_d$~\cite{Lucini:2012wq,Berg:2017tqu,Bonanno:2025eeb}.\footnote{More precisely, lattice results give $\chi_d(0)\approx0.16\,T_{\rm conf}^4$, where $T_{\rm conf}$ is the confinement temperature. Here we make the coarse identification $T_{\rm conf}\approx\Lambda_d$, appropriate for the accuracy of our analytic estimate.}
For definiteness, we take $N_d=4$ in this estimate and the numerical estimates in \cref{sec:three_axions_numerical}. Using the corresponding high-temperature scaling of $\chi_d(T)$ (see \cref{app:topological_susceptibility}), we find
\begin{align}
    \tau_{3,2}\approx\zeta\left(\frac{\Lambda_{\rm KS}^4}{0.16\Lambda_d^4}\right)^{3/16}\,.
\end{align}
We take the residual misalignment of $\theta_{{\rm KS}\perp}$ relative to the nearby dark-potential minimum to be $\theta_{{\rm KS}\perp}(\tau_{\rm rel})={\cal O}(1)$, where $\tau_{\rm rel}$ is the time when the monodromy has relaxed to within one period of the global minimum.

Once the large monodromy oscillation has relaxed, the light mode $\theta_{\rm KS\perp}$ has mass squared $2\chi_d/(3f^2)$ and dimensionless sphaleron friction $2D/3$. Its oscillations begin at a time $\tau_{d,\rm osc}<\tau_{3,2}$, which we estimate by (again, restricting to $N_d = 4$)
\begin{align}
    \label{eqn:light_osc}
    \sqrt{\frac23}\left(\frac{\tau_{d,\rm osc}}{\tau_{3,2}}\right)^{8/3}\sim\frac{3}{2\tau_{d,\rm osc}}+\frac23D(\tau_{d,\rm osc})\,.
\end{align}
Before oscillations begin, the light mode can already relax toward its minimum while remaining overdamped. This early relaxation is generally a small correction when Hubble friction dominates, but can appreciably suppress the misalignment when dark sphalerons delay oscillations. 
Neglecting $\theta_{\rm KS\perp}$'s acceleration and anharmonic corrections during this overdamped period, we estimate its displacement at the $\tau_{d,\rm osc}$ as
\begin{align}
    \label{eqn:light_relaxation}
    \theta_{\rm KS\perp}^2(\tau_{d,\rm osc})\approx{}&\theta_{\rm KS\perp}^2(\tau_{\rm rel})\exp\!\left[-\frac43\int_{\tau_{\rm rel}}^{\tau_{d,\rm osc}}{\rm d}\tau\,\frac{(\tau/\tau_{3,2})^{16/3}}{\frac{3}{2\tau}+\frac23D(\tau)}
    \right]\,.
\end{align}
The lower limit of integration is appropriate to the restriction where $\theta_{\rm KS\perp}$ remains approximately frozen until the monodromy has unwound. 

After the light-mode oscillations begin, its amplitude evolves in the WKB approximation. Since its mass grows as $\tau^{8/3}$, the oscillation-averaged squared amplitude at the monodromy-dark crossing is
\begin{align}
    \label{eqn:light_WKB}
    \langle\theta_{{\rm KS}\perp}^2(\tau_{3,2})\rangle\approx{}&\frac12\theta_{{\rm KS}\perp}^2(\tau_{d,\rm osc})\left(\frac{\tau_{d,\rm osc}}{\tau_{3,2}}\right)^{25/6}\exp\!\left[-\frac23\int_{\tau_{d,\rm osc}}^{\tau_{3,2}}
    {\rm d}\tau\,D(\tau)
    \right]\,.
\end{align}
The power $25/6=3/2+8/3$ accounts for dilution with volume and the growing mass for $N_d=4$, while the exponential accounts for sphaleron damping during the oscillatory regime.

At the dark crossing, the initially heavy monodromy mode follows the heavy eigenstate and is identified post-crossing with $\theta_3$. This mode continues to gain energy from the plasma as the dark instanton potential continues to grow towards confinement.\footnote{Note that we have assumed dark sphaleron friction has shut off before the crossing, so no additional energy is lost to dissipation.}
The resulting relic abundance is then
\begin{align}
    \frac{\Omega_d}{\Omega_{\rm DM}}\sim \frac{\frac12\Lambda_{\rm KS}^4\langle\theta_{\rm KS}^2(\tau_{3,2})\rangle}{\frac54\Heq^2\Mpl^2}\left(\frac{\tau_{3,2}}{\tau_{\rm eq}}\right)^{3/2}\left(\frac{\chi_d(0)}{3\Lambda_{\rm KS}^4}\right)^{1/2}\,,
\end{align}
where the last term accounts for the growth of $\chi_d$.

On the other hand, the pre-crossing light mode $\theta_{\rm KS\perp}$ follows the light eigenstate through the crossing, and is identified post-crossing with $\theta_{3\perp}$, which gets its mass from the monodromy potential. Neglecting order-one matching factors through the crossing, $\theta_{3\perp}$'s energy density subsequently dilutes as cold matter:
\begin{align}
    \label{eqn:rho_mon}
    \rho_{3\perp}\sim\Lambda_{\rm KS}^4\langle\theta_{\rm KS\perp}^2(\tau_{3,2})\rangle\left(\frac{\tau}{\tau_{3,2}}\right)^{-3/2}\,.
\end{align}

Next, at a time $\tau_{2,1}$, the remaining monodromy mode $\theta_{3\perp}$ undergoes an avoided crossing with the QCD mode. The crossing occurs when
\begin{align}
    \chi_c\approx\Lambda_{\rm KS}^4\,,
\end{align}
giving
\begin{align}
    \tau_{2,1}\approx2.7\times10^4\left(\frac{f}{10^{8.3}\,{\rm GeV}}\right)^{1/2}\,.
\end{align}

The resulting late-time QCD axion abundance, expressed as a fraction of the observed dark matter abundance, is approximately
\begin{align}
    \frac{\Omega_c}{\Omega_{\rm DM}}\sim\frac{\Lambda_{\rm KS}^4\langle\theta_{{\rm KS}\perp}^2(\tau_{3,2})\rangle}{\frac54\Heq^2\Mpl^2}\left(\frac{\tau_{\rm eq}}{\tau_{3,2}}\right)^{-3/2}\left(\frac{\tau_{c,0}}{\tau_{2,1}}\right)^2\,,
\end{align}
where the factor $\tau_{c,0}^2/\tau_{2,1}^2$ accounts for the growing QCD instanton potential, $\tau_{\rm eq}$ is the dimensionless time of matter--radiation equality, $\rho_{\rm DM}(t_{\rm eq})$ is the observed dark matter density evolved back to equality, and $\tau_{c,0}$ is the time when the QCD susceptibility reaches its constant value, and we have taken dark matter to be $5/6$ the total matter density.

The late-time monodromy-like mode $\theta_1$, on the other hand, inherits its energy from the pre-crossing QCD-like mode $\theta_{2}-\theta_1$. Before the crossing, this mode's energy density grows with the QCD topological susceptibility. The crossing suppresses its final abundance relative to ordinary QCD misalignment. For an ${\cal O}(1)$ misalignment, the ordinary QCD axion accounts for the dark matter at $f\sim10^{12}\,{\rm GeV}$, so $f\sim10^9\,{\rm GeV}$ would typically yield only a few $\times10^{-4}$ of the observed abundance. The crossing further suppresses this already small contribution.

\begin{table*}[t]
    \centering
    \renewcommand{\arraystretch}{1.15}
    \setlength{\tabcolsep}{5.5pt}
    \begin{tabular}{cccccccc}
        \toprule
        $f$
        & $\Lambda_d$
        & $\Lambda_{\rm KS}$
        & $R_{\rm th}$
        & $N_i$
        & $\Omega_{\rm KS}/\Omega_{\rm DM}$
        & $\Omega_c/\Omega_{\rm DM}$
        & $\Omega_d/\Omega_{\rm DM}$
        \\
        \midrule
        $5\times10^7\,\GeV$
        & $0.3\,\GeV$
        & $1.8\,\MeV$
        & $24.75$
        & $2.8\times10^7$
        & $10^{-8}$
        & $10^{-3}$
        & $1$
        \\
        \bottomrule
    \end{tabular}
    \caption{Benchmark parameter point and resulting analytic relic abundances. This parameter point is indicated by the red `$\times$'s in \cref{fig:hard_cutoff_money}.}
    \label{tab:benchmark}\label{tbl:benchmark_analutical}
\end{table*}
We provide an example set of parameters that satisfies the assumptions of this estimate and gives the observed dark matter abundance within the level of accuracy of our approximations in \cref{tbl:benchmark_analutical}. We find that the heavy axion $\theta_3$ aligned with the dark confining potential dominates the relic abundance. In the subsequent section, we will find that numerical integration of the full set of equations agrees in the case that the friction is taken to be the high-temperature asymptotic expression and then cut off prior to confinement. On the other hand, if the friction is taken to be the lattice extrapolation, we will find that the dynamics are substantially different, and the QCD axion dominates the dark matter abundance.

\subsection{Parameter space numerical results}
\label{sec:three_axions_numerical}

The discussion in \cref{sec:three_axions_parameter_space_estimate} provides an example of the evolution of the full three-axion system from the electroweak phase transition through the QCD phase transition. The WKB approximation allowed us to efficiently evolve the mode amplitudes as the growing instanton contributions and sphaleron damping changed the instantaneous field basis. At the transitions between WKB regimes, we assumed adiabatic mode crossings, an approximation that was accurate under the specific conditions considered there.

In \cref{app:fast_three}, we generalize this approach to perform a limited parameter scan for both the high-temperature asymptotic friction with a cutoff and the lattice-extrapolated friction discussed in \cref{subsec:sphaleron_rate_summary}. To accurately resolve the generation of the baryon asymmetry, we numerically integrate the full multi-axion--Boltzmann system down to $60\GeV$ before switching to the inexpensive WKB-based solver. The resulting baryon and axion evolution are visible in the right insets labeled (c) in \cref{fig:hard_cutoff_money,fig:monodromy_crossing_money} evaluated at benchmark points in the parameter space (four colorful dots in the main panel). For the parameter scans, this solver uses simplified prescriptions for the mode crossings, which we describe below.

The low computational cost of the approximate solver allows us to explore departures from the equal-$C_{ij}$, equal-$f_i$ benchmark studied in \cref{sec:three_axions_parameter_space_estimate}. We use a large language model (LLM) agent to perform a semantic optimization (see \cref{app:semantic}) of the scan parameters, identifying low-dimensional parameter slices whose projection covers a relatively broad region of the $(f_{\rm eff,i}^{-1},m_i)$ plane visible as the colored contours in \cref{fig:hard_cutoff_money,fig:monodromy_crossing_money}. Here, $f_{\rm eff,i}^{-1}$ are the effective inverse QCD coupling decay constants, defined below in \cref{eq:QCD_coupling}. The LLM uses the physical interpretation of the scan parameters and the outcome of each trial point to propose parameter adjustments motivated by the dynamics of the system. We report the resulting parameter slices for each friction prescription below.

Because the inexpensive solver treats the mode crossings approximately, we validate the resulting parameter regions with a set of more costly simulations that explicitly resolve the crossings. For each representative case, we initialize the resolved calculation at a parameter point selected from the approximate scan. Resolving the crossings generally shifts the predicted dark matter relic abundance slightly. We then use the same LLM-guided optimization procedure to adjust the parameters until the relic abundance lies within a specified tolerance of the observed value. The proximity of these resolved points to those obtained from the approximate scan provides a direct check that the simplified crossing prescriptions locate the viable parameter regions with reasonable accuracy. The resolved axion energy density versus time curves are visible for selected parameter points in the left (b) insets of \cref{fig:hard_cutoff_money,fig:monodromy_crossing_money}, with corresponding $(f_{\rm eff,i}^{-1},m_i)$ parameter points indicated on the plots.

The final relic abundance and its partition among the three axions depend sensitively on the treatment of the dark-sector friction $\Upsilon_d$ in the poorly understood regime above confinement. We therefore present results for both prescriptions introduced in \cref{subsec:sphaleron_rate_summary}: the high-temperature asymptotic friction rate terminated at a phenomenological cutoff $T_{\rm cut}=R_{\rm th}\Lambda_d$ (\cref{fig:hard_cutoff_money}), and the near-confinement lattice result extrapolated to higher temperatures (\cref{fig:monodromy_crossing_money}).

\subsection*{Asymptotic friction with cutoff}

\begin{figure}
    \centering
    \includegraphics[width=\columnwidth]{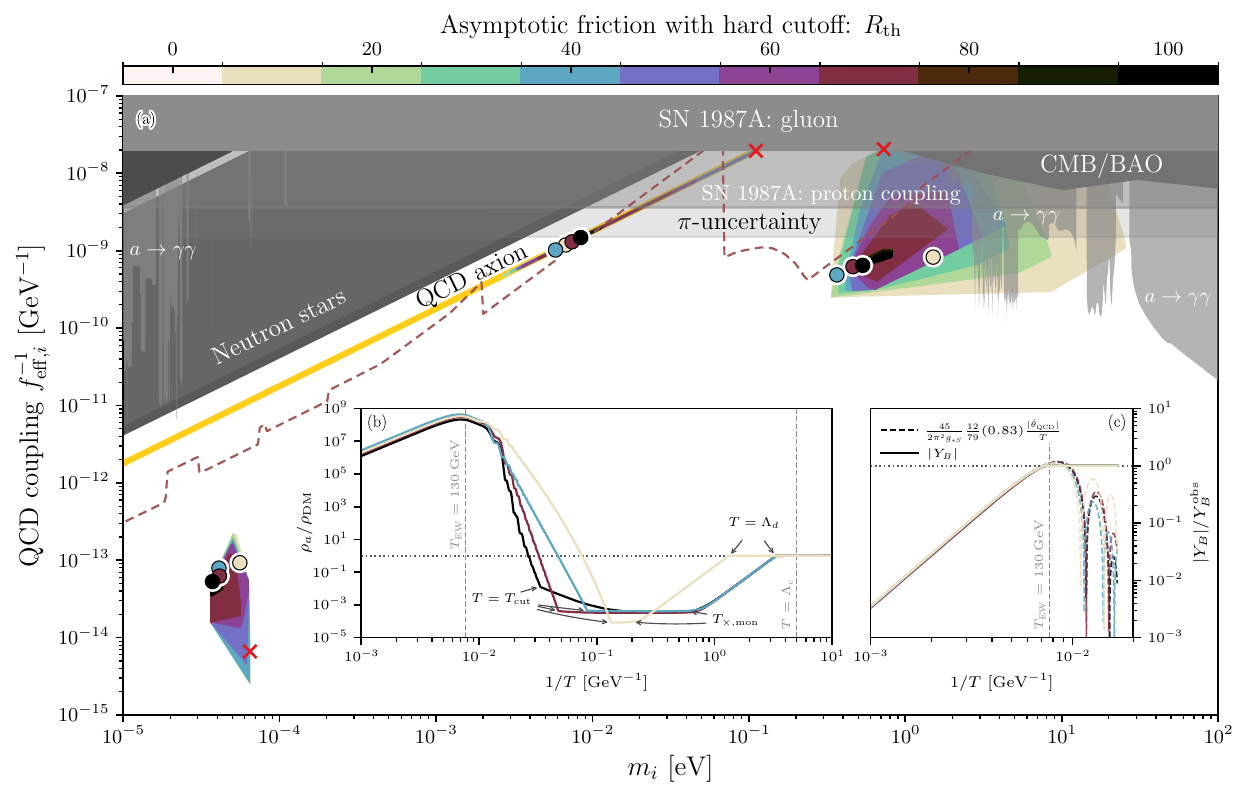}
    \caption{
    Available parameter space in the late-time mass--QCD-coupling plane, with dark sphaleron friction modeled by the high-temperature asymptotic form of \cref{eqn:sphaleron_rate_with_coefficient} and cut off at $T_{\rm cut}=R_{\rm th}\Lambda_d$. For each mass eigenstate $i$, the axes show $m_i$ and the effective inverse QCD decay constant $f_{{\rm eff},i}^{-1}$ defined in \cref{eq:QCD_coupling}. Colored contours show the approximate parameter space obtained with the low-cost solver: each sample is evolved exactly through the electroweak era to $T=60\,{\rm GeV}$, with the initial monodromy displacement calibrated to reproduce the observed baryon asymmetry (\cref{sec:monodromy}), then continued schematically to $T=0.01\,{\rm GeV}$ using the WKB/crossing prescription of \cref{sec:three_axions_numerical}. We require $\Lambda_d\gtrsim 0.3\,{\rm GeV}$ for consistency with glueball decay limits (\cref{sec:depletion_of_DS}). Shading indicates points with final axion abundance below the observed dark matter abundance. Only the heavy axion, corresponding to the rightmost colored regions, constitutes an ${\cal O}(1)$ fraction of the dark matter. Colored markers show representative points with $R_{\rm th}=10,40,70,$ and $100$ for which we perform fully resolved numerical evolutions of the level crossings. Insets show the total axion abundance (b) and baryon yield versus inverse temperature (c), with curve colors corresponding to the marked points. In (c), solid curves show $|Y_B|/Y_B^{\rm obs}$ and dashed curves the axion-velocity equilibrium prediction of \cref{eq:fin_B_val}, where $\dot\theta_{\rm QCD}$ is the velocity projected onto the QCD direction. The vertical line marks sphaleron freeze-out and horizontal dotted lines the observed abundances.
    In (a), the yellow diagonal shows the canonical QCD-axion coupling. Solid gray regions show existing cosmological and astrophysical exclusions on the axion-gluon coupling~\cite{Hook:2017psm,Gomez-Banon:2024oux,Kumamoto:2024wjd,Blum:2014vsa,Springmann:2024ret,Balkin:2022qer,Caloni:2022uya}. The bands labeled ``SN1987A: proton coupling'' and ``$\pi$-uncertainty'' show, respectively, the reach when the proton coupling takes its natural value and the uncertainty from the unknown pion abundance in the supernova~\cite{Lella:2023bfb}. Transparent gray regions labeled $a\to\gamma\gamma$ show cosmological and astrophysical bounds (though see \cref{sec:axion_signatures}) with $g_{a\gamma\gamma}=\alpha_{\rm EM}/2\pi f_{\rm eff}$ for axions composing all the dark matter~\cite{Gemmell:2026xaa,Saha:2025any,Yin:2024lla,Regis:2020fhw,Todarello:2023hdk,Wang:2023imi,Grin:2006aw,Todarello:2024qci,Carenza:2023qxh,Nakayama:2022jza,Porras-Bedmar:2024uql,Bernal:2022xyi,Wadekar:2021qae,Liu:2023nct,Capozzi:2023xie,Bolliet:2020ofj} (or a thermally produced subcomponent~\cite{Cadamuro:2011fd}) with haloscope exclusions~\cite{Wuensch:1989sa,ADMX:2021mio,ADMX:2018ogs,MADMAX:2024sxs,GigaBREAD:2025lzq,Grenet:2021vbb,Ahyoune:2024klt,CAST:2020rlf,QUAX:2024fut,QUAX:2023gop,Alesini:2022lnp,Alesini:2020vny,Alesini:2019ajt,Quiskamp:2024oet,Quiskamp:2023ehr,Quiskamp:2022pks,McAllister:2017lkb,TASEH:2022vvu,Adair:2022rtw,Bae:2024kmy,Kim:2023vpo,Yang:2023yry,Yoon:2022gzp,Lee:2022mnc,CAPP:2020utb,Jeong:2020cwz,HAYSTAC:2024jch,HAYSTAC:2023cam,HAYSTAC:2020kwv,HAYSTAC:2018rwy,Brubaker:2016ktl}. Red dashed contours show projected sensitivities~\cite{Schutte-Engel:2021bqm,Qiu:2025bbi,McAllister:2017lkb,Beurthey:2020yuq,Baryakhtar:2018doz,Alesini:2017ifp,Alesini:2023qed,Fan:2024mhm,Heinze:2024bdc,Obata:2018vvr,DeMiguel:2023nmz,Aja:2022csb,BREAD:2021tpx,Ahyoune:2023gfw,ALPHA:2022rxj,Lawson:2019brd,Nagano:2019rbw,Graham:2015ouw,Stern:2016bbw,Liu:2018icu}. Red `$\times$' from \cref{tbl:benchmark_analutical}.
    }
    \label{fig:hard_cutoff_money}
\end{figure}

The parameter space available under the high-temperature asymptotic friction prescription of \cref{subsec:sphaleron_rate_summary} is illustrated in \cref{fig:hard_cutoff_money}. We perform an inexpensive scan over a restricted parameter slice. We define
\begin{align}
\Lambda_{\rm ref}(f_3) \equiv 4.38\times10^{-3}\,{\rm GeV} \sqrt{\frac{f_3}{3\times10^8\,{\rm GeV}}}\,,
\end{align}
and fix $N_d=4$, with no dark sector charged fermions in the plasma, $\chi_d(0)=0.16\Lambda_d^4$, $\Lambda_c=0.2\,{\rm GeV}$, and $\chi_c(0)=(0.075\,{\rm GeV})^4$. The initial conditions are $\theta_{\rm QCD}=0.3$ and $\theta_d=0$, with all velocities vanishing. For each point, the initial monodromy displacement is adjusted to reproduce the observed baryon asymmetry, $|Y_B|_{\rm obs}$. We retain points satisfying $\rho_a\leq\rho_{\rm DM}$ at $T=0.01\,{\rm GeV}$. The remaining parameters are scanned as summarized in \cref{tab:hard_cutoff_scan}.

\begin{table}[t]
\centering
\small
\renewcommand{\arraystretch}{1.15}
\begin{tabular}{@{}ll@{}}
\toprule
Parameter & Values or relation \\
\midrule
Monodromy scale
  & $\Lambda_{\rm KS}=\Lambda_{\rm ref}(f_3)$ \\
Decay constants
  & $(f_1,f_2)=f_3(1.2^a,10^a)$ \\
Coefficients
  & $(C_{12},C_{13},C_{23})
     =(10^a,(16/10)^a,16^{-a})$ \\
Range of $a$
  & $0\leq a\leq1.4$ \\
Range of $f_3$
  & $(2.2\times10^7\text{--}2.6\times10^8)\,{\rm GeV}$ \\
Range of $\Lambda_d$
  & $(0.3\text{--}2.2)\,{\rm GeV}$ \\
$R_{\rm th}$
  & $\{10,20,\ldots,100\}$ \\
\midrule
Additional choices at large $R_{\rm th}$
  & $(f_1,f_2)=f_3(1.2q,10)$ \\
  & $(C_{12},C_{13},C_{23})
     =(10,10 q/16,1/(16q))$ \\
Refined range of $q$
  & $0.85\leq q\leq1.50$ \\
Other variations
  & Correlated changes in $f_2,C_{12},C_{23}$, \\
  & changes in $C_{12}$ alone, and interpolation \\
  & between nearby parameter choices \\
\bottomrule
\end{tabular}
\caption{Parameter choices for the hard-cutoff scan, obtained
by semantic optimization (\cref{app:semantic}). The ranges
include subsequent searches near the boundaries of the
allowed regions. Sampling is nonuniform, and the two
parametrized families are explored separately rather than
as a Cartesian product. Of 2857 tested points, 1544 are
accepted. Stochastic monodromy--cosine crossings occur in
13 accepted points ($0.84\%$), all at $R_{\rm th}=10$;
their post-crossing energy partition is chosen a posteriori
to satisfy the abundance and radiation bounds.}
\label{tab:hard_cutoff_scan}
\end{table}

The high-temperature asymptotic friction [\cref{eqn:sphaleron_rate_with_coefficient}] can produce strong dissipation at early times. Consequently, throughout much of the parameter space that avoids overclosure, the initial monodromy winding relaxes shortly after oscillations begin. The left inset (b) of \cref{fig:hard_cutoff_money} illustrates this behavior as the energy-density curves fall rapidly before $T=10\,{\rm GeV}$. Dissipation does not completely erase the monodromy energy, however, because the friction shuts off at $T=T_{\rm cut}=R_{\rm th}\Lambda_d$, which is visible as the first kink in the energy density versus temperature curves. The surviving monodromy component then continues to dominate until its avoided level crossing with the dark-instanton mode at $T_{\times,{\rm mon}}$, visible as the second kink in the energy density curves. 
The comoving energy density of the axion system asymptotes near $T \sim 
\Lambda_d$ as indicated in the inset (b) of the figure.

Each of the four fully resolved numerical trajectories corresponds to three points of the same color in the $(f_{{\rm eff},i}^{-1},m_i)$ plane, one for each late-time mass eigenstate. Here, $f_{{\rm eff},i}^{-1}$ refers to the effective QCD coupling decay constant. For the canonically normalized fields $\phi_i=f_i\theta_i$,
\begin{align}\label{eq:QCD_coupling}
f_{{\rm eff},i}^{-1} =|\bm q^{T}\hat{\bm e}_i|,
\end{align}
where $\bm q^T=(0,1/f_2,C_{23}/f_3)$ is the coupling vector and $\hat{\bm e}_i$ is its unit-normalized mass eigenvector of the zero-temperature mass matrix. 

As in \cref{sec:three_axions_parameter_space_estimate}, the heavy state constitutes most of the dark matter and may lie within reach of LAMPOST~\cite{Baryakhtar:2018doz}, as well as searches for axion decays~\cite{Sun:2023acy,Bessho:2022yyu,Carenza:2023qxh,Capozzi:2023xie}. The QCD axion typically contributes an ${\cal O}(10^{-3})$ fraction of the dark matter. Although the highlighted regions overlap the projected experimental reach of laboratory probes (dashed red), its suppressed abundance makes astrophysical processes that directly produce the axion, such as SN1987A~\cite{Lella:2023bfb} and future supernovae, the most promising probes. The monodromy state has an even smaller abundance and is more weakly coupled, owing to the mass-ratio-suppressed mixing induced through QCD~\cite{Demirtas:2021gsq,Agrawal:2022lsp,Agrawal:2024ejr,Agrawal:2025rbr,Gendler:2023kjt,Baryakhtar:2026oun}.

\subsection*{Lattice-extrapolated friction}

\begin{table}[t]
\centering
\small
\renewcommand{\arraystretch}{1.15}
\begin{tabular}{@{}ll@{}}
\toprule
Parameter & Values or relation \\
\midrule
Decay constants
 & $(f_1,f_2)=f_3(r\,1.2^a,10^a)$ \\
Coefficients
 & $(C_{12},C_{13},C_{23})
    =(A\,10^a,D,b\,16^{-a})$ \\
Monodromy scale
 & $\Lambda_{\rm KS}=\ell\,\Lambda_{\rm ref}(f_3)$ \\
Range of $f_3$
 & $[5\times10^7,3\times10^8]\,{\rm GeV}$ \\
Range of $\Lambda_d$
 & $[1.2,7]\,{\rm GeV}$ \\
Range of $a$
 & $[\log_{10}6,1]$ \\
Ranges of $A,D$
 & $A\in[0.75,1.25]$, $D\in[1.1,2.1]$ \\
\midrule
Initially fixed ratios
 & $r=b=\ell=1$ \\
Additional variations
 & $r\in[0.95,1.10]$, $b\in[0.80,1.05]$ \\
 & $\ell\in[0.85,1.15]$ \\
Values of $R$
 & $-10,-8,\ldots,10$ \\
\bottomrule
\end{tabular}
\caption{Parameters used in the lattice-extrapolation scan, obtained by semantic optimization (\cref{app:semantic}), for which the corresponding $(f_{\rm eff,i}^{-1}, m_i)$ parameter space is presented in \cref{fig:monodromy_crossing_money}. The initial search fixed $r=b=\ell=1$; these ratios were then varied around selected examples. Additional points were tested near the edges of the allowed regions, including small departures from the listed ranges. Each displayed bound on $|R|$ includes points satisfying the relic upper bound for at least one tested $R$ within that bound.}
\label{tab:lattice_extrapolation_scan}
\end{table}

\begin{figure}
    \centering
    \includegraphics[width=\columnwidth]{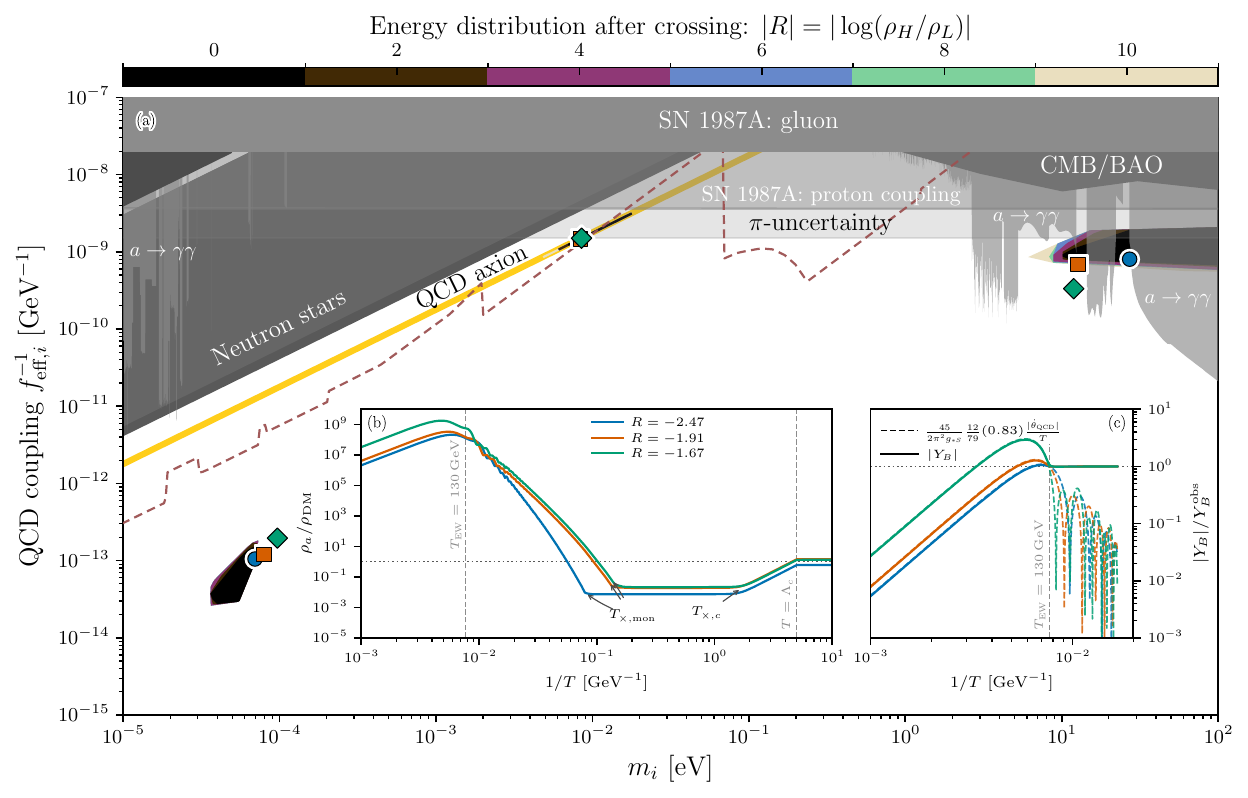}
    \caption{Available parameter space in the late-time mass--QCD-coupling plane using the lattice-extrapolated dark sphaleron friction prescription~\cref{subsec:sphaleron_rate_summary}. Compared to \cref{fig:hard_cutoff_money}, the weaker friction leaves substantial monodromy energy at the dark-instanton--monodromy level crossing, leading to the effectively stochastic crossing dynamics discussed in \cref{subsec:monodromy_crossing}. Post crossing, both the monodromy and dark states carry ${\cal O}(1)$ of the energy density, but the heavy dark-aligned state is strongly depleted by dark sphaleron damping. However, the heavy state does have a significant though still subdominant thermal relic abundance, possibly putting it in reach of axion decay searches. The monodromy state then crosses with QCD: the outgoing QCD axion state acquires the monodromy energy, and thus constitutes the dominant dark matter component. The colored contours show the approximate parameter space obtained with the low-cost solver, and the corresponding post-crossing energy partition required to avoid overproducing the dark matter, while the colored markers indicate representative points evolved with fully resolved numerical simulations. 
    The inset plots show the total axion abundance (b) and baryon yield versus inverse temperature (c). Different color curves of insets (b) and (c) correspond to the points marked in the three axion parameter space by the same color. The resolved trajectories in (b) account for an ${\cal O}(1)$ fraction of the observed dark matter abundance, but do not reproduce it exactly because small changes in the initial conditions and model parameters can produce order-one changes in the post-crossing energy partition, making precise numerical convergence difficult. Details of the exclusions and prospective sensitivity contours are given in the caption of \cref{fig:hard_cutoff_money}.}
    \label{fig:monodromy_crossing_money}
\end{figure}

The parameter scan using the lattice-extrapolated friction prescription of \cref{subsec:sphaleron_rate_summary} is shown in \cref{fig:monodromy_crossing_money}. We use the same fixed values as before: $N_d=4$, with no charged fermions in the plasma, $\chi_d(0)=0.16\Lambda_d^4$, $\Lambda_c=0.2\,{\rm GeV}$, and $\chi_c(0)=(0.07546\,{\rm GeV})^4$. The initial conditions are $\theta_{\rm QCD}=0.3$ and $\theta_d=0$, with all velocities vanishing. For each point, the initial monodromy displacement is adjusted to reproduce the observed baryon asymmetry, $|Y_B|_{\rm obs}$, and we retain points satisfying $\rho_a\leq\rho_{\rm DM}$ at $T=0.01\,{\rm GeV}$. The remaining parameters are chosen differently from the high-temperature asymptotic scan and are summarized in \cref{tab:lattice_extrapolation_scan}.

The inset curves in \cref{fig:monodromy_crossing_money} show the weaker dissipation produced by the lattice-extrapolated friction relative to the high-temperature asymptotic expression: the energy densities decay much more slowly than in \cref{fig:hard_cutoff_money}. By the avoided level crossing between the dark-instanton and monodromy states, the monodromy has not yet fully unwound. The system therefore undergoes the effectively stochastic monodromy--cosine crossing dynamics discussed in \cref{subsec:monodromy_crossing}.

For the resolved example shown, $R=\log(\rho_H/\rho_L)<0$, so the monodromy state carries most of the energy density after the dark crossing. In contrast to the setup of \cref{subsec:monodromy_crossing}, dark sphaleron damping remains active and continues to dissipate the $\chi_d$-aligned state after the crossing. The monodromy state is effectively undamped and therefore continues to dominate the energy density until the QCD crossing. After that crossing, the QCD axion inherits most of the energy density, which then grows as $\chi_c$ approaches its confinement value. The final dark matter abundance is consequently dominated by the QCD axion. The homogeneous component of the heavy dark-aligned state becomes a negligibly small subcomponent because of the continued dark-sector damping, though the heavy axion does obtain a small thermal component which we discuss in \cref{sec:axion_signatures}, though we leave its precise determination to future work. The final monodromy state carries approximately the same small fraction of the energy density as in the high-temperature asymptotic case. Its late-time abundance is ultimately sourced by the QCD axion, whose dynamics are similar in the two friction prescriptions.

The resolved trajectories lie approximately within the regions identified by the coarse parameter scan, although the green point falls noticeably outside them. In addition, the blue trajectory does not converge precisely to the observed dark matter abundance. We attribute both discrepancies to the effectively stochastic level-crossing dynamics. Small changes in the initial conditions or model parameters can substantially alter the post-monodromy-crossing partition of the energy density, making precise numerical convergence difficult.

The dominant QCD axion dark matter is accessible to astrophysical probes, including SN1987A~\cite{Lella:2023bfb}, as well as to the BREAD experiment~\cite{BREAD:2021tpx}. The heavy dark-aligned axion has no appreciable relic abundance arising from misalignment. However, because its mass is quite large, the heavy axion has a significant though still-subdominant thermally-produced abundance, and therefore its decay signatures are possibly within reach of future searches~\cite{Sun:2023acy,Bessho:2022yyu,Carenza:2023qxh,Capozzi:2023xie}. Its parameter space also lies close to existing astrophysical bounds, so it may be within reach of future supernova observations. See \cref{sec:axion_signatures} for further discussion of the thermal subcomponent. \\

Taken together, \cref{fig:hard_cutoff_money,fig:monodromy_crossing_money} show that both the viable parameter space and the identity of the dominant dark matter component—whether the heavier dark-aligned axion or the QCD axion—depend sensitively on the dark-sector friction near confinement. This uncertainty motivates further lattice studies of the sphaleron rate in this regime. A comprehensive numerical scan of the full viable parameter space is also left for future work. Nevertheless, our central conclusion is robust across the friction prescriptions considered here: our multi-axion framework can simultaneously account for the observed baryon asymmetry and dark matter abundance while predicting signals in a broad range of forthcoming searches. In the next section, we examine these phenomenological signatures in greater detail.

\section{Phenomenology}\label{sec:phenomenology}

In this section we outline the main phenomenological signatures of the model described in \cref{sec:three_axions}. Achieving successful baryogenesis while avoiding overclosure and satisfying QCD axion quality pushes two of the axions into detectable regions of parameter space, as shown in \cref{fig:hard_cutoff_money,fig:monodromy_crossing_money}. This gives rise to a variety of potentially observable signatures. These signatures fall into two broad classes. First, the confining dark sector itself can be probed through its relic glueball population, its decay products, and possible collider signatures; we discuss these in \cref{sec:depletion_of_DS}. Second, the axions give rise to direct-detection, cosmological, and astrophysical signals; we discuss these in \cref{sec:axion_signatures}.

\subsection{Dark glueball and meson signatures}
\label{sec:depletion_of_DS}

In this section, we review relevant constraints on dark glueballs and dark mesons. 
In particular, we argue that ensuring glueballs evade BBN bounds~\cite{Jedamzik:2006xz,Kawasaki:2020qxm} puts a lower bound on the dark confinement scale. 
Furthermore, achieving sufficiently rapid glueball decays requires dark quarks and mesons with masses not far above current LHC limits, thereby giving rise to promising signals in upcoming searches.

We solved the axiogenesis overclosure problem with the friction from a new confining dark sector. 
We assumed that the dark sector maintains a temperature comparable to that of the SM thermal bath. 
In the absence of additional interactions between the dark sector and the SM, the resulting dark glueballs would be stable and overclose the Universe~\cite{Forestell:2016qhc,Halverson:2016nfq}.\footnote{In Ref.~\cite{Asadi:2025btr}, it was argued that the glueball overclosure problem can be circumvented in models with a sufficiently low reheat temperature $T_{\mathrm{RH}}$, typically below the electroweak phase transition scale. This is not a viable solution in our setup, since the generation of the baryon asymmetry relies on high enough temperatures in the early Universe where electroweak sphalerons are active.}
This can be resolved by introducing new degrees of freedom to enable glueball decays.

\begin{figure}
    \centering
    \includegraphics[width=0.4\linewidth]{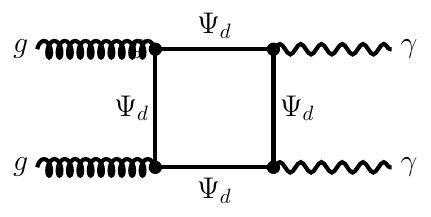}
    \caption{The loop diagram allowing the lightest dark glueballs to decay to two SM photons via bi-fundamental dark quark loops. 
    }
    \label{fig:GB_decay}
\end{figure}

We augment the model by adding dark quarks $\Psi_d$ of mass $m_\Psi$ that are bifundamentals of the dark confining group and the SM SU(2)$_L$. 
The existing LEP bounds force $m_\Psi \gtrsim 100$~GeV. 
For the dark-quark masses and values of $\Lambda_d$ considered here, the quarks do not appreciably modify the friction during the relevant axion evolution, leaving the baryogenesis and relic-abundance analyses of the preceding sections unchanged.\footnote{For the dark-quark masses considered here, the friction coefficient has already approached its pure Yang--Mills value by the onset of appreciable axion motion, which occurs not too far above $\Tew$. The heavy quarks therefore have a negligible effect on the subsequent evolution.}
It has been shown that this portal allows CP-even glueballs in the spectrum to decay to the SM via dimension-8 operators \cite{Juknevich:2009gg} once the dark quarks are integrated out, see \cref{fig:GB_decay}.\footnote{Heavier glueballs have exponentially suppressed abundances relative to the lightest state \cite{Forestell:2016qhc}, so we impose the decay constraint only on the lightest CP-even glueball. The dimension-eight portal does not mediate decays of C-odd glueballs protected by dark G-parity, which may also stabilize dark mesons \cite{Bai:2010qg,Asadi:2026mip}. We assume that UV physics breaks this symmetry sufficiently to make these potentially overabundant relics decay before BBN, $\tau \lesssim 0.1\,\mathrm{s}$. } 

The lifetime for the lightest glueball is given by
\cite{Juknevich:2009ji}
\begin{equation}
\tau_{\mathrm{GB}} \simeq 3600\, \frac{32 \pi}{(N_d^2-1) \, \mathfrak{q}_d^4 \, \alpha_{\mathrm{em}}^2\alpha_d^2(m_{\mathrm{GB}})}  \, \left(\frac{m_{\Psi}^3}{F_{\mathrm{GB}}^S}\right)^2 \, \left(\frac{m_\Psi}{m_{\mathrm{GB}}}\right)^2 \frac{1}{m_{\mathrm{GB}}},
    \label{eq:GB_lifetime}
\end{equation}
where $\alpha_{\mathrm{em}}$ ($\alpha_d (x)$) is the electromagnetic coupling (dark gauge coupling evaluated at the scale $x$), $\mathfrak{q}_d$ is the dark quark charge, $N_d$ is the number of colors of the dark gauge group, $m_\Psi$ is the new heavy dark quark mass, $m_{\mathrm{GB}} \simeq 7 \Lambda_d$ \cite{Morningstar:1999rf,Mathieu:2008me} is the lightest glueball mass, and 
\begin{equation}
    F_{\mathrm{GB}}^S \simeq \frac{3}{4\pi \, \alpha_d (m_{\mathrm{GB}})}  m_{\mathrm{GB}}^3 
    \label{eq:FS0}
\end{equation}
is the lightest glueball decay constant \cite{Chen:2005mg}. 
With dark quarks transforming as the fundamental of SU(2)$_L$, and reasonably small $N_d$ values, we find
\begin{equation}
    \tau_{\mathrm{GB}} \simeq 10^{-22}~\mathrm{s} \, \times \left( \frac{m_\Psi}{\Lambda_d}\right)^8 \, \times \, \left( \frac{1~\mathrm{GeV}}{\Lambda_d}\right) \,,
    \label{eq:lifetime_numerical}
\end{equation}
Requiring $\tau_{\mathrm{GB}} \lesssim 0.1$s introduces a lower bound on the dark confinement scale $\Lambda_d$. 
To estimate this bound, we should first discuss the existing bounds on the dark quark mass $m_\Psi$ more in detail.

The scalar dark meson can manifest as a microscopic quirk~\cite{Kang:2008ea}. The phenomenology of such states has recently been revisited in Ref.~\cite{Forsyth:2025wks}, where it was shown that existing LHC di-photon searches constrain masses up to $m_\Psi \sim 100~\text{GeV}$, i.e. comparable to LEP bounds. 
In the quirky region, this is the most stringent existing bound, while future LHC searches, combined with novel search techniques \cite{Curtin:2025ngf}, can further constrain this parameter space.

While other studies have obtained stronger bounds on dark bi-fundamental quarks in theories with the same field content, these bounds apply to scenarios in which the dark confinement scale is comparable to $m_\Psi$, i.e. outside the quirky regime. For instance, although the dark mesons can form multiplets of SU(2)$_L$, the low confinement scale in our setup makes long-range dark strong-force effects important for their production. As a result, the EFT analysis of Ref.~\cite{Asadi:2025vfr} cannot be directly applied to predict their collider production cross sections. Consequently, the stringent disappearing-track bounds derived in Ref.~\cite{Asadi:2025vfr} can be avoided in our setup.
Similarly, bounds from resonant production of the dark $\rho$ meson through its mixing with the SM photon \cite{Kribs:2018ilo} are significantly alleviated in the quirky regime. These bounds rely on sizable $\rho$--photon mixing, which is proportional to the dark meson decay constant, $f_\rho \sim 4\pi \Lambda_d$.\footnote{This estimate is consistent with the observed decay constants of heavy-quark mesons in the SM, such as $\Upsilon(1S)$.} In Ref.~\cite{Kribs:2018ilo}, the authors assumed $\Lambda_d \sim m_\Psi$, leading to a large mixing and hence an appreciable collider production rate. In our quirky setup, by contrast, $\Lambda_d \ll m_\Psi$ suppresses production through this channel, thereby substantially weakening the bounds.

We emphasize that direct production of dark glueballs at the LHC is highly suppressed and does not yield observable signals. Similarly, contributions to the electroweak oblique parameters $S$ and $T$ from vector-like dark quarks with degenerate masses within each SU(2)$_L$ multiplet are negligible, see, e.g., Refs.~\cite{Lavoura:1992np,Albergaria:2022jne,Albergaria:2023nby}.

Using the bound $m_\Psi \gtrsim 100$~GeV, and imposing $\tau_{\mathrm{GB}} \lesssim 0.1\mathrm{s}$ to avoid BBN constraints~\cite{Jedamzik:2006xz,Kawasaki:2020qxm}, \cref{eq:lifetime_numerical} implies:
\begin{equation}
    \Lambda_d \gtrsim 0.28\GeV\left(\frac{m_\Psi}{100\GeV}\right)^{8/9}\left(\frac{\tau_{\rm GB}}{0.1\,{\rm sec}}\right)^{-1/9}\, .
    \label{eq:Lambda_bound_GBt}
\end{equation}

Finally, let us mention that dark quarks can form stable neutral dark baryons that would constitute a very small fraction of the DM abundance today \cite{Bai:2010qg,Mitridate:2017oky}.\footnote{If dark quarks were charged under the SM U(1)$_Y$ (instead of the SU(2)$_L$), the lightest baryon would be electrically charged, which would be ruled out for the mass ranges of interest, even if it constitutes a very tiny fraction of DM abundance today.} 
The direct detection signal of these hadrons is suppressed by the $\mathcal{H}$-parity and selection rules in such a sector \cite{Asadi:2024bbq,Asadi:2024tpu,Asadi:2026iot} and their tiny relic abundance \cite{Mitridate:2017oky}.\footnote{Note that the non-zero $\dot{\theta}_i$ does not generate any asymmetry for the dark quarks since there is no equivalent of SU(2)$_L$ sphalerons on the dark side to convert the chiral asymmetry into a $B+L$ asymmetry and this dark chiral asymmetry is washed out by the heavy quark's bare mass when $T \lesssim m_\Psi$.}

\subsection{Axion signatures}
\label{sec:axion_signatures}

As introduced in \cref{sec:three_axions_numerical}, the axion relics predicted by this baryogenesis scenario give rise to signatures that are largely within reach of near-future probes. Here we review these signatures in more detail, including the main astrophysical and cosmological constraints and their associated uncertainties. In minimal QCD-coupled axion models, axion production in SN1987A provides the strongest constraints of the axion--nucleon coupling~\cite{Lella:2023bfb}. CMB and baryon acoustic oscillation (BAO) bounds on the gluon and photon couplings become nearly as restrictive at larger masses~\cite{Caloni:2022uya}. Additionally, the strongest bound on the QCD axion-photon coupling is from its influence on the
cooling of stars within globular clusters \cite{Ayala:2014pea}. Although either the nucleon or photon coupling can be suppressed in nonminimal QCD axion models, suppressing both simultaneously requires additional model building; in particular, nucleophobic constructions generally retain a sizable photon coupling~\cite{DiLuzio:2017ogq}. These probes therefore offer complementary and particularly promising sensitivity to the heavy axion in our model. We discuss these bounds in detail, together with possible signatures from the decay of the heavy-axion subcomponent into photons and prospects for direct detection of the QCD axion in the meV mass range.

\subsection*{Supernova 1987A} 
\label{sec:pheno_astro}
After RG evolution and matching onto the low-energy chiral Lagrangian, the QCD axion acquires derivative couplings to nucleons~\cite{GrillidiCortona:2015jxo},
\begin{align}
    {\cal L}\supset\frac{\partial_\mu a}{2f_a}\left(C_{ap}\bar p\gamma^\mu\gamma^5 p+C_{an}\bar n\gamma^\mu\gamma^5 n + \cdots\right)\,.
\end{align}
Here $C_{ap}$ ($C_{an}$) is the dimensionless low-energy axion coupling to the proton (neutron). The ellipsis denotes the corresponding interactions with pions and heavier hadronic states that arise in the low-energy chiral description after matching the axion's quark and gluon couplings.

The extreme temperatures and densities reached in supernova cores make them powerful probes of new weakly coupled particles, and in particular of axion--nucleon interactions. The observation of neutrinos from SN1987A, including events in Kamiokande-II~\cite{Kamiokande-II:1987idp}, showed that the proto-neutron star cooled over a timescale of order seconds. Axions coupled to nucleons would provide an additional cooling channel: if produced in the proto-neutron-star core and able to free-stream out, they would carry away energy, thus changing the observed neutrino flux. The dominant axion-production processes are nucleon bremsstrahlung~\cite{Raffelt:1993ix,Carenza:2019pxu,Carena:1988kr,Brinkmann:1988vi,Raffelt:1996wa,Carenza:2023lci},
\begin{align}
    N+N\to N+N+a\,,
\end{align}
and pion conversion~\cite{Carenza:2020cis,Fischer:2021jfm},
\begin{align}
    \pi+N\to a+N\,.
\end{align}
The relative importance of these channels depends on the abundance of pions in the hot, dense core. In particular, pion conversion can significantly enhance the axion luminosity when a thermal pion population is present.

The recent analysis of Ref.~\cite{Lella:2023bfb} constrains axion--nucleon couplings using one-dimensional simulations of an $18\,M_\odot$ progenitor. Rather than applying a fixed luminosity criterion independent of the supernova model, the axion luminosity is compared with the neutrino luminosity obtained from the same simulation. This comparison is important because both the absolute neutrino and axion luminosities are sensitive to the core temperature and density. Hotter and denser profiles enhance axion production, while also increasing the neutrino luminosity against which the axion luminosity is compared.

The bounds are typically reported in terms of the couplings $g_{ap}$ and $g_{an}$, defined by
\begin{align}\label{eqn:gap_gan}
    g_{ap}=C_{ap}\frac{m_p}{f_a}\,,\qquad g_{an}=C_{an}\frac{m_n}{f_a}\,,
\end{align}
where $C_{ap}$ and $C_{an}$ are typically $\mathcal{O}(0.01$--$1)$ coefficients whose precise values depend on the details of the UV completion. As a benchmark, in our analysis we assume only a coupling to protons, as approximately realized in KSVZ axion models. Deviating from this assumption can change the resulting bounds by at most $\mathcal{O}(1)$ factors. With this prescription, Ref.~\cite{Lella:2023bfb} finds a characteristic bound
\begin{align}
    g_{ap}\lesssim 10^{-9}\,,
\end{align}
for axion masses below the supernova temperature scale, $m_a\lesssim\MeV$, with the precise boundary depending on the relative proton and neutron couplings and on the treatment of pion-induced production.

Two main known uncertainties enter this bound. The first is the dependence on the supernova model. The baseline calculation of Ref.~\cite{Lella:2023bfb} uses the \texttt{GARCHING SFHO-18.8} simulation~\cite{GarchingCCSNArchive}, while also comparing with a spherically symmetric \texttt{AGILE-BOLTZTRAN} simulation of an $18\,M_\odot$ progenitor~\cite{Mezzacappa:1993gn,Liebendoerfer:2002xn}. The latter has lower core temperatures and densities, which reduce axion production, especially in the free-streaming regime, while also reducing the neutrino luminosity. Because the bound is obtained by comparing the axion luminosity with the neutrino luminosity from the same simulation, these effects partially cancel. The resulting change in the inferred bound on $g_{ap}$ is roughly $25$--$30\%$.

The second, and more important, uncertainty is the pion abundance in the proto-neutron-star core. The population of negatively charged pions depends on the dense-matter equation of state, and a fully self-consistent treatment remains unavailable. Ref.~\cite{Lella:2023bfb} therefore compares two limiting constraints: one including their estimate of the thermal pion abundance and one in which the pion abundance is set to zero. In the latter case, axion production proceeds only through nucleon bremsstrahlung. This gives their most conservative free-streaming cooling bound, weakening the sensitivity to $g_{ap}$ by roughly a factor of two relative to the pion-enhanced calculation. The weakening becomes more pronounced at larger axion masses, where bremsstrahlung is more strongly Boltzmann suppressed than pion conversion.

We illustrate these bounds and their associated astrophysical uncertainty in \cref{fig:hard_cutoff_money,fig:monodromy_crossing_money}. A significant fraction of the viable parameter space lies close to the current SN1987A cooling sensitivity, while the precise boundary of the excluded region remains uncertain. The curve labeled ``SN1987A Proton Coupling'' corresponds to the conservative free-streaming bound obtained by setting the pion abundance to zero, following the right panel of Fig.~4 of Ref.~\cite{Lella:2023bfb}. The lower, lighter-gray region labeled ``$\pi$-uncertainty'' shows the stronger bound obtained when the estimated pion abundance is included. This region represents the dominant dense-matter uncertainty identified in Ref.~\cite{Lella:2023bfb} for the part of parameter space relevant to our benchmarks and should not be interpreted as a statistical confidence interval. Since both simulations considered in Ref.~\cite{Lella:2023bfb} are spherically symmetric and make different simplifying assumptions about proto-neutron-star dynamics, the true astrophysical uncertainty may be larger than the spread shown by this comparison.

Additional uncertainty arises because SN1987A is a single event whose detailed properties cannot be directly reconstructed. For example, Ref.~\cite{Bar:2019ifz} pointed out that a sufficiently rapidly rotating progenitor could have collapsed into a black hole surrounded by an accretion disk. Since that work appeared, JWST observations have found signatures consistent with the presence of a neutron-star remnant~\cite{Fransson:2024csf}. Nevertheless, this and other possible single-event confounding factors have motivated proposals to use the diffuse supernova neutrino background (DSNB) to calibrate axion sensitivity~\cite{Mori:2026nye}. The DSNB has so far been tentatively measured at the $2.6\sigma$ level~\cite{Sekiya:2026skdsnb}, so the corresponding axion bound remains tentative. The resulting tentative $1\sigma$ constraint is already comparable to the SN1987A bounds quoted above and may be more robust against several classes of single-event systematic uncertainty.

As is visible in both \cref{fig:hard_cutoff_money,fig:monodromy_crossing_money}, the viable parameter space lies close to the current SN1987A cooling bound. This proximity is a generic feature of this class of models and is largely independent of the mechanism responsible for generating the axion velocity at the electroweak phase transition, as illustrated in the companion paper Ref.~\cite{Asadi:2025cvm}. A future nearby supernova could therefore provide a decisive test of this scenario across a broad range of implementations. A neutrino light curve consistent with standard cooling would substantially strengthen constraints on axion emission in this coupling range, while a statistically significant cooling anomaly, particularly if accompanied by an independent axion-induced signal, could instead provide evidence for the scenario.

The axion--nucleon couplings entering these bounds are, however, model dependent and can in principle be suppressed~\cite{DiLuzio:2017ogq}. It is therefore useful to consider supernova constraints that follow directly from the axion coupling to QCD and are insensitive to such tuning. After matching onto the low-energy chiral theory, the QCD axion possesses model-independent shift-symmetry-breaking interactions fixed by its role in solving the strong-CP problem. Ref.~\cite{Springmann:2024ret} identified a next-to-leading-order operator that generates a tree-level contribution to nucleon bremsstrahlung and that persists even when the nucleon couplings are absent, resulting in the weaker but robust bound
\begin{align}
    f_a\gtrsim 1.1_{-0.6}^{+0.4}\times 10^8\GeV\,,\qquad(68\%\quad {\rm C.L.})\,.
\end{align}
The quoted uncertainty accounts for uncertainties in the EFT input parameters and for truncation of the chiral expansion at NLO. In our plots in \cref{fig:hard_cutoff_money,fig:monodromy_crossing_money}, we therefore adopt the $68\%$ confidence bound $f_a\gtrsim 5\times 10^7\GeV$, shown as the dark-gray region labeled ``SN1987A: gluon''. We note that the quoted limit itself does not account for astrophysical uncertainties.

\subsection*{Cosmology}
\label{sec:pheno_cosmo}
Because the axions in our model mix with the QCD axion, they are thermally produced in the early Universe and, if the reheating temperature is sufficiently high, may have entered thermal equilibrium with the SM. The resulting thermal relic population provides some of the strongest constraints at larger axion masses, where the relic energy density becomes more important, while even a subdominant thermal population can remain relevant for direct and indirect-detection experiments.

The thermal production rate of axions in the quark-gluon plasma has been studied in Refs.~\cite{Graf:2010tv,Salvio:2013iaa,DEramo:2021lgb,DEramo:2021psx,Bouzoud:2024bom}. In particular, Ref.~\cite{Caloni:2022uya} uses these rates to derive cosmological bounds on the QCD axion. Assuming $C_{ij}=1$ and equal $f_i$, thermal production of the heavy axion through QCD dominates over production through the dark ${\rm SU}(N_d)$ sector. Although one might expect the large number of dark gauge degrees of freedom to enhance the latter rate, there is no parametric enhancement at large $N$. The thermal axion production rate per unit volume scales as~\cite{Braaten:1991dd,Manohar:1998xv,Graf:2010tv,Bouzoud:2024bom}
\begin{align}
    \gamma_a^{\rm HTL}\propto \alpha_d^3N(N^2-1)\frac{T^6}{f^2}\propto N^0+{\cal O}(N^{-1})\,,
\end{align}
since $\alpha_d N\to{\rm const}$ in the large-$N$ limit. Charged fermions further enhance the production rate by a factor of order $1+n_f/N$. This enhancement is important for QCD, where $n_f=6$ above the electroweak scale. Since the dark-charged fermions lie at or above the electroweak scale and may have fewer active degrees of freedom, we expect QCD production to dominate over dark-sector production by an ${\cal O}({\rm few})$ factor in this benchmark limit.

Once thermally produced, an axion that later becomes non-relativistic retains a relic abundance approximately given by~\cite{Caloni:2022uya}
\begin{align}\label{eq:overclosure_thermal}
    \Omega_a h^2 \simeq 7.2\times 10^{-3}
    \left(\frac{m_a}{10\,\eV}\right)
    \left(\frac{g_{\star,s}(T_{\rm dec})}{106.75}\right)^{-1}\,.
\end{align}
Thus, once the axion mass and decoupling temperature are specified, the thermal relic abundance is fixed up to assumptions about entropy production and the relative temperature of any decoupled sectors. In particular, later decoupling corresponds to a smaller $g_{\star,s}(T_{\rm dec})$ and therefore a larger relic abundance. Requiring that this population not overclose the Universe consequently places an upper bound on the mass of thermally connected axions.

The benchmark estimate above does not apply uniformly across the parameter space shown in \cref{fig:hard_cutoff_money,fig:monodromy_crossing_money}, where neither $C_{ij}=1$ nor equal $f_i$ is generally realized. In fact, the coupling to the dark confining sector is often substantially stronger than the coupling to QCD, as required to efficiently damp coherent axion oscillations over otherwise viable regions of parameter space. In the more extreme examples shown in \cref{tab:hard_cutoff_scan,tab:lattice_extrapolation_scan}, the axion can remain coupled to the dark bath until temperatures comparable to $\Lambda_d$. As a simple estimate, taking the dark and SM baths to have the same temperature, even at the lower end of our range, $\Lambda_d\sim0.3\,\GeV$, the SM bath contains only $g_{\star s}\sim61.75$ relativistic degrees of freedom at axion decoupling. The corresponding thermal relic abundance is therefore enhanced relative to the $g_{\star, s}(T_{\rm dec})=106.75$ benchmark in \cref{eq:overclosure_thermal}.\footnote{If the dark and SM baths decouple before axion decoupling, the axion abundance also depends on the dark-to-visible temperature ratio. Before dark confinement, this ratio typically differs from unity by only tens of percent, since the SM entropy degrees of freedom change only modestly over this interval.} Possible later decoupling may further increase the thermal axion abundance if the axion remains in equilibrium through part of the QCD crossover, where $g_{\star, s}$ decreases rapidly toward $\sim17.25$.

At lower masses, the thermal relic can remain relativistic during the CMB epoch and instead contribute to $\Delta N_{\rm eff}$. The corresponding constraints studied in Ref.~\cite{Caloni:2022uya} combine Planck measurements of $N_{\rm eff}$~\cite{Planck:2018vyg} with BAO measurements from BOSS DR12~\cite{BOSS:2016wmc}, 6dFGS~\cite{Beutler:2011hx}, and SDSS-MGS~\cite{Ross:2014qpa}. For a single pseudoscalar degree of freedom decoupling from the SM bath at temperature $T_{\rm dec}$,
\begin{align}\label{eq:Neff_thermal}
    \Delta N_{\rm eff}\simeq 0.027\left(\frac{g_{\star,s}(T_{\rm dec})}{106.75}\right)^{-4/3}\,,
\end{align}
assuming the axion remains relativistic during the CMB epoch. An axion coupled to QCD with a decay constant above a few $\times 10^8\,\GeV$ decouples before the electroweak phase transition and consequently contributes too little to $\Delta N_{\rm eff}$ to be excluded by current bounds~\cite{Caloni:2022uya}.

The overclosure bound from \cref{eq:overclosure_thermal} is most relevant for $m_a\gtrsim100\eV$, while the $\Delta N_{\rm eff}$ bound from \cref{eq:Neff_thermal} applies for $m_a\lesssim0.1\eV$. In the intermediate mass range, the thermal axion becomes non-relativistic during cosmologically relevant epochs, and a full CMB+BAO analysis is required; see Refs.~\cite{Caloni:2022uya,DEramo:2022nvb,Notari:2022ffe} for a detailed treatment.

In our model, both $\theta_2$ and $\theta_3$ may contain an order-one fraction of the thermally produced interaction eigenstate, so the single-species bounds of Ref.~\cite{Caloni:2022uya} do not apply directly and are modified by order-one mixing factors, as well as by the presence of the dark bath. If the QCD and dark interactions decouple at parametrically different temperatures, the interaction that remains efficient to the lower temperature sets the final thermal abundance. Because the viable parameter space tends to suppress the QCD coupling relative to the dark coupling, it will often be the dark gauge group that decouples last. For $\theta_3$ masses of order $10\,\eV$, as relevant for our baryogenesis benchmarks, the thermal population behaves as a partially free-streaming relic whose cosmological effects interpolate between those of extra radiation at early times and matter at late times. This provides a characteristic signature of the model, and forthcoming measurements from ACT/SPT and weak-lensing surveys, and a future CMB-S4 mission could improve sensitivity to this region of parameter space~\cite{Caloni:2022uya}. A detailed treatment of the coupled two-axion decoupling dynamics, including interactions with the dark sector, is beyond the scope of this work and is left for future study. We therefore treat the cosmological bounds shown in \cref{fig:hard_cutoff_money,fig:monodromy_crossing_money} as indicative.

\subsection*{Decay to photons}
\label{sec:pheno_decay}

The axion generically possesses a model-dependent coupling to photons and can therefore decay through $a\to\gamma\gamma$. For the relatively large masses and small decay constants favored by this scenario, the decay rate can be large enough for the resulting photons to be observable with spectroscopic telescopes as a narrow line. MUSE, DESI, and HST provide some of the strongest sensitivity to $\theta_3$ decays in the eV to $\sim 20\,\eV$ mass range~\cite{Regis:2020fhw,Todarello:2023hdk,Wang:2023imi,Todarello:2024qci,Todarello:2025rff}.

These searches are particularly relevant for the lattice-extrapolated friction scenario of \cref{subsec:sphaleron_rate_summary}. In this case, the heavy-axion misalignment component is efficiently damped away, leaving the thermal relic of \cref{eq:overclosure_thermal} as its dominant remaining population. Although subdominant in the total dark matter abundance, this population can still give rise to observable decay signatures. The decay bounds shown in \cref{fig:hard_cutoff_money,fig:monodromy_crossing_money} generally assume that the axion constitutes all of the dark matter. In our scenario, the $\theta_3$ abundance depends on the thermal history of the Universe, particularly the reheating temperature, and may therefore be substantially smaller than unity. Further, given its partly-thermal origin, its distribution in the halo may be considerably different than the majority of the cold dark matter. The corresponding limits should be rescaled according to the relic abundance and warm dark matter distribution, so the gray regions in \cref{fig:monodromy_crossing_money} should be interpreted as indicative reaches. Nevertheless, accounting for the thermal relic abundance, decay searches may still probe this otherwise depleted heavy-axion population~\cite{Dror:2024ibf}.

At larger masses, particularly strong constraints on axion decays to photons can be derived from heating of the dwarf galaxy Leo T~\cite{Wadekar:2021qae}. Leo T is both dark-matter dominated and gas rich, making it especially sensitive to energy injection from dark matter decays. For axion masses above $2\times13.6\,\eV$, the decay photons can ionize hydrogen. A fraction of the resulting photoelectron kinetic energy is subsequently deposited as heat through interactions with the gas. By comparing this heating rate with the astrophysical heating and cooling rates, Ref.~\cite{Wadekar:2021qae} derived the bound shown at $m_{\theta_3}\gtrsim{\cal O}(27.2)\,\eV$. The boundary is nearly vertical because of the hydrogen ionization threshold at $13.6\,\eV$, above which the gas becomes optically thick to the decay photons. The strength of the Leo T bound makes it an excellent probe of small heavy-axion subcomponents.

\subsection*{Direct detection}
\label{sec:pheno_direct}

Depending on the form of the friction, either the QCD axion or the heavier dark-aligned axion can constitute the dominant dark matter component. For the lattice-extrapolated friction scenario of \cref{subsec:sphaleron_rate_summary}, the QCD axion makes up most of the dark matter. A distinctive feature of this axiogenesis mechanism is that it populates QCD axion masses above the canonical misalignment target, placing them within the range of proposed haloscopes designed to probe heavier QCD axion dark matter. In particular, BREAD~\cite{BREAD:2021tpx} is sensitive to part of this parameter space, as illustrated in \cref{fig:monodromy_crossing_money}. The heavier state is still thermally populated and can therefore give rise to complementary decay signatures. Observation of both a heavy QCD axion and a second, heavier axion coupled to the SM through its mixing with QCD would provide a distinctive signature of the multi-axion dynamics underlying this mechanism.

For friction that remains closer to the high-temperature extrapolation discussed in \cref{subsec:sphaleron_rate_summary}, the relic composition is instead reversed: the QCD axion is only weakly populated, while the heavy axion constitutes nearly all of the dark matter. In this case, the heavy state may lie within reach of direct-detection experiments such as LAMPOST~\cite{Baryakhtar:2018doz}, as illustrated in \cref{fig:hard_cutoff_money}. The QCD axion coupling to the SM nevertheless remains close to existing astrophysical limits, so a future nearby supernova could provide a complementary probe of the QCD axion even when its relic abundance is small.

Together, these two friction scenarios illustrate the sensitivity of axion haloscopes spanning the ${\rm meV}$ to ${\rm eV}$ mass range to this mechanism. Combined with astrophysical and decay searches, they offer the opportunity of probing both axion eigenstates and potentially discovering the multi-axion structure associated with axiogenesis.

\section{Outlook}\label{sec:outlook}

We have shown that multi-axion theories provide a viable framework for axiogenesis comprising only technically natural ingredients, while preserving axion quality. The additional axion directions provide sufficient freedom to satisfy the physical and observational requirements, while their mixing correlates the resulting baryon asymmetry, relic abundances, masses, and couplings.

The viable parameter space points to comparatively narrow regions of late-time axion masses and couplings, as anticipated in the companion paper~\cite{Asadi:2025cvm}. Much of this space lies close to current supernova cooling bounds~\cite{Lella:2023bfb,Springmann:2024ret}, while cosmological measurements, axion-decay searches, and haloscopes provide complementary probes~\cite{Caloni:2022uya,Todarello:2025rff,BREAD:2021tpx,Baryakhtar:2018doz}. In particular, the heavier dark-aligned axion can remain observable as a thermally produced subcomponent even when the QCD axion constitutes most of the dark matter. The confining sector required by the mechanism also gives rise to complementary collider signatures through the heavy states that mediate the interactions with the SM~\cite{Forsyth:2025wks,Curtin:2025ngf}. 

The behavior of the dark-sphaleron friction at temperatures a factor of 10–100 above the confinement scale remains poorly determined, with different extrapolations yielding qualitatively different dark matter compositions. For this reason, further lattice calculations of the sphaleron rate in this intermediate regime are crucial to determine which axion dominates the relic abundance and to sharpen the corresponding experimental targets.

Whether the dynamics presented here open a path towards resolving the matter coincidence problem remains an important open question. The baryon-to-dark-matter ratio is very sensitive to the model parameters, with dark matter overproduced by many orders of magnitude across broad regions of parameter space. 

An important future direction is the construction of an explicit compactification that realizes the required monodromy scale and winding number. The warped double Klebanov--Strassler throat discussed in this paper offers a possible starting point for the small monodromy scale; accommodating the large winding remains the principal challenge.

\begin{acknowledgements}
    We are especially indebted to Jakob Moritz and Matt Reece. We thank Bobby Acharya, Asimina Arvanitaki, Masha Baryakhtar, Austin Batz, Claudio Bonanno, Raymond Co, Basudeb Dasgupta, Savas Dimopoulos, Joshua Forsyth, Joshua Foster, Ella Henry, Anson Hook, Junwu Huang, Graham Kribs, Mehrdad Mirbabayi, Guy Moore, Rudin Petrossian-Byrne, Davide Racco, Stephen Sharpe, Eva Silverstein, Jedidiah Thompson, Chris Verhaaren, Giovanni Villadoro, and Lawrence Yaffe for useful discussions and feedback on the manuscript. 

    The research of P.A. and S.G. is supported in part by the U.S. Department of Energy grant number DE-SC0010107. S.G. also acknowledges support from the Miller Institute for Basic Research in Science, at the University of California Berkeley.
    D.C.\ acknowledges the receipt of a grant from the Abdus Salam International Centre for Theoretical Physics (ICTP), Trieste, Italy, and the support of the Istituto Nazionale di Fisica Nucleare (INFN). D.C.\ also acknowledges support by the U.S.\ Department of Energy Office of Science under Award Number DE-SC0024375 and the Department of Physics and College of Arts and Science at the University of Washington.

    Experimental limits and prospects presented in this paper were in part compiled using the GitHub repository associated with Ref.~\cite{AxionLimits}.
    
    This work was initiated and performed in part at Aspen Center for Physics, which is supported by National Science Foundation grant PHY-2210452.
    For facilitating portions of this research, D.C.\ wishes to acknowledge the Center for Theoretical Underground Physics and Related Areas (CETUP*), the Institute for Underground Science at Sanford Underground Research Facility (SURF), and the South Dakota Science and Technology Authority for hospitality and financial support, as well as for providing a stimulating environment.
    
    Generative AI tools, including ChatGPT versions 4 through 6 and Claude Fable, were used for language editing, code assistance, and to confirm mathematical results. The authors take full responsibility for the content of the manuscript.
\end{acknowledgements}
\appendix

\section{Boltzmann equations in a rotating axion background}
\label{app:boltzmann}

In this appendix we review the basics of Boltzmann equations governing various SM asymmetries in a rotating scalar field background. 
This includes a review of the contributions of SM perturbative processes, whether as a source term or the wash-out term, as well as non-perturbative contributions. 
The former consists of the contribution from all SM Yukawa processes, while the latter arises from sphaleron \cite{Klinkhamer:1984di} field configurations. 
The rotation of an axion field in the early Universe contributes as a source term to the Chern-Simons number of the SU(2)$_L$, which in turn acts as an asymmetry source term for all anomalous charges. 
We review these contributions in detail to derive the final Boltzmann equations in \cref{eq:final_boltz} for the U(1) symmetries associated with each SM fermion field and their corresponding net charge asymmetry. 
Along the way, we review some basics of the sphaleron processes and the vacuum of non-abelian gauge fields that are relevant for the study of the friction term on an axion.

\subsection{Perturbative processes: Yukawas}
\label{subapp:perturbative}

Let us begin by reviewing how various perturbative processes contribute to the evolution of a charge asymmetry.
It is useful to define the \textit{symmetric} and \textit{asymmetric} parts of the matrix element as follows:
\begin{align}
     |\Sigma\M^\alpha_\beta|^2 &\equiv |\M^\alpha_\beta|^2 + |\M^{\alpha^\dagger}_{\beta^\dagger}|^2  \ ,
    \label{eq:defSigmas}
    \\
    |\Delta \M^\alpha_\beta|^2 &\equiv |\M^\alpha_\beta|^2 - |\M^{\alpha^\dagger}_{\beta^\dagger}|^2 \ ,
    \label{eq:defDeltaM}
\end{align}
where $\mathcal{M}^\alpha_\beta$ denotes the matrix element for a process from an initial state $\alpha$ to a final state $\beta$
\begin{equation}
    \M(\alpha\rightarrow\beta) \equiv \M^{\alpha}_{\beta} \ .
    \label{eq:defMnotation}
\end{equation}
Note that a non-zero asymmetric matrix element $\Delta \mathcal{M}$ requires both a source of CP violation and interference between multiple diagrams for the process.
This involves the presence of both \textit{weak} \cite{Weinberg:1979bt,Kolb:1979qa} and \textit{strong} \cite{Cutkosky:1960sp} phases in the amplitude.
The weak phase arises from physical complex parameters in the Lagrangian, e.g., from complex Yukawa couplings.
The strong phase, on the other hand, originates from kinematic factors such as the imaginary parts of propagators associated with intermediate on-shell states in a Feynman diagram. 

To calculate macroscopic quantities such as the net number densities and free energies, we use   
\begin{equation}
    f_{i} = \frac{1}{e^{\frac{E_i-\mu_i}{T}}\pm 1}\simeq e^{\mu_i/T}e^{-E_i/T} \ \equiv \ e^{\mu_i/T}f_{i}^0
\ ,
\label{eq:def_f}
\end{equation}
where $\mu_i$ is the chemical potential for the particle species $i$. The last approximate equality is valid for $E_i-\mu_i\gg T$, and corresponds to Maxwell-Boltzmann distribution for all particles.
The net asymmetry in the number density of a particle can be expressed in terms of its chemical potential as
\begin{equation}
    n_{i} \simeq \frac{1}{6}g_ik_i\mu_iT^2 \ ,
    \label{eq:def_mu}
\end{equation}
where $g_i$ is the number of internal degrees of freedom, and $k_i$ accounts for particle statistics, with $k_i=1$ ($k_i=2$) for fermions (bosons). 

The collision term arising from the process $1+2 \rightarrow 3+4$ where the numbers label different particle species, contributes to the Boltzmann equation for a quantum number $Q$ as
\begin{align}
[{{C}}_Q]^{12}_{34} & \equiv 
\left(\Delta Q\right)^{12}_{34} \int\!d\Pi_1\!\int\!d\Pi_2
\;\mathcal{W}^{12}_{34}\, 
f_1(T)f_2(T) \, ,
\label{eq:cnterm1}
\end{align}
where $\left(\Delta Q\right)^{12}_{34}$ denotes the change in charge $Q$ during this process, $d\Pi_i$ stands for a Lorentz-invariant phase space element of particle $i$, and we assume all particles are in the same thermal bath with temperature $T$.
The \textit{scattering kernel} $\mathcal{W}^{12}_{34}$ is defined as
\begin{align}
    \mathcal{W}^{12}_{34}
\ \equiv \ {S}\!\int\!{d\Pi_3}\!\int\!{d\Pi_4}\;(2\pi)^4\delta^{(4)}(p_1+p_2-p_3-p_4)|\,\mathcal{M}^{12}_{34}|^2 \ ,
\label{eq:wscatt}
\end{align}
where $S$ is a symmetry factor.

In this notation, if we put together the contribution of a process $1+2\rightarrow 3+4$ and its reverse $3+4 \rightarrow 1+2$ to the net charge $Q$, and after expanding the exponential in \cref{eq:def_f} to the leading order in $\mu_i$, we arrive at
\begin{align}
\label{eq:delcqnonself}
[{{C}}_Q]^{12}_{34}/\left(\Delta Q\right)^{12}_{34} = \int\!d\Pi_1\!\int\!d\Pi_2
\Bigg(
&2\Delta\mathcal{W}^{12}_{34}f_1^0f_2^0
\  \\
&+ \ \frac{\Sigma\mathcal{W}^{12}_{34}}{T}f_1^0f_2^0
\left(\mu_1+\mu_2 - \mu_3 -\mu_4\right) + \mathcal{O}(\mu_i^2)
\Bigg)
\ , \nonumber
\end{align}
where $\lbrace \Delta \mathcal{W}, \Sigma \mathcal{W} \rbrace$ are defined analogous to \cref{eq:wscatt}, with the simple replacement $\M \rightarrow \lbrace \Delta\M , \Sigma \M\rbrace$, respectively. We have also used energy conservation to write $f_3^0f_4^0 = f_1^0f_2^0$. 
Note that we implicitly assume all asymmetries are small, so that a leading-order expansion in the chemical potentials $\mu_i$ is valid. 
The first term in \cref{eq:delcqnonself} acts as a source term for the $Q$ asymmetry arising from this process, while the second term describes the backreaction from the existing asymmetries, i.e. the wash-out term.

We now apply this formalism to various fermion number asymmetries in the SM.
The only relevant perturbative processes in the SM arise from the Yukawa interactions.\footnote{The flavor-changing SU(2)$_L$ interactions contribute as well. However, their contributions cancel when we sum over different generations.}
In the SM, the contribution of these processes to the source terms is small and can be neglected.
This leaves only the wash-out terms from \cref{eq:delcqnonself}, which are proportional to the symmetric matrix element $\Sigma \M$. 
Note that if these Yukawa interactions are the sole processes affecting the comoving number densities, and in the absence of source term ($\Delta \mathcal{W}\rightarrow 0$), \cref{eq:delcqnonself} implies the system reaches equilibrium when
\begin{equation}
    \sum_i \mu_i =0 \ .
    \label{eq:chemical_eqb}
\end{equation}
This corresponds to the usual condition for a process in chemical equilibrium.
As we will see in the next subsection, the rotation of an axion field introduces a source term for various SM charges. As a result, \cref{eq:chemical_eqb} will no longer hold for all Yukawa processes in the SM.

At high temperature, all SM fields involved in Yukawa interactions are massless, and thus the corresponding $2\rightarrow 1$ processes are kinematically forbidden. While finite-temperature effects generate thermal masses that open up these phase spaces, the dominant perturbative contributions typically arise from $2 \rightarrow 2$ scatterings involving a background gauge boson from the thermal bath. As a result, the leading-order finite-temperature contribution to the symmetric matrix element is given by
\begin{equation}
    \Sigma\mathcal{M} \propto g y T,
    \label{eq:M_finiteT}
\end{equation}
where $g$ is the gauge coupling associated with the background gauge fields and $y$ is the relevant SM Yukawa coupling
\begin{equation}
    \mathcal{L} \supset y H^{(\dagger)} f_1 f_2 \, .
    \label{eq:yukawa_fake}
\end{equation}
After integrating over the phase space and using \cref{eq:def_mu}, as well as including finite-temperature effects from the thermal plasma \cite{Besak:2012qm,Bodeker:2019ajh} as described in the main body of the text,
the contribution of Yukawa interactions to the wash-out term in \cref{eq:delcqnonself} becomes
\begin{equation}
    [{{C}}_{f_1,f_2}]^{H}_{f_1f_2} \sim  \kappa \, N_x \min (\alpha ,\, y^2 )\, T \left( \pm \sum_f \frac{n_f}{g_f} \pm \frac{n_H}{4} \right),
    \label{eq:CQ_yukawa}
\end{equation}
where subscripts $f_i$ denote the fermion number currents of the fermion species $i$, $\alpha$ is the background gauge structure constant of the gauge group SU($N_x$), $\kappa$ captures the multiplicative finite-temperature effects, the sum over $f$ runs over the fermions involved in the Yukawa interaction.
The quantity $n_H$ is the net number density of the Higgs boson, with the sign for each particle chosen appropriately depending on the specific process.
We will neglect $\mathcal{O}(1)$ finite-temperature prefactors ($\kappa \rightarrow 1$) and take \cref{eq:CQ_yukawa} as the contribution of the various Yukawa interactions to the final Boltzmann equations for fermion numbers.

\subsection{Non-perturbative processes: sphalerons}
\label{subapp:nonperturbative}

The sphalerons of both the SU(2)$_L$ and SU(3) gauge groups of the SM contribute to the evolution of various fermion charge asymmetries - both as source terms (in the presence of a rotating axion background) and as wash-out terms.
In this appendix, we briefly review the relevant physics for a general SU($N_x$) gauge group; further details can be found in, e.g., Refs.~\cite{Klinkhamer:1984di,Kuzmin:1985mm,Rubakov:1996vz,Riotto:1998bt,Srednicki:2007qs,Servant:2014bla}.

Let us consider the Lagrangian density 
\begin{equation}
\mathcal{L} \supset \frac{\alpha}{8\pi} \zeta \ W^{a}_{\mu\nu} \tilde{W}^{\mu\nu \ a},
    \label{eq:zeta_L}
\end{equation}
where $\zeta$ can be time-dependent and $W^{\mu\nu}$ is the field strength tensor of an SU($N_x$) gauge group. 
We will show in this appendix that this term, and the transition between different vacua of the SU($N_x$) gauge, have important implications for fermion asymmetries.

A pure SU($N_x$) gauge group has a non-trivial vacuum manifold in 3+1 dimensions.
There exist local gauge field configurations that interpolate between different SU($N_x$) vacua. 
Sphalerons \cite{Klinkhamer:1984di} are one such field configuration that can interpolate between different vacua with a non-negligible rate at high enough temperatures. 
These configurations, and the corresponding vacua, are characterized by their Chern-Simons (CS) number, $N_{\mathrm{CS}}$. 
The CS number is a discrete quantity associated with the divergence of the CS current (see, e.g., Ref.\cite{Srednicki:2007qs}):
\begin{equation}
    \partial_\mu j_{\mathrm{CS}}^\mu = \frac{\alpha}{8\pi} W_{\mu\nu}^a \tilde{W}^{\mu\nu, \ a} \ .
    \label{eq:changeCS_app}
\end{equation}

\begin{figure}
    \centering
    \includegraphics[width=0.6\linewidth]{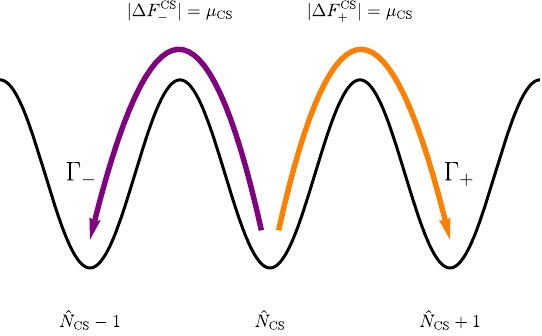}
    \caption{A schematic depiction of transitions between vacua with different Chern-Simons numbers $\hat{N}_{\mathrm{CS}}$ showing hopping to neighboring vacua that either increase (orange) or decrease (purple) $N_{\mathrm{CS}}$ via sphaleron transitions. The rates for these processes are denoted by $\Gamma_+$ and $\Gamma_-$, respectively. The change in free energy associated with each transition receives contributions from the CS chemical potential $\mu_{\mathrm{CS}}$ and from fermions charged under the gauge group. An axion rotation, characterized by $\dot{\theta}$, acts as a source for $\mu_{\mathrm{CS}}$, while the fermion contributions to the free energy act as a wash-out term in the sphaleron contributions to the Boltzmann equations.}
    \label{fig:hopping}
\end{figure}

A schematic illustration of these vacuum transitions via sphalerons is shown in \cref{fig:hopping}. 
The transitions between different $N_{\mathrm{CS}}$ vacua are dominated by nearest-neighbor hops, i.e. processes with $\Delta N_{\mathrm{CS}} = \pm 1$.
The rate of change of $N_{\mathrm{CS}}$ within a volume $V$ is given by
\begin{equation}
\partial_t N_{\mathrm{CS}} = R_+ - R_- \ ,
    \label{eq:dNCS_dt}
\end{equation}
where $R_{\pm}$ denotes the rate of transitions to vacua with $N_{\mathrm{CS}} = \pm 1$. 
From basic thermodynamics, we know that
\begin{equation}
    \frac{R_+}{R_-} = e^{-\Delta F /2T} \ ,
    \label{eq:ratio_F}
\end{equation}
where $\Delta F = \Delta F_+ - \Delta F_-$ denotes the difference in free energy changes associated with the respective hoppings. 
As a result, to calculate the transition rate between different vacua, we have to find their free energy difference. 
The free energy receives contributions both from (i) a non-zero Chern-Simons chemical potential $\mu_{\mathrm{CS}}$, and from (ii) a change in the net number of fermions charged under the gauge group. 
Below we review these contributions.

Let us start our study of the free energy difference from \cref{eq:zeta_L}. 
When integrated, it contributes to the action as
\begin{align}
    \label{eq:def_muCS}
    \mathcal{S} = \int d^4x \ \mathcal{L} & \supset  \int d^4x \ \frac{\alpha}{8\pi} \ \zeta \ W^{a}_{\mu\nu} \tilde{W}^{\mu\nu \ a} \\
    & =  \int d^4x \ \zeta \ \partial_\mu j_{\mathrm{CS}}^\mu \nonumber \\
    & =  \int d^4x \ \zeta \ \partial_t j_{\mathrm{CS}}^0 \nonumber \\
    & =  - \int d^4x \ \partial_t\zeta \ j_{\mathrm{CS}}^0 \nonumber \ .
\end{align}
In the second line, we used \cref{eq:changeCS_app}; in the third line, we assumed that the spatial divergence of the current is negligible on the boundary of the integration region; and in the last line, we integrated by parts.
The resulting expression takes the form of a chemical potential coupled to the Chern-Simons current \cite{Turok:1990in}.
It is therefore standard to define
\begin{equation}
    \mu_{\mathrm{CS}} \equiv - \partial_t \zeta \ .
    \label{eq:muCS_derivative}
\end{equation}
This suggests that if the Lagrangian contains a term like that in \cref{eq:zeta_L}, the time variation of the prefactor $\zeta$ acts as a chemical potential for the associated Chern-Simons number, biasing transitions between different Chern-Simons vacua.
In particular, a time-dependent axion field $\theta=a/f_a$ acts as such a chemical potential, thereby inducing transitions between topologically distinct Chern-Simons vacua.
In our specific model, the chemical potential of the strong force is equal to the time derivative of its associated strong phase, i.e. $\dot{\theta}_i$. 
The electroweak CS chemical potential is proportional to the same derivative, via the weak anomaly coefficient $c_W$.

In presence of a CS chemical potential, the contribution to the system’s free energy is
\begin{equation}
    \Delta F^\mathrm{CS} = - \mu_{\mathrm{CS}}  \ \Delta N_{\mathrm{CS}} \ .
    \label{eq:F_muCS}
\end{equation}
In this equation, and throughout the following derivation, we focus on a spatial region of volume $V$.
The expression quantifies the change in the free energy within this volume when transitioning between two vacua with different Chern-Simons numbers.
For the transition depicted in \cref{fig:hopping}, $\Delta N_{\mathrm{CS}}=2$ between the two vacua adjacent to the initial vacuum the system starts in.
Thus, the CS chemical potential contributes to the free energy difference $\Delta F$ as
\begin{equation}
 \Delta F^\mathrm{CS} = -2\mu_{\mathrm{CS}} \ .
    \label{eq:muCS_toF}
\end{equation}

The free energy of the system in different CS vacua gets a contribution from the net asymmetry of different fermions charged under the gauge group as well. 
To understand this, let us focus on the symmetry U(1)$_i$ associated with the fermion number of Weyl fermion $i$. 
Since the fermion is chiral, this symmetry is anomalous under an SU($N_x$) gauge group.
The divergence of the associated current receives a non-vanishing contribution from the following triangle diagram:
\begin{center}
    \includegraphics[width=0.3\textwidth]{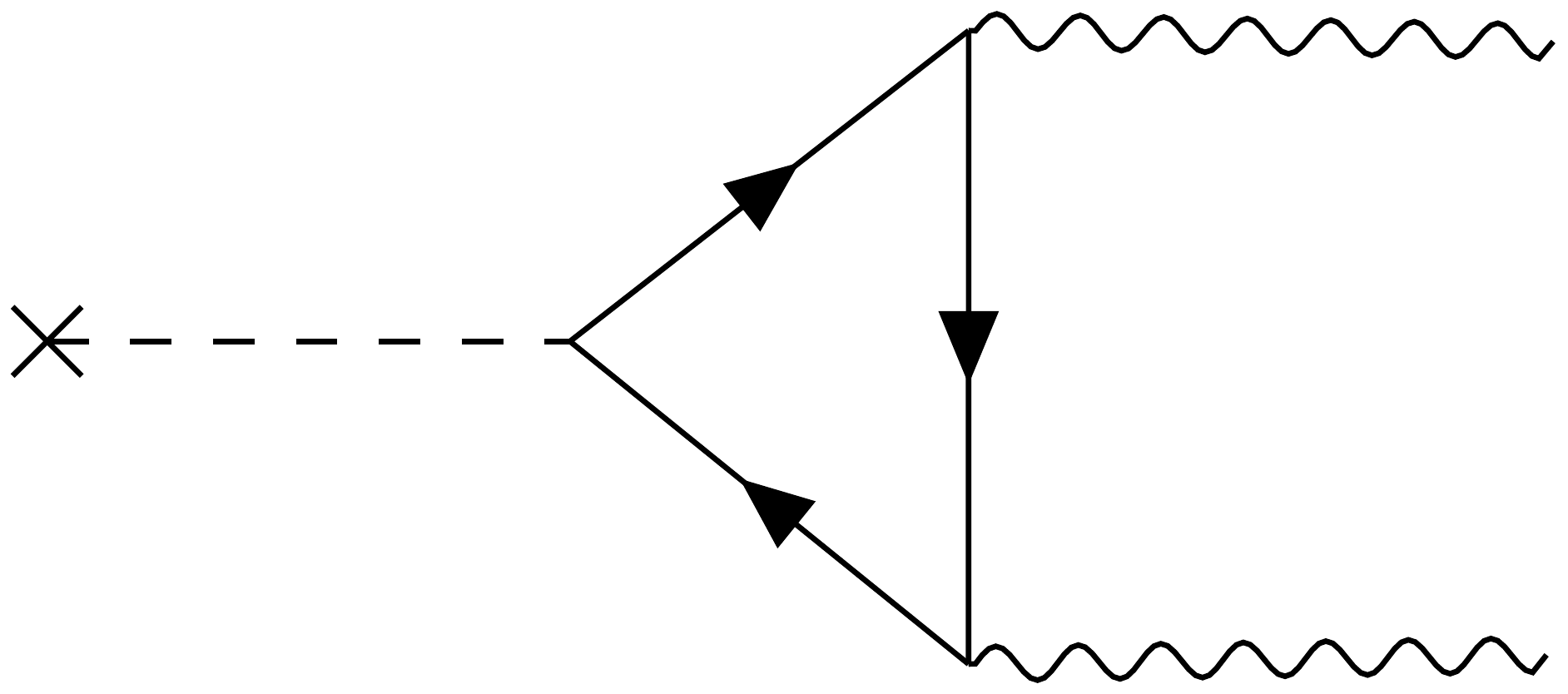} 
\end{center}

\noindent where all fermions charged under both SU($N_x$) and the anomalous global U(1)$_i$ symmetry contribute in the loop.
This anomaly implies that the divergence of the U(1)$_i$ current is given by
\begin{equation}
\partial_\mu j_Q^\mu = \mathcal{N}_i^x \frac{\alpha}{8\pi} W_{\mu\nu}^a \tilde{W}^{\mu\nu, \ a},
    \label{eq:div_Q_FF}
\end{equation}
where $a$ is the SU($N_x$) adjoint index, $\alpha$ is its structure constant, and $\mathcal{N}_i^x$ is the anomaly multiplicity factor for fermion $i$. For a chiral fermion in the fundamental representation of $SU(N_x)$, $\mathcal{N}_i^x$ simply counts the internal states not associated with the gauge group; it is equal to the total number of internal degrees of freedom of the fermion, $g_i$, divided by $N_x$ 
(since $N_x$ of those degrees of freedom arise from the fermion's gauge index). 
This equation can be integrated to compute the change in the net $i$ charge within a bounded region, yielding
\begin{equation}
\delta N_i = \mathcal{N}_i^x \frac{\alpha}{8\pi} \int d^4x W_{\mu\nu}^a \tilde{W}^{\mu\nu, \ a} \ ,
    \label{eq:deltaNQ}
\end{equation}
where we have neglected any flux of the spatial current $\vec{j}_Q$ entering or leaving the region.

Equations~\eqref{eq:div_Q_FF} and \eqref{eq:changeCS_app} imply the standard result
\begin{equation}
    \delta N_i = \mathcal{N}_i^x \times \delta N_{\mathrm{CS}} \ .
    \label{eq:NQNCS}
\end{equation}
Thus, if a gauge field configuration interpolates between two different vacua of SU($N_x$), i.e. it changes the Chern-Simons number, then the net fermion number associated with U(1)$_i$ changes in that region.

In a space of volume $V$, the free energy contribution of a fermion species $i$, with $g_i$ internal degrees of freedom at temperature $T$, is given by 
\begin{equation}
    F_i = g_i T V \int \frac{d^3k}{(2\pi)^3} \ \left[ \log \left(  1+ e^{- (E-\mu_i)/T} \right) + \left( \mu_i \rightarrow -\mu_i \right) \right] \ ,
    \label{eq:F_fermion_def}
\end{equation}
where the first term captures the fermion's contribution and the second term (with $\mu \rightarrow -\mu$) captures its anti-particle's contribution. (See Ref.~\cite{Quiros:1999jp} for a review of finite temperature potential and free energy.) 
Here $\mu_i$ is the fermion's chemical potential. 
In the massless limit, and assuming small asymmetries, this expression can be expanded in $\mu_i$ as a power series:
\begin{equation}
    F_i \simeq \frac{7}{360} g_i \pi^2 T^4 V + \frac{1}{12} g_i \mu_i^2 T^2 V + \mathcal{O}(\mu^4) \ .
    \label{eq:Fi_expanded}
\end{equation}
Using \cref{eq:def_mu} we can rewrite this equation as
\begin{equation}
    F_i \simeq \frac{7}{360} g_i \pi^2 T^4 V + \frac{3 V}{g_i T^2} n_i^2 + \mathcal{O}(n_i^4) \ ,
    \label{eq:Fi_withni}
\end{equation}
where $n_i$ is the net fermion number density of fermion species $i$.

When we hop between neighboring CS vacua with a sphaleron, \cref{eq:NQNCS} suggests the net number of fermions $i$ change by $\mathcal{N}_i^x$ units.
Thus the net fermion density changes by $\mathcal{N}_i^x/V$. 
If we denote the net fermion density in the original CS vacua by $n_{i \, 0}$, \cref{eq:Fi_withni} implies
\begin{align}
    \label{eq:changeFi_Nf}
    \Delta F_+^i & =  \frac{3V}{g_i T^2} \left[ \left(n_{i \, 0} + \frac{\mathcal{N}_i^x}{V}\right)^2 - n_{i \, 0}^2 \right] \, , \\
    \Delta F_-^i & =  \frac{3V}{g_i T^2} \left[ n_{i \, 0}^2 - \left(n_{i \, 0} - \frac{\mathcal{N}_i^x}{V}\right)^2 \right]  \, , \nonumber \\
    \Longrightarrow \Delta F^i & =  12 \mathcal{N}_i^x \frac{n_{i \ 0}}{g_i T^2} \, . \nonumber
\end{align}

Putting \cref{eq:muCS_toF} and \eqref{eq:changeFi_Nf} together, and summing over all quark species $i$ charged under SU($N_x$), we find the free energy change between the two sphalerons shown in \cref{fig:hopping}:
\begin{equation}
    \Delta F = \Delta F^\mathrm{CS} + \Delta F^i = - 2 \mu_{\mathrm{CS}} + 12 \sum_i \frac{\mathcal{N}_i^x}{g_i T^2} n_i\,, 
    \label{eq:F_total}
\end{equation}
where the sum runs over all fermion species $i$ charged under the gauge group.

We can now use these results in \cref{eq:ratio_F} and define
\begin{equation}
    \hat\Gamma_{\mathrm{sph}} \equiv R_+ + R_- \  ,
    \label{eq:Rsum_sph}
\end{equation}
to find $R_\pm$, where $\hat\Gamma_{\mathrm{sph}}$ is the rate of Chern-Simons number diffusion (related to the rate at which the Chern-Simons number changes in the background of a chemical potential as $2\, \hat{\Gamma}_\mu=\hat\Gamma_{\mathrm{sph}}$ \cite{Moore:1996qs}). 
With a little algebra, we find
\begin{equation}
    R_+ - R_- =  \hat\Gamma_{\mathrm{sph}} \frac{e^{-\Delta F/2T}-1}{e^{-\Delta F/2T} + 1} \ .
    \label{ea:Rdif}
\end{equation}
After expanding in $\mu_{\mathrm{CS}}$ and $n_i$, we find 
\begin{equation}
    \partial_t N_{\mathrm{CS}} = R_+ - R_- = \hat\Gamma_{\mathrm{sph}} \frac{\mu_{\mathrm{CS}}}{2T} - 3 \frac{\hat\Gamma_{\mathrm{sph}}}{T^3} \sum_i \frac{\mathcal{N}_i^x}{g_i} n_i  \,.
    \label{eq:master_eq_fin_almost}
\end{equation}
For a gauge group SU($N_x$), the ratio $g_i/\mathcal{N}_i^x$ corresponds to the number of internal degrees of freedom associated with SU($N_x$); for example, $g_i/\mathcal{N}_i^x = N_x$ for fermions in the fundamental representation of SU($N_x$).
Thus, 
\begin{equation}
    \partial_t N_{\mathrm{CS}} = \hat\Gamma_{\mathrm{sph}} \frac{\mu_{\mathrm{CS}}}{2T} - \frac{3}{N_x} \frac{\hat\Gamma_{\mathrm{sph}}}{T^3} \sum_i n_i \ .
    \label{eq:master_eq_fin}
\end{equation}

We can now express this equation as a contribution to the number density of any fermion species $f$ in a volume $V$. 
Using \cref{eq:NQNCS} we have
\begin{equation}
    \partial_t n_{f} = \frac{\mathcal{N}_f^x}{V} \partial_t N_{\mathrm{CS}} = \frac{3\mathcal{N}_f^x}{N_x} \frac{\Gamma_{\mathrm{sph}}}{T^3} \left( \frac{N_x}{6} \mu_{\mathrm{CS}} T^2 - \sum_i n_i \right) \ ,
    \label{eq:master_eq_Ni}
\end{equation}
where $\Gamma_{\mathrm{sph}}=\hat\Gamma_{\mathrm{sph}}/V$ is the sphaleron rate per unit volume for the gauge group SU($N_x$). 
This equation is the main result of this appendix.

As a cross-check, we can compare this equation to different standard results in the literature.
The first term in \cref{eq:master_eq_fin} corresponds to the standard source term describing the sphaleron-induced change in the Chern-Simons number; see, for example, Ref.\cite{Moore:1996qs} and references therein.
We can similarly recover the standard wash-out term for e.g., $n_{B+L}$ from \cref{eq:master_eq_Ni}.
To do so, observe that when the SM Yukawa interactions are out of equilibrium, and all fermion asymmetries originate from the left-handed fermions $q$ and $\ell$, for each generation of fermions we have 
\begin{equation}
    n_{B \pm L} = \frac{1}{3} n_q \pm n_\ell \ ,
    \label{eq:B+-L}
\end{equation}
which can be used in \cref{eq:master_eq_Ni} to yield (for $N_g=3$ generations of $q$ and $\ell$ fermions)
\begin{equation}
    \dot{n}_{B+L} \supset - 6 N_g \frac{\Gamma_{\mathrm{sph}}}{T^3} n_{B+L} \ .
    \label{B+L_check}
\end{equation}
This is in agreement with the standard wash-out term for $n_{B+L}$ from electroweak sphalerons, e.g. see Ref.~\cite{Bochkarev:1987wf,Rubakov:1996vz} and references therein.

We use \cref{eq:master_eq_Ni} to capture the contribution of the strong and the electroweak sphalerons to the Boltzmann equation of various fermion asymmetries. Putting \cref{eq:master_eq_Ni} and \cref{eq:CQ_yukawa} together, we find the final form of the relevant Boltzmann equations in the main body of the manuscript, see \cref{eq:final_boltz}.\footnote{Our Boltzmann equations are slightly different than those in Ref.~\cite{Co:2019wyp}. The difference arises from the factor $3\mathcal{N}_f^x/N_x$ in \cref{eq:master_eq_Ni}, i.e. an $\mathcal{O}(1)$ prefactor multiplying the sphaleron rates. While the final baryon asymmetry is independent of this prefactor, the friction term $\Upsilon$ scales at most linearly with this prefactor. } It should be noted that the only asymmetry source term in the equations is proportional to $\dot{\theta}$, thus all fermion asymmetries will be proportional to this quantity.

The discussion above can be used to calculate the friction term $\Upsilon$ felt by $\theta$ as well. 
To see that, we should start from the equation of motion of $\theta$. Neglecting the topological susceptibility, $\chi$, (i.e. the second derivative of the energy density with respect to $\theta$) and assuming the $\theta$ has no potential, the equation of motion is 
\begin{equation}
\ddot\theta + 3H \dot\theta  - \frac{\nabla^2}{\asf^2}\theta= \frac{\alpha}{8\pi f_a^2} W^a_{\mu\nu}\tilde{W}^{\mu\nu,\,a}\, ,
    \label{eq:EOM_theta_appx}
\end{equation}
where $\asf$ is the scale factor and $f_a$ is the PQ breaking scale for the associated axion $a=\theta f_a$. 
Neglecting the spatial variation of $\theta$ and using \cref{eq:changeCS_app} and \eqref{eq:master_eq_fin}, we find
\begin{equation}
\ddot\theta + 3H \dot\theta = \Gamma_{\mathrm{sph}} \frac{\mu_{\mathrm{CS}}}{2f_a^2 T} - \frac{3}{N_x} \frac{\Gamma_{\mathrm{sph}}}{f_a^2 T^3} \sum_i n_i \equiv - \frac{\Upsilon}{f_a^2} \dot{\theta} \, ,
    \label{eq:friction_EOM_appx}
\end{equation}
where in the last equality we have used the fact that all fermion asymmetries will be linearly proportional to $\dot\theta$. 
This term acts similar to the Hubble, i.e. as a friction term, and dissipates the axion energy due to its interaction with chiral fermions in the bath; see Ref.~\cite{Berghaus:2020ekh} for further discussions. 
In the main text, we apply the results of this appendix to find the friction terms in the visible and the dark confining sector.

In deriving \cref{eq:friction_EOM_appx}, we have neglected the possible gauge-helicity screening identified in Ref.~\cite{Broadberry:2025ggb}: the rolling axion biases the two gauge-boson helicities, while the resulting helicity asymmetry feeds back to screen the sphaleron drive. 
This effect was not quantified for a non-Abelian plasma in Ref.~\cite{Broadberry:2025ggb}, and subsequent work has either treated it qualitatively or studied related helicity-changing channels. 
A recent transport analysis \cite{Dejrah:2026ltr} finds only an $\mathcal{O}(1)$ reduction of the friction---rather than a near-cancellation---although a first-principles non-Abelian determination is still lacking.

\section{Further details on solving the degenerate system of Boltzmann equations}
\label{app:linear_sys}

Here we discuss in more detail the linear system of equations from the equilibrium condition on the Boltzmann equations in \cref{eq:final_boltz}. 
The system has 16 unknowns ($n_X$ with $X=(q_g,\ell_g ,\bar{u}_g,\bar{d}_g,\bar{e}_g,H)$ with generation index $g=1,2,3$). 
In addition to the fermion asymmetries, we should include the Boltzmann equation for Higgs asymmetry, which only involves the wash-out terms from the $2\to 2$ scatterings:
\begin{align}
    \dot{n}_H + 3 H n_H = & \sum_g
    3\kappa_{{\rm ch},u}\min(\alpha_3 ,y_{u_g}^2)T\left( -\frac{n_{q_g}}{6} - \frac{n_{\bar{u}_g}}{3} - \frac{n_H}{4}  \right) \\ 
    \label{eq:boltz_H}
    &- 3\kappa_{{\rm ch},d}\min(\alpha_3 ,y_{d_g}^2)T\left( -\frac{n_{q_g}}{6} - \frac{n_{\bar{d}_g}}{3} + \frac{n_H}{4}  \right) \nonumber \\ 
    &- 2\kappa_{{\rm ch},e}\min(\alpha_2 ,y_{e_g}^2)T\left( -\frac{n_{\ell_g}}{2} - n_{\bar{e}_g} + \frac{n_H}{4}  \right) \ . \nonumber
\end{align}
By setting the righthand side of all these Boltzmann equations to zero, i.e. imposing the equilibrium condition, we are at a system of 16 linear equations with 16 unknowns. 

However, we find the matrix multiplying the unknown vector $\vec{n}_X$ is singular. This simply follows from different conservation relations, which imply some rows are not linearly independent. 
These conservation laws can in turn be added to the system as new equations. Two such relations arise from conservation of hypercharge and $B-L$ number. 

One can check that with these equations, the rank of the coefficient matrix is 12, which is still smaller than the number of unknowns (16). 
This suggests the solutions to the system make up a 4 dimensional subspace and the system has more than a single solution. 
Further conservation laws respected by the processes involved can be included to single out a specific solution.

That said, we are only interested in the net Baryon asymmetry, which is a sum of a subset of unknowns (different quark asymmetries). 
This combination of unknowns is orthogonal to the kernel of the coefficient matrix, thus it has a unique solution for the net comoving baryon number $Y_B$ and $n_{\mathrm{CS}}$.

Although these equations possess an equilibrium solution with $\dot\theta\sim Y_B T$, one must be careful when considering the backreaction of a fermion asymmetry on the axion. The energy stored in the asymmetry itself is a quantity which may only decrease. For instance, suppose the SM reheats and is initially populated with some $Y_B$. Ignoring the limit of the initial energy stored in the asymmetry, one may erroneously conclude that $\dot\theta(T_{\rm RH})\sim Y_B T_{\rm RH}$. However, in this configuration, the asymmetry would have an energy density $\frac12f_a^2\dot\theta^2\sim f_a^2Y_B^2 T_{\rm RH}^2$, much in excess of the bath's initial asymmetry energy density $\sim Y_B^2 T_{\rm RH}^4$. Thus, accounting for the full dynamics, one would find that the bath itself partly equilibrates as the axion accelerates, resulting in $f_a^2\dot\theta^2\sim Y_B^2 T_{\rm RH}^6/f_a^2$ and an asymmetry in the bath $Y_{B}^{\rm(final)}\sim Y_B T_{\rm RH}^2/f_a^2$, as required by decoupling as $f_a\to\infty$.

\section{Running couplings and masses}
\label{app:running}

In this appendix, we summarize the renormalization-scale dependence of the gauge couplings and quark masses used in the finite-temperature calculation. In the thermal bath we evaluate these running quantities at a scale $\mu \sim T$, up to the usual order-one ambiguity.
With this prescription, the fine structure constant $\alpha_{c,d}$ and quark masses $m_{c,d}$ are functions of temperature at one-loop. The renormalization group equations are \cite{kapusta2007finite,Luthe:2016ima}
\begin{align}
    -\frac{2\pi}{\alpha_{c,d}(\mu)}\frac{\partial\log\alpha_{c,d}(\mu)}{\partial\log \mu} &= \frac{11}{3}N_{c,d} - \frac{2}{3}\sum_f B_0\left(\frac{m_f^2(\mu)}{\mu^2}\right)\,,\\
    -\frac{2\pi}{\alpha_{c,d}(\mu)}\frac{\partial \log m_f(\mu)}{\partial\log\mu}&=3\frac{N_{c,d}^2-1}{2N_{c,d}}C_0\left(\frac{m_f^2(\mu)}{\mu^2}\right)\,,
\end{align}
where the functions $B_0$ and $C_0$ are
\begin{align}
    B_0(x) &= 1 - 6 x + 12\left(\frac{x^2}{\sqrt{1+ 4x}}\right)\log\left(\frac{\sqrt{1+ 4x} + 1}{\sqrt{1+ 4x} - 1}\right)\,,\\
    C_0(x)&=1 - x\log(1 + x^{-1})\,,
\end{align}
and we have assumed that the fermions are charged under exactly one of SU($N_c$) or SU$(N_d)$ and transform in the fundamental representation.
We define the confinement scale $\Lambda_{c,d}$ by the boundary condition $\alpha_{c,d}(\Lambda_{c,d}) = 1$ and $m_f(\Lambda_{c,d}) = m_f$ at the confinement scale. This choice of boundary condition is to avoid the formal blowup at the confinement scale in the usual prescription. With this boundary condition, the general behavior of the fine structure constants is well-described by
\begin{align}\label{eqn:running_fine_structure}
    \alpha_{c,d}^{-1}(T) &= 1 + \frac{11 N_{c,d} - 2 n_{c,d}(T)}{6\pi}\log\left(\frac{T}{\Lambda_{c,d}}\right)\,,
\end{align}
where the effective number of active quarks is approximated by~\cite{kapusta2007finite}
\begin{align}
    n_{c,d}(T) \approx \frac{1}{\log\left[\left(T/\Lambda_{c,d}\right)^2\right]}\sum_f\log\left[\frac{T^2 + m_{f}^2(T)}{\Lambda_{c,d}^2 + m_{f}^2(T)}\right]\,.
\end{align}
In our calculations, we are concerned only with the mass of the lightest dark quark $m_d$. In practice, its running has a marginal effect, since it is only significant once $m_d(T) \lesssim  T$.

\section{Topological susceptibility and sphaleron rate}\label{app:topological_susceptibility}

Here we review the temperature scaling of the axion potential in the dilute instanton gas approximation. Specifically, the axion potential is proportional to the topological susceptibility $\chi_{c,d}(T)$ of the gauge field  
\cite{Gross:1980br,Laine:2016hma}
\begin{align}
    \chi(T)=\int_0^{1/T} d\tau \int d^3x\langle Q(\tau,\mathbf{x})Q(0)\rangle_{\theta=0}\,.
\end{align}
where
\begin{align}
    Q(x)\equiv \frac{\alpha}{8\pi}G^a_{\mu\nu}\tilde G^{a\mu\nu}\,,
\end{align}
Using the periodicity of $\theta$ and assuming the dominance of the leading harmonic,
\begin{align}
    \frac{\alpha}{4\pi}\langle {\rm Tr}\tilde G_{c,d} G_{c,d}\rangle_T \supset - \chi_{c,d}(T)\sin\theta\,.
\end{align}
The high temperature limit of the topological susceptibility can be estimated via a semi-classical approximation \cite{Gross:1980br}
\begin{align}\label{eqn:DIGA}
    \chi_{c,d} \approx T^4 e^{-2\pi/\alpha_{c,d}}\prod_{f:m_f< T}\frac{m_f}{T}\,,
\end{align}
where the $T^4$ scaling may have been anticipated from naive dimensional analysis; the exponential factor $e^{-2\pi/\alpha_{c,d}}$ comes from the Euclidean action of the instanton which interpolates between neighboring $\theta$-vacua; and the factors of fermion mass arise from the fermion zero-modes in the background of the instanton. 

In the thermal bath, the typical energy scale is $T$ up to $\calO(1)$ factors, so we evaluate $\alpha_{c,d}$ at $\mu =  T$ in \cref{eqn:DIGA}. As $T$ approaches $\Lambda_{c,d}$, our perturbative expansion breaks down badly, and ultimately we must make a choice of how to connect between the perturbative and non-perturbative regime. Thus, for $T < \Lambda_{c,d}$ we write
\begin{align}\label{eqn:vacuum_topological_susceptibility}
    \chi_{c,d}(0) = X_{c,d} \Lambda_{c,d}^4\min\left\{1,\frac{m_f}{\Lambda_{c,d}}\right\}\,,
\end{align}
where $X_{c,d}$ is a dimensionless number characterizing the relative size of $\chi_{c,d}$ compared to the naive dimensional analysis captured by the remaining factors, and where $m_f$ denotes the lightest fundamental fermion mass (the ``minimum''
term is omitted for pure Yang-Mills). Approximately independent of the gauge group, lattice results suggest $X\approx 0.16$, which we assume in our calculations~\cite{Lucini:2012wq,Berg:2017tqu,Yamada:2024vsk,Bonanno:2025eeb}. Note that in treating $\Lambda_{c,d}$ as a ``confinement temperature'' we are making a coarse approximation, though it is accurate to the level of approximation in this paper.

In the case of the SM, the value of the zero-temperature topological susceptibility $\chi_c(0)$ is known from the measured value of the pion masses and decay constants \cite{GrillidiCortona:2015jxo},
\begin{align}
    \chi_c(0) = \left[\frac{m_u/m_d}{(1 + m_u/m_d)^2} + \cdots\right]m_{\pi^0}^2 f_\pi^2  = [75.46(29) \MeV]^4\,,
\end{align}
where $\cdots$ indicates next-to-leading order corrections which shift the leading order result at the percent level, 
corresponding to $X_c \approx 2$ with $\Lambda_c = 200\MeV$ and $m_u = 2 \MeV$, where $m_u/m_d = 0.472(11)$, and $m_{\pi^0} = 135\MeV$ and $f_\pi = 92.3\MeV$ are the neutral pion mass and decay constant. The fact that $X_c$ is order one in the SM is an indication that \cref{eqn:vacuum_topological_susceptibility} should capture the important behavior of $\chi_d$, though we will be applying \cref{eqn:vacuum_topological_susceptibility} in the limit of $m_{q_d}> \Lambda_d$ where only lattice gauge theory may provide accurate insight into the size of $X_d$.

To connect the vacuum topological susceptibility \cref{eqn:vacuum_topological_susceptibility} to the high temperature scaling \cref{eqn:DIGA}, we straightforwardly patch the temperature scaling of \cref{eqn:DIGA} onto the zero-temperature susceptibility \cref{eqn:vacuum_topological_susceptibility}:
\begin{align}\label{eqn:topological_susceptibility_scaling}
    \chi_{c,d}(T) \approx \chi_{c,d}(0) \left(\frac{\Lambda_{c,d}}{T}\right)^{\frac{11}{3}N_{c,d}-4}\prod_{f:m_f < T}\left(\frac{\max\left\{m_f,\Lambda_{c,d}\right\}}{T}\right)^{1/3},\hspace{1cm} T > \Lambda_{c,d}\,,
\end{align}
where we have dropped the additive constant in $\alpha^{-1}$ in \cref{eqn:running_fine_structure} to extract the large-temperature power law, and we have approximated the number of active quarks as a step function. The fractional power in the fermion mass threshold factor comes from combining the fermion zero-mode factor $m_f/T$ with the change in the instanton action from the fermion contribution to the beta function above the quark threshold.
While the approximation does capture the leading scaling at high temperature, it misses behavior of $\chi_{c,d}(T)$ around the transition between the confined and de-confined phases, though such errors are unavoidable in a perturbative treatment near confinement. The dynamics we discuss in the main body of the paper are not sensitive to the precise temperature dependence of the potential, only that $\chi_{c,d}(T)$ turn on rapidly as $T\to\Lambda_{c,d}$, so these uncertainties have minimal impact on our results. Therefore, we utilize \cref{eqn:topological_susceptibility_scaling} in all of our numerical axion evolution.

The diffusion coefficient $\Gamma_{\rm sph}$ is the rate of sphaleron transitions per spacetime volume, and has been computed in Ref.~\cite{Arnold:1996dy,Bodeker:1998hm,Moore:2010jd}:
\begin{align}\label{eqn:sphaleron_rate_with_coefficient}
    \Gamma_{\rm sph} &= C_{\rm sph}(N\alpha)^5 T^4\,,\\
    C_{\rm sph} &\equiv (0.21\pm0.01)\frac{6 N}{2N + N_f}W\left(e^{3.041}\sqrt{\frac{2\pi}{3\alpha}\frac{2 N + N_f}{N^2}}\right)\frac{N^2 - 1}{N^2}\,.
\end{align}
Here, $N_f$ is the number of light Dirac fundamental fermions contributing to Debye screening at temperature $T$, and $W$ is the Lambert $W$ function. We emphasize that this expression is only expected to hold in the regime of weak couplings.

\section{Relic abundance from adiabatic invariants}
\label{app:relic_abundance_from_adiabatic_invariant}

At temperatures well above the QCD phase transition, the topological susceptibility of QCD is vanishingly small and the axion potential negligible. The exactness of the axion shift symmetry then implies that the yield
\begin{align}
    Y_\theta\equiv \frac{n_\theta}{s}\,,
\end{align}
is constant to very high accuracy, where
\begin{align}
    s = \frac{2\pi^2}{45}g_{\star,s} T^3\,,
\end{align}
is the entropy density of the Universe. The large axion velocity means that Hubble friction, which in the canonical misalignment story sets the onset of axion oscillations when $3H \approx m_a$, does not set the relevant timescale. Instead, the axion switches from rolling to oscillating when the energy density stored in its rotation is comparable to the height of its potential barriers. This defines the oscillation temperature via
\begin{align}
    \frac12\frac{n_\theta^2(\Tosc)}{f_a^2} = 2\chi_c(\Tosc)\,.
\end{align}
Below the temperature $\Tosc$, the charge density $n_\theta$ is no longer approximately conserved, though the oscillating axion system does admit an approximately conserved quantity: the axion number density $n_a$, which is conserved up to non-adiabaticity in the changing mass $m_a^2(T) = \chi_c(T)/f_a^2$ and anharmonicity of the potential.

The Hamiltonian of the axion system corresponding to the Lagrangian density
\begin{align}
    \calL = \sqrt{-g}\left(\frac12 f_a^2\dot\theta^2 - \frac12 m_a^2f_a^2\theta^2\right)\,,
\end{align}
is
\begin{align}
    H = \frac12\int{\rm d}^3 x\, \left(\frac{1}{f_a^2 \asf^3} p^2 + f_a^2 \asf^3m_a^2q^2\right)\,.
\end{align}
where the canonical coordinates $q,p$ are defined as
\begin{align}
    q \equiv \theta\,,\hspace{1cm} p \equiv \frac{\partial\calL}{\partial\dot q} = \asf^3 f_a^2 \dot q\,.
\end{align}
The system admits periodic solutions both in the massless and massive limits, corresponding to rolling on the compact field space and oscillations about the potential minimum respectively. Therefore, the adiabatic invariant
\begin{align}
    J\equiv\int_0^T{\rm d}t\, p \dot q\,,
\end{align}
is well defined in both regimes, where $T$ is the period. In the massless regime, we have 
\begin{align}
    J = \int_0^{2\pi}{\rm d}\theta\, a^3 f_a^2 \dot \theta = 2\pi a^3 f_a^2\dot\theta = 2\pi a^3 n_\theta\,.
\end{align}
In the massive case,
\begin{align}
    J=\int_0^{2\pi/m_a}{\rm d}t\, a^3 f_a^2 \dot \theta^2 = 2\pi \frac{a^3 f_a^2\langle\dot\theta^2\rangle}{m_a} = 2\pi a^3 n_a\,,
\end{align}
where we have used the standard relation $n_a = f_a^2 \langle\dot\theta^2\rangle/m_a$ for coherent harmonic oscillations. Neglecting non-adiabaticity across the transition between these two regimes, we obtain
\begin{align}\label{eqn:adiabatic_matching}
    n_a = n_\theta\,.
\end{align}
Therefore, the final relic abundance can be estimated as
\begin{align}
    \rho_a = m_a(0) n_\theta = m_a(0) Y_\theta s.
\end{align}
Solving the full equation of motion across the transition at $\Tosc$ to account for both anharmonicity and nonadiabaticity, we find a factor of $\sim 2$ enhancement of the relic abundance, in agreement with Ref.~\cite{Co:2019jts}, so in practice the relationship in \cref{eqn:adiabatic_matching} should be replaced by $n_a \approx 2\,n_\theta$.

Taking $n_\theta = f_a^2 \omega_{\rm ew}$ at $\Tew$ and matching onto the dark matter energy density at matter–radiation equality (i.e{.} $\approx 5/6$ of the matter density), we find
\begin{align}\label{eqn:axiogenesis_relic_abundance}
    \frac{\Omega_a}{\Omega_{\rm cdm}} = \frac{72}{\pi^2}\frac{g_{\star,s}(T_{\rm eq})}{g_\star(T_{\rm eq})g_{\star,s}(\Tew)}\frac{\chi_c(0)}{T_{\rm eq}\Tew^3}\frac{\omega_{\rm ew}}{m_a(0)} \approx 49\left(\frac{f_a}{\fSN}\right)\,,
\end{align}
where the numerical estimate assumes the benchmark value of $\omega_{\rm ew}$ corresponding to the observed baryon asymmetry. This result illustrates the remarkable fact that the kinetic misalignment mechanism axion relic abundance is, up to ${\cal O}(1)$ corrections from anharmonicity and nonadiabaticity, independent of the temperature dependence of the QCD topological susceptibility. This follows from the near-conservation of PQ charge above $\Tosc$ and axion number density below $\Tosc$, or equivalently from the conservation of the adiabatic invariant $J$.

The adiabatic invariant approach can also be generalized to the case where the axion experiences dissipation. The appropriate Lagrangian density which encodes the dissipation from \cref{sec:review} is
\begin{align}\label{eqn:Lagrangian_dissipation}
    \calL = \sqrt{-g}\exp\left\{\int_{t_i}^t{\rm d}t \frac{\Upsilon_c}{f_a^2}\right\}\left(\frac12 f_a^2 \dot\theta^2 - \frac12 m_a^2 f_a^2\theta^2\right)\,.
\end{align}
The remaining steps to compute the relic abundance are analogous to those above. While this construction is perfectly well suited to a single axion, it does not straightforwardly generalize to multiple axions with a general dissipative matrix, where the appropriate effective Lagrangian is not obvious. In what follows we will instead analyze the equations of motion for $\theta$ directly.

\section{Baryon abundance from an oscillating axion}
\label{app:rapid_oscillation_YB}

Here we discuss the scaling of the $B + L$ yield depending on the behavior of the axion field, i.e. whether it is rolling or oscillating slowly or oscillating rapidly at temperatures around $\Tew$. While the full system of Boltzmann equations is quite involved, we can gain some insight using two key facts. First, since $B + L$ violation is mediated by electroweak sphalerons, the rate of $B + L$ violation must be proportional to $\Gamma_w$. Second, as we have observed in \cref{subsec:sph_boltzman}, the equilibrium value for $Y_B$ is independent of the weak sphaleron rate and in fact is approximately independent of the strong sphaleron rate. Thus, we can write a toy model that captures the key physics as
\begin{align}
    \dot Y_{B + L} = - \gamma_{B + L}\left(Y_{B + L} - Y_{B + L}^{\rm eq}\right)\,,
\end{align}
where $Y_{B + L}^{\rm eq}\sim c T^2\dot\theta/s$, $c$ is approximately a constant independent of the weak and strong sphaleron rates, and $\gamma_{B + L}\propto \Gamma_w/T^3$. The sign has been chosen so that $Y_{B + L}$ relaxes toward $Y_{B + L}^{\rm eq}$. To make the equations more readable, we drop the $B + L$ subscript going forward. We can directly integrate this equation to give
\begin{align}
    Y(t)=\int_{-\infty}^t{\rm d}t'\gamma(t')Y^{\rm eq}(t')e^{-\int_{t'}^t{\rm d}t''\gamma(t'')}\,.
\end{align}
As the electroweak sphaleron freezes out, the sphaleron rate decreases exponentially~\cite{DOnofrio:2014rug}. As a toy model, we take
\begin{align}
    \gamma = \gamma_0 e^{-r t}\,.
\end{align}
where $\gamma_0$ and $r$ are taken to be constants with units of mass. Writing
\begin{align}
    Y^{\rm eq} = \bar Y^{\rm eq} \cos(\omega t)\,,
\end{align}
the final abundance is
\begin{align}
    Y_\infty&=\bar Y^{\rm eq}\Re\left[\left(\frac{\gamma_0}{r}\right)^{{i \omega }/{r}}\Gamma \left(1-\frac{i \omega }{r}\right)\right]\,,
\end{align}
where $\Gamma$ is the gamma function. We can evaluate this function in the limiting cases:
\begin{align}
    \Re\left[\left(\frac{\gamma_0}{r}\right)^{{i \omega }/{r}}\Gamma \left(1-\frac{i \omega }{r}\right)\right] \to
    \begin{cases}
        1&\omega\ll r\,,\\
        e^{-\pi\omega/2r}\sqrt{\frac{2\pi \omega}{r}}\cos\left[\frac{\omega}{r}\left(1 + \log\frac{\gamma_0}{\omega}\right) - \frac{\pi}{4}\right]&\omega\gg r\,.
    \end{cases}
\end{align}
Thus the yield is exponentially suppressed in the large $\omega/r$ limit:
\begin{align}
    \frac{Y(\infty)}{\bar Y^{\rm eq}}\propto \sqrt{\frac{\pi\omega/r}{\sinh(\pi\omega/r)}}\,,
\end{align}
where we have dropped the rapid oscillations at large $\omega$ to illustrate the envelope's scaling. In this toy model, the duration of freeze-out is controlled by $1/r$. Rapid oscillations of the axion-induced equilibrium value therefore cancel across the freeze-out interval when $\omega > r$. The parameter $\gamma_0$ only sets the location of this interval, and thus appears in the oscillatory phase rather than in the envelope.

\section{Mode mixing in the decoupling limit}
\label{app:mode_mixing_decoupling}

The discussion in \cref{sec:damping_from_dark_sphalerons} focused on identifying the quasinormal modes of the two-axion system and their corresponding decay rates. In general, however, these modes are not perfectly orthogonal solutions to the equations of motion. The late-time evolution is therefore controlled not only by the decay rates, but also by how efficiently the slowly decaying mode is populated at early times. We illustrate this point with a simple limiting example.

Consider the following reduced version of \cref{eqn:two_axion_EOM} written in terms of the variables \cref{eqn:A} and~\eqref{eqn:B}, and with $C = 1$ for simplicity:
\begin{align}\label{eqn:AB_EOM_Reduced}
    -\alpha B' &=  A'' + \alpha\bar\beta^2 A' + A\,,\\
    -\alpha\beta^2\bar\beta^2 A' &= B'' + \alpha\beta^2 B'\,.
\end{align}
Here we have removed the mass term for $B$, expanded the sinusoidal potential for $A$ to quadratic order, moved to the dimensionless time coordinate $\tau=m_A t$, with primes denoting derivatives with respect to $\tau$, and neglected Hubble friction. The latter can be restored by the usual WKB rescaling.

We take initial conditions
\begin{align}
    A(0)=A_0\,,\qquad A'(0)=0\,,\qquad B'(0)=0\,,
\end{align}
while the constant mode of $B$ is arbitrary. We also work in the limit $\beta\ll1$, so that $\bar\beta\simeq1$. This corresponds to the decoupling limit $f_1\to\infty$, where one might expect the dynamics to be well described by a single-axion effective theory.

Indeed, at early times this expectation is correct. Since $B'$ is only weakly sourced by $A'$, the leading dynamics are obtained by treating $A$ as approximately independent,
\begin{align}
    0 \simeq A''+\alpha\bar\beta^2 A' + A\,.
\end{align}
For small $\alpha$, this gives
\begin{align}
    A(\tau)\simeq A_0 e^{-\alpha\bar\beta^2\tau/2}\cos\tau\,.
\end{align}
The field $B$ is then sourced by these damped oscillations. Most of the induced response consists of oscillatory terms that inherit the relatively fast decay rate $\alpha\bar\beta^2/2$. However, the system also populates a non-oscillatory rolling mode whose velocity decays only at the rate $\alpha\beta^2$:
\begin{align}
    B'(\tau)\sim A_0\,\alpha\beta^2\bar\beta^2\,e^{-\alpha\beta^2\tau}\,.
\end{align}
At sufficiently late times, the original oscillatory $A$ component has decayed away. Nevertheless, the slowly decaying $B'$ mode continues to source $A$, preventing it from fully relaxing to the minimum of its potential. Parametrically,
\begin{align}
    A(\tau)\sim A_0\,\alpha^2\beta^2\bar\beta^2\,e^{-\alpha\beta^2\tau}\,.
\end{align}

This example highlights an important subtlety of the two-axion decoupling limit. Taking one decay constant large does not simply remove the corresponding degree of freedom at all times. Rather, the decoupling is dynamical: at early times the system is well approximated by the independent motion of $A$, but at late times the nearly decoupled rolling mode dominates because it is also only weakly damped. Equivalently, although the two-axion system may appear reducible to a one-axion effective theory over intermediate timescales, any finite ratio of decay constants generically leaves the slowly decaying mode populated, which ultimately controls the asymptotic evolution.

\section{Monodromy oscillation fragmentation}
\label{app:Monodromy_oscillation_fragmentation}

In general, the monodromy potential of \cref{sec:monodromy} will introduce axion self-interactions that cause certain momentum modes to grow, possibly becoming nonperturbatively large and spoiling homogeneity, invalidating the assumptions used to derive the axion-Boltzmann equations. Here, we perform a simple analytical estimate of the fragmentation time as a function of the phenomenological monodromy potential parameters. The resulting fragmentation timescale should then be viewed as a cutoff on the regime of validity of \cref{fig:monodromy_evolution_YB,fig:monodromy_evolution_osc} in the rapidly oscillating regime.

Consider the axion equation of motion \cref{eq:single_axion_monodromy_EOM} written in dimensionless units of time and a re-scaled field in a non-expanding Universe:
\begin{align}
    \Theta\equiv\frac{\langle\theta\rangle}{\theta_0}\,,\quad\delta\vartheta\equiv\frac{\theta - \langle\theta\rangle}{\theta_0}\,,\quad\zeta\equiv\mu\theta_0^{(p-2)/2}t\,,\quad\kappa\equiv \theta_0^{(2-p)/2}\frac{k_{\rm phys}}{\mu}\,,\quad\varepsilon\equiv\frac{\alpha}{\theta_0}\,,
\end{align}
where $k_{\rm phys}$ is the physical wavenumber, 
in which case the equations of motion become
\begin{align}
    0&=\Theta'' + \Theta(\varepsilon^2 + \Theta^2)^{p/2 - 1}\,,\\
    0&=\delta\vartheta_\kappa'' + \left[\kappa^2 + \frac{\varepsilon^2 + (p-1)\Theta^2}{(\varepsilon^2 + \Theta^2)^{2 - \frac{p}{2}}}\right]\delta\vartheta_\kappa\,,
\end{align}
where $'$ denotes differentiation with respect to $\zeta$.
To simplify the analysis, we take the limit $\varepsilon \to 0$.

The (in)stability of long wavelength modes is easy to understand from the perspective of the effective equation of state, which determines the dispersion relation of long wavelength modes~\cite{Cembranos:2015oya}
\begin{align}
    \Omega^2 = c_{\rm eff}^2\kappa^2
\end{align}
where the physical-time frequency is related to the dimensionless time frequency by $\omega = \mu\theta_0^{(p-2)/2}\Omega$. Using
\begin{align}
    c_{\rm eff}^2= w_p = \frac{p-2}{p+2}\,,
\end{align}
we immediately find that a long-wavelength instability exists for $p < 2$.

To understand higher frequencies, we need to study the principal instability band using the Mathieu equation. For this purpose, it is useful to further adapt the units so that the oscillation frequency is unity to avoid unphysical secular terms. The equation of motion is integrable, so the frequency shift owing to self-interactions can be calculated by direct integration
\begin{align}
    \varpi_p =\frac{\pi\sqrt{p/2}}{B(1/p,1/2)}\,,
\end{align}
where $B$ is the $\beta$-function. We then define
\begin{align}
    \zeta_p = \varpi_p\zeta\,,\qquad\kappa_p = \varpi_p^{-1}\kappa\,,
\end{align}
in terms of which the equation of motion for the perturbations take the form
\begin{align}
    0&=\partial_{\zeta_p}^2\delta\vartheta + \left[\kappa_p^2 + \frac{(p-1)}{\varpi_p^2}|\Theta|^{p-2}\right]\delta\vartheta\,.
\end{align}
The homogeneous mode may then be written
\begin{align}
    \Theta(\zeta_p) = \sum_{n\in\rm odd} a_{n,p}\cos n\zeta_p\,.
\end{align}

\begin{figure}
    \centering
    \includegraphics[width=\columnwidth]{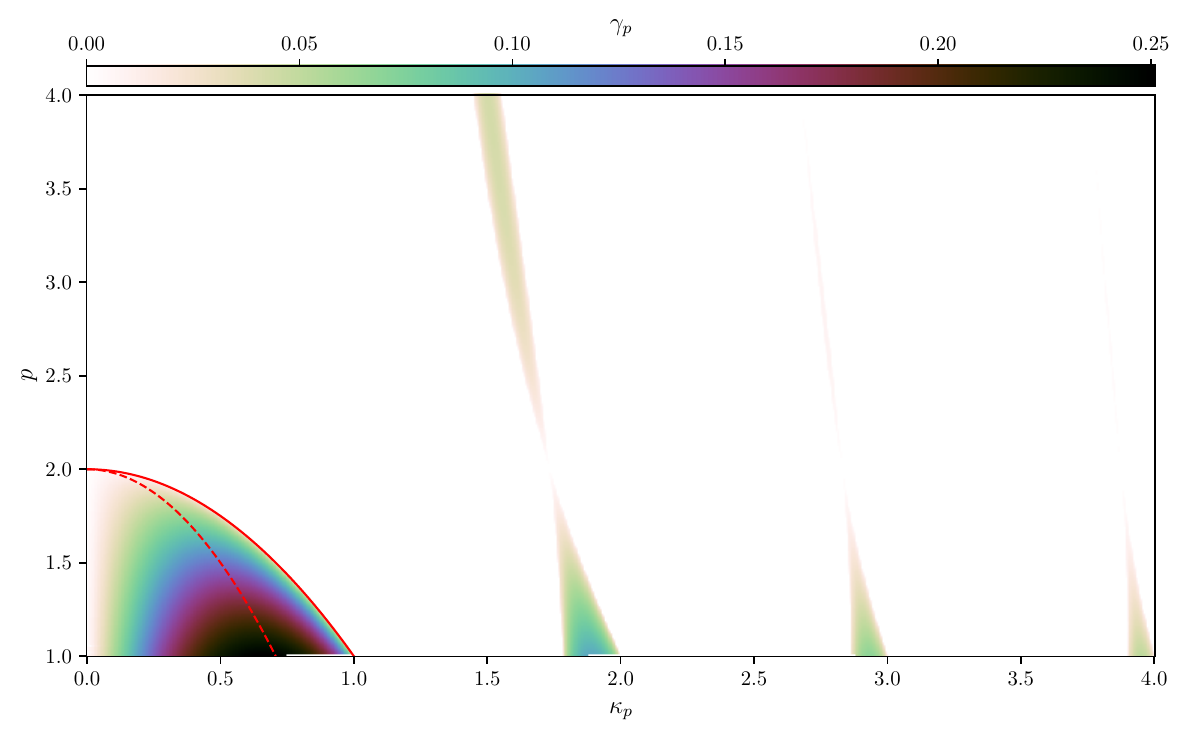}
    \caption{Growth rate of perturbations of the axion field oscillating in the monodromy potential \cref{eq:mono-potential} in the limit $\alpha/\theta_0\to 0$ as a function of the re-scaled wavenumber $\kappa_p$ and the asymptotic monodromy power law $p$. The principal Mathieu instability is visible, encompassed by a solid red curve indicating the boundary of the band as calculated in the $\eta\to 0$ limit of this appendix. The dashed red curve indicates the fastest growing wavenumber $\kappa_{p,\rm max}$ calculated in the same approximation.}
    \label{fig:floquet}
\end{figure}

We now solve for the perturbations in the $p\to 2^{-}$ limit analytically.
We define the expansion parameter $\eta\equiv2-p\ll1$. In this limit, the homogeneous mode takes the form
\begin{align}
    \Theta(\zeta_p) = \cos\zeta_p\,.
\end{align}
to zeroth order in $\eta$. The perturbation equation then assumes the form
\begin{align}
    0=\partial_{\zeta_p}^2\delta\vartheta_k+\left[\kappa_p^2+\frac{(1-\eta)}{\varpi_p^2}|\Theta|^{-\eta}\right]\delta\vartheta_k\,.
\end{align}
We illustrate the growth rate of the perturbations as a function of $\kappa_p$ and $p$ in \cref{fig:floquet}.
The principal Mathieu instability band is controlled by the constant and frequency $2$ term. Dropping higher harmonics from the effective frequency of the perturbation
\begin{align}
    0=\partial_{\zeta_p}^2\delta\vartheta_k+\left[a +   2q\cos2\zeta_p\right]\delta\vartheta_k\,,
\end{align}
with
\begin{align}
    a\equiv 1-\frac{\eta}{2}+\kappa_p^2\,,\qquad q\equiv-\frac{\eta}{2}\,.
\end{align}
The principal Mathieu instability growth rate is then
\begin{align}
    \gamma_p = \frac12\sqrt{q^2 - (a - 1)^2} = \frac12\kappa_p\sqrt{\eta-\kappa_p^2}\,.
\end{align}
The boundary of the principal instability band is indicated with the solid red contour in \cref{fig:floquet}. 
The fastest growing mode has wavenumber and growth rate
\begin{align}
    \kappa_{p,\rm max} = \sqrt{\frac{\eta}{2}}\,,\qquad\gamma_{p,\rm max} = \frac{\eta}{4}\,.
\end{align}
$\kappa_{p,\rm max}$ is indicated by the dashed red contour in \cref{fig:floquet}.
The total number of oscillations before the onset of nonlinearity, when $\langle\delta\vartheta^2\rangle\sim 1$, is then
\begin{align}
    N_{\rm osc}\approx \frac{2}{\pi\eta}\log [\delta\vartheta_{\kappa_{p,\rm max}}^{-1}(0)] \sim \frac{7.3}{\eta}\left(1 + 0.087\log\left[\frac{10^{-5}}{\delta\vartheta_{\kappa_{p,\rm max}}(0)}\right]\right)\,.
\end{align}

\section{Monodromy crossing fragmentation}
\label{app:Monodromy_Crossing_Fragmentation}

Here we test to what extent the results of the homogeneous treatment in \cref{subsec:monodromy_crossing} extend to the case with spatial perturbations. As in \cref{subsec:monodromy_crossing} we treat the case of equal decay constants. Writing $\tau=\mu t$, where $\mu^2 = \Lambda_{\rm mon}^4/f^2$, $\bm y=\mu\bm x$, and $\asf=\sqrt{\tau}$, the spatial equations follow by adding $-\asf^{-2}\nabla^2\theta_i$ to the corresponding homogeneous equations~\cref{eq:two-axion-monodromy-crossing-eom}. We assume radiation-dominated expansion and no dissipation aside from Hubble damping $3H\dot\theta_i$. The susceptibility is $\chi_{\rm conf}/(f^2\mu^2)=10^4\min\{1,(\tau/500)^4\}$.

At $\tau_i=0.01$, we take $\Theta_1=500x$, $\Theta_2=500(1-x)$, and vanishing velocities, sampling 49 initial partitions from the uniform-$x$ ensemble. Each field receives independent, initially frozen Gaussian perturbations with variance $d\langle\delta\theta_i^2\rangle/d\ln\kappa=10^{-16}$ on the available lattice modes in $0.5\leq\kappa\leq80$, where $\kappa=k/\mu$. We evolve the perturbations linearly until the latest of $\tau=75,80,85,90$ preceding the crossing for which $\max|\delta\theta_i|<10^{-4}$, and then evolve the full nonlinear equations. The $2+1$ dimensional ensemble uses a periodic $256^2$ grid with side length $L_y=\pi/2$, momentum spacing $\Delta\kappa=4$, and radial cutoff $\kappa_{\max}=508$. Runs are terminated at $\tau = 170$.

The reason for performing $2+1$ as opposed to $1+1$ or $3+1$ dimensional simulations is as follows: $3+1$ is simply too costly, while in $1+1$ stable domains form upon axion fragmentation.
The potential has degenerate minima $(\theta_1,\theta_2)=(-2\pi n,2\pi n)$, and adjustment of both fields removes the volume-scaling of any monodromy-induced bias between wells. In a simulation with only one spatially varying coordinate, the interfaces are planar and there is no curvature to drive collapse.

\begin{figure}
    \centering
    \includegraphics[width=\columnwidth]{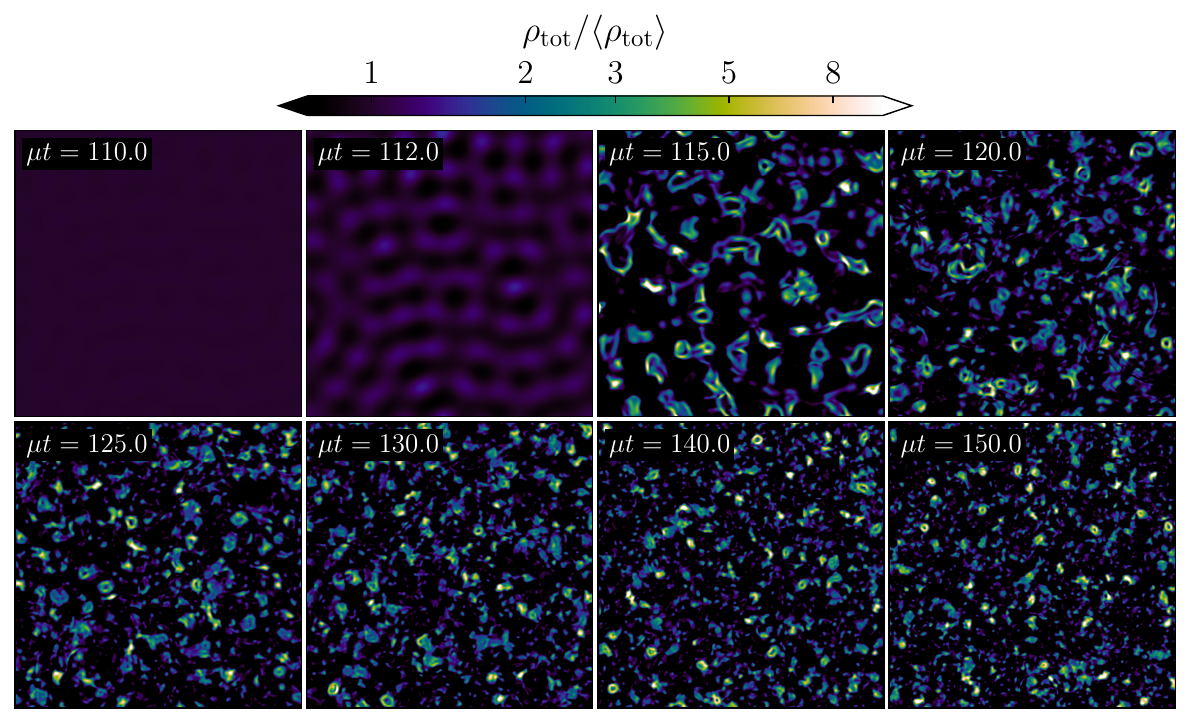}
    \caption{Evolution of the total axion energy density during monodromy--cosine crossing in a $2+1$ dimensional simulation, illustrating axion fragmentation, the collapse of domains and oscillon formation.}
    \label{fig:monodromy_fragmentation_snapshots}
\end{figure}

Therefore, two dimensions offers a minimal testing ground for non-perturbative effects in monodromy--crossing. In this case, the simulations are still small enough to run within a reasonable time, while curvature causes domains to collapse.
\Cref{fig:monodromy_fragmentation_snapshots} plots the total energy density of the system in one such simulation, and illustrates the growth of perturbations, followed by fragmentation and subsequent oscillon formation.

Our goal is to use these $2+1$ dimensional simulations to compare against our homogeneous simulations in \cref{subsec:monodromy_crossing}.
We extend the definition \cref{eq:R} of $R$, the log ratio of the heavy to light mass eigenstates WKB-inverted back to crossing, to the non-perturbative case as follows. Define 
\begin{align}
    \frac{\rho_{\rm light}}{f^2\mu^2}&=\frac12(\partial_\tau\theta_1)^2+\frac{1}{2\tau}(\nabla_{\bm y}\theta_1)^2+\frac12(\theta_1+\theta_2)^2,\\
    \frac{\rho_{\rm heavy}}{f^2\mu^2}&=\frac12(\partial_\tau\theta_2)^2+\frac{1}{2\tau}(\nabla_{\bm y}\theta_2)^2+\frac{\chi_{\rm conf}}{f^2\mu^2}(1-\cos\theta_2)\,.
\end{align}
The total energy density of the system is then $\rho_{\rm tot} = \rho_{\rm heavy} + \rho_{\rm light}$, though the identification of each term with the heavy and light state is only approximate, and becomes accurate when $\chi_{\rm conf}\to\infty$\,. Using the homogeneous crossing time $\tau_\times$ defined in \cref{subsec:monodromy_crossing}, we construct
\begin{align}\label{eq:Rapp}
    R \equiv \log\frac{\langle\rho_{\rm heavy}(\tau_{\times})\rangle}{\langle\rho_{\rm light}(\tau_\times)\rangle}\,,
\end{align}
where $\langle\cdot\rangle$ denotes the spatial average over the box. Since our simulations are short, the nonlinear structures still have not dissipated by the time we measure $R$. Therefore, to increase numerical stability, we infer $R$ by reversing WKB for 31 times, spaced by $1$, between $140$ and $170$ and take the resulting average when reporting our results. We denote the resulting average as $\bar R$. 

\begin{figure}
    \centering
    \includegraphics[width=0.8\columnwidth]{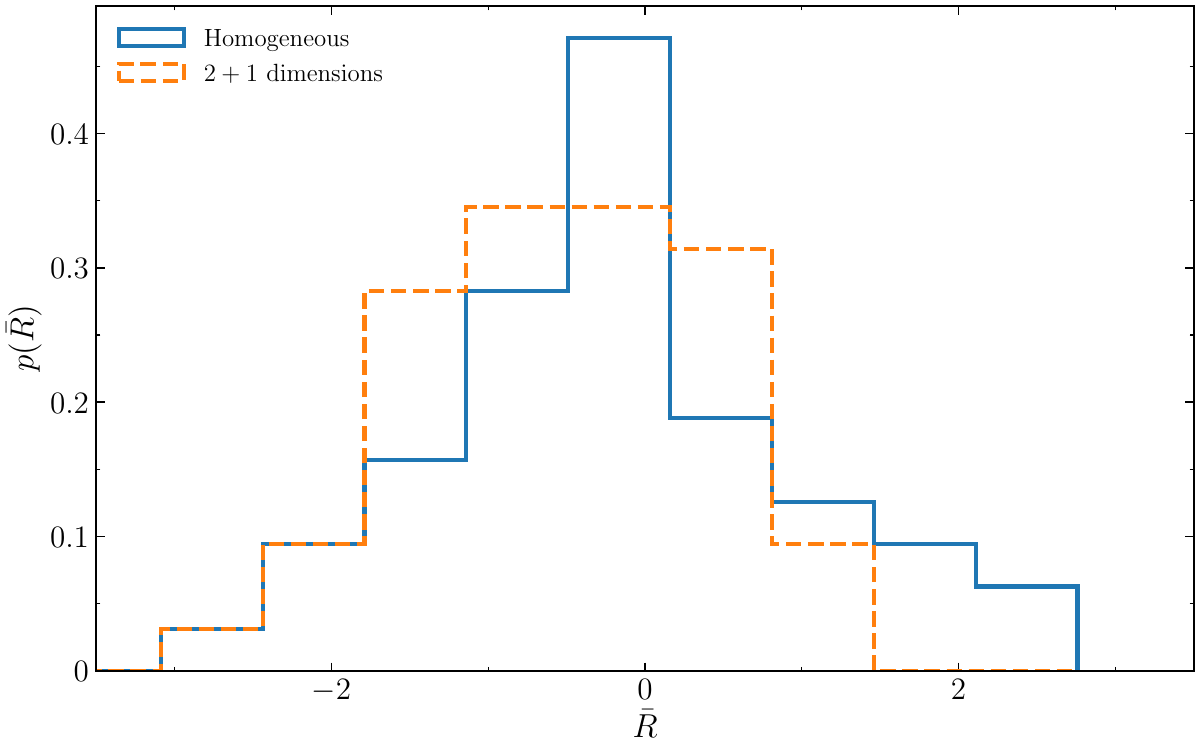}
    \caption{Distribution of the energy partition $\bar R$ defined in \cref{eq:Rapp}, comparing 49 $2+1$ dimensional simulations with homogeneous controls at $\theta_0=500$.}
    \label{fig:monodromy_fragmentation_histogram}
\end{figure}

\Cref{fig:monodromy_fragmentation_histogram} compares 49 $2+1$ dimensional simulation realizations with homogeneous controls evaluated at the same initial conditions, using the same temporal averaging to compute $\bar R$. The distributions are qualitatively similar despite fragmentation and oscillon formation, though notably the fragmented $2+1$ dimensional result is broader around $\bar R = 0$ and does not have as extended a tail towards large $\bar R$. While this result is encouraging that the homogeneous calculations continue to capture the main physics even after including fragmentation, a larger set of $3+1$ dimensional simulations remains necessary.

\section{Fast three-axion evolution}
\label{app:fast_three}

\begin{table}[t]
\centering
\renewcommand{\arraystretch}{1.15}
\begin{tabular}{@{}lccc@{}}
    \toprule
     Asymptotic cutoff & $\quad\Delta f_i\quad$ & $\quad\Delta\theta_{\rm KS}\quad$ & $\quad\Delta\theta_3\quad$ \\
    \midrule
    $R_{\rm th}=10$  & $f_3:\ +0.89\%$     & $+6.96\%$   & $0$ \\
    $R_{\rm th}=40$  & $f_3:\ +0.90\%$     & $+7.11\%$   & $0$ \\
    $R_{\rm th}=70$  & $f_3:\ +1.72\%$     & $+8.73\%$   & $0$ \\
    $R_{\rm th}=100$ & $f_2,f_3:\ -6.44\%$ & $-0.94\%$   & $0$ \\
    \bottomrule
    \toprule
     Lattice-extrapolated & $\quad\Delta \Lambda_{\rm KS}\quad$ & $\quad\Delta\theta_{\rm KS}\quad$ & $\quad\Delta\theta_3\quad$ \\
    \midrule
    Blue circle   & $0$                          & $0$         & $\pi/8$ \\
    Orange square & $0$                          & $0$         & $0$ \\
    Green diamond & $ +30.00\%$ & $+128.41\%$ & $0$ \\
    \bottomrule
\end{tabular}
\caption{
Net parameter changes between fast and resolved trajectories in \cref{fig:hard_cutoff_money,fig:monodromy_crossing_money}. The initial monodromy angle refers to the linear combination $\theta_{\rm KS}=\theta_1+C_{12}\theta_2+C_{13}\theta_3$. All unlisted parameters remain fixed. For example, in the blue circle lattice-extrapolation example, the $\theta_3$ shift is compensated by shifts of $\theta_1$ and $\theta_2$ that preserve both the monodromy and QCD angles.
}
\label{tab:resolved_benchmark_shifts}
\end{table}

The cheap numerical scans in \cref{sec:three_axions_numerical} are performed according to the following methodology: from an initial temperature above $\Tew$ down to $T = 60\GeV$, we evolve the full system of equations including the SM asymmetry Boltzmann equations. This ensures that every point in the tested parameter space reproduces the observed baryon abundance. After this point, we switch to the following approximate procedure.

We define the following vectors
\begin{align}
    \bm k^T\equiv\left(\frac{1}{f_1},\frac{C_{12}}{f_2},\frac{C_{13}}{f_3}\right)\,,\quad\bm q^T\equiv\left(0,\frac{1}{f_2},\frac{C_{23}}{f_3}\right)\,,\quad\bm d^T\equiv\left(0,0,\frac{1}{f_3}\right)\,,
\end{align}
in terms of which the mass and damping matrices take the form
\begin{align}
    \bm M^2\equiv \Lambda_{\rm KS}^4\bm k\bm k^T + \chi_c\bm q\bm q^T + \chi_d\bm d\bm d^T\,,\quad\bm\Gamma\equiv\Upsilon_c\bm q\bm q^T + \Upsilon_d\bm d\bm d^T\,.
\end{align}
As the mass eigenvalues of $\bm M^2$ can be extremely hierarchical at any given time, it is useful to decompose $\bm M^2 = \bm B\bm B^T$ with
\begin{align}
    \bm B\equiv\left(\Lambda_{\rm KS}^2\bm k,\sqrt{\chi_c}\bm q,\sqrt{\chi_d}\bm d\right)\,.
\end{align}
Taking the singular value decomposition of $\bm B$ computes the mass eigenvalues $m_i$ of $\bm M^2$ as its singular values, and the eigenvectors $\hat{\bm e}_i$ of $\bm M$ as its left singular vectors. This procedure avoids possible imprecision arising from the approximate cancellation of large terms when directly computing the eigenvalues and eigenvectors of $\bm M^2$.

To keep track of states throughout the adiabatic evolution, we maximize the eigenvector overlap at adjacent temperatures:
\begin{align}
    O_{ij} = |\bm e_i(T_n)^T\bm e_j(T_{n + 1})|\,.
\end{align}
All six possible assignments are tested, and the assignment which maximizes the overlap is chosen at each temperature. Near crossings, sufficiently fine temperature spacing is necessary to avoid losing track of which vector corresponds to which physical state.

Adiabaticity of a crossing is checked by computing the ratio of the rate at which the basis orientation changes relative to the mass splitting:
\begin{align}
    \epsilon_{{\rm rot},i} \equiv \frac{\arccos|\bm e_i(T_n)^T\bm e_i(T_{n + 1})|}{\Delta t \min_{j\neq i}|m_i - m_j|}\,,
\end{align}
where $\Delta t$ is the time step corresponding to the temperature step $\Delta T = T_n - T_{n + 1}$. We define values of $\epsilon_{{\rm rot},i}$ in excess of $0.3$ to be non-adiabatic for the purpose of classification.

Within the oscillatory WKB regime, each state's energy density is evolved accounting for the changing mass and the projected dissipation rate. Adiabatic crossings are assumed to produce no energy transfer between these branches. Trajectories violating the adiabatic criterion are rejected, except for the explicitly modeled monodromy–cosine crossing described in \cref{subsec:monodromy_crossing}. In that case, the total energy of the participating pair is redistributed according to $R=\log(\rho_H/\rho_L)$.

The final abundance is evaluated at $T=0.01\,{\rm GeV}$. For the lattice-extrapolated friction scan, prescribed values of $R$ are tested and points are retained when the total axion abundance does not exceed the observed dark-matter abundance. For the asymptotic friction with a hard-cutoff, adiabatic points have a fixed predicted abundance, and are retained if they likewise avoid overclosure.

To validate the fast WKB method, we perform a number of simulations that resolve the crossings where the WKB approximations no longer apply. The evolution of the axion energy density over time for these examples is illustrated in the insets (b) of \cref{fig:hard_cutoff_money,fig:monodromy_crossing_money}. Initially, these runs are seeded by the parameters obtained from the fast WKB method, and are therefore not guaranteed to produce the observed relic density. We shift the parameters and initial conditions to bring the dark matter abundance back within tolerance of the observed relic density. The required shifts are presented in \cref{tab:resolved_benchmark_shifts}.

\section{LLM semantic optimization}
\label{app:semantic}

Semantic optimization provides a practical way to navigate a high-dimensional parameter space subject to a mixture of quantitative constraints and qualitative preferences. It is particularly useful when individual simulations are computationally expensive, so that reducing the number of evaluations required to identify viable regions of parameter space is important.

An agent is provided with the simulation software---in our case, the axion--Boltzmann solver---together with access to its input parameters and outputs. The prompt specifies a set of hard constraints on the inputs and outputs, such as requiring the axion decay constants to satisfy astrophysical bounds and the resulting dark matter abundance not to exceed the observed abundance. It also specifies softer preferences that guide the search among otherwise viable points. In our application, these include favoring values of $C_{ij}$ and ratios $f_i/f_j$ closer to unity.

The procedure does not provide a mathematical guarantee of convergence, nor does it guarantee that the resulting parameter point is globally or locally optimal. If desired, conventional local optimization methods can subsequently be applied once semantic optimization has identified a suitable region of parameter space. Its practical advantage is that the search can use the physical meaning of the parameters and the outcomes of previous simulations when selecting subsequent trial points. It can therefore balance local refinement against broader exploration according to the stated preferences, including in parameter spaces containing multiple disconnected viable regions.

\bibliographystyle{apsrev4-1}
\bibliography{bibliography}

\end{document}